\documentclass[letterpaper,12pt]{report}	

\usepackage[authoryear, round]{natbib}
\usepackage{styles/utformat}  		

\usepackage{pdflscape}
\usepackage{graphicx}
\usepackage{amsmath,amsthm,amsfonts,amscd}
\usepackage{eucal} 	 	
\usepackage{verbatim}      	
\usepackage{styles/citesort}         	%
\usepackage{algorithm}
\usepackage{bibentry}
\usepackage{algorithmic}
\usepackage{xcolor}
\usepackage{markdown}
\usepackage{booktabs}
\usepackage{url}

\usepackage{multirow}
\usepackage{makecell}
\usepackage{tabularx}
\usepackage{bbm}
\usepackage{mathtools}
\usepackage{hyperref}
\hypersetup{colorlinks=false}

\oneandonehalfspacing

\theoremstyle{definition}

\theoremstyle{remark}

\newcommand{\latexe}{{\LaTeX\kern.125em2%
                      \lower.5ex\hbox{$\varepsilon$}}}

\chardef\bslash=`\\	

\makeatletter		
\def\square{\RIfM@\bgroup\else$\bgroup\aftergroup$\fi
  \vcenter{\hrule\hbox{\vrule\@height.6em\kern.6em\vrule}%
                                              \hrule}\egroup}
\makeatother		

\nobibliography*

\usepackage[utf8]{inputenc}
\usepackage{graphicx}
\usepackage{amsmath}
\usepackage{physics}
\usepackage{xcolor}
\usepackage{svg}

\usepackage{amsfonts}
\usepackage{subfig}
\usepackage[export]{adjustbox}
\usepackage{xcolor}
\usepackage{amssymb}
\usepackage{physics}
\usepackage{xcolor}
\usepackage{hyperref}
\usepackage{stmaryrd}

\usepackage{tensor}
\usepackage{wrapfig}
\newcommand{\gke}{\texttt{Gkeyll}}
\newcommand{\eqr}[1]{Eq.\thinspace(#1)}
\newcommand{\pfrac}[2]{\frac{\partial #1}{\partial #2}}

\newcommand{\pfraca}[1]{\frac{\partial}{\partial #1}}

\newcommand{\mvec}[1]{\mathbf{#1}}

\newcommand{\gcs}{\nabla_{\mvec{x}}}

\newcommand{\basis}[1]{\mvec{e}_{#1}}

\newcommand{\dbasis}[1]{\mvec{e}^{#1}}

\setkeys{Gin}{width=0.9\linewidth,keepaspectratio}

\newcommand{\buni}{\mvec{b}}

\author{Akash Shukla}  	

\address{akashukla@utexas.edu}  

\title{Gyrokinetic Simulations for Spherical Tokamak Divertor Design} 

\supervisor[Richard Fitzpatrick]
	{David Hatch}

\committeemembers
	[Swadesh Mahajan]
	{Anna Tenerani}

\degree{Doctor of Philosophy}
\degreeabbr{PhD} 

\dissertation  

\begin{document}


\copyrightpage  

\commcertpage   

\titlepage      



\begin{acknowledgments}		
During my PhD, I have been lucky to have many mentors and collaborators to teach me, introduce me to other scientists, and help me navigate my PhD.

First I would like to thank professor David Hatch. David took me on as a student during my undergraduate degree and began to teach me about plasma physics, gyrokinetics, programming, and supercomputers. He always trusted me to present my work to others and introduced me to collaborators in other departments and at other institutions. I vividly remember my first time present our work outside of UT - David sent me to Oakridge National Laboratory in 2018 to collaborate with the math department. I was nervous to present at the seminar there, and David talked to me on the phone every day to help me with my slides. In the end, the seminar went very well, and I got a lot of good feedback. This was a really exciting experience that made me excited to continue working as a scientist and gave me the confidence to present my work. Since then, David has always given me the option to choose what I want to work on. When I started my PhD he gave me the option to continue working on core turbulence or work on the tokamak edge with the Gkeyll group. I opted to work on the edge, and David connected me with Greg Hammett and Ammar Hakim at PPPL.

I would also like to thank professor Swadesh Mahajan for his advice and guidance. Swadesh was the first person I came to at IFS when I wanted to start research as an undergrad. Swadesh welcomed me and immediately introduced me to David to get started on a project. Since then Swadesh has always supported and advised me. He has helped me with all of the hard decisions I have had to make during my PhD and always looks out for my best interest. Thank you, Swadesh.

I would like to thank Mike Kotschenreuther and Jonathan Roeltgen for working tirelessly with me on the physics problems that make up my thesis. David introduced me to Mike after the second year of my PhD and I have worked very closely with him and Jonathan since. Mike has led me to work on novel ideas, which are sometimes contrary to popular opinions in the community. Mike has held me to a very high standard, and my work has improved dramatically because of it. We always investigate every facet of a problem, no matter how difficult or time consuming, so we can be confident in our results. Mike has taught me about divertor physics and the big picture of tokamak design. Mike's encyclopedic knowledge of tokamaks has allowed me to ask questions and learn more quickly than I ever could have otherwise. I want to thank Mike for investing so much time in my education, proposing novel research that can really make an impact, and holding my work to a high standard. I want to thank Jonathan for working with me on all of the projects we have done under Mike's supervision. We have spent countless hours making sense of confusing data and writing code and have done so much good work together.

I want to thank Greg Hammett for welcoming me into the Gkeyll group at PPPL. I spent the first summer of my PhD at PPPL working with Greg. He taught me about gyrokinetics and more importantly about how to think through problems. From Greg I began to learn how to break problems into smaller pieces and look at which components could be responsible for the physics (or errors) I was seeing. Working with Greg also taught me how to come to a meeting or seminar prepared. In the beginning, Greg would ask about relevant parameters for a model or problem we were looking at, and I would not know the answers. Since then I have learned to think about and make note of relevant information that will be useful in a discussion with other scientists.

I have worked very closely with Ammar Hakim, the leader of the Gkeyll group. Ammar made great efforts to bring me to PPPL as often as possible for exhausting but exciting and extremely productive visits. I remember one of my first discussions with Ammar in the hallways of the old CSD department at PPPL. He drew diagrams on the blackboard and taught me about the sheath boundary condition in Gkeyll at a time when I knew almost nothing about numerical methods. Ammar taught me about numerical methods and programming and was willing to dive into the enormous and uncertain project of including the X-point in Gkeyll with me. I have been able to learn so much from Ammar because he has always encouraged me to ask questions - I never felt afraid to look stupid and always asked every question no matter how basic, and he would always explain in detail. I want to thank Ammar for spending so much time teaching me, supporting me, and guiding my education.

I also want to thank Manaure Francisquez and Jimmy Juno for mentoring me as I worked on the Gkeyll code. I have spent countless hours on zoom with Mana planning projects, discussing algorithms, and debugging problems. Mana took the time to go into detail about different numerical methods with me, writing down the equations on his iPad. This must have been painstaking for him when I was a novice, but it has taught me so much.  I would not be where I am today if I had not had Mana as an example. His attention to detail and careful analysis of problems has taught me how to be rigorous and isolate issues. I want to thank Jimmy for teaching me how to program on GPUs, which is an extremely valuable skill that I would not have been able to learn without him. It was very difficult, and I asked a lot of silly questions in the beginning, but Jimmy took the time to walk me through countless examples of how to do write and optimize GPU code. I also want to thank Jimmy for always quickly giving me helpful big picture feedback on my papers and manuscripts.

I also want to thank my teachers and advisors. Thank you professor Waelbroeck and professor Tenerani for teaching the plasma physics courses and going through all the nitty gritty details of coulomb collisions and Landau damping. Thank you professor Fitzpatrick for the detailed textbooks and online resources you provide - they have been crucial to my understanding of plasma physics - for answering all of my questions, and helping me choose classes and navigate my PhD.

Finally, I want to thank my wife, parents, and brother for supporting me through my PhD. They have listened to me practice talks and worry over problems for countless hours. Sometimes they have helped me figure out how to proceed, but more often and more importantly, they help me stop worrying, feel better, and take a break when I need to. I couldn't have done it without you.

\end{acknowledgments}

\begin{preface}		
My contribution to the co-authored papers is as follows:
\begin{itemize}
\item Chapter~\ref{chap:2}: All simulations, tests, and analysis were performed by me. The algorithms described in the paper to allow for X-point geometry were developed and implemented in the Gkeyll code by me with help from Ammar Hakim, Manaure Francisquez, and James Juno. The paper~\cite{Shukla2025Xpt} (submitted to Journal of Plasma Physics) was written by me.
\item Chapter~\ref{chap:3}: All simulations were performed by me. Jonathan Roeltgen and I worked together to write the code required to couple Gkeyll to EIRENE. All analysis was performed by me and the paper~\cite{shukla2025LR} (planned submission to Nuclear Fusion) was written by me. 
\item Chapter~\ref{chap:4}: All Gkeyll simulations were performed by me. I developed the Gkeyll code required to be able to use numerical equilibrium from EFIT in simulations. All analysis was performed by me and the paper~\cite{Shukla25} (published in AIP Advances) was written by me.
\end{itemize}
\end{preface}

%
\utabstract
Nuclear fusion is an appealing source of energy because of the abundance of the fuel and the low levels of carbon emissions it produces. The tokamak, which confines a plasma using magnetic fields, is the most mature nuclear fusion reactor concept. Net energy production has not yet been achieved in a tokamak, but companies and governments across the world are leading a push towards commercially viable fusion. Producing net energy in a tokamak has proven much more difficult than initially expected because of anomalous heat leakage caused by turbulence, which makes maintaining the temperature and density of the plasma costly. Maximizing energy confinement by minimizing turbulent heat leakage and minimizing damage to the reactor by reducing or accommodating large heat loads is essential for producing efficient fusion reactors. Achieving good confinement and reactor survivability at minimal cost will be essential for the commercial viability of fusion energy.

Theory and modeling have been essential tools for understanding and designing tokamaks. Reactors are expensive to build, so insight that can be gained from simulations is very valuable. We have used simulations to develop an understanding of heat transport out of the core plasma in tokamaks, but have found that the conditions in the edge of the reactor greatly influence energy confinement. Thus, it is essential to model the boundary of the plasma, called the scrape-off layer (SOL), to find designs that maximize fusion performance and result in tolerable heat loads on material surfaces called divertors in the device. Modeling the edge of a tokamak plasma is difficult because it involves large fluctuations, magnetic X-points, and interactions between the plasma and material surfaces. 

Fluid rather than kinetic or gyrokinetic models have mostly been used to model the SOL for divertor design. This can be appropriate when a conventional divertor design that has a low SOL density and high SOL temperature is used. However, in alternative approaches which aim to improve energy confinement, the SOL temperature can be high and the density low. In these regimes, called low-recycling regimes, fluid modeling is no longer appropriate. In order to study these regimes, we have developed the \gke\ code's gyrokinetic model into a code suitable for divertor design. We have improved \gke's geometric flexibility to allow simulations in realistic X-point tokamak geometry and have also coupled \gke\ to a monte carlo code EIRENE for modeling plasma wall interactions and neutral particle evolution.
With \gke\ as our primary tool, we have investigated the feasibility of low recycling regimes with numerical simulations. 

Low-recycling regimes are appealing because they entail a high edge temperature and low edge density which are good for core confinement. 
However, due to considerably enhanced heat flux, the exhaust problems become severe. In addition, in the low-recycling regime, the conventional fluid simulations may not capture the physics of the Scrape-Off Layer (SOL) plasma which is in the long mean free path regime; kinetic calculations become necessary. In this thesis, by performing both Kinetic and fluid simulations, we explore the feasibility of a low-recycling regime in the magnetic geometry of the Spherical Tokamak for Energy Production (STEP); kinetic effects come out to be crucial  determinants of the SOL dynamics.
The simulation results indicate that a high SOL temperature and low SOL density could be achieved
even when the divertor target is not made of a low recycling material. This can be done by using a low recycling material as a wall material.
This is an important step towards demonstrating the feasibility of a low-recycling scenario.
Lithium, a commonly used low recycling material, tends to evaporate at high heat fluxes which counteracts the desired high temperature, low density regime, and materials that can handle high heat fluxes are generally high recycling. 
Comparisons of gyrokinetic and fluid simulation results indicate that one can take advantage of kinetic effects to address some of the issues associated with a low-recycling SOL. Specifically, kinetic simulations show better confinement of impurities to the divertor region and greater broadening of the heat flux width due to drifts when compared with fluid simulations. Impurity confinement would help prevent core contamination from sputtering, and a broader heat flux width would help reduce the peak heat load at the target in the absence of detachment.    

\tableofcontents   

\listoftables      
\listoffigures     

%
%

\chapter{Introduction}
\label{chap:introduction}
\section{Motivation}
In order to mitigate the effects of climate change, we must replace fossil fuel consumption with alternative energy sources that do not produce greenhouse gases such as carbon dioxide. A combination of renewable energy sources such as wind and solar and a more consistent energy source such as nuclear fusion or fission to provide baseload power will be required to support the world's energy needs.

Nuclear power is an appealing source of energy when compared to fossil fuels primarily because of the abundance of the fuel and the low levels of carbon emissions it produces~\citep{Cowley2016}. Nuclear power also offers advantages over renewable energy because it does not suffer from intermittency and does not require large amounts of space compared with wind and solar power~\citep{Dunlap21}. Because nuclear energy does not suffer from intermittency, it does not require the development of efficient energy storage technology that would be required for renewable energy to produce consistent electricity~\citep{Dunlap21}. These advantages are true for both nuclear fission and nuclear fusion. However, fission produces long-lived radioactive waste and poses security concerns due the potential for proliferation~\citep{Meschini2023,Dunlap21}, so fusion is preferable if it can be made economically viable. The difficulty of waste disposal and damage caused by past accidents has made fission unpopular~\citep{Kim2014,Meschini2023} and expensive due to regulatory burdens, so nuclear fusion has become a promising option for fulfilling our future energy needs.


The great appeal of fusion energy has led to a push towards commercially viable fusion (CVF) across the world. There are many different fusion concepts including intertial confinement  fusion (ICF) and magnetic confinement fusion (MCF) concepts. 
In 2022, an ICF experiment at the National Ignition Facility (NIF) achieved a scientific gain (ratio of energy produced by the fuel to energy that hit the fuel) of 1.5~\citep{NIF2024}. However, using ICF as an energy source requires repeating a pulse like this many times per second~\citep{Betti2025}, which is operationally very difficult and would require significant engineering advances. Additionally, the fuel pellets used for ICF are currently time consuming and expensive to make, and a power plant would need to use upwards of 500,000 pellets per day~\citep{Goodin2006}. The manufacturing cost would need to be reduced from thousands of dollars to 25-30 cents per capsule for laser ICF to be cost-effective~\cite{Goodin2004}. Another major challenge for ICF is that it is very energetically costly to power the lasers, resulting in overall efficiencies (engineering gain) of $\sim$1\%~\cite{Betti2025}. This `wall-plug efficiency' is a much less severe problem for MCF devices like the tokamak; there is a much smaller difference between scientific and engineering gain.
While ICF is pulsed in nature, some MCF reactor types are intended to operate in a steady state ideal for electricity generation. The stellarator in particular is ideal for steady state operation because it does not require toroidal current in the plasma. Tokamak plasmas feature a toroidal current which is typically driven by ramping up current in a central solenoid. This makes tokamaks pulsed in nature~\citep{Meschini2023} but they have a much higher duty cycle than ICF concepts, making them more suitable for electricity generation. The current in a tokamak could be driven by neutral beam injection, radio frequency  current drive, or bootstrap current rather than the central solenoid to enable steady state operation, but these options can be difficult so some tokamaks such as SPARC plan for pulsed operation~\citep{SPARCDutyCycle}.
Tokamaks are the most mature type of MCF device and have come the closest to achieveng fusion gain (Q) greater than 1 (net power production). SPARC plans to demonstrate Q$\approx$11 in 2027 which will be a huge step forward towards commercially viable fusion~\citep{SPARCGain,SPARCGain2, Crownhart2024}. However, further developments will be required to make devices with low enough capital cost and enough longevity to be commercially viable. 
Additionally, further developments in other areas such as the tritium fuel cycle, blanket technology,
structural materials with adequate lifetime in a fusion neutron environment, and Plasma Facing Components (PFCs) with adequate lifetime upon exposure to the plasma and neutrons simultaneously
will be needed to construct a fully integrated power plant that can actually provide electricity for commercial use.

\section{Fusion \& Tokamak Basics}
Fusion requires forcing nuclei close enough together that the attractive nuclear force overpowers the repulsive electrostatic force. This requires producing a plasma with a very high temperature and pressure -  the nuclei must have enough energy to overcome electrostatic repulsion. The most accessible reaction is fusing deuterium and tritium to produce an alpha particle and a high energy neutron
\begin{equation}
    D + T \rightarrow ^4He\, (3.52 MeV) + n\, (14.07 MeV) .
    \label{eq:fusionreaction}
\end{equation}

A fusion reactor would collect these high energy neutrons and convert their energy into thermal energy. In the sun, extreme gravitational pressures create the conditions necessary for fusion, but other approaches are necessary on earth in a manmade device. The most promising method is to heat a gas of deuterium and tritium to around 10keV (~6 times as hot as the sun) so that collisions between the particles can overcome the electrostatic repulsion and result in a fusion reaction. In order for fusion reactions to occur frequently, the high temperature and density of the plasma must be maintained, which means both the energy and particles must be confined. At these temperatures, the gas cannot simply be confined by a vessel made of any known material (any vessel would be destroyed), so instead we attempt to confine the gas with magnetic fields. Confinement via magnetic fields is possible because the gas is completely ionized (a plasma) at these temperatures and the particles are charged; charged particles spiral around magnetic fields due to the Lorentz force.

The simplest magnetic confinement solution would be to have a ring configuration; a torus with a magnetic field pointing along the axis in the toroidal direction. Naively one would think that this would confine charged particles: the particles would move along the direction of the magnetic field but their perpendicular motion would be constrained as they would gyrate around the magnetic field. However, the presence of magnetic drifts will cause the electrons and ions to be lost in the vertical direction rather than confined in this configuration. Drifts of the guiding center particle orbits occur as a result of (1) an electric field perpendicular to the magnetic field, (2) a gradient in the magnetic field perpendicular to the magnetic field, (3) curvature of the magnetic field and (4) time variation in the electric field. One solution to this particle loss is to wrap the magnetic field lines around the torus in a helical configuration. The simplest version of this concept, the tokamak, is what we will focus on.

\begin{figure}[h]
\includegraphics[width=\textwidth]{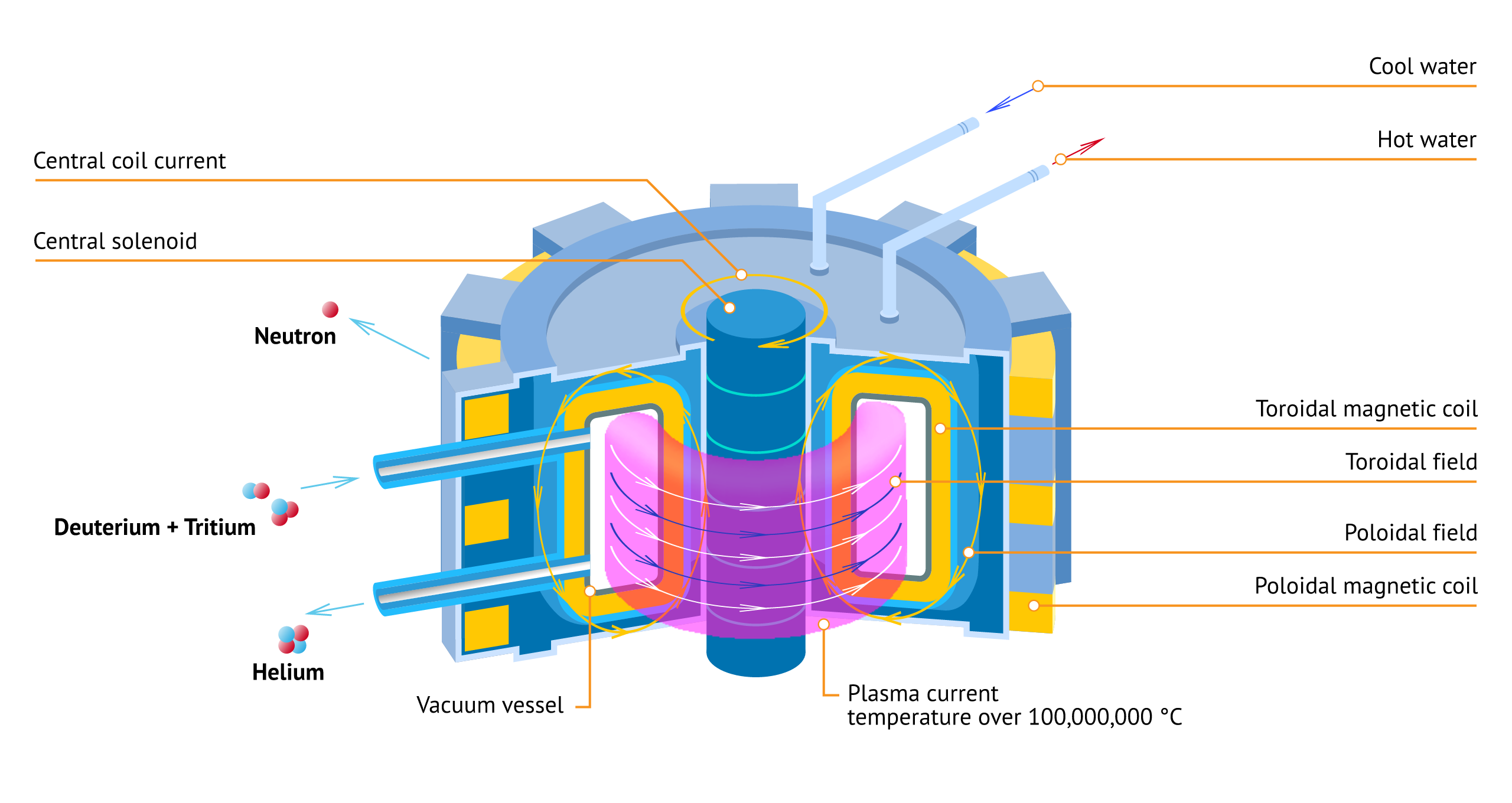}
\caption{\label{fig:tokamakdiagram}
Schematic of tokamak.}
\end{figure}

A schematic of a tokamak is shown in Fig.~\ref{fig:tokamakdiagram}. The toroidal magnetic field is created by the toroidal field coils, which are like a solenoid that is bent into a torus. To achieve the helical magnetic configuration, a poloidal magnetic field (the poloidal direction is the short way around the torus) is produced by inducing a toroidal current in the plasma. The plasma current is induced by the central solenoid; the central solenoid and plasma act as a transformer where the plasma serves as the secondary winding. The poloidal field is also supplemented by the poloidal field coils.

\section{Power Balance \& The Triple Product}
As mentioned in the previous section, confining the energy of the plasma is essential to producing fusion conditions; if the plasma is well confined it will stay hot and dense with less auxiliary heating power than if it was poorly confined. Here we will lay out the power balance equations for a fusion reactor and the conditions under which energy can be produced.

Consider a plasma made up of deuterium ions, tritium ions, and electrons, and let us assume that all species have the same temperature, $T$. The rate of fusion reactions occurring per unit volume is 
\begin{equation}
f=n_D n_T\langle\sigma v\rangle_{D T}(T)
\end{equation}
where $n_D$ is the deuterium ion density, $n_T$ is the tritium ion density, $\sigma$ is the cross section for the D-T fusion reaction in Eq.~\ref{eq:fusionreaction}, $v$ is the relative velocity of the ions, and $\langle\quad \rangle$ indicates an average over the distribution functions of the  ions. $\langle\sigma v\rangle_{D T}(T)$ increases rapidly with temperature in  the range 1-30 keV and reaches $\approx 10^{-22}\, m^3/s $ at $T \approx 10\, keV$~\citep{FitzReact}.

The density of each species in this plasma will be $n_D = n_T = n_e/2$ (as demanded by  quasineutrality), where $n_e$ is the electron density. Thus, the rate of fusion reactions can be expressed in terms of the electron density as~\citep{FitzLawson}
\begin{equation}
f \equiv n_D n_T\langle\sigma v\rangle_{D T}\left(T\right)=\frac{n_e^2}{4}\langle\sigma v\rangle_{D T}\left(T\right) .
\label{eq:rate}
\end{equation}

The total energy, W, of this plasma with volume V, is given by
\begin{equation}
    W \equiv V (\frac{3}{2} n_D T+\frac{3}{2} n_T T+\frac{3}{2} n_e T )=3 n_e T V
    \label{eq:storedenergy}
\end{equation}
where a uniform density and temperature have been assumed for simplicity.

In the steady state, the energy balance equation for this plasma is given by 
\begin{equation}
    \pdv{W}{t} = P_{heating} + P_{\alpha} - P_{loss} = 0
    \label{eq:balance}
\end{equation}
where $P_{heating}$ is the auxiliary heating power injected into the plasma, $P_{loss}$ is the lost power, and $P_\alpha$ is the power transferred to the plasma by fusion reactions. Not that $P_{alpha}$ includes only the power of the alpha particles and not of the neutrons, since the neutrons are lost and do not heat the plasma.

The energy confinement time is defined as $\tau_E = W/P_{loss}$ and is a measure of the rate at which the plasma loses energy. In order for the energy to be  produced, the fusion power, $P_{fusion} = P_\alpha + P_{neutron}$, which includes the power of the alpha particles and neutrons produced by the fusion reaction, must exceed the losses. Break-even occurs when the fusion power is equal to the loss power (when the power from the neutrons is equal to the input heating power):
\begin{equation}
P_{fusion} = P_{losses} \quad \textrm{for break-even}
\end{equation}
For the plasma to become self sustaining and not require any auxiliary heating power, a stricter condition, the Lawson criterion, must be met:
\begin{equation}
    P_{\alpha} \geq P_{loss}\quad \textrm{for ignition}
\end{equation}

Using Eq.~\ref{eq:rate} and writing the energy of the alpha particle (3.502 MeV) as $E_\alpha$, we get $P_\alpha = \frac{n_e^2}{4} \langle \sigma v\rangle E_\alpha V$ and the condition for ignition becomes
\begin{equation}
\frac{n_e^2}{4} \langle \sigma v\rangle E_\alpha \geq \frac{3 n_eT}{\tau_E}
\end{equation}
The break-even condition becomes
\begin{equation}
\frac{n_e^2}{4} \langle \sigma v\rangle (E_\alpha + E_n) = \frac{3 n_eT}{\tau_E}
\end{equation}
where $E_n = 14.07\, MeV$ is the energy of the neutrons produced by  the fusion reaction. Rearranging the ignition and break-even condition  we can write them in a typical form:
\begin{align}
    &n_e \tau_E \geq \frac{12T}{ (E_\alpha) \langle \sigma v \rangle} \quad \textrm{ ignition, }\label{eq:ignition} \\
    &n_e \tau_E = \frac{12T}{ (E_n +E_\alpha) \langle \sigma v \rangle} \quad \textrm{ break-even.} \label{eq:breakeven}
\end{align}

The right hand side of Eq.~\ref{eq:ignition} is a function only of temperature and reaches a minimum at of $1.49\times 10^{20} s/m^3$ at $T=25.67 keV$~\citep{FitzLawson,Wesson}. This gives the famous Lawson criterion for ignition : $n\tau_E \geq 1.49 \times 10^{20} s/m^3$. Since $\tau_E$ is itself a function of temperature, the actual ignition temperature is likely to be lower, $T = 13.54keV$~\citep{FitzLawson, Wesson}. In the range 10-20 keV, the reaction rate can be approximated as 
\begin{equation}
    \langle\sigma v\rangle=1.1 \times 10^{-24} T^2 \mathrm{~m}^3 \mathrm{~s}^{-1}
    \label{eq:approxrate}
\end{equation} 
where $T$ is in keV. Multiplying both sides of Eq.~\ref{eq:ignition} by T we get the more commonly used triple producet formulation of the Lawson criterion:
\begin{equation}
    n_e T \tau_E \geq \frac{12T^2}{ E_\alpha \langle \sigma v \rangle}.
    \label{eq:tripleproduct}
\end{equation}
Substituting the value of $E_\alpha$ and Eq.~\ref{eq:approxrate} into Eq.~\ref{eq:tripleproduct}, the temperature dependence on the right hand side cancels, and we get
\begin{equation}
    n_e T \tau_E \geq  2.76 \times 10^{21} \mathrm{keV} \mathrm{~s} \mathrm{~m}^{-3}
    \label{eq:tripleproductnumerical}
\end{equation}
for ignition.


From Eq.~\ref{eq:tripleproductnumerical} it becomes clear that increasing the density of the plasma, increasing the temperature, or improving the confinement can help achieve net energy gain. In a tokamak, the density cannot be increased arbitrarily. The Greenwald limit~\citep{Greenwald1988} gives a limit on  the density beyond which a disruption (an abrupt breakdown in confinement) will typically occur. The limit is given by $n_G = I_p/\pi a^2$ where $n_G$ is  the line averaged density in units of $10^{20} m^{-3}$, $I_p$ is the plasma current in  MA, and a is the major radius in m. The reaction cross section rises rapidly up to around 20-30 keV and then flattens out, so increasing the temperature beyond 30keV yields diminishing returns. Increasing the plasma temperature also increases its pressure and thus the plasma beta ($\beta$ = the ratio of plasma pressure to magnetic pressure) which can drive instabilities. Stiff temperature gradient driven transport also limits the effectiveness of increasing the temperature beyond a certain point; temperature gradient driven turbulence will cause heat loss and make increasing the plasma temperature inefficient.  Thus, assuming a tokamak operates in the optimal density and temperature ranges, increasing confinement time is the main avenue for reaching net energy production in tokamaks.

\section{Challenges of Tokamak Design}
Two of the most pressing challenges for tokamaks on the pathway to fusion energy are confining the plasma (maximizing $\tau_E$) and managing heat exhaust. 
Confining the plasma is essential for producing energy and managing the heat exhaust is essential for having a long-lived reactor. 
Plasma turbulence has made confinement much more difficult than initially anticipated; minimizing turbulent heat transport in the core (closed field-line region) is essential to producing energy in a tokamak. Heat that is exhausted from the plasma tends to be deposited in a very small area, which destroys the plasma facing materials. Therefore, finding ways to spread out this heat or develop new strategies for exhausting it is also necessary for a functional reactor. The challenges of heat exhaust and confinement are not independent; the methods for managing heat exhaust affect the conditions at the edge of the plasma which greatly affect confinement. 

Beyond demonstrating energy production, a reactor must also be commercially viable in order to replace fossil fuels. Thus, it is essential to consider affordable methods for simultaneously reducing turbulent heat transport and managing heat exhaust. The work presented in this thesis is aimed at using numerical simulations to find cost effective solutions for maximizing confinement and managing heat exhaust.

\section{Fusion Performance and Cost}
Fusion performance is measured by the triple product $n T \tau_E$ where $n$ is the plasma density, $T$ is the temperature, and $\tau_E$ is the energy confinement time. The confinement time measures the effectiveness of heating the plasma, so it is an essential performance metric.

The confinement time is defined as the stored energy in the plasma divided by the input power $\tau_E = 3nTV/P$ where $V$ is the plasma volume and $P$ is the input power. In the steady state, the input power is equal to the heat leaving the device. The power leaving the device is a sum of the radiated power and the power deposited on the material surfaces of the device which is equal to the heat flux, Q, times the cross sectional area of the device, A. In the absence of radiated power, the expression for the power is $P = QA$. 
The heat flux is often conceptualized as a diffusive process following Fick's law: $Q = n\chi\grad T$. The heat flux is determined by turbulent transport, which is extremely difficult and complex to model, generally requiring large HPC simulations of the gyrokinetic equations. However, the fundamental underlying scaling of the turbulent diffusivity follows the so called gyro-Bohm scaling, $\chi = \rho^2v_t/L_T$~\citep{ITER}, 
where $\rho$ is the gyroradius, $v_t$ is the thermal velocity, and $L_T = T/\nabla T$ is the temperature gradient length scale. Putting this all together we get the following scaling law for the energy confinement time
\begin{equation}
\tau_E \propto \frac{RL_T^2}{\rho^2 v_{t}^3}
\label{confinement}
\end{equation}
where $R$ is the major radius of the reactor. 

An empirical scaling law for the confinement time in terms of machine parameters has been constructed using experimental data from many machines ~\ref{tau_98}~\citep{Becker04}.
\begin{equation}
\tau_{H 98(y, 2)}=0.0562 I_p^{0.93} B_t^{0.15} \bar{n}_e^{0.41} P^{-0.69} R^{1.97} \kappa_a^{0.78} A^{-0.58} A_i^{0.19}
\label{tau_98}
\end{equation}
whe $I_p$ is the plasma current, $B_t$ is the toroidal field, $\kappa$ is the elongation, $A$ is the aspect ratio, and $A_i$ is the hydrogenic atomic mass number.
In fusion literature, confinement performance is measured by the $H$ factor which reflects the improvement in confinement beyond this scaling law; $H = \tau_E / \tau_{H98}$.

According to a commonly cited estimate, for a high fusion gain tokamak, the required device major radius is proportional to one over the H factor; $R \sim 1/H$~\citep{Zohm10,Wade2021}. So, if confinement can be doubled, the required major radius is reduced by a factor of 2. Device cost scales roughly like the area or volume of the device; $cost \sim R^{2-3}$. Thus, doubling $H$ reduces cost by a factor of 4 conservatively. From Eq.~\ref{confinement} we can see that increasing $R$ improves confinement, but this does not provide any cost benefit. Increasing the field strength $B$ also improves confinement but it is expensive. Thus, improving confinement in some way other than increasing the device size or increasing the magnetic field strength is a route to cheaper, commercially viable reactors~\citep{Wade2021}.

\section{Confinement \& H-Mode}
The energy loss observed in tokamaks is much higher than the losses predicted by the theory of collisional particle and heat transport for plasma in a torus (neoclassical transport). Plasma instabilities result in turbulent heat transport which can be orders of magnitude larger than predicted by neoclassical theory~\citep{Wesson}, which has made achieving net energy gain in tokamaks much more difficult than anticipated. In experiments it was found that, in some scenarios, as the heating power is increased, there is an abrupt jump in confinement time. This was surprising because normally increasing the heating power decreases confinement due to stiff transport. The low confinement regime is referred to as L-mode and the enhanced confinement regime is called H-mode. H-mode is caused by the development of an edge transport barrier (ETB) near the edge of the core plasma in which perpendicular heat transport is reduced. The mechanism for the formation of ETBs is not yet completely understood and is an active area of research. In H-mode, the transport barrier results in the formation of a temperature and density `pedestal', a narrow radial region over which there is a sharp rise in temperature and density. This pedestal allows easier access to high temperature and density in the core. Without a transport barrier, these gradients would be flattened by turbulent transport removing the pedestal.

In order to try to predict the confinement time for ITER, a database of data from L-mode plasmas was used to construct an empirical scaling law (ITER89-P) for the confinement time in terms of machine parameters:
\begin{equation}
\tau_{\mathrm{E}}^{\mathrm{IT} \mathrm{ER} 89-\mathrm{P}}=0.048 \frac{I^{0.85} R^{1.2} a^{0.3} \kappa^{0.5}\left(n / 10^{20}\right)^{0.1} B^{0.2} A^{0.5}}{P^{0.5}} \mathrm{~s}.
\label{eq:L-mode}
\end{equation}
After H-mode was discovered, a database and empirical scaling law were developed for H-mode, the H98 scaling in Eq.~\ref{tau_98}. The H-mode scaling law offers more favorable scaling than L-mode.

\section{The Importance of the Boundary Layer}
\label{sec:boundarylayer}
\begin{figure}[h]
   \centering
   \subfloat[\label{fig:edge_temp}]{
   \includegraphics[width=0.45\textwidth]{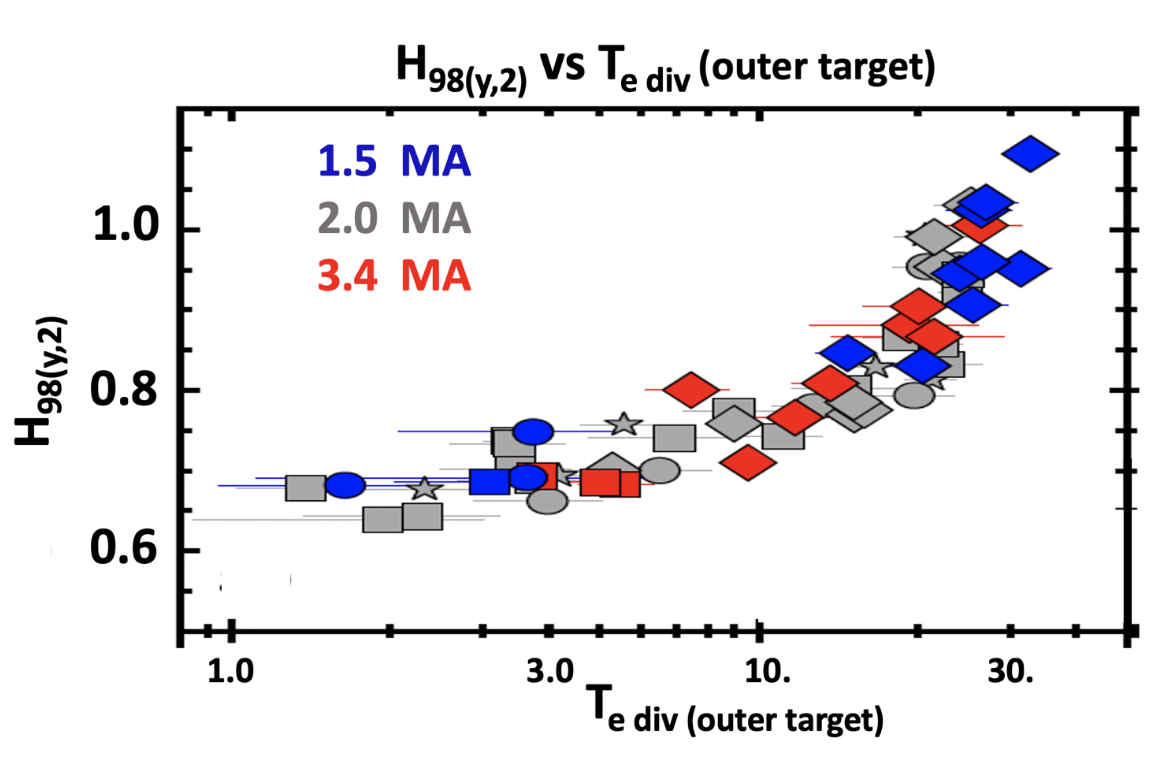}
    \makebox[0pt][r]{
      \raisebox{0.5em}{
         $ [eV] $
      }\hspace*{1em}
    }
   }
   \subfloat[\label{fig:edge_density}]{
   \includegraphics[width=0.45\textwidth]{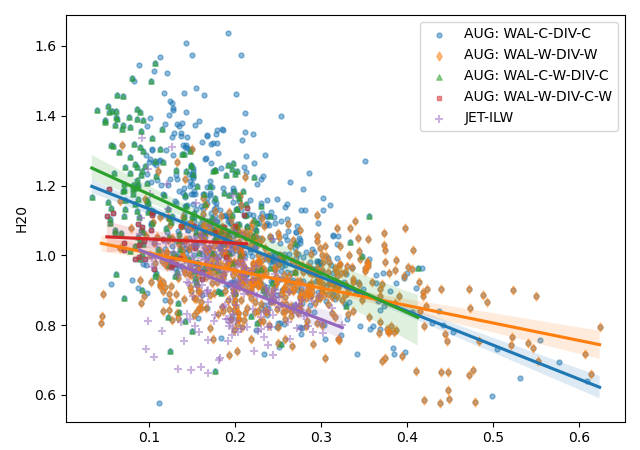}
    \makebox[0pt][r]{
      \raisebox{-1.5em}{
         $ n_{sep}/\bar{n} $
      }\hspace*{6em}
    }
   }
   \caption[H factor vs. electron temperature at the divertor plate]{ (a) H factor vs. electron temperature at the divertor plate. Data from JET-ILW~\citep{Lomanowski22} and (b) H factor vs. $n_{sep}/\bar{n}$, where $\bar{n}$ is the line averaged density, from experiments in JET (Joint European Torus) and Asdex Upgrade~\citep{Mike}.
   }
\end{figure}

Introducing transport barriers in tokamak plasmas to reduce turbulent heat transport is thought to be necessary for an energy producing reactor. Transport barriers reduce the turbulent heat flux leaving the device and therefore improve confinement. The  exact mechanism for introducing transport barriers is not is not fully understood, but experimental data shows us certain regimes that produce transport barriers. One way to introduce a transport barrier is by having a large density gradient. In ~\cite{Mike}, it is shown that a constraint on the radial charge flux results in the formation of transport barriers when there are large density gradients in the plasma. This is demonstrated in Fig.~\ref{fig:edge_density}; the $H$ factor improves with decreasing edge density.

Heat and particle transport in the core of a tokamak have been well studied and we have developed powerful predictive tools in the form of gyrokinetic simulations, which will be described in section~\ref{sec:numerical_simulations}, to aid in reactor design. However, the behavior of the boundary layer of the plasma that comes in contact with the vessel determines the temperature and density gradients at the edge of the core plasma, and thus strongly affects the heat and particle transport in the core.


The plasma conditions at the edge of the reactor are essential to energy confinement and maximizing H-mode performance. High core temperatures are necessary for fusion, but large temperature gradients drive microinstabilities including Ion Temperature Gradient (ITG) and Electron Temperature Gradient (ETG) turbulence causing heat loss. Thus, maintaining a high edge temperature improves confinement; a high edge temperature allows the tokamak to have a high core temperature with smaller temperature gradients which results in less heat loss and better confinement. Fig.~\ref{fig:edge_temp} shows one example of this trend~\citep{Lomanowski22}, and there are many more in the literature~\citep{MikeArxiv2024}.

\section{Heat Exhaust: High and Low Recycling}
\begin{figure}[h]
    \centering
    \includegraphics[width=0.5\textwidth]{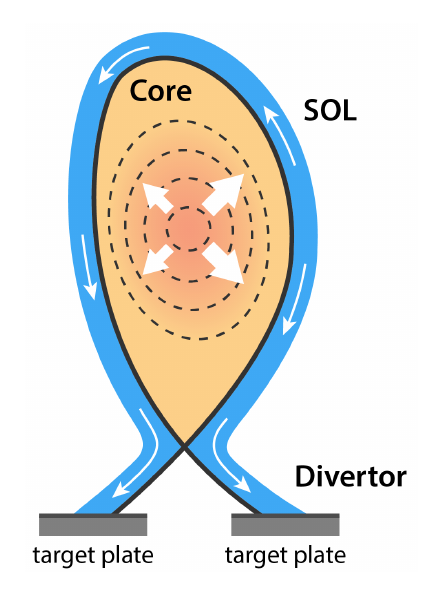}
    \caption[Poloidal cross section of a single-null tokamak.]{Poloidal cross section of a single-null tokamak. The thick white arrows indicate particle and heat transport in the radial direction from the core to SOL and the thin white arrows indicate parallel particle and heat transport along field lines in the SOL to the divertor plates}
    \label{fig:poloidaldiagram}
\end{figure}

In order to manage heat exhaust in tokamaks, we must study the edge region of the plasma where heat is exhausted, the scrape-off layer (SOL). In the core, magnetic flux surfaces are closed, but in the SOL, magnetic field lines terminate on material surfaces called divertor plates. The plasma in the SOL streams along the field lines and is incident on the divertor plates made of a special material able to handle the heat. Fig.~\ref{fig:poloidaldiagram} shows a diagram of a tokamak showing the core (orange) and SOL (blue). The thick white arrows indicate particle and heat transport in the radial direction from the core to SOL and the thin white arrows indicate parallel particle and heat transport along field lines in the SOL to the divertor plates (aka target plates). Radial transport across flux surfaces is much slower than the parallel transport along field lines. This results in an SOL that is typically very narrow (the width in Fig.~\ref{fig:poloidaldiagram} is exaggerated). See chapter 1, sections 1.5-1.7 in~\cite{Stangeby2000} for more information about the length and time scales associated with parallel and perpendicular particle motion in tokamaks.

Recycling is the process by which ions incident on the divertor plate recombine with electrons in the metal and are re-emitted into the plasma as cold neutral atoms. When a recycled neutral re-enters the plasma it can be ionized by electron impact ionization or charge-exchange with an ion, both of which will cool the plasma. If the recycled neutral passes through the plasma without reacting  and collides with the wall or divertor target it can either be recycled again or absorbed. The fraction of incident particles that are re-emitted as neutrals is called the recycling coefficient, R. Sputtering is a process by which ions incident on the divertor plate knock atoms off of the plate and release them into the plasma. The amount of sputtering that occurs is dependent on the incident ion temperature and sputtering erodes the divertor plate. Chapter 1 and section 3.3 in~\cite{Stangeby2000} provide a quantitative description of sputtering and recycling for various materials.

There are two basic categories in which a tokamak SOL can exist, the sheath-limited and conduction-limited regimes. Here, `Limited' can  be confusing, it actually indicates which physical effect is determining the SOL density and temperature - the sheath or the parallel heat conduction. The amount of heat leaving the plasma will be fixed, but whether it leaves mostly via conduction or convection will determine the parallel profile of the density and temperature. The conducted parallel heat flux is given by $q_{\parallel,cond} = -\kappa_\parallel\nabla_\parallel T$. The parallel  heat conductivity $\kappa_\parallel$ is proportional to the $T/\nu$, and $\nu$, the collision frequency is proportional to $n/T^{3/2}$. So, $q_{\parallel,cond}\propto T^{7/2}$. The heat conduction is also inversely proportional to mass, so electron heat conduction is much stronger than ion heat conduction. The parallel convected heat flux, $q_{\parallel,conv}$ is proportional to $nTu_\parallel$ where $u_\parallel$ is the parallel flow velocity. Chapter 2 and sections 1.9, 4.10, and 5.3 in~\cite{Stangeby2000} give a detailed description of the sheath and the conduction and sheath limited regimes.

In most situations, when the temperature is large, the heat conductivity is very large, so large temperature gradients cannot exist - the rapid conduction quickly flattens the temperature profile along the field line. In this case, most heat leaves by convection and we call the regime `sheath limited' because the effect of the sheath boundary condition at the divertor plate will be the dominant effect on the parallel profiles. Sheath-limited regimes are produced by low collisionality SOLs.

If the temperature is low, then the conductivity is lower, and large parallel temperature gradients can exist in the SOL, which will lead to a large conducted heat flux. In this case, we refer to the regime as conduction-limited. The conduction-limited regime is typically  associated with a more colisional SOL, and the downstream temperature (near the divertor plate) is significantly lower than the upstream  temperature (near the core plasma).

\subsection{Historical Background and Key Considerations}
In 2011,~\cite{Goldston2011} proposed a heuristic drift-based model (the HD model) for determining the heat flux width, $\lambda_q$ in terms of machine parameters. The heat flux width, also called the power scrap-off width, is the width of the region of the divertor over which most of the heat is deposited. For a given amount of power loss, a smaller value of $\lambda_q$ (narrower heat flux width) will entail a larger peak heat flux.~\cite{Eich2011} and~\cite{Eich2013} used a multi-machine database including experimental data from JET, DIII-D, ASDEX Upgrade, C-Mod, NSTX and MAST to fit an empirical scaling law for $\lambda_q$ in terms of machine parameters. The empirical scaling law showed strong agreement with the HD model. Two significant features of both the HD and empirical scaling law are 1) An inverse scaling with the poloidal magnetic field ($~\lambda_q\sim 1/B_p$) and 2) a weak  scaling with device size. 

Based on these scaling laws, the heat flux width predicted for ITER is very small, $\lambda_q\sim1$ mm. The predicted heat flux width for SPARC with its high field and compact size is also very narrow, $\lambda_q\sim 0.15$ mm~\citep{Lore2024}. These narrow heat flux widths imply unacceptably large unmitigated peak heat fluxes for these devices.
Peak heat fluxes which are too large are extremely problematic; they  will melt or erode the surface of the divertor. If this melting or erosion happens on a fast time scale, it will contaminate the plasma with impurities, ruining the fusion power output. If it happens on a longer time scale, then it will prevent the tokamak from running for long before the divertor needs to be replaced, which could take months~\citep{SuperXT}. What value of peak heat flux is `too large' depends on the machine and divertor configuration, but a typical maximum acceptable value is $10 MW/m^2$. In light of these narrow heat flux widths, approaches for lowering the peak heat fluxes to acceptable values have been developed.

\subsection{Conventional Approach: High Recycling Regime}
\label{sec:HR}
When mitigating the damage to  the divertor plate there are two important factors to be considered, the peak heat flux and the temperature of the plasma near the plate. As discussed above, large peak heat fluxes can destroy the divertor. Another  factor influencing how fast the divertor is eroded and the level of contamination in the core plasma is sputtering.
Sputtering is more severe if the temperature of the incident ions is higher. A conventional approach for mitigating damage to the divertor involves lowering the temperature of the SOL plasma and radiating a significant fraction of the power that enters the SOL before it reaches the divertor plates.
High recycling regimes typically employ walls and divertor plates with R$\sim$0.99 which allows recycling to cool the plasma. This results in a conduction limited regime where there is a significant temperature drop along the SOL which allows the plasma near the divertor to be cold which can alleviate the issue of sputtering.


High recycling regimes typically also inject impurities into the plasma in the divertor region to radiate power and cool the plasma, which lowers the total amount of heat hitting the divertor plate and thus reduces the peak  heat flux. The power radiated is proportional to the product of the electron and impurity densities, so having a high SOL plasma density is required for effective power removal. Impurity seeding can also be used to facilitate detachment, which refers to the plasma adjacent to the divertor becoming cold enough that it recombines. In a detached divertor scenario, there is a region of neutral gas between the plasma and the divertor. Impurity seeding can be detrimental for the fusion power output if the core plasma becomes overly contaminated by impurities. Low Z impurities can dilute the core plasma resulting in lower fusion power output, and high  Z impurities can be even more hazardous as they will radiate power from the main plasma. So, confining impurities to the SOL to prevent core contamination is an important consideration when using impurity seeding to  control the heat flux. The shape of the divertor plasma as well as the shape of the machine wall can affect how many impurities travel upstream and contaminate the core plasma.

High recycling regimes also facilitate an important aspect of exhaust in tokamaks: helium exhaust. Helium ions are one of the products of D-T fusion, and if the core plasma becomes dominated by helium rather than the deuterium and tritium fuel, the fusion reaction will be `choked.' Helium that hits the divertor target is recycled and then cryogenic pumps positioned in the divertor region are typically used to remove the neutral helium gas from the machine. The pumping duct has to be long enough that it is shielded from the neutrons emitted by the fusion in the main plasma; the neutrons will destroy the pump. Pumping neutral gases is easier when the density is high and the mean free path is short relative to the  duct entrance of  the pump. So, the high SOL density associated with a high recycling regime are conducive to exhausting the helium produced by fusion. 

In summary, high recycling regimes have many appealing properties for heat exhaust; they provide effective solutions for reducing peak heat fluxes to manageable levels, for mitigating damage caused by sputtering, and for pumping helium. However, high recycling regimes make it more difficult to get high SOL temperatures and high confinement, which motivates alternative approaches to heat exhaust.

\subsection{Alternative Approach: Low Recycling Regime}
As discussed in section~\ref{sec:boundarylayer}, a low edge density and high edge temperature improve core confinement. The high recycling divertor scenarios described in the previous section~\ref{sec:HR} are effective at protecting the divertor plates, but result in a low edge temperature and high edge density which is not good for confinement. Thus, alternative divertor solutions which aim to keep the SOL at a high temperature and low density are being explored. Scenarios with a high SOL temperature low SOL density are said to be in the low recycling regime.
In  the low-recycling regime, something is done to absorb the ions incident on the plate - the divertor region could be coated with lithium to absorb hydrogen or hydrogen could be pumped out of the device. This lowers recycling and prevents the associated cooling and increase in density. In this regime, the edge of the plasma will be high temperature and low density, which is good for confinement but requires new methods for managing particle and heat exhaust.

In the early 2000s, researchers at Princeton Plasma Physics Laboratory (PPPL) began to propose strategies for low recycling regimes in tokamaks~\citep{Krasheninnikov2003, Zakharov2007}. They outlined the issues with the high recycling regime. High recycling regimes cool the plasma edge which increases temperature gradients resulting in turbulent energy losses and poor confinement. High recycling regimes also drive edge-localized modes (ELMS) which degrade plasma performance. This early work on low recycling regimes proposed using lithium coated plasma facing components (PFCs) to reduce recycling and improve confinement, laying out initial concepts for low recycling divertors. The benefits of low recycling for confinement were confirmed in experiments on CDX-U~\citep{Kaita2007} and NSTX~\citep{Maingi2012} at PPPL which demonstrated improved energy confinement when using lithium-coated walls. More recent work lays out a design for flowing liquid lithium divertors and details the expected performance on a JET-sized tokamak~\citep{Zakharov2019}.

Designing feasible low recycling regimes for fusion pilot plants is an active area of research - low recycling regimes will offer enhanced confinement if heat exhaust issued can be managed. The low SOL density associated with this regime means that it is difficult to radiate power effectively (remember the radiated power is proportional to the electron and impurity density). Since significant power cannot be radiated, the total heat flux incident on the divertor and peak heat flux will be larger than in the high recycling regime.
The low recycling coefficients also mean that the SOL plasma will be much hotter and will be sheath-limited, unable to support large parallel temperature gradients. So, the plasma hitting the divertor will be hot and sputtering is a more severe issue than in the high recycling regime. In addition, the low SOL density makes helium exhaust more difficult. In spite of these heat exhaust issues, low recycling regimes are appealing because of the enhanced core confinement they offer.

The research conducted in this thesis focuses on developing solutions for heat exhaust issues in the low recycling regime that will allow tokamaks to access the confinement advantages it offers. 
The high recycling regime limits the parameter space in which the SOL can operate (the density will be high). Low recycling divertors open up a larger parameter space for optimization; they give us the opportunity to explore higher temperature and lower density SOLs.
We have conducted several simulations that begin to address the heat exhaust issues in low recycling regimes. 
Simulation results indicate that the peak heat flux can be reduced by the interaction of mirror trapping and drifts which can drastically increase $\lambda_q$ at high SOL temperatures as we will see in section~\ref{sec:plasma_only_drifts}. Simulation results also indicate that the problem of core contamination from sputtering may be manageable because impurities are better confined to the divertor region in low recycling regimes~\citep{Shukla25}. The problem of erosion would persist if a solid divertor target was used, but liquid metal targets~\citep{SuperXT,ARPAE2025,INFUSE2024} targets could be used to avoid detrimental erosion. We have not yet explored the problem of helium exhaust with numerical simulations, but long legged divertors could make pumps with a large opening and short duct feasible, which would make pumping helium easier in the long mean free path regime. Additionally, a low SOL plasma density would reduce the helium ionization source in the SOL and reduce its affect on the plasma.

\subsection{Relevance and Impact of PFCs}
The choice of material for different PFCs can have a large impact on the edge conditions, confinement, and fusion performance. Many different materials have been considered in the context of different regimes. Here we outline the advantages and disadvantages of some of those materials and which target regimes they are compatible with.

Some important considerations are whether the material is high or low recycling, how much heat flux the material can handle, and the Z of the material. Clearly low and high recycling materials will help facilitate their respective regimes. Refractory metals which can handle large heat fluxes without detrimental erosion are desirable because will have a longer lifetime in a reactor and have high melting points. If the divertor or wall materials suffer too much erosion or evaporation, the main plasma can be heavily polluted by impurities. Regardless of the choice of material, some of impurities will be sputtered into the plasma, so it is important to consider the effects of high and low Z impurities. High Z impurities are more hazardous for fusion performance because they radiate large amounts of power in the core while low Z impurities are less hazardous - they will dilute the plasma which degrades performance but do not radiate as much power as high Z impurities.

For high recycling regimes, carbon and tungsten walls have been considered. Carbon is high recycling and has low Z but suffers from tritium retention~\citep{Stangeby1990,Haasz2005,Tanabe2003}. Tritium retention is undesirable in a reactor because it makes the tritium fuel cycle more difficult and can pose environmental hazards in the case of a leak or accident~\cite{Pitts2011}.
Tungsten has a very high melting point and can handle large heat fluxes which will increase the lifetime of PFCs~\citep{Pitts2019, Scholte2025, YOU2022}. However, tungsten is high Z, so sputtering can be detrimental to fusion performance~\citep{Pitts2019}. 
In high recycling regimes, sometimes low Z impurity seeding of gases such as argon can be used to purposely radiate power from the SOL and facilitate detachment. 
Lithium is low recycling but evaporates easily, and so cannot handle high heat fluxes without rapid evaporation that would cool the plasma.

Liquid metals PFCs offer an advantage over solid PFCs because they can be replenished to mitigate erosion~\citep{Tabares2017, MikeArxiv2024}. Liquid tin has been considered as PFC because it is capable of high temperature operation ($\geq$ 1000 $^\circ C$ unlike lithium), but it is high Z and could result in core radiation issues similar to tungsten~\cite{MikeArxiv2024}. Recent experiments in ASDEX Upgrade have demonstrated that liquid tin ejects droplets when exposed to hydrogen plasma which could result in unacceptable radiative lossesin the core~\cite{Scholte2025}. 
Recent advances indicate that there could be suitable liquid metal alloys for the divertor target which have low Z sputtering but are capable of high temperature operation~\citep{MikeArxiv2024} .

The use of some combination of different materials for different PFCs as well as the development of new materials will be essential for improving plasma performance. 
Some low recycling scenarios envisioned for FPPs involve using novel liquid metal alloys~\citep{ARPAE2025,INFUSE2024,SuperXT} to coat the divertor plates (regions of high heat flux) and low recycling materials such as lithium to coat the side walls (regions of low heat flux).

\section{Numerical Simulations for Reactor Design}
\label{sec:numerical_simulations}
Tokamaks are expensive to build, so predicting how new devices in untested regimes will behave is extremely important; numerical modeling guides reactor design. Realistic modeling of tokamak plasmas is very complex; tokamak plasmas are a complicated nonlinear system involving a wide range of length and time scales and multiple species with very different masses (ions and electrons). Beyond this, tokamaks have complicated geometries and interactions with the wall materials must also be modeled. The dynamics in the core and SOL effect have great influence on each other, so modeling both is essential to accurately predict conditions in a reactor and to understand and interpret phenomena observed in present day experiments. There are a wide range of models used for tokamak plasmas varying in complexity and regime of validity.

The most accurate description of a plasma would be a fully kinetic description involving 3 spatial dimensions, 3 velocity dimensions, and time. A kinetic model must resolve very small length and times scales (compared to the size of a tokamak and a particle confinement time respectively). This makes fully kinetic modeling poorly suited for tokamaks. Fluid modeling instead evolves the fluid moments of the plasma and involves only 3 spatial dimensions and time which greatly reduces the complexity of the problem. However, fluid modeling assumes that the plasma is maxwellian, which is only valid for a collisional (low temperature, high density) plasma. The fluid assumption of high collisionality is not valid at the high temperatures present in the core of the tokamak and is valid in the SOL only in some scenarios. A model of intermediate complexity is gyrokinetics which averages out the gyro-motion of the particles around the magnetic field. Gyrokinetics has one fewer dimension than kinetics (3 space and 2 velocity dimensions) and does not have to resolve the electron gyroradius or the gyrofrequency (gyrokinetics can resolve the gyroradius if those length scales are important but cannot resolve the gyrofrequency). This makes gyrokinetic models much less costly than kinetic models.

Gyrokinetics has been a powerful tool for modeling plasma turbulence in the tokamak core, but until recently, fluid models were typically used for modeling the SOL. In light of interest in the low-recycling regime, in which SOL would not be cold or dense enough to warrant the fluid assumptions, gyrokinetic modeling of the SOL plasma has become desirable. Gyrokinetic simulations of the SOL pose several challenges not present in simulations of the core such as modeling the interaction between the plasma and the material wall, including complex magnetic geometries with magnetic X-points, and modeling the dynamics of impurity species. Here we will focus on modeling of the SOL, specifically gyrokinetic modeling.

\subsection{Challenges of SOL Modeling}
The first major difference between core and SOL modeling is that while the domain of a core simulation lies entirely inside the separatrix in the closed field line region, SOL simulation domains typically include a portion of the closed and open field line regions. In diverted tokamak geometries, this means that a magnetic X-point will be present in the domain. Core simulations typically take advantage of field aligned coordinate systems which allow for a reduction in resolution in the direction aligned with the magnetic field. However, field aligned coordinate systems have a coordinate singularity at magnetic X-points so special approaches and coordinate systems have had to be developed to handle magnetic geometry in SOL simulations.

The second major difference is the interaction of the plasma with the wall materials. One aspect of the wall interaction is modeling the rate of recycling and sputtering of wall materials caused by plasma incident on the divertor plates and machine walls. This involves not only calculating the sputtering and recycling rates, but modeling the dynamics of the neutral particles ejected from the wall as well as modeling their reactions with the plasma(ionization, recombination, and charge-exchange). The second aspect of the wall interaction is the modeling of the sheath. The sheath is a thin region of net positive charge space (the rest of the plasma is quasineutral) that develops at the divertor plate. The sheath is very thin, only a few debye lengths, and thus poses a challenge to gyrokinetic modeling which assumes the plasma is quasineutral and does not resolve the debye length. Boundary conditions including the logical and conducting sheath~\citep{shithesis,Shi17} have been developed to enable gyrokinetic simulations of the SOL.

The last major difference between core and SOL modeling is the treatment of plasma fluctuations. In core simulations, delta-$f$ gyrokinetics is typically used. In this treatment, the fluctuations of the plasma are assumed to be small fluctuations on top of a background equilibrium which is not evolved. The fluctuations are driven by fixed gradients of the background equilibrium. In local delta-$f$ codes, the radial domain is very narrow and the background gradients are taken to be radially constant. In global delta-$f$ the gradients come from a provided density and temperature profile and thus vary radially across the domain. In SOL modeling, the fluctuations are not small compared to the background equilibrium, so full-$f$ gyrokinetics must be employed. In full-$f$ gyrokinetics the entire plasma is evolved, so the background gradients change in time.

For SOL modeling and divertor design, simulation codes often reduce the complexity of the problem by assuming an axisymmetric plasma. Core codes are primarily geared at turbulence, which necessitates doing a modeling a full 3D (in space) plasma. SOL codes are often used for divertor design and sometimes operate in a 2D axisymmetric mode to reduce computational cost. In these axisymmetric simulations, a perpendicular particle and heat diffusivity, $D_\perp$ and $\chi_\perp$, are inserted to mimic turbulent heat transport. The diffusivities can be chosen based on a 3D turbulent calculation or to target an SOL width expected in an experiment.  

\subsection{Summary of Present Tokamak Modeling Codes}
It is helpful to have an idea of the different available tokamak simulation codes and their capabilities. Here we present a non-exhaustive list of commonly used codes in table~\ref{tab:codelist} including which model they employ, the domain that can be simulated, and whether gyro-averaging is included for the gyrokinetic codes. Gyrokinetics can be full-$f$ or delta-$f$ (local or global), contiunuum or particle-in-cell (PIC), and can include gyro-averaging in some cases.
Of the full-$f$ codes, only XGC includes gyro-averaging. The axisymmetric fluid codes SOLPS and UEDGE simulate a thin portion of the core, but they are generally not considered valid for predictive modeling in the core; the simulated portion of the core serves to infer the particle and heat source entering the SOL. Other fluid codes such as BOUT++ can capture pedestal dynamics in the core. Because the temperature in the core of the tokamak is high, gyrokinetic rather than fluid codes are better suited to capturing all of the relevant effects and micro-instabilities in the core.

SOL gyrokinetic and fluid codes typically either evolve a fluid neutral species or are coupled to a kinetic monte-carlo solver that evolves the neutral particle dynamics. B2.5 is tightly coupled to a neutral solver called EIRENE and the composite software package is called SOLPS. In this work, Gkeyll has also been loosely coupled to EIRENE for the evolution of neutrals. Another monte-carlo neutral solver is DEGAS2 which XGC is coupled to. UEDGE has the option to couple to either DEGAS2 or EIRENE. UEDGE and B2.5 both also have the option to evolve fluid neutral species.

\begin{table}[]
\centering
\small
\begin{tabular}{|c|c|c|c|c|c|}
     \hline
     Name &  Model & Domain & Gyro-     & Spatial    & Velocity\\
          &        &        & averaging & Dimensions & Dimensions\\
     \hline
     GENE &  Continuum Delta-f GK& Core & Yes & 3 & 2\\
         &   (Local/Global)       &      &     &   &  \\
     \hline
     CGYRO &  Continuum Delta-f GK& Core & Yes & 3 & 2\\
           &   (Local)            &      &     &   &  \\
     \hline
     GENE-X &  Continuum Full-f GK& Core \& SOL & No & 3 & 2\\
     \hline
     XGC &  PIC Full-f GK & Core \& SOL & Yes & 2 or 3 & 2 \\
     \hline
     Gkeyll &  Continuum Full-f GK& Core \& SOL & No & 2 or 3 & --\\
     \hline
     B2.5 &  Braginskii Fluid & Core \& SOL & -- & 2 & --\\
     \hline
     UEDGE &  Fluid & Core \& SOL & -- & 2 & --\\
     \hline
     BOUT++ &  Fluid & Core \& SOL & -- & 3 & --\\
     \hline
\end{tabular}
 \caption[List of commonly used codes for tokamak modeling.]{List of commonly used codes for tokamak modeling. Gyrokinetic is abbreviated as GK in the table.}
 \label{tab:codelist}
\end{table}

\subsection{Gkeyll's Gyrokinetic Model}
\label{sec:intro_Gkeyll_model}
The simulations conducted as part of this thesis use the \gke code.
Gkeyll is a full-f, long-wavelength gyrokinetic code using a Discontinuous Galerkin (DG) method for spatial discretization and Runge-Kutta for discretization in time. We use the electrostatic version of the code which solves the gyrokinetic equation
\begin{eqnarray}
\frac{\partial f_s}{\partial t}+\dot{\boldsymbol{R}} \cdot \nabla f_s+\dot{v}_{\|} \frac{\partial f_s}{\partial v_{\|}} - \nabla \cdot (\mathbf D \cdot \nabla f_s) = \nonumber \\
C\left[f_s\right]+S_s+C^{iz}_s+C^{rec}_s +  C^{rad}_s,
\label{eq:intro_gkeq}
\end{eqnarray}
with $\dot{\mathbf{R}}=\{\mathbf{R}, H\}, \dot{v}_{\|}=\left\{v_{\|}, H\right\}, \textrm{ and } H_s=\frac{1}{2} m_s v_{\|}^2+\mu B+q_s \phi$
along with the gyrokinetic Poisson equation
\begin{equation}
-\nabla_{\perp} \cdot\left(\sum_s\frac{m_sn_{0s}}{B_0^2} \nabla_{\perp} \phi\right) = \sum_sq_s n_s(\boldsymbol{R}),
\label{eq:intro_poisson}
\end{equation}
where  $f_s = f_s(\mathbf{R}, v_\parallel,\mu.t)$ is the gyrocenter distribution function for species $s$, $\mathbf R$ is the guiding center position, $\mathbf D$ is the particle diffusivity, $v_\parallel$ is the velocity parallel to the magnetic field, $\mu = \frac{mv_\perp^2}{2B}$ is the magnetic moment, $v_\perp$ is the velocity perpendicular to the magnetic field, $B$ is the magnetic field magnitude, $B_0$ is the magnetic field magnitude at the center of the simulation domain, $H_s$ is the gyrocenter center Hamiltonian of species s, $\phi$ is the electrostatic potential, $n_s$ is the guiding center density of species s, $n_{0s}$ is a reference density for species s, $q_i$ is the ion charge, and $e$ is the elementary charge.

The gyrokinetic Poisson bracket is given by~\citep{Noah21}
\begin{equation}
\{F, G\}=\frac{\boldsymbol{B}^*}{m B_{\|}^*} \cdot\left(\nabla F \frac{\partial G}{\partial v_{\|}}-\frac{\partial F}{\partial v_{\|}} \nabla G\right)-\frac{\mathbf{b}}{q B_{\|}^*} \times \nabla F \cdot \nabla G,
\end{equation}
with $\mathbf{B^*} = \mathbf{B} + (mv_\parallel/q)\nabla\times\mathbf{b}$ and $B_\parallel^* = \mathbf{b} \cdot \mathbf{B^*} \approx B$ where $\mathbf{b} = \mathbf{B}/B$ is the unit vector along the magnetic field.

The right-hand side of Eq.~\ref{eq:intro_gkeq} contains the collision term~\citep{Mana22} $C[f_s]$, volumetric source terms $S_s$, ionization and recombination terms $C^{iz}_s$ and $C^{rec}_s$~\citep{Tess22}, and the radiation term $C^{rad}_s$~\citep{radiation}. 
The radiation term removes energy from the electrons based on the electron distribution function and the densities of the electrons and radiating impurities~\citep{Roeltgen25}.
Gkeyll uses the Dougherty collision operator~\citep{Dougherty1964, Mana22}.

Note that Gkeyll implements the drift-kinetic limit of the gyrokinetic equation, neglecting all gyroaveraging operations. However, first order finite Larmor radius (FLR) effects are present in the ion polarization term in the gyrokinetic poisson equation, which distinguishes the long-wavelength gyrokinetic model from a drift-kinetic model~\citep{Noah21}.

\section{Thesis Overview}
Using numerical simulations to inform reactor design is essential to realizing commercially viable fusion. Exploring novel regimes with simulations can help us find methods for increasing fusion performance while keeping reactor cost low. 
The low recycling regime is appealing because it offers confinement advantages, but it presents challenges for heat exhaust. In addition to physical challenges, the low recycling regime presents modeling challenges.
Fluid simulations are typically used to study the Scrape-Off Layer (SOL) in tokamaks ~\citep{Hudoba2023, Osawa2023, Rozhansky2021, Zhang2024}, but modeling the collisionless SOL of a low recycling regime in which the fluid assumptions are not valid requires a kinetic treatment. 

The aim of this thesis is to use numerical simulations to demonstrate the feasibility of low recycling regimes in fusion pilot plants.
To this end, we have developed \gke 's 2D gyrokinetic solver into a code suitable for divertor design; we have improved \gke 's geometric flexibility allowing simulations in X-point geometry and coupled \gke\ to EIRENE for modeling wall interactions and the evolution of neutral particles.
We have then used \gke\ simulations to study low recycling scenarios in the magnetic geometry of the Spherical Tokamak for Energy Production (STEP).


In chapter~\ref{chap:2}, we introduce the gyrokinetic equation and the methods developed for simulating tokamaks with X-point geometry using a field-aligned coordinate system. In the remaining chapters, we begin to demonstrate the feasibility of a low recycling regime using gyrokinetic simulations. In chapter~\ref{chap:3} we show Gkeyll-EIRENE simulations that indicate a low-recycling (high temperature) SOL can be maintained without a lithium target and also note some of the heat exhaust challenges associated with this regime. In chapter~\ref{chap:4} we address these challenges by comparing gyrokinetic and fluid simulations. We show that, in a low recycling regime, impurities can be confined downstream to avoid core contamination. We also demonstrate how kinetic effects can be used to broaden the SOL and reduce the peak heat flux. Finally, in chapter~\ref{chap:conclusion}, we summarize the main results and outline future research that will extend this work. 


\chapter{Constructing Field Aligned Coordinate Systems for Gyrokinetic
  Simulations of Tokamaks in X-point Geometries}
\label{chap:2}
\section{Introduction}
    \footnote{
    All simulations, tests, and analysis were performed by me. The algorithms described in the paper to allow for X-point geometry were developed and implemented in the Gkeyll code by me with help from Ammar Hakim, Manaure Francisquez, and James Juno. The paper~\cite{Shukla2025Xpt} (submitted to Journal of Plasma Physics) was written by me. 
    }
    Structures in tokamak plasmas are anisotropic: they are elongated along the field line but short perpendicular to it. Many tokamak simulation codes, especially core codes such as GS2~\citep{Dorland2000,GS2-zenodo}, GENE~\citep{Jenko2000,Gorler2011} and GYRO~\citep{Candy2010, Candy2009}, take advantage of this by using a field aligned coordinate system; the field aligned coordinate system allows for coarse resolution along the field line reducing computational expense~\citep{beer95}. However, field aligned coordinate systems have a coordinate singularity at magnetic X-points where the poloidal magnetic field vanishes, so using field-aligned coordinate systems in edge codes which simulate  a portion of the core and scrape-off layer (SOL) simultaneously is more difficult~\citep{STEGMEIR2016139, leddy2017}. Grids using field-aligned coordinates also suffer from cell deformation in the presence of strong magnetic shear which can be addressed using the shifted metric procedure~\citep{Scott2001,Scott1998,Dimits1993}.

    There have been a variety of approaches to handling the coordinate singularity at the X-point. BOUT++ handles it by using multiple blocks, each with a field aligned coordinate system,
    and avoiding the calculation of geometric quantities at the X-point~\cite{leddy2017, Dudson2024}.
    Flux-aligned coordinates  do not suffer from the same shear as field-aligned coordinates since they typically use the toroidal angle as cone coordinate, but they are also singular at X-points. This singularity can be cured numerically~\cite{Mattor1995594}, but still have resolution imbalances near the X-point.
    COGENT uses multiple blocks each with a coordinate system that is flux aligned except near the X-point where they overlap. A high order interpolation scheme is used to transfer information between the overlapping regions of each block~\citep{MCCORQUODALE2015181, Dorf16}. The 2D fluid codes SOLPS~\cite{WIESEN2015} and UEDGE~\cite{ROGNLIEN1992} as well as the 3D fluid code SOLEDGE3X~\citep{BUFFERAND2024} also employ flux-aligned coordinates. 

    Other edge gyrokinetic codes such as GENE-X~\citep{Michels21} and fluid codes such as GRILLIX~\citep{STEGMEIR2026}, which use the same framework~\citep{STEGMEIR2016139}, have abandoned field and flux aligned coordinates in favor of the Flux-Coordinate-Independent (FCI) approach because of the difficulty of dealing with the singularity at the X-point. The gyro-fluid code FELTOR~\citep{Wiesenberger2024} uses a finite volume formulation of the FCI approach~\cite{WIESENBERGER2023}.  The FCI approach breaks the simulation domain into a series of poloidal planes which do not have a field aligned coordinate system and employs a field-line following discretization of the parallel derivative operator to minimize the number of poloidal planes needed. Interpolation within the poloidal plane is required to compute the parallel derivatives~\citep{HARIRI20132419, STEGMEIR2016139, Stegmeir2018}. Another alternative is to forego the resolution advantages offered by FCI and field-aligned coordinate systems and use a cylindrical coordinate system as is done in the fluid code GBS~\cite{GIACOMIN2022}.
    
    Here we present an algorithm for computing geometric quantities in a standard field aligned coordinate system that avoids the singularity at the X-point. We employ a multi-block approach where each block conforms to the separatrix leaving no gap around the X-point. Our numerical scheme allows us to avoid calculation of any geometric quantities or fluxes at the X-point while still having block corners at the X-point. We implement and test this algorithm in the gyrokinetic model in the \gke\ simulation framework\citep{Mana25, Shukla25, Mandell2020, Ammar2019}.

    The rest of the paper is organized as follows: in section~\ref{sec:clebsch} we give background on the Clebsch representation of magnetic fields and field aligned coordinates,
    in section~\ref{sec:transforms} we present the equations of our gyrokinetic model in a field aligned coordinate system, 
    in section~\ref{sec:coordinates} we detail the coordinate system we employ, 
    in section~\ref{sec:discretization} we show how the spatial discretization of our algorithm avoids the cooordinate singularity at the X-point,
    and in section~\ref{sec:computation} we describe how we generate simulation grids and calculate geometric quantities and also show example grids. 
    Finally, in section~\ref{sec:tests},  we conduct convergence and consistency tests and show an example 2D axisymmetric gyrokinetic simulation of STEP~\citep{Karhunen2024} using our algorithm.
    Consistent with findings in~\cite{WIESENBERGER2018}, we find that convergence is reduced in the vicinity of the X-point. However, we find that our scheme converges faster than first order for all tests and maintains exact particle conservation even in domains with an X-point.  

\section{Coordinate Systems for Magnetized Plasma Simulations}
\label{sec:clebsch}

As is well known, in certain situations (described below), we can
write the magnetic field in the Clebsch representation~\citep{DHaessler}
\begin{align}
  \mvec{B} = \nabla\psi\times\nabla\alpha \label{eq:clebsch}
\end{align}
where $\psi(\mvec{x})$ and $\alpha(\mvec{x})$ are
scalar functions of the position vector $\mvec{x}$. 
However, not all magnetic field configurations can be described by the
Clebsch representation: the field-lines of Clebsch-representable
magnetic fields are \emph{integrable} and hence enforce some stringent
constraints on the type of fields that can be described in this
way\footnote{A generalized Clebsch representation of the form
  $\mvec{B} = \nabla\psi_1\times\nabla\alpha_1 +
  \nabla\psi_2\times\nabla\alpha_2$ allows representing arbitrary
  magnetic fields, including ones in which the field-lines are not
  integrable. However, these are not useful to construct coordinate
  systems.}. Thankfully, for tokamaks, where the fields are
axisymmetric, or in regions of stellarators with nested flux surfaces,
such representations can be found. Hence, in this paper we will
restrict ourselves to such magnetic configurations.

The importance of the Clebsch representation (when it exists) is that
the two vectors $\nabla\psi$ and $\nabla\alpha$ can be used as the
\emph{dual basis vectors} (contravariant basis) of a field-line
following coordinate system. To understand what this means and
establish notation for rest of the paper consider an arbitrary
coordinate transform given by the \emph{invertible} map
\begin{align}
  \mvec{x} = \mvec{x}(z^1, z^2, z^3)
\end{align}
where $(z^1, z^2, z^3)$ are computational coordinates. This maps a
rectangular region in $\mathbb{R}^3$ to a (generally non-rectangular)
region of physical space. Once this mapping is known then we can compute the
\emph{tangent vectors}
\begin{align}
  \basis{i} = \pfrac{\mvec{x}}{z^i}
\end{align}
and the \emph{dual vectors} $\dbasis{i}$ defined implicitly by the
relation
\begin{align}
  \dbasis{i}\cdot\basis{j} = \delta\indices{^i_j}.
\end{align}
If the inverse mapping $z^i = z^i(\mvec{x})$ is known, then we can
show that $\dbasis{i} = \nabla z^i$. At each point $\mvec{x}$ either
the tangents or duals form a linearly independent set of vectors and
hence can be used to represent vector and tensor quantities at that
point. For example, a vector $\mvec{a}$ can be written as
\begin{align}
  \mvec{a} = a^i \basis{i} = a_i \dbasis{i}
\end{align}
where $a^i = \mvec{a}\cdot\dbasis{i}$, $a_i = \mvec{a}\cdot\basis{i}$
and we have assumed the summation convention over repeated
indices. Once the tangent and dual vectors are determined we can
compute the covariant and contravariant components of the metric
tensor as
\begin{align}
  g_{ij} &= \basis{i}\cdot\basis{j} \\
  g^{ij} &= \dbasis{i}\cdot\dbasis{j}.
\end{align}

Defining the \emph{Jacobian} (volume element) of the transform
$J_c = \basis{1}\cdot(\basis{2}\times\basis{3})$ we can
easily derive the explicit expressions for the duals:
\begin{align}
  \dbasis{1} &= \frac{1}{J_c} \basis{2}\times\basis{3} \\
  \dbasis{2} &= \frac{1}{J_c} \basis{3}\times\basis{1} \\
  \dbasis{3} &= \frac{1}{J_c} \basis{1}\times\basis{2}.
\end{align}
From this we also see that
$J_c^{-1} = \dbasis{1}\cdot(\dbasis{2}\times\dbasis{3}) =
\nabla z^1\cdot(\nabla z^2\times\nabla z^3)$. We assume that the basis
are arranged such that $J_c > 0$. 

As we need the mapping to be invertible we must ensure that
$J_c(\mvec{x})$ does not vanish anywhere in the domain. At
the $X$- and $O$-points of a tokamak configurations, however, we have
$J_c = 0$ for field-line following coordinates, that is, the
coordinate system is non-invertible. We get around this issue by
ensuring that we \emph{do not} compute any geometrical quantities or
numerical fluxes at these isolated singular points in the domain. The
use of a high-order scheme (we use the discontinuous Galerkin scheme)
that uses interior (to surfaces and volumes) quadrature nodes where
numerical fluxes are computed, automatically ensures this, allowing us
to work with coordinate systems that have singularities at a finite
set of isolated points. However, despite not computing any geometric or physical quantity at the $X$- or $O$-points, we ensure a corner node lies exactly there, producing an accurate representation of the geometry, without any ``holes''. As will be discussed in section~\ref{sec:tests}, avoiding evaluation at the X-point does not entirely remove the effect of the coordinate singularity.

Identifying the dual vectors as $\dbasis{i} = \nabla z^i$, we can see why the Clebsch
form \eqr{\ref{eq:clebsch}} is useful: once we find the Clebsch form
we can \emph{construct} a coordinate system (as described later in
this paper) such that the resulting mapping has
$\dbasis{1} = \nabla\psi$ and $\dbasis{2} = \nabla\alpha$. With this,
the two scalar function $z^1 = \psi$ and $z^2 = \alpha$ would be two
of the three computational coordinates. The choice of the third
coordinate, $z^3 = \theta$, called the field-line coordinate, can then
be made independently.

Now, as
$\mvec{B} = \nabla\psi\times\nabla\alpha =
\dbasis{1}\times\dbasis{2}$ we must have
\begin{align}
  \mvec{B}\cdot\dbasis{1} = \mvec{B}\cdot\dbasis{2} = 0
\end{align}
and hence
\begin{align}
  \mvec{B} = (\mvec{B}\cdot\dbasis{3})\basis{3} = \frac{1}{J_c} \basis{3}.
\end{align}
From this we get a relation between the Jacobian, the magnitude of the
magnetic field and the $g_{33}$:
\begin{align}
  J_c B = \sqrt{g_{33}}.
\end{align}

In these \emph{field-line following} coordinates the magnetic field
always points in the direction of $\basis{3}$. The unit vector in the
direction of the magnetic field is
\begin{align}
  \buni = \frac{\basis{3}}{\| \basis{3} \|}
\end{align}

The choice of these field-line following coordinates, is \emph{not},
in general, global or unique, and depends on the topologically
distinct regions that need to be included in a simulation. 
In general, a single mapping is not usually enough to cover all of the physical region of interest, and hence 
several maps are needed that between them cover the physical domain~\cite{BUFFERAND2024,leddy2017}. 
For simple
devices, like the magnetic mirror, a single coordinate map is enough
to grid the complete domain. However, for tokamaks we usually have to
divide the physical domain into multiple regions, at least one for
each topologically distinct region, and construct field-line following
coordinates specific to each region. For example, for a double-null
configuration we have to construct separate coordinate systems in the
outer and inner scrape-off-layers (SOLs), the upper and lower private
flux (PF) regions and the core region. In our implementation, in fact,
for double-null configurations, we generate 
five maps to ensure a reasonably smooth grid that includes the core, the
SOLs and the private-flux regions. We refer to the assembly of grids that covers the entire physical region of interest as a multi-block grid.

\section{Transforms of the Gyrokinetic Equation}
\label{sec:transforms}
\subsection{The Gyrokinetic Equations}

The electrostatic gyrokinetic equation can be written as a Hamiltonian
system
\begin{align}
  \pfrac{f}{t} + \{f,H\} = 0 \label{eq:gen-ham}
\end{align}
where $f$ is the distribution function and $H$ is the Hamiltonian. In conservative form we can write this
as
\begin{align}
  \pfrac{(J_vf)}{t} + 
  \gcs\cdot(J_v\dot{\mvec{x}} f)
  +
  \pfraca{v_\parallel}
  (J_v\dot{v}_\parallel f) = 0
  \label{eq:gen-ham-cons}
\end{align}
where $v_\parallel$ is the velocity parallel to the magnetic field, $\mu$ is the magnetic moment, $\dot{\mvec{x}} = \{\mvec{x},H\}$, $\dot{v}_\parallel = \{v_\parallel,H\}$ and
$J_v = B_\parallel^*/m$. Further, for any two phase-space functions
$f(\mvec{x},v_\parallel,\mu)$ and $g(\mvec{x},v_\parallel,\mu)$ the
Poisson bracket given by
\begin{align}
  \{f,g\}
  =
  \frac{\mvec{B}^*}{m B_\parallel^*}
  \cdot
  \left(
  \gcs f \pfrac{g}{v_\parallel}
  -
  \pfrac{f}{v_\parallel}
  \gcs g
  \right)
  -
  \frac{\buni}{q B_\parallel^*}
  \times
  \gcs f \cdot \gcs g
\end{align}
where $\mvec{B}^* = \mvec{B} + (m v_\parallel/q) \gcs\times\buni$ and
$B_\parallel^* = \buni\cdot\mvec{B}^* \approx B$. The gyrocenter Hamiltonian is
\begin{align}
  H = \frac{1}{2} m v_\parallel^2 + \mu B + q \phi,
\end{align}
where $m$ is the species' mass, $q$ is the species' charge, and $\phi$ is the electrostatic potential. 
Here we have taken the long-wavelength (drift-kinetic) limit to neglect gyroaveraging of $\phi$~\citep{Mandell2020}.
Substituting the Hamiltonian into the Poisson bracket, we get get
\begin{align}
  \{f,H\}
  &=
  \frac{\mvec{B}^*}{m B_\parallel^*}
  \cdot
  \left(
  m v_\parallel
  \gcs f
  -
  \pfrac{f}{v_\parallel}
  \gcs H
  \right)
  -
  \frac{\buni}{q B_\parallel^*}
  \times
  \gcs f \cdot \gcs H \\
\end{align}
where
\begin{align}
  \gcs H
  =
  \mu\gcs B + q\gcs\phi.
\end{align}
The characteristics are
\begin{align}
  \dot{\mvec{x}} = \{\mvec{x},H\} =
  \frac{\mvec{B}^*}{B_\parallel^*} v_\parallel
  +
  \frac{\buni}{q B_\parallel^*}
  \times
  \gcs H
\end{align}
and
\begin{align}
  \dot{v}_\parallel = \{{v}_\parallel,H\} =
  -\frac{\mvec{B}^*}{m B_\parallel^*}\cdot\gcs H.
\end{align}

The electrostatic potential $\phi$ is determined by the gyrokinetic Poisson equation (also sometimes called the quasineutrality equation)
\begin{align}
  -\gcs\cdot
  \left(
  \varepsilon_\perp \nabla_\perp \phi
  \right)
  =
  \sum_s q_s \int J_v f_s \thinspace d^3\mvec{v}
  \label{eq:poisson1}
\end{align}
where $\varepsilon_\perp(\mvec{x})$ is a polarization tensor and $d^3\mathbf v = d\mu dv_\parallel$ indicates integration of velocity space. The
operator $\nabla_\perp$ is defined as
\begin{align}
  \nabla_\perp = \gcs - \buni(\buni\cdot\gcs).
\end{align}
Even in the long-wavelength limit with no gyroaveraging, the first-order polarization charge density on the left-hand side of Eq.~\ref{eq:poisson1} incorporates some finite-Larmor-radius effects~\citep{Mandell2020}. 
Note that as will be discussed later in section~\ref{sec:gkfieldaligned}, the flute ordering approximation ($k_\parallel \ll k_\perp)$ can be employed to drop derivatives parallel to the magnetic field in the gyrokinetic Poisson equation.

Until this point we have written all equations in an coordinate
independent form. Now we introduce coordinates. Consider transforming
the configuration space coordinates as
\begin{align}
  \mvec{x} = \mvec{x}(z^1,z^2,z^3)
\end{align}
where $(z^1,z^2,z^3)$ are computational coordinates. From this
mapping, as we discussed above, we can compute the tangent vectors
, the duals, and the
co- and contravariant components of the metric-tensor.

One we have the tangents and duals, we can construct the fundamental
vector derivative operator
\begin{align}
  \gcs  = \dbasis{i}\frac{\partial}{\partial z^i}.
\end{align}
This operator is enough now to write the equations in aribitrary
coordinate systems. To ease the derivations we need the identities
\begin{align}
  \gcs\cdot\mvec{U}
  =
  \frac{1}{J_c}\frac{\partial}{\partial z^i}\left(J_c
  \dbasis{i}\cdot\mvec{U} \right)
\end{align}
and
\begin{align}
  \gcs\times\mvec{U}
  &=
  \frac{1}{J_c}\frac{\partial}{\partial z^i}
  \left(
  \epsilon^{ijk} U_j
  \right) \basis{k} \\
  &=
  \frac{1}{J_c}\left( \pfrac{U_3}{z^2} - \pfrac{U_2}{z^3} \right) \basis{1}
  +
  \frac{1}{J_c}\left( \pfrac{U_1}{z^3} - \pfrac{U_3}{z^1} \right) \basis{2}
  +
  \frac{1}{J_c}\left( \pfrac{U_2}{z^1} - \pfrac{U_1}{z^2} \right) \basis{3}
\end{align}
where $\mvec{U}$ is any vector field and $\epsilon^{ijk}$ is the Levi-Civita tensor.

\subsection{Gyrokinetic Equation in Field Aligned Coordinates}
\label{sec:gkfieldaligned}
The GK equation in computational coordinates becomes
\begin{align}
  \pfrac{(J_vf)}{t} 
  + 
  \frac{1}{J_c}\frac{\partial}{\partial z^i}
  (
  J_c J_v(\dbasis{i}\cdot\dot{\mvec{x}}) f
  )
  +
  \pfraca{v_\parallel}
  (J_v\dot{v}_\parallel f) = 0.
  \label{eq:gk-zi}  
\end{align}
Further, in these coordinates we have
\begin{align}
  \mvec{B}^*
  =
  (\mvec{B}\cdot\dbasis{3})\basis{3}
  +
  \frac{m v_\parallel}{q} 
  \frac{1}{J_c}
  \frac{\partial}{\partial z^i}
  \left(
  \epsilon^{ijk} b_j
  \right) \basis{k}
\end{align}
where $b_j = \basis{j}\cdot\buni$.  Hence, we have
\begin{align}
  \dbasis{i}\cdot\mvec{B}^*
  =
  (\mvec{B}\cdot\dbasis{3})\delta\indices{^i_3}
  +
  \frac{m v_\parallel}{q} 
  \frac{1}{J_c}
  \frac{\partial}{\partial z^k}
  \left(
  \epsilon^{kji} b_j
  \right).
  \label{eq:em-dot-B*}
\end{align}
Further, we can compute
\begin{align}
  \dbasis{i}\cdot(\buni\times\gcs H)
  =
  \dbasis{i}\cdot(\dbasis{j}\times\dbasis{k}) b_j \pfrac{H}{z^k}
  =
  \frac{\epsilon^{ijk}}{J_c} b_j \pfrac{H}{z^k}.
\end{align}
Hence, we have
\begin{align}
  \dbasis{i}\cdot\dot{\mvec{x}}
  =
  \frac{v_\parallel}{B^*_\parallel} (\dbasis{i}\cdot\mvec{B}^*)
  +
  \frac{1}{q B^*_\parallel}
  \frac{\epsilon^{ijk}}{J_c} b_j \pfrac{H}{z^k}.
  \label{alpha config}
\end{align}
Further, we have
\begin{align}
  \dot{v}_\parallel
  =
  -\frac{(\dbasis{k}\cdot\mvec{B}^*)}{m B_\parallel^*}\pfrac{H}{z^k}.
  \label{alpha phase}
\end{align}
We can again use \eqr{\ref{eq:em-dot-B*}} to compute $\dbasis{k}\cdot\mvec{B}^*$.

The gyrokinetic Poisson equation in computational coordinates becomes
\begin{align}
  -\frac{1}{J_c}
  \frac{\partial}{\partial z^i}
  \left(
  J_c \varepsilon_\perp \dbasis{i}\cdot\nabla_\perp \phi
  \right)
  =
  \sum_s q_s \int J_v f_s \thinspace d^3\mvec{v}.
  \label{poisson}
\end{align}
We can compute
\begin{align}
  \dbasis{i}\cdot\nabla_\perp \phi
  &=
  \dbasis{i}\cdot\dbasis{j}\pfrac{\phi}{z^j}
  -
  (\dbasis{i}\cdot \buni) (\buni\cdot\dbasis{m})\pfrac{\phi}{z^m} \\
  &=
  g^{ij} \pfrac{\phi}{z^j}
  -
  \delta\indices{^i_3}\frac{1}{\lVert \basis{3} \rVert^2 }
    \pfrac{\phi}{z^3}.
  \label{ei dot gradperp}
\end{align}
Note that in gyrokinetics we typically drop the derivatives in $z^3$ in the gyrokinetic Poisson
equation due to the assumption that gradient scale lengths in the parallel direction are much longer than those in the perpendicular direction. This is the commonly used "flute approximation"~\citep{Scott2010,Seto2019}. 

\subsection{Simplifications in Axisymmetric Limit}

For divertor design, axisymmetric simulations which are 2D rather than 3D in configuration space are often used~\cite{Dominski2025,Shukla25,shukla2025LR}. In these simulations, 
cross-field transport is modeled with ad-hoc diffusive terms.
If we take the second computational coordinate $z^2$ as our ignorable coordinate assuming $\partial F/\partial z^2 = 0$ for all quantities $F$, we get the following equations of motion by taking $i=1,3$ in Eq.~\ref{alpha config}
\begin{equation}
\mathbf e^1 \cdot \mathbf{\dot x} = \dot z^1 = -\frac{mv_\parallel^2}{qJ_cB^*_\parallel}\pdv{b_2}{z^3} + \frac{b_2}{qJ_cB_\parallel^*}  \pdv{H}{z^3}
\label{eq:z1}
\end{equation}

\begin{equation}
\mathbf e^3 \cdot \mathbf{\dot x} =  \dot z^3 = \frac{Cv_\parallel}{J_cB_\parallel^*} + \frac{mv_\parallel^2}{qJ_cB^*_\parallel}\pdv{b_2}{z^1} + \frac{b_2}{qJ_cB_\parallel^*}  \pdv{H}{z^1}
\label{eq:z3}
\end{equation}

and Eq.~\ref{alpha phase}
\begin{equation}
\dot v_\parallel = -\frac{C}{mJ_cB_\parallel^*}\pdv{H}{z^3} + \frac{v_\parallel}{qJ_cB_\parallel^*} \left(\pdv{b_2}{z^3}\pdv{H}{z^1} - \pdv{b_2}{z^1}\pdv{H}{z^3}\right)
\label{eq:vpar}
\end{equation}

Neglecting derivatives in $z^2$ in Eq.~\ref{poisson} by gives the axisymmetric limit of the gyrokinetic poisson equation.
After dropping the $z^3$ derivatives in Eq.~\ref{poisson}, as is typically justified because of the long parallel and short perpendicular wavelengths present in tokamaks, neglecting derivatives in $z^2$ gives the axisymmetric limit of the gyrokinetic poisson equation

\begin{equation}
\rho = -\frac{1}{J_c} \pdv{}{z^1} \Big[ J_c\epsilon_\perp g^{11} \pdv{\phi}{z^1} \Big]
\end{equation}

\section{Coordinate System}
\label{sec:coordinates}
\subsection{Coordinate Definitions}
Tokamak equilibrium magnetic fields are axisymmetric and can be written as~\citep{Cerfon2010}
\begin{equation}
\mathbf{B}=\frac{F(\psi)}{R} \mathbf{\hat e}_\phi+\frac{1}{R} \nabla \psi \times \mathbf{\hat e}_\phi
\end{equation}
where $\mathbf{\hat e}_\phi$ is a unit vector and $\mu_0$ is the vacuum permeability.
The poloidal flux $\psi$ will satisfy the Grad-Shafranov equation shown here in cylindrical $(R,Z,\phi)$ coordinates.
\begin{equation}
R \frac{\partial}{\partial R}\left(\frac{1}{R} \frac{\partial \psi}{\partial R}\right)+\frac{\partial^2 \psi}{\partial Z^2}=-\mu_0 R^2 \frac{d p}{d \psi}-F \frac{d F}{d \psi}
\label{eq:GS}
\end{equation}
where $F$ is the poloidal current and $p$ is the pressure. There are equilibrium codes such as the Python package FreeGS~\citep{freegs-docs, amorisco2024}, that solve Eq.~\ref{eq:GS} for $\psi(R,Z)$ and provide the solution in the commonly used G-EQDSK format~\cite{lao1997geqdsk}.

Given $\psi(R,Z)$ we choose to use field-aligned coordinates $(z^1,z^2,z^3) = (\psi,\alpha, \theta)$ where $\alpha$ is the field line label and $\theta$ is the poloidal projection of the length along the field line normalized to $2\pi$. We choose these coordinates such that our field can be represented in the Clebsch form  as
\begin{equation}
\boldsymbol{B} = \nabla \psi \times \nabla \alpha
\label{clebsch}
\end{equation}

Note that there are many possible choices of the parallel coordinate $\theta$ depending on the topology one wishes to represent. For example in the core of a tokamak where flux surfaces are closed, one could choose the actual poloidal angle as the parallel coordinate $\theta$. However, as discussed in~\cite{leddy2017}, in the scrape-off layer, the actual poloidal angle is not a suitable choice because typically more than one point on the same field would have the same value of $\theta$. (In the poloidal plane, a line of constant poloidal angle will intersect the same flux surface twice in the SOL). For the outer SOL of a double-null tokamak configuration, the cylindrical coordinate $Z$ could be a suitable choice but that would of course not work for the core and would restrict the divertor plates to be horizontal in the R-Z plane which is undesirable. The choice of poloidal arc length as the parallel coordinate $\theta$ which we make here is suitable for both the open and closed field line regions of a tokamak and allows for flexible divertor plate shapes. More details on the derivation of our coordinate system can be found in~\cite{MandellThesis} and other possible choices of the parallel coordinate can be found in~\cite{Jardin}. Another similar coordinate system that aims to reduce cell deformation is described in ~\cite{Riberio2010}. 

In order to have a generalized poloidal angle that sweeps out equal poloidal arc-lengths, we choose the Jacobian to be
\begin{equation}
J_c =s(\psi) \frac{R}{|\nabla \psi|}.
\label{eq:jacexpr}
\end{equation}
Note here that the Jacobian is proportional to $1/|\nabla \psi|$. This will be true, regardless of the choice of parallel coordinate, for field aligned coordinate systems. The coordinate singularity discussed earlier results from the fact that $\nabla\psi$ vanishes at X-points and O-points, which causes the Jacobian to diverge. 

With this Jacobian, the $\theta$ coordinate, parameterized in terms of the cylindrical $Z$ coordinate, is given by
\begin{equation}
\theta(R, Z)=\frac{1}{s(\psi(R, Z))} \int_{Z_{\text {lower }}(\psi)}^{Z(\psi)} \sqrt{1+\left(\frac{\partial R\left(\psi, Z^{\prime}\right)}{\partial Z^{\prime}}\right)^2} \mathrm{~d} Z^{\prime} - \pi
\label{eq:theta}
\end{equation}
where the normalization factor is 
\begin{equation}
s(\psi)=\frac{1}{2 \pi} \oint \mathrm{d} \ell_p=\frac{1}{2\pi} \int_{Z_{\text {lower }}(\psi)}^{Z_{\text {upper }}(\psi)} \sqrt{1+\left(\frac{\partial R\left(\psi, Z^{\prime}\right)}{\partial Z^{\prime}}\right)^2} \mathrm{~d} Z^{\prime}
\label{eq:normalization}
\end{equation}
Now we define the last coordinate such that Eq.~\ref{clebsch} is satisfied
\begin{equation}
\alpha(R, Z, \phi)= \phi - F(\psi) \int_{Z_{\text{lower }}(\psi)}^{Z(\psi)} \frac{1}{|\nabla \psi| R\left(\psi, Z^{\prime}\right)} \sqrt{1+\left(\frac{\partial R\left(\psi, Z^{\prime}\right)}{\partial Z^{\prime}}\right)^2} \mathrm{~d} Z^{\prime}
\label{eq:alpha}
\end{equation}
where $F(\psi) = RB_\phi$.

The choice of computational coordinates, $(z^1, z^2, z^3) = (\psi, \alpha, \theta)$, along with Eqs.~\ref{eq:normalization},~\ref{eq:theta}, and ~\ref{eq:alpha} define the mapping of computational coordinates $(\psi, \alpha, \theta)$ to physical $(R,Z,\phi)$ coordinates where $\theta \in [-\pi, \pi]$ and $\alpha \in [-\pi, \pi]$. From the mapping we can compute tangent vectors, dual vectors, and then metric coefficients, which are written explicitly in Eqs.~\ref{eq:metrics1}-~\ref{eq:metrics6}.

\begin{figure}
    \subfloat[\label{fig:dnarrows}
    Schematic for field line tracing for a double null tokamak configuration. There are 5 distinct regions: the outboard scrape-off-layer, the inboard scrape-off-layer, the lower private flux region, the upper private region, and the core. We plot one flux surface in black in each region. The tracing of the flux surface starts in each region at $Z_{lower}(\psi)$  (marked in blue) and stops at $Z_{upper}(\psi)$ (marked in red) in accordance with Eq.~\ref{eq:normalization}. The green arrows indicate the direction of the tracing in each region.]{
    \includegraphics[width=0.5\linewidth, valign=t]{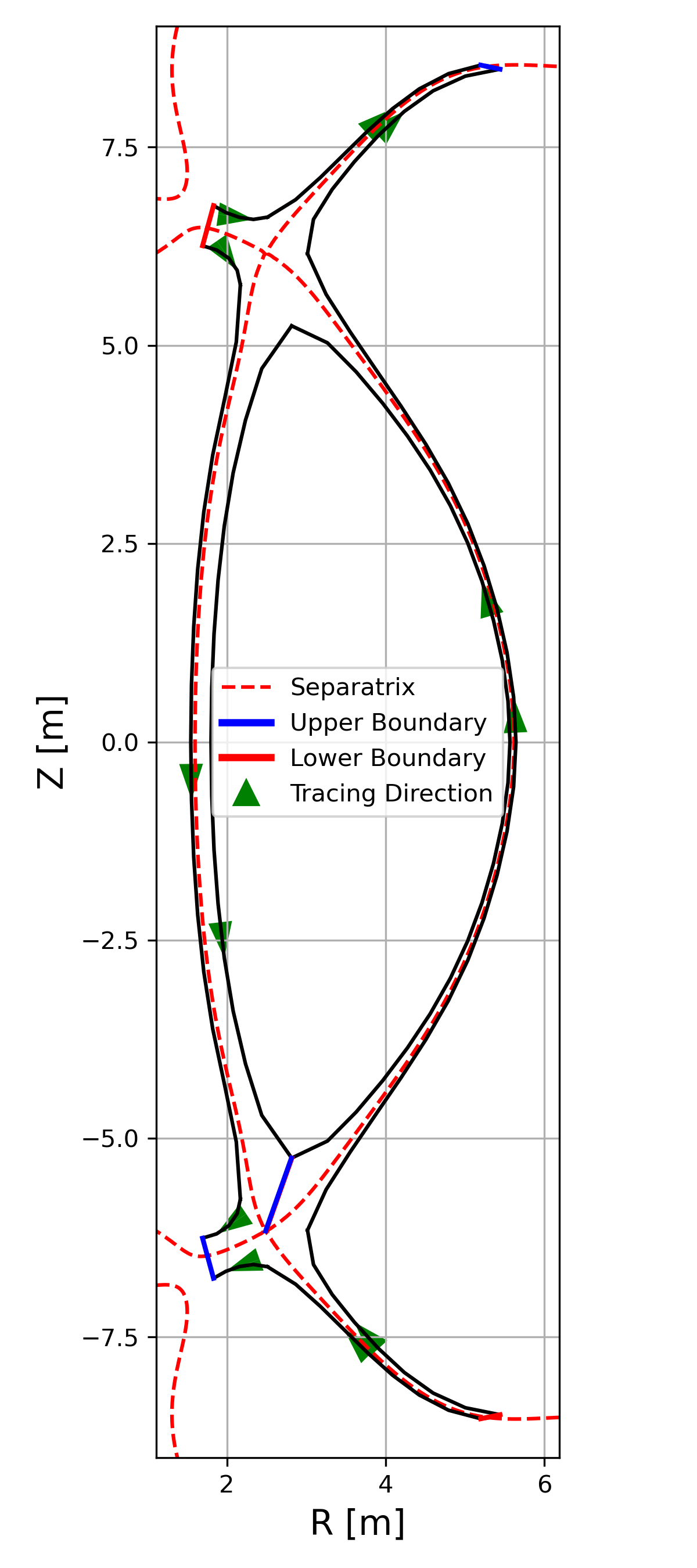}
    }
    \subfloat[\label{fig:snarrows}
    Schematic for field line tracing for a single null tokamak configuration. There are 3 distinct regions: the scrape-off-layer, the private flux region, and the core. We plot one flux surface in black in each region. The tracing of the flux surface starts in each region at $Z_{lower}(\psi)$  (marked in red) and stops at $Z_{upper}(\psi)$ (marked in blue) in accordance with Eq.~\ref{eq:normalization}. The green arrows indicate the direction of the tracing in each region.]{
    \includegraphics[width=0.5\linewidth, valign=t]{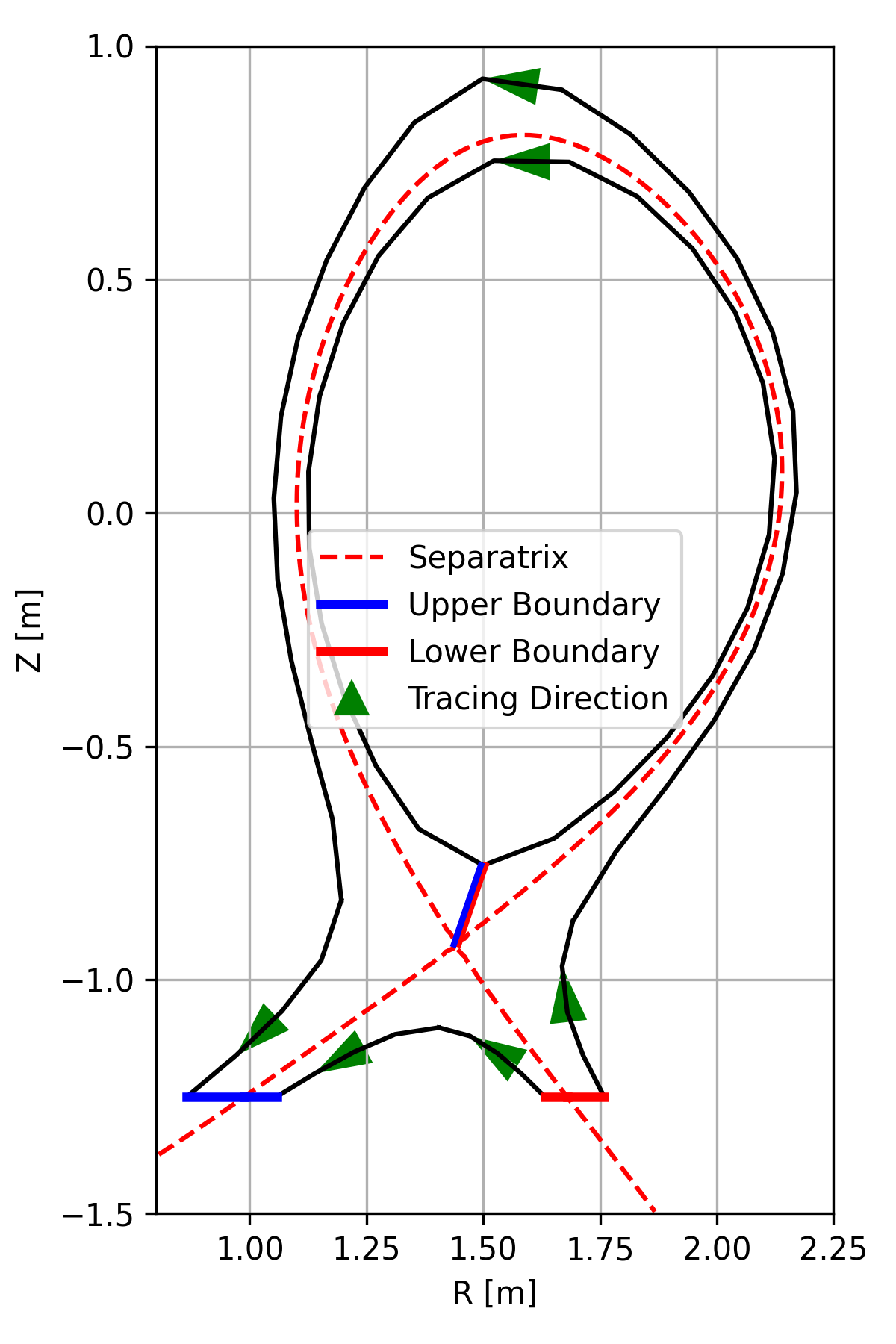}
    }
    \caption{Schematic for field line tracing in a double null (a) and single null (b) configuration.
    \label{fig:arrows}}
\end{figure}

The integrals in ~\ref{eq:theta},~\ref{eq:normalization}, and~\ref{eq:alpha} are along contours of constant $\psi$. The Z limits of the integration can be chosen based on the part of the poloidal plane one wishes to trace. For example integral in Eq.~\ref{eq:normalization} traces from divertor plate to divertor plate in the SOL but makes a complete poloidal circuit in the core. For a double null tokamak configuration there are 5 distinct topological regions: the outboard SOL, the inboard SOL, the lower private flux region, the upper private region, and the core. In Fig.~\ref{fig:dnarrows} we show how each region is traced for a double null tokamak.
For a single null tokamak configuration there are 3 distinct topological regions: the SOL, the private flux region, and the core. In Fig.~\ref{fig:snarrows} we show how each region is traced for a lower single null tokamak.

\section{Discretization of the Gyrokinetic Equation: Avoiding the X-point }
\label{sec:discretization}

The gyrokinetic equation, Eq.~\ref{eq:gk-zi} for the evolution of $F = J_v J_c f$ in the axisymmetric limit becomes
\begin{equation}
\pdv{F}{t} + \dot{z}^1\pdv{F}{z^1} + \dot{z}^3\pdv{F}{z^3} + \dot{v_\parallel}\pdv{F}{v_\parallel} = 0.
\label{eq:gk_axi}
\end{equation}

We use a Discontinuous Galerkin (DG) scheme to discretize this equation as described in ~\cite{Mana25, Ammar2019, Mandell2020}. The discrete approximation of $F$ in each cell $K_i$ is given by
\begin{equation}
F_i=\sum_{k=1}^{N_b} F_i^{(k)} \psi_i^{(k)}
\label{eq:expansion}
\end{equation}
where $\psi_i$ are the phase-space basis functions and $N_b$ is the number of basis functions.
The discrete form of Eq.~\ref{eq:gk_axi} can be obtained by projecting it onto the phase space basis $\psi_j^{(k)}$ in cell $K_j$ and integrating by parts
\begin{equation}
\begin{aligned}
& \int_{K_j} d\mathbf{z} dv_\parallel d\mu \psi_j^{(\ell)} \frac{\partial F}{\partial t}+\oint_{\partial K_j} \mathrm{~d} \mathbf{S}_{\mathrm{i}} dv_\parallel d\mu \psi_{j \pm}^{(\ell)} \dot{z}_{ \pm}^i \widehat{F}_{ \pm}+\oint_{\partial K_j} d \mathbf{z} d\mu \psi_{j \pm}^{(\ell)} \dot{v}_{\| \pm} \widehat{F_{ \pm}} \\
& \quad-\int_{K_j} d\mathbf{z} dv_\parallel d\mu\left(\frac{\partial \psi_j^{(\ell)}}{\partial z^i} \dot{z}^i+\frac{\partial \psi_j^{(\ell)}}{\partial v_\parallel} \dot{v}_{\|}\right) F=0 .
\end{aligned}
\end{equation}
where $d\mathbf S_i$ is the surface element perpendicular to the i-th direction and $\widehat{F_\pm}$ is the upwind flux evaluated at the upper and lower edge of the cell in direction i. 

Substituting in the expansion of $F$ in the first term of Eq.~\ref{eq:expansion} and making use of the orthonormality relation $\int_{K_j} d\mathbf z dv_\parallel d\mu \psi_j^{(\ell)}\psi_j^{(k)} = \delta_{lk}$, we get the time evolution of each expansion coefficient of $F$
\begin{equation}
\begin{aligned}
 &\frac{\partial F_j^{(\ell)}}{\partial t}+\oint_{\partial K_j} \mathrm{~d} \mathbf{S}_{\mathrm{i}} dv_\parallel d\mu \psi_{j \pm}^{(\ell)} \dot{z}_{ \pm}^i \widehat{F}_{ \pm}+\oint_{\partial K_j} d \mathbf{z} d\mu \psi_{j \pm}^{(\ell)} \dot{v}_{\| \pm} \widehat{F_{ \pm}} \\
 &\quad-\int_{K_j} d\mathbf{z} dv_\parallel d\mu\left(\frac{\partial \psi_j^{(\ell)}}{\partial z^i} \dot{z}^i+\frac{\partial \psi_j^{(\ell)}}{\partial v_\parallel} \dot{v}_{\|}\right) F=0 .
 \label{eq:evolution}
\end{aligned}
\end{equation}

We evaluate the integrals in Eq.~\ref{eq:evolution} analytically using DG expansions of the characteristics  $\dot z^i$ and $\dot v_\parallel$ on the phase basis in the volume term (the fourth term) and DG expansions of the fluxes ($\dot{z}^i\hat{F}_\pm$ and $\dot{v_\parallel}\hat{F}_\pm$) in the second and third terms (the surface terms). In the last term (the volume term) the expansion of the characteristic velocities ($\dot{z}^i$ and $\dot{v_\parallel}$) are constructed by evaluating the characteristics at interior Gauss-Legendre quadrature points and converting to a modal representation. In the second and third terms, the expansion of the fluxes are calculated by evaluating the flux at surface Gauss-Legendre quadrature points and converting to a modal representation. 
Labeling the characteristics and fluxes based on whether they are calculated by evaluation at interior or surface quadrature points with a subscripts $int$ and $surf$ respectively, we can rewrite Eq.~\ref{eq:evolution} as

\begin{equation}
\begin{aligned}
 &\frac{\partial F_j^{(\ell)}}{\partial t}+\oint_{\partial K_j} \mathrm{~d} \mathbf{S}_{\mathrm{i}} dv_\parallel d\mu \psi_{j \pm}^{(\ell)} (\dot{z}^i \widehat{F})_{\pm , surf} +\oint_{\partial K_j} d \mathbf{z} d\mu \psi_{j \pm}^{(\ell)} (\dot{v}_{\|} \widehat{F})_{\pm , surf} \\
 &\quad-\int_{K_j} d\mathbf{z} dv_\parallel d\mu\left(\frac{\partial \psi_j^{(\ell)}}{\partial z^i} \dot{z}^i_{int}+\frac{\partial \psi_j^{(\ell)}}{\partial v_\parallel} \dot{v}_{\|,int}\right) F=0 .
 \label{eq:evolution_quad}
\end{aligned}
\end{equation}

\begin{figure}
    \centering
    \subfloat[
    Gauss-Legendre quadrature points on the surface (blue and green) and interior (red) points of a computational cell along with cell corners (orange). 
    \label{fig:quad_points}
    ]{
    \includegraphics[width=0.45\linewidth]{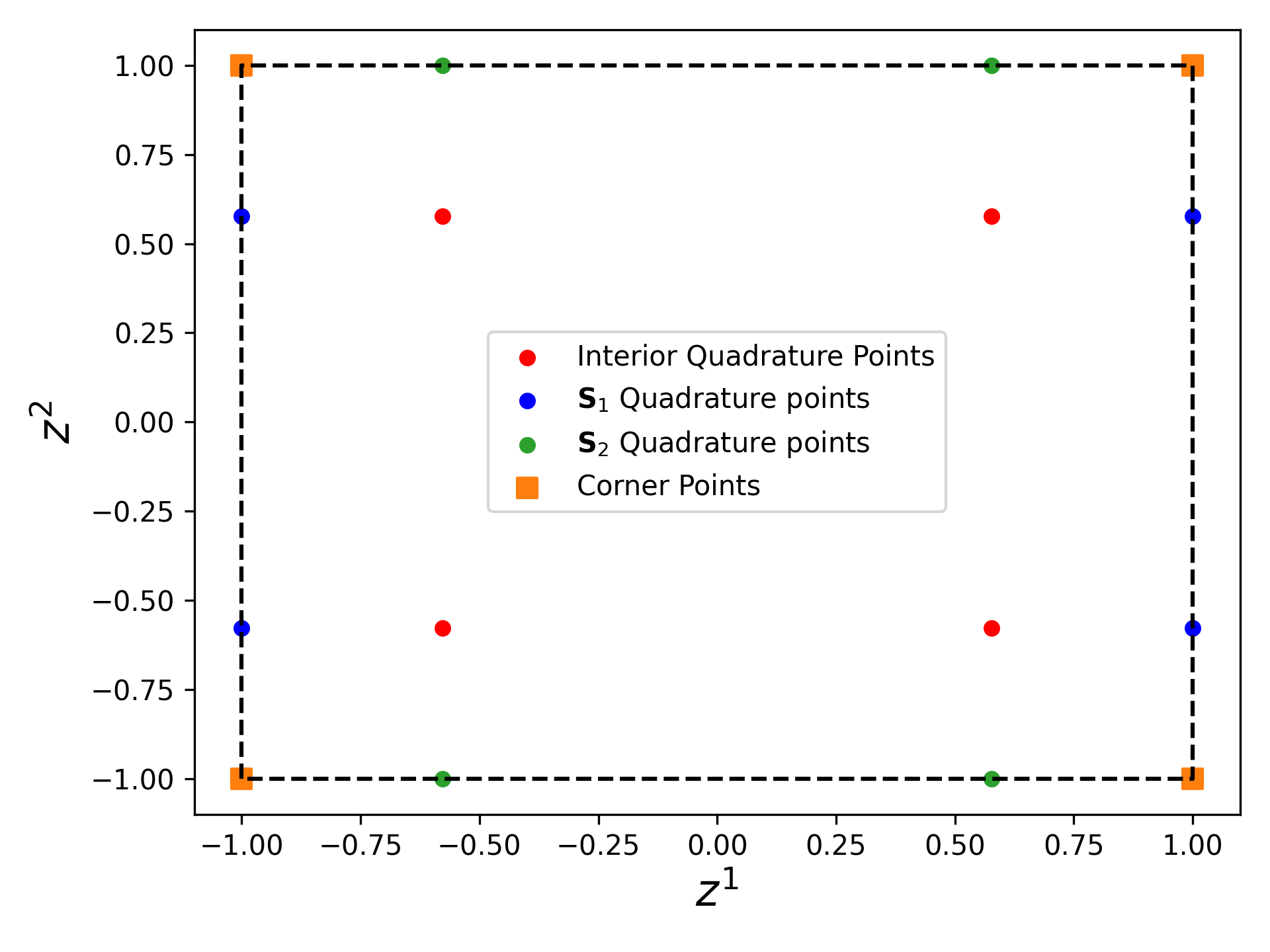}
    }\hspace{0.2cm}
    \subfloat[
    Gauss-Legendre quadrature points on the surface (blue and green) and interior (red) points along with cell corners (orange) mapped to physical cells abutting the X-point. 
    \label{fig:xpt_quad_points}
    ]{
    \includegraphics[width=0.45\linewidth]{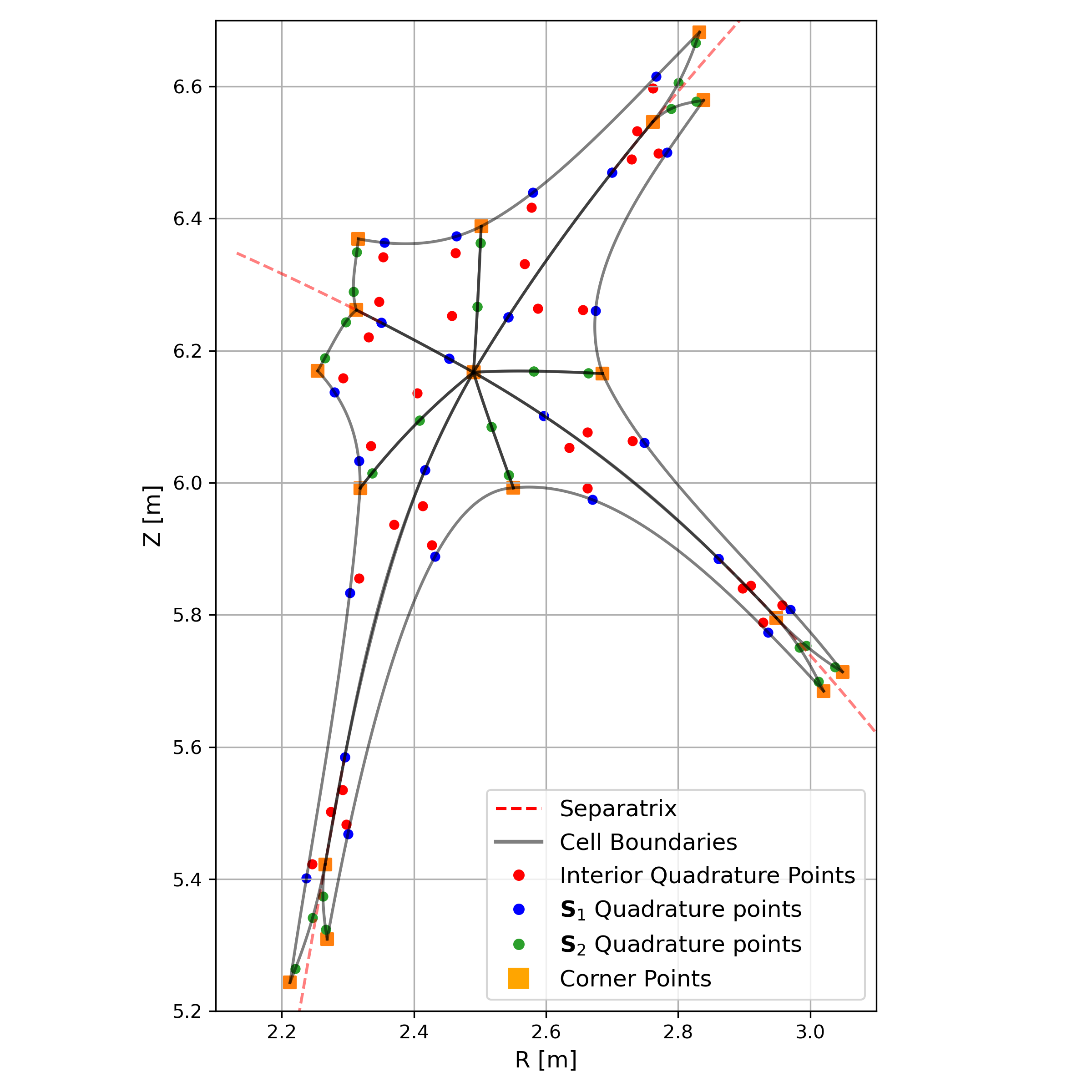}
    }
    \caption[In (a) we show the interior, surface, and corner points on the unit cell]{In (a) we show the interior, surface, and corner points on the unit cell. In (b) we show these points mapped to the physical domain for cells abutting the X-point. The cell in physical space is \emph{not rectangular}, allowing for an accurate representation of the flux-surface geometry.  The surface and interior nodes used for the evaluation of geometric quantities do not lie directly on the X-point and are thus well defined. 
    \label{fig:xpt_quad_points_mapped}
    }
\end{figure}

An example of the quadrature points used in 2D is depicted in Fig.~\ref{fig:quad_points}. For example to construct the volume representation $\dot{z^i_{int}}$ in this cell, we evaluate $\dot z^i$ at the 4 red points and convert to a modal representation. To calculate the surface representation $(\dot z^1\hat{F})_{+,surf}$ at the upper $z^1$ edge of this cell we would evaluate $\dot z^1\hat{F}$ at the two blue points at $z^1=1$ and convert to a modal representation.
The use of an orthonormal, modal representation for the DG fields allows us to significantly reduce the computational cost of DG~\citep{HakimandJuno2020} while respecting the need to eliminate aliasing errors in DG discretizations of kinetic equations~\citep{Juno18}.

This method of evaluating the characteristics is the key feature of our algorithm that allows us to simulate magnetic geometries with an X-point as shown in Fig.~\ref{fig:xpt_quad_points_mapped}. The geometric quantities such as the Jacobian, $J_c$, contained in the characteristics written in Eqs.~\ref{eq:z1},~\ref{eq:z3}, and~\ref{eq:vpar} diverge at the X-point as mentioned below Eq.~\ref{eq:jacexpr}. However, since we evaluate the characteristics at either interior quadrature points or surface quadrature points and not corner points, we can avoid evaluating any geometric quantities at the X-point as long as cell corners lie at the X-point, which our multi-block grid generation routine ensures. 
The gyrokinetic poisson equation, Eq.~\ref{poisson}, also benefits from the distinction between corner and interior evaluations. Our solution, described in detail in ~\cite{Mana25}, makes use of the interior geometric quantities to avoid the coordinate singularity. 
Note that while our method allows including the X-point in the domain, the effect of the coordinate singularity is not completely avoided; it affects convergence as described in section~\ref{sec:tests} and previously observed in~\cite{WIESENBERGER2018}.

At block boundaries, one must be careful to ensure consistency; the surface normal vectors must be consistent at block boundaries
For example at a radial block boundary (the $z^1$ direction), the surface normal $\mathbf{\hat{e}^1} = \mathbf{e^1}/|\mathbf{e_1}|$ must match. Our method for calculating the tangent vectors and metric coefficients described in section~\ref{sec:metrics} ensures that the  surface nodes used to calculate fluxes across block boundaries are common and that the normal vectors are consistent at either side of the boundaries.
Because our algorithm evolves $F=J_vJ_cf$ rather than $f$, we must also be careful to correctly handle the change in the normalization of $J_c$ (Eq.~\ref{eq:normalization}) at radial block boundaries. Blocks sharing a parallel boundary share the same normalization, so there is a change only at radial block boundaries. To address this, the 2nd term in eq.~\ref{eq:evolution_quad} must be modified at radial block boundaries. The flux $(\dot{z}^i\hat{F})_{\pm, surf}$ is divided by $J_c$ of the block the flux is leaving and multiplied by $J_c$ of the block it is entering. The values of $J_c$ used for the division and multiplication are the value of $J_c$ at the surface quadrature point.

If the normal vectors are not consistent at block boundaries or at cell boundaries, particle conservation will broken~\cite{Ammar2019}. In section~\ref{sec:example} we conduct a test of particle conservation which demonstrates consistency of the normal vectors and the correct handling of the re-normalization at block boundaries.
\section{Grid Generation and Geometric Quantities}
\label{sec:computation}

In order to conduct simulations, we need to generate a physical simulation grid and then calculate all of the geometric quantities appearing in the equations of motion on that grid. All of the geometric quantities required can be extracted from two basic quantities: the magnitude of the magnetic field $B(\psi, \alpha, \theta)$ and the tangent vectors.

\subsection{Representation of Magnetic Field}
The starting point for our grid generation is a tokamak equilibirum provided by the commonly used G-EQDSK format~\citep{lao1997geqdsk}. The G-EQDSK format provides $\psi(R,Z)$ on an $N_R \times N_Z$ grid and from that we can construct a DG expansion of $\psi(R,Z)$ on the same grid. 
G-EQDSK files also give the toroidal magnetic field by providing $F(\psi) = RB_\phi$ on a grid of length $N$ from which we can construct a DG expansion of $F(\psi)$. 
We use either a biquadratric or bicubic representation of $\psi(R,Z)$ for the field line tracing described in section~\ref{sec:exact_mapping} and for calculating the magnitude of the magnetic field at each grid point. The biquadratic representation offers a speedup over the bicubic representation in the grid generation process because it enables a simple and fast root finding procedure. The magnetic field components and magnitude can be calculated from $\psi(R,Z)$ and $F(\psi)$ in cylindrical coordinates as
\begin{subequations}
\begin{align}
B_R & =\frac{1}{R} \frac{\partial \psi}{\partial Z}=\frac{\partial}{\partial Z}\left(\frac{\psi}{R}\right) \label{eq:BR}\\
B_Z & = -\frac{1}{R} \frac{\partial \psi}{\partial R}=-\frac{\partial}{\partial R}\left(\frac{\psi}{R}\right)-\frac{\psi}{R^2} \label{eq:BZ}\\
B_\phi &= \frac{F(\psi)}{R}\\
B &= \|\mathbf B \| = \sqrt{B_R^2+B_Z^2+B_\phi^2}.
\end{align}
\end{subequations}

The poloidal magnetic field is $\mathbf B_{pol} = B_R \hat{\mathbf R} + B_Z\hat{\mathbf Z}$. Looking at Eq.~\ref{eq:jacexpr} and Eqs.~\ref{eq:BR} and~\ref{eq:BZ}, one can now see the connection between a vanishing poloidal field and the coordinate singularity at the X-point; when $\mathbf B_{pol}$ vanishes, $|\nabla\psi| = 0$, and the Jacobian, $J_c$, diverges.

\subsection{Grid Generation Algorithm}
\label{sec:exact_mapping}

We use a rectangular computational grid with extents $(L_\psi, L_\alpha, L_\theta)$ and number of cells $(N_\psi, N_\alpha, N_\theta)$. The grid spacing is $(\Delta\psi, \Delta\alpha, \Delta\theta) = (L_\psi/N_\psi, L_\alpha/N_\alpha, L_\theta/N_\theta)$. This computational grid has $(N_\psi+1)(N_\alpha+1)(N_\theta+1)$ nodes. 

In order to lay out a physical grid for our simulation and to calculate the geometric factors appearing in Eq.~\ref{alpha config} and Eq.~\ref{alpha phase} we calculate mapping $\mathbf x(\psi, \alpha, \theta)$ at each point on our computational grid. For each point, $(\psi_0, \alpha_0, \theta_0)$, on our grid, we calculate the mapping using the following algorithm:

\begin{itemize}
    \item {\bf Step 1:} Pick an initial $Z$ and find $R$ such that $\psi(R,Z) = \psi_0$. In practice this is done by inverting our piecewise polynomial representation of $\psi(R,Z)$ to get a polynomial $R(\psi,Z)$.
    \item {\bf Step 2:} Calculate $\theta(R,Z)$ using Eq.~\ref{eq:theta}. The integral is done with a double exponential method~\citep{Bailey} and will require doing Step 1 and evaluating the derivative of the polynomial $R(\psi,Z)$ at each quadrature point to remain on the flux surface.
    \item {\bf Step 3:} Repeat Steps 1 and 2 choosing $Z$ using a root-finder (we use Ridders method~\citep{Ridders}) until we find $R$ and $Z$ such that $\theta(R,Z) = \theta_0$.
    \item {\bf Step 4:} Calculate $\phi$ using Eq.~\ref{eq:alpha}.
    \item {\bf Step 5:} Calculate the Cartesian coordinates from the cylindrical coordinates: $X = R\cos\phi, Y=R\sin\phi$, $Z=Z$.
\end{itemize}

The method described here requires the inversion of $\psi(R,Z)$. Although this inversion is simple with our biquadratic representation of $\psi$, an alternative employ streamline integration~\citep{WIESENBERGER2017} could be used in the future to avoid the inversion and root finding.~\cite{WIESENBERGER2018} describes how to use streamline integration in domains including an X-point.

\subsection{Multi-Block Grids}
\label{sec:grids}
\begin{figure}
    \vspace{-3cm}
    \subfloat[\label{fig:stepgridfull}]{
    \includegraphics[width=0.55\textwidth, valign=t]{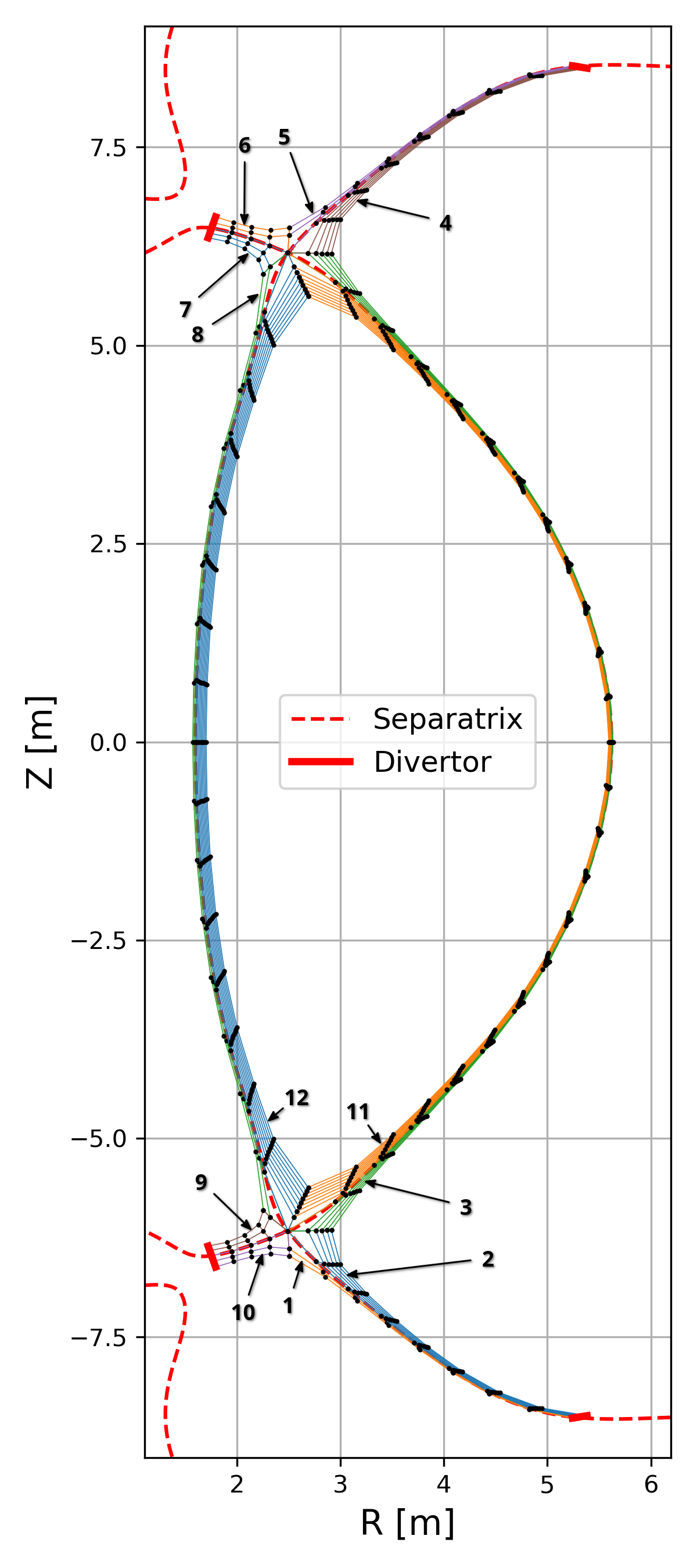}
    }
    \subfloat[\label{fig:stepgridzoom}]{
    \includegraphics[width=0.45\textwidth, valign=t]{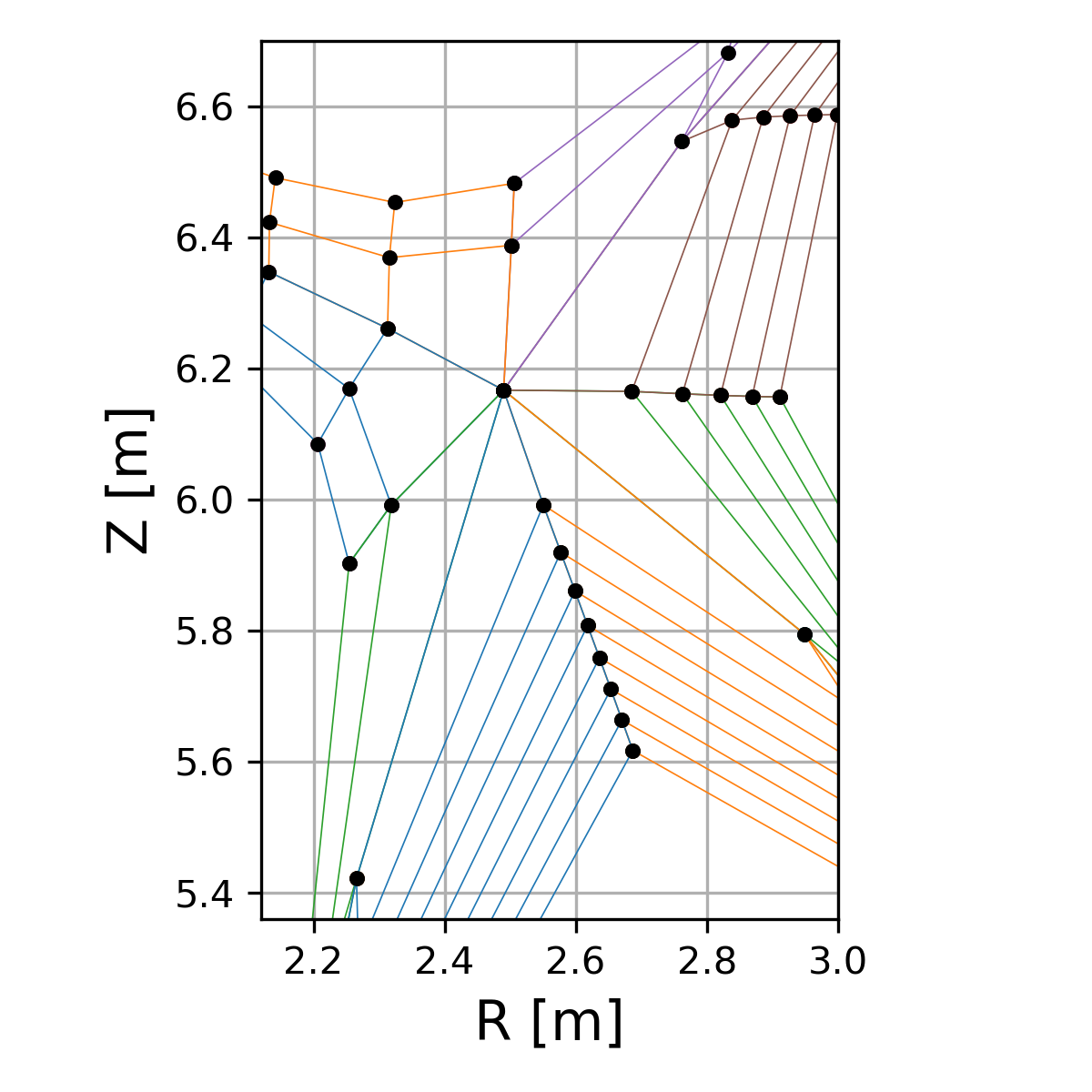}
    }\\
    \begin{flushright}
    \vspace{-10cm}
    \subfloat[\label{fig:stepgridzoomdivertor}]{ 
    \includegraphics[width=0.4\textwidth]{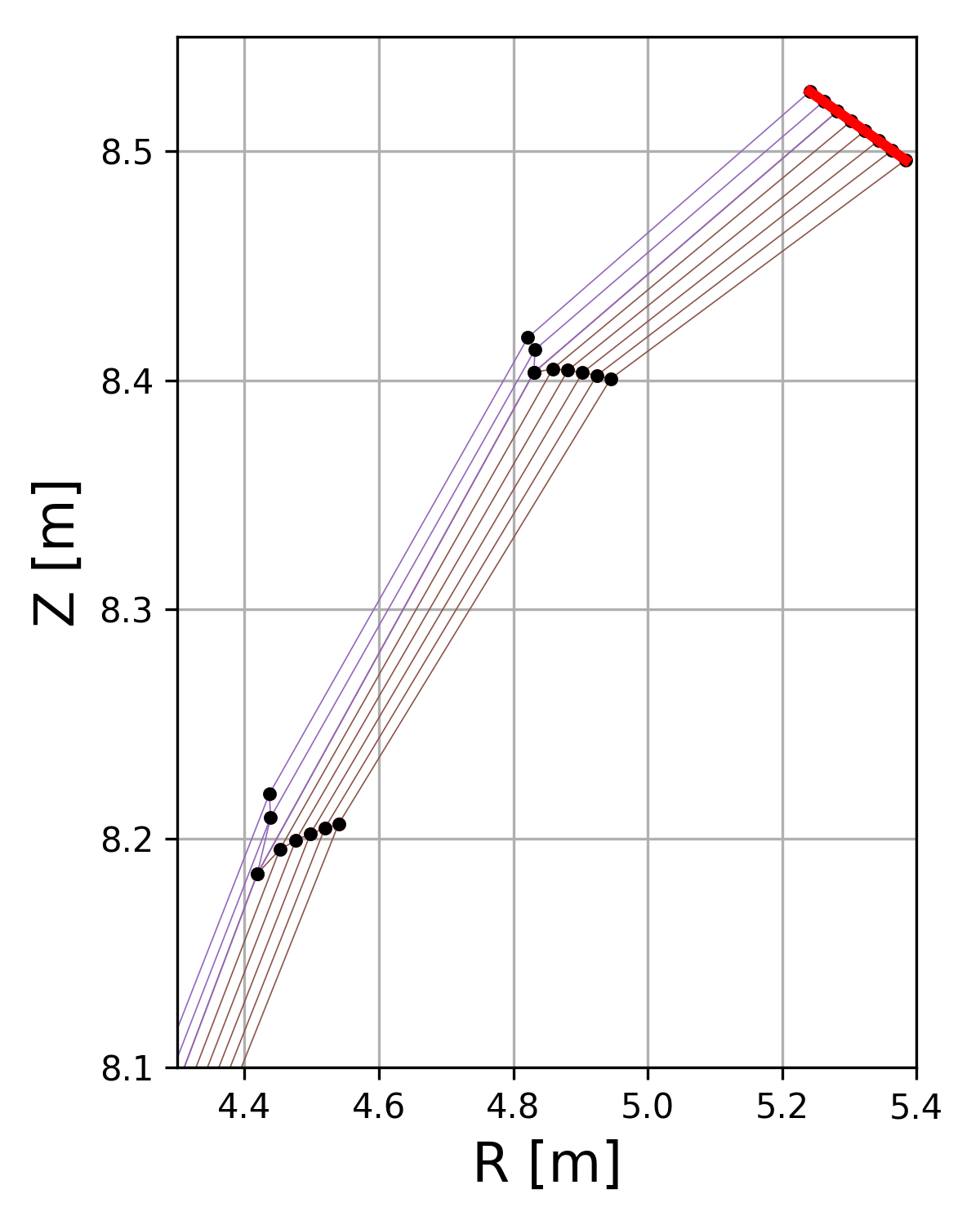}
    }
    \vspace{1cm}
    \end{flushright}
    \caption[Block layout and grid for the Spherical Tokamak for Energy Production]{Block layout and grid for the Spherical Tokamak for Energy Production in a double null configuration with different colors indicating different blocks and a number 1-12 labeling each block. The full grid is shown in (a), (b) shows a close-up of the grid near the upper X-point, and (c) shows a close-up of the grid near the upper outer divertor plate (red). 
    \label{fig:stepgrid} }
\end{figure}

\begin{figure}
    \subfloat[\label{fig:asdexgridfull}]{
    \includegraphics[width=0.45\textwidth, valign=t]{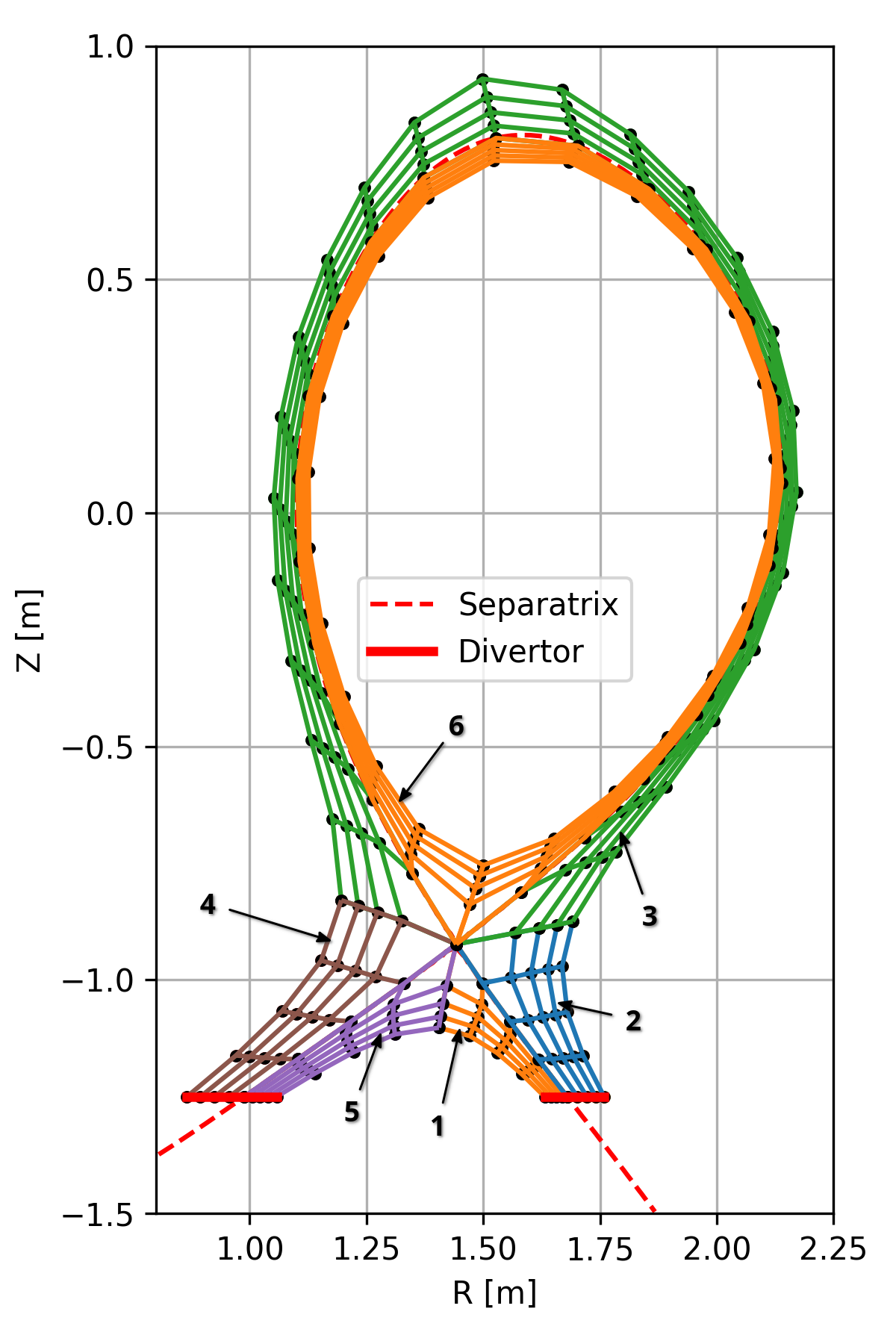}
    }
    \subfloat[\label{fig:asdexgridzoom}]{
    \includegraphics[width=0.45\textwidth, valign=t]{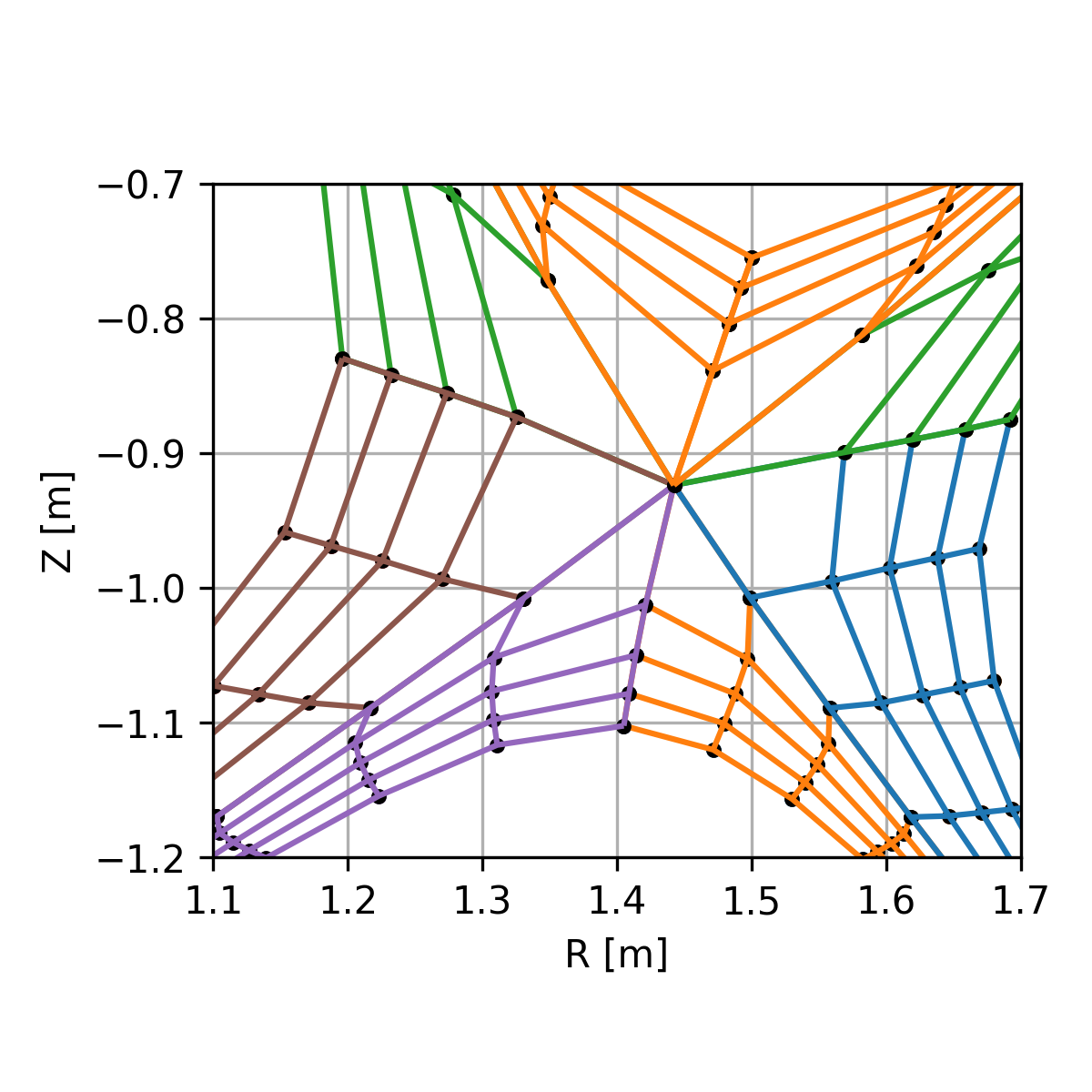}
    }
    \caption[Grid for ASDEX-Upgrade in a single null configuration]{Grid for ASDEX-Upgrade in a single null configuration with different colors indicating different blocks and a number 1-6 labeling each block. The full grid is shown in (a) and (b) shows a close-up of the grid near the X-point. 
    \label{fig:asdexgrid} }
\end{figure}

To enable simulations of domains including the core, private flux, and SOL, we break the domain up into blocks. We first break the domain up into the distinct topological regions (5 for double null and 3 for single null) described in Sec.~\ref{sec:coordinates} and then split each region at the X-point. As shown in Fig.~\ref{fig:stepgridfull}, a double null tokamak has 12 blocks each of which has one edge along the separatrix and at least one corner at the X-point. Fig.~\ref{fig:asdexgridfull} shows a lower single null tokamak with 6 blocks.

Once the domain has been split into blocks, we can generate a uniform computational grid within each block. In Fig.~\ref{fig:stepgrid} we show the multi-block grid generated for a double null configuration of STEP, and in Fig.~\ref{fig:asdexgrid} we show the multi-block grid generated for ASDEX-Upgrade~\citep{Stroth2022} in a lower single null configuration. The input files used to generate these grids can be found at \url{https://github.com/ammarhakim/gkyl-paper-inp/tree/master/2025_JPP_Xpt}.

\subsection{Metric Coefficients}
\label{sec:metrics}
One we have generated mapping from computational to physical coordinates at all of the grid nodes, we can calculate the metric coefficients associated with this coordinate transformation. 

Using the definitions of transformation from cartesian coordinates $(x,y,z)$ to cylindrical coordinates $(R, Z, \phi)$ 
\begin{align}
x &= R(\psi, \theta)\cos\phi(\psi, \theta, \alpha)\\
y &= R(\psi, \theta)\sin\phi(\psi, \theta, \alpha)\\
z &= Z(\psi, \theta)
\end{align}
we can express the metric coefficients, $g_{ij}$, in terms of the derivatives of the cylindrical coordinates $(R, Z, \phi)$ with respect to the computation coordinates $(\psi, \alpha, \theta)$ as follows
\begin{align}
g_{11} &= \Big(\pdv{R}{\psi}\Big)^2 + R^2\Big(\pdv{\phi}{\psi}\Big)^2 + \Big(\pdv{Z}{\psi}\Big)^2 \\
\label{eq:metrics1}
g_{12} &= R^2 \pdv{\phi}{\psi} \\
g_{13} &= \pdv{R}{\psi}\pdv{R}{\theta} + R^2\pdv{\phi}{\psi}\pdv{\phi}{\theta} + \pdv{Z}{\psi}\pdv{Z}{\theta} \\
g_{22} &= R^2 \\
g_{23}  &= R^2\pdv{\phi}{\theta} \\
g_{33} &= \Big(\pdv{R}{\theta}\Big)^2 + R^2\Big(\pdv{\phi}{\theta}\Big)^2 + \Big(\pdv{Z}{\theta}\Big)^2 .
\label{eq:metrics6}
\end{align}




The derivatives $(R,Z,\phi)$ with respect to the computational coordinate $\theta$ can be calculated directly from our representation of $\psi(R,Z)$ and the derivatives with respect to $\alpha$ are trivial:

\begin{align}
\pdv{R}{\theta} &= \sin\Big[\arctan(\pdv{R}{Z})\Big] s(\psi)\\
\pdv{Z}{\theta} &= \cos\Big[\arctan(\pdv{R}{Z})\Big] s(\psi)\\
\pdv{\phi}{\theta} &= \frac{F(\psi)}{R|\grad\psi|} s(\psi) \\
\pdv{R}{\alpha} &= 0 \\
\pdv{Z}{\alpha} &= 0 \\
\pdv{\phi}{\alpha} &= 1
\label{exact_derivs}
\end{align}
The derivative $\pdv{R}{Z}$ appearing in the derivatives with respect to $\theta$ can be calculated easily from our biquadratic representation of $\psi(R,Z)$. To calculate the remaining 3 derivatives with respect to $\psi$, $\pdv{R}{\psi}$, $\pdv{Z}{\psi}$, and $\pdv{\phi}{\psi}$, we use second order finite differences.



%

\section{Convergence and Consistency Tests}
\label{sec:tests}
In previous work on grids with X-points, it has been observed that the effects of the X-point singularity cannot be avoided entirely by avoiding evaluation at the X-point~\citep{WIESENBERGER2018}. Here we conduct several convergence checks and an example simulation including a consistency test. In section~\ref{sec:volume} we test the convergence of the enclosed volume, in section~\ref{sec:poissonconvergence} we test the convergence of the poisson solver, and in section~\ref{sec:advection} we test convergence for the advection of a gaussian bump.
In section~\ref{sec:example} we conduct an example gyrokinetic simulation and demonstrate geometric consistency at multi-block boundaries by showing that our simulation conserves particles to machine precision.
In line with previous work, we see a reduced order of convergence for dynamics in the vicinity of the X-point. The convergence of our second order scheme is not completely destroyed or reduced to 1, but we observe an order of ~1.5.
The input files used for the convergence tests and simulations described in this section can be found at \url{https://github.com/ammarhakim/gkyl-paper-inp/tree/master/2025_JPP_Xpt}.

\subsection{Enclosed Volume}
\label{sec:volume}
\begin{figure}
    \subfloat[\label{fig:volaway}]{
    \includegraphics[width=0.45\textwidth, valign=t]{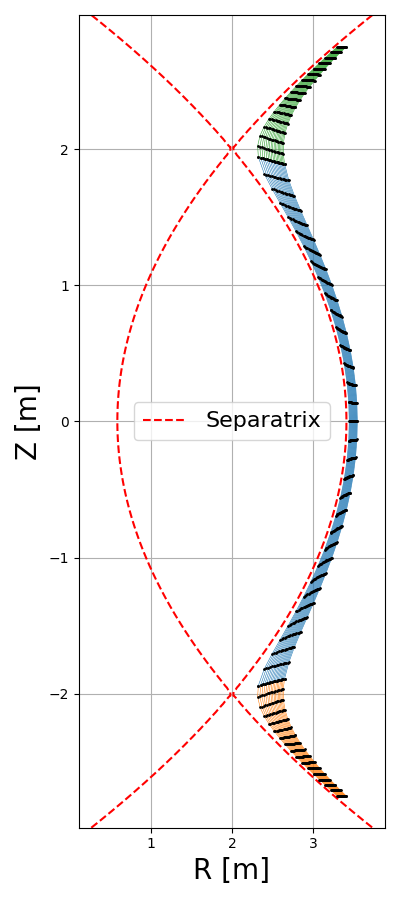}
    }
    \subfloat[\label{fig:volon}]{
    \includegraphics[width=0.45\textwidth, valign=t]{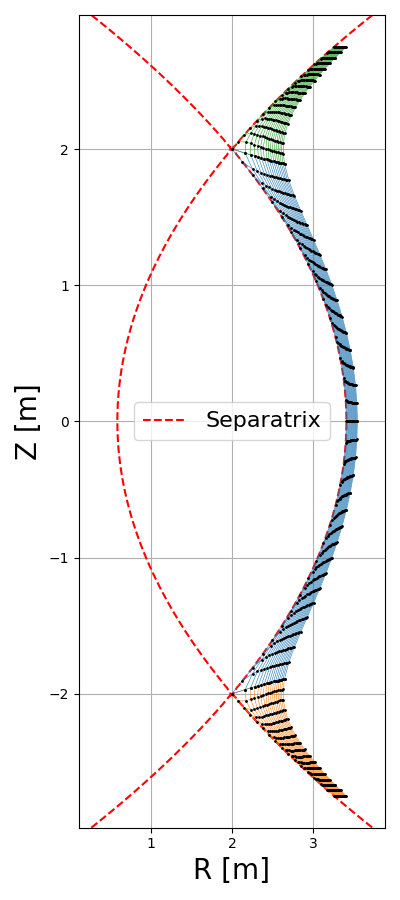}
    }
    \caption[Analytical bump solutions]{Analytical bump solutions away from the X-point (a) and on the X-point (b) for the potential, $\phi$.
    \label{fig:vol} }
\end{figure}

As a first step, we check the order at which the enclosed volume of a multi-block grid not containing the X-point converges. As a second step we check the order at which the enclosed volume of a similar multi-block grid including the X-point converges. For these tests we use an analytical double null equilibrium with $\psi(R,Z) = (R-2)^2 + Z^2 - Z^4/8$. The grids without and with the X-point are shown in Figures~\ref{fig:volaway} and~\ref{fig:volon} respectively.

 We define the relative error for the volume as $\mathcal{E} = |V_{grid} - V_{ref}|/V_{ref}$ where $V_{grid}$ is the volume of our grid and $V_{ref}$ is the reference enclosed volume. We calculate the reference volume between the two flux surfaces (labeled with subscripts outer and inner) in python as
\begin{equation}
    V_{ref}=\pi \int_{-2.75}^{2.75}\Big[ (\psi_{outer}-\psi_{inner} + 4\Big(\sqrt{\psi_{outer} - Z^2 + Z^4/8} - \sqrt{\psi_{inner} - Z^2 + Z^4/8} \Big)\Big]dZ
\end{equation}
using an adaptive double quadrature scheme (scipy.integrate.dblquad). The reference volume has some error, but this is much smaller than the difference between $V_{grid}$ and $V_{exact}$.  We define the order of convergence between two successive tests with increasing resolution as 
\begin{equation}
\mathcal{O} = -\frac{\ln(\mathcal{E}_{2}/\mathcal{E}_{1})}
                    {\ln(\sqrt{(N_{\psi,2}N_{\theta,2}) / (N_{\psi,1}N_{\theta,1})})}
\end{equation}
where the subscript 1 indicates the lower resolution test and the subscript 2 indicates the higher resolution test.

The numerical results of the convergence test are shown in table~\ref{tab:volorder}. We observe an average convergence order of 3.31 for the grid without the X-point and a reduced convergence order of 1.42 when the X-point is included in the domain. The presence of the X-point reduces the convergence order of the enclosed volume, which we can attribute to the diverging Jacobian. 

\begin{table}
\begin{center}
\subfloat[Convergence table for the volume of the grid away from the X-point. \label{tab:volaway}]{
\begin{tabular}{lcccc} 
$N_\psi$ & $N_\theta$ & Relative Error & Order \\
 1  & 4    & 8.109e-03&      \\
 2  & 8    & 5.428e-03&  0.58\\
 4  & 16   & 1.671e-04&  5.02\\
 8  & 32   & 3.734e-05&  2.16\\
 16 & 64   & 8.410e-07&  5.47\\
\end{tabular}
} \quad\quad\quad\quad
\subfloat[Convergence table for the volume of the grid touching the X-point. \label{tab:volon}]{
\begin{tabular}{lcccc} 
$N_\psi$ & $N_\theta$ & Relative Error & Order \\
1  & 4    & 4.20e-02&      \\
2  & 8    & 9.22e-03&  2.19\\
4  & 16   & 3.51e-03&  1.39\\
8  & 32   & 1.62e-03&  1.11\\
16 & 64   & 8.28e-04&  0.97\\
\end{tabular}
}
\caption[Relative error for enclosed volume of the double null]{Relative error for enclosed volume of the double null outer SOL grid away from the X-point (a) and touching the the X-point (b). The average order of convergence without the X-point is 3.37 and with the X-point is 1.42. \label{tab:volorder}}
\end{center}
\end{table}

\subsection{Poisson Solve}
\label{sec:poissonconvergence}
In this section we used the method of manufactured solutions to test the convergence of the gyrokinetic poisson equation, Eq.~\ref{poisson}, on our multi-block grids. For these tests we use an analytical single null equilibrium with  $\psi(R,Z) = (R-2)^2 + Z^2 + Z^3/3$. We conduct tests similar to the tests conducted in~\cite{WIESENBERGER2018}. We use a function of the following form 
\begin{equation}
\phi(R, Z)= \begin{cases}e^{1+\left(\frac{\left(R-R_0\right)^2}{\sigma_R^2}+\frac{\left(Z-Z_0\right)^2}{\sigma_Z^2}-1\right)^{-1}} & \text { for }\left(R-R_0\right)^2/\sigma_R^2+\left(Z-Z_0\right)^2/\sigma_Z^2< 1 \\ 0 & \text { else }\end{cases}
\label{eq:phibump}
\end{equation}
for the potential $\phi$. Note that we do not employ the flute approximation for this test because the potential we have chosen to test is not elongated in the direction of the magnetic field. We insert Eq.~\ref{eq:phibump} into Eq.~\ref{poisson} to compute the corresponding charge density analytically in cylindrical coordinates and project this charge density onto the grid. We then solve Eq.~\ref{poisson} on our multi-block grid with homogeneous Dirichlet boundary conditions to get a numerical solution $\phi_{num}$. Finally, we project the analytical solution onto the grid to get $\phi_{ref}$ and compute the relative error as
\begin{equation}
    \mathcal{E} = \Big(\frac{\int J_c d\psi d\theta (\phi_{num}-\phi_{ref})^2} {\int J_c d\psi d\theta \phi_{ref}^2} \Big)^{1/2}.
    \label{eq:poissonerror}
\end{equation}

\begin{figure}
    \subfloat[\label{fig:phianaaway}]{
    \includegraphics[width=0.45\textwidth, valign=t]{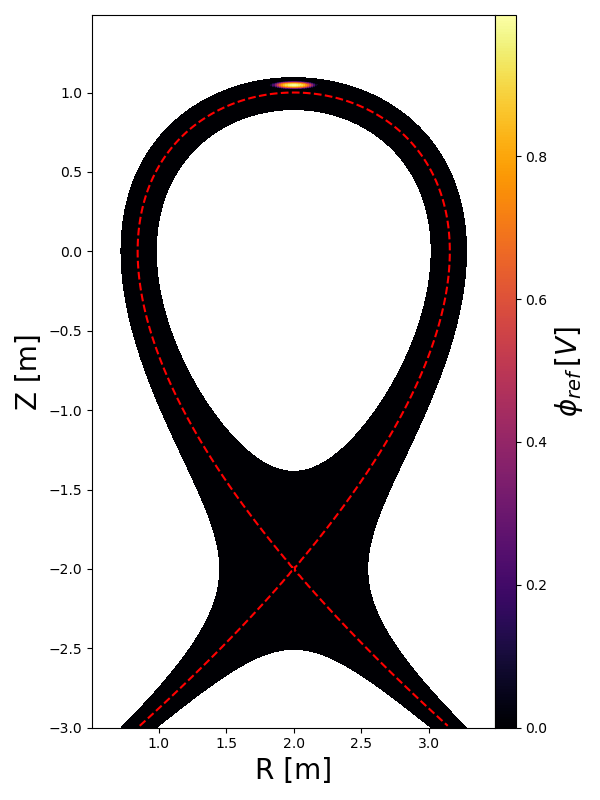}
    }
    \subfloat[\label{fig:phianaon}]{
    \includegraphics[width=0.45\textwidth, valign=t]{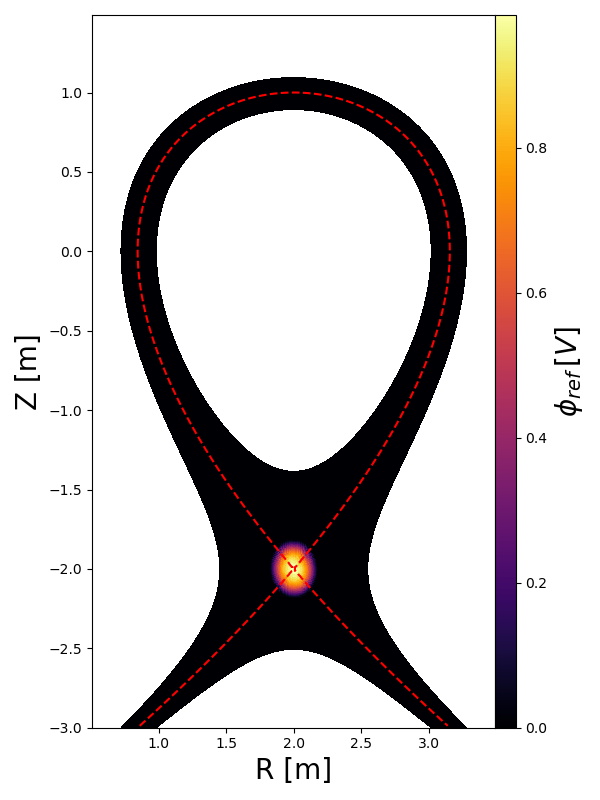}
    }
    \caption[Projections of the analytical bump solutions]{Projections of the analytical bump solutions away from the X-point (a) and on the X-point (b) for the potential, $\phi$.
    \label{fig:phiana} }
\end{figure}

We conducted one test with the potential far away from the X-point with parameters $R_0=2m, Z_0=1.05m, \sigma_R=0.2m$, and $\sigma_Z = 0.03m$. The projection of the analytical solution for this test is shown in Fig.~\ref{fig:phianaaway} and the numerical results are shown in table~\ref{tab:orderaway}. We conducted another test with the potential centered on the X-point with parameters $R_0=2m, Z_0=-2m, \sigma_R=0.2m$, and $\sigma_Z = 0.2m$. The projection of the analytical solution for this test is shown in Fig.~\ref{fig:phianaon} and the numerical results are shown in table~\ref{tab:orderon}.

\begin{table}
\begin{center}
\subfloat[Convergence table for the bump solution away from the X-point. \label{tab:orderaway}]{
\begin{tabular}{lcccc} 
$N_\psi$ & $N_\theta$ & Relative Error & Order \\
8 & 24    & 0.678&    \\
16 & 48   & 0.227&   1.58 \\
32 & 96   & 0.150&   0.60 \\
48 & 144  & 0.065&  2.05 \\
64 & 192  & 0.041&  1.59 \\
80 & 240  & 0.034&  0.88 \\
120 & 360 & 0.016&  1.84 
\end{tabular}
} \quad\quad
\subfloat[Convergence table for the bump solution on top of the X-point. \label{tab:orderon}]{
\begin{tabular}{lcccc} 
$N_\psi$ & $N_\theta$ & Relative Error & Order \\
8 & 24    & 14.596&   \\
16 & 48   & 1.312&  3.48 \\
32 & 96   & 1.016&  0.370 \\
48 & 144  & 0.322&  2.83 \\
64 & 192  & 0.191&  1.81 \\
80 & 240  & 0.156&  0.90 \\
120 & 360 & 0.109&  0.89 
\end{tabular}
}
\caption[Relative error from the poisson solve]{Relative error from the poisson solve for a potential located away from the X-point (a) and on the the X-point (b). The average order of convergence is 1.42 for the potential centered away from the X-point and 1.71 for the potential centered on the X-point.
\label{tab:order}}
\end{center}
\end{table}

For both cases (with and without the X-point) we observe somewhat irregular convergence with an abnormally low order going from $(N_\psi,N_\theta)=(16,48)$ to $(N_\psi,N_\theta) = (32,96)$. The average order of convergence is 1.42 for the test case away from the X-point and 1.71 for the test case on the X-point. For both tests the average order is lower than the expected order of 2 and the average order for the the potential centered on the X-point is unexpectedly higher than for the potential away from the X-point. This could be because the aspect ratio of the cells near the X-point is much closer to 1 than the aspect ratio of the cells near the top of the grid; the cells near the top of the equilibrium are much longer in the R direction than the Z direction. Note that the reference potential, $\phi_{ref}$ is a projection so it has some discretization error as well, which makes it harder to measure convergence.

\subsection{Advection}
\label{sec:advection}
For this test case we advect a gaussian bump in the $\hat{Z}$ direction with velocity $\mathbf{v} = v_0\hat{Z}$ in the vicnity of the X-point. This means that we set the characteristic velocities (Eqs.~\ref{eq:z1}-~\ref{eq:vpar}) to $\dot{z}^1 = v_0\mathbf{e^1} \cdot \hat{Z}$, $\dot{z}^3 = v_0 \mathbf{e^3} \cdot \hat{Z}$, and $\dot{v}_\parallel = 0$. So, this test is not a test of the gyrokinetic equations, but a test of the quality of the grid and geometric quantities near the X-point.

\begin{figure}
    \subfloat[\label{fig:blobanaaway}]{
    \includegraphics[width=0.45\textwidth, valign=t]{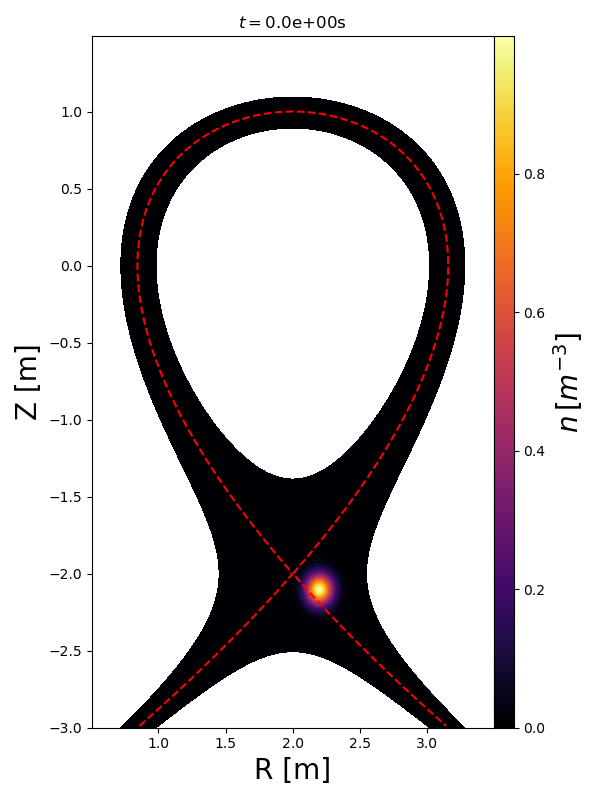}
    }
    \subfloat[\label{fig:blobanaon}]{
    \includegraphics[width=0.45\textwidth, valign=t]{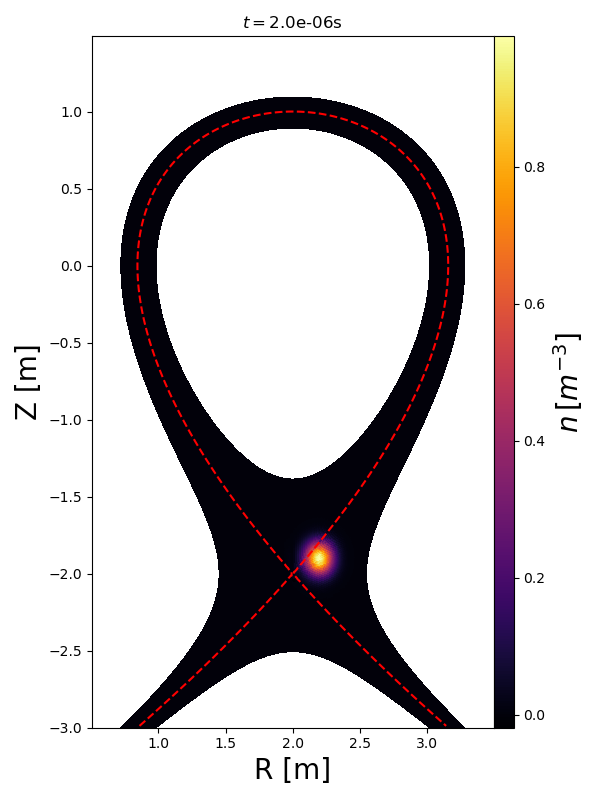}
    }
    \caption[Projection of the initial condition]{Projection of the initial condition (a) and the analytical final solution (b) for the gaussian bump advection test.
    \label{fig:blobana} }
\end{figure}

 We chose an initial condition for this test of $n = e^{[(R-2)^2 + (Z-2.1)^2]/0.1^2}$. We chose a speed $v_0 = 10^5 \textrm{ m/s}$ and evolve the simulation for $2.0 \times 10^6\textrm{ s}$, which makes the final solution $n=e^{[(R-2)^2 + (Z-1.9)^2]/0.1^2}$. The initial condition and analytical final solution are shown in Fig.~\ref{fig:blobana}.
 
We compare the density at the end of the simulation, $n_{num}$ with the projection of the analytical solution, $n_{ref}$, and calculate the error as 
\begin{equation}
    \mathcal{E} = \Big(\frac{\int J_c d\psi d\theta (n_{num}-n_{ref})^2} {\int J_c d\psi d\theta n_{ref}^2} \Big)^{1/2}.
    \label{eq:bloberror}
\end{equation}
The results of the convergence test are shown in table~\ref{tab:bloborder}. The average order of convergence is 1.55 with lower order convergence at low resolutions. Note that, as was true in the test of the poisson solver, the reference solution, $n_{ref}$, is a projection so it has some discretization error as well, which makes it harder to measure convergence.

\begin{table}
\begin{center}
\begin{tabular}{lcccc} 
$N_\psi$ & $N_\theta$ & Relative Error & Order \\
8 & 24    & 0.664&      \\
16 & 48   & 0.431&  0.62\\
32 & 96   & 0.165&  1.38\\
48 & 144  & 0.078&  1.85\\
64 & 192  & 0.045&  1.94\\
80 & 240  & 0.029&  1.94\\
120 & 360 & 0.015&  1.55 
\end{tabular}
\caption[Relative error from the advection test for a gaussian bump]{Relative error from the advection test for a gaussian bump advected in the $\hat{Z}$ direction just to the right of the X-point. The average order of convergence is 1.55.
\label{tab:bloborder}}
\end{center}
\end{table}

\subsection{Multi-Block Demonstration \& Consistency Test}
\label{sec:example}
\begin{figure}
    \subfloat[\label{fig:step_density}]{
    \includegraphics[width=0.33\textwidth, valign=t]{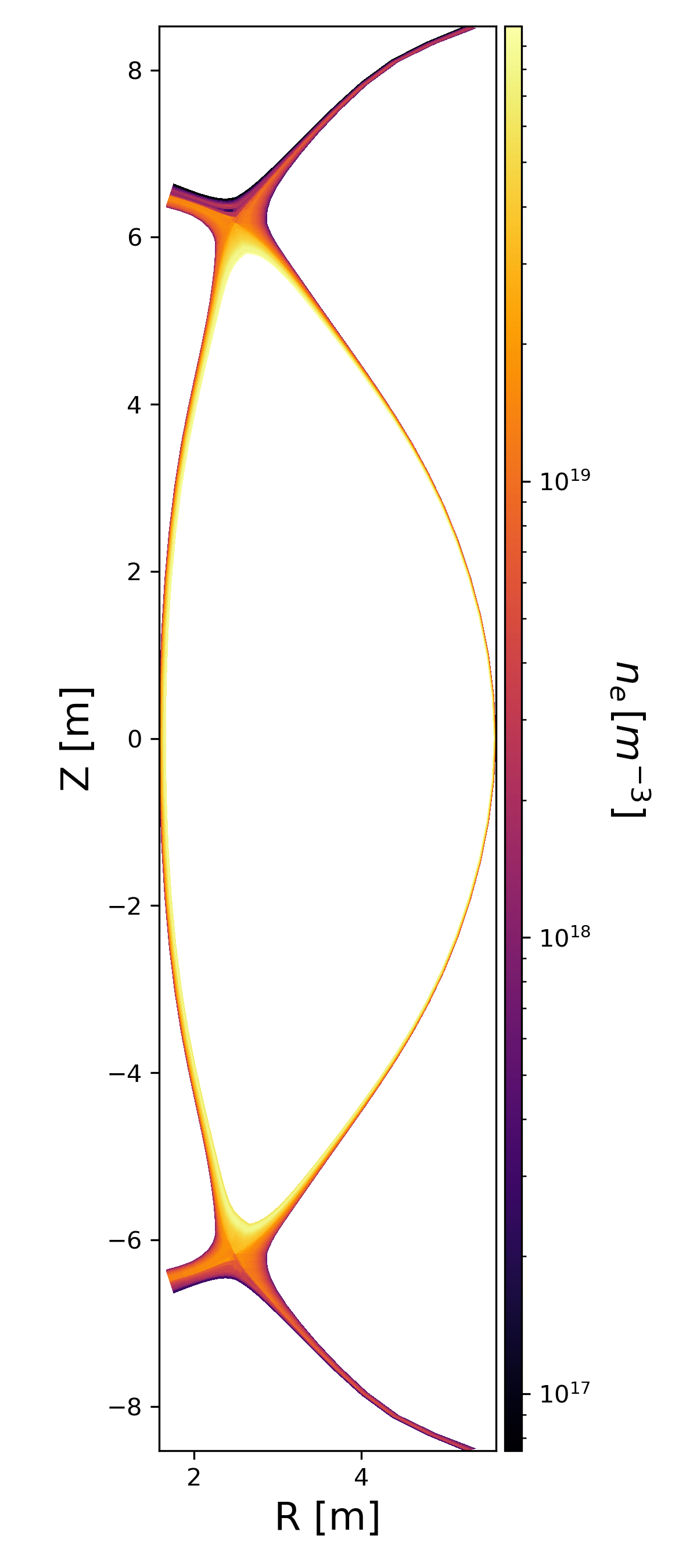}
    }
    \subfloat[\label{fig:step_temp}]{
    \includegraphics[width=0.33\textwidth, valign=t]{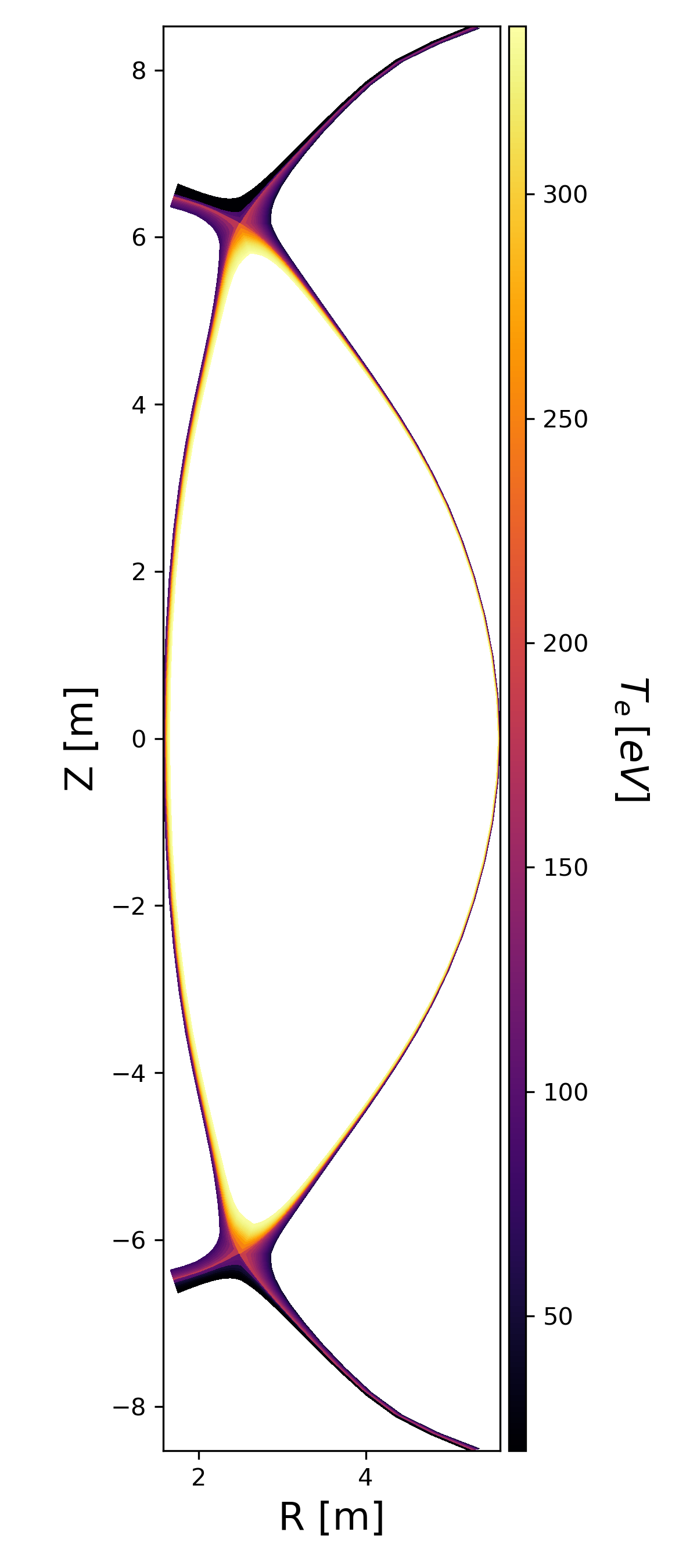}
    }
    \vspace{-11cm}
    \begin{flushright}
    \subfloat[\label{fig:step_densityzoom}]{
        \includegraphics[width=0.33\textwidth, valign=t]{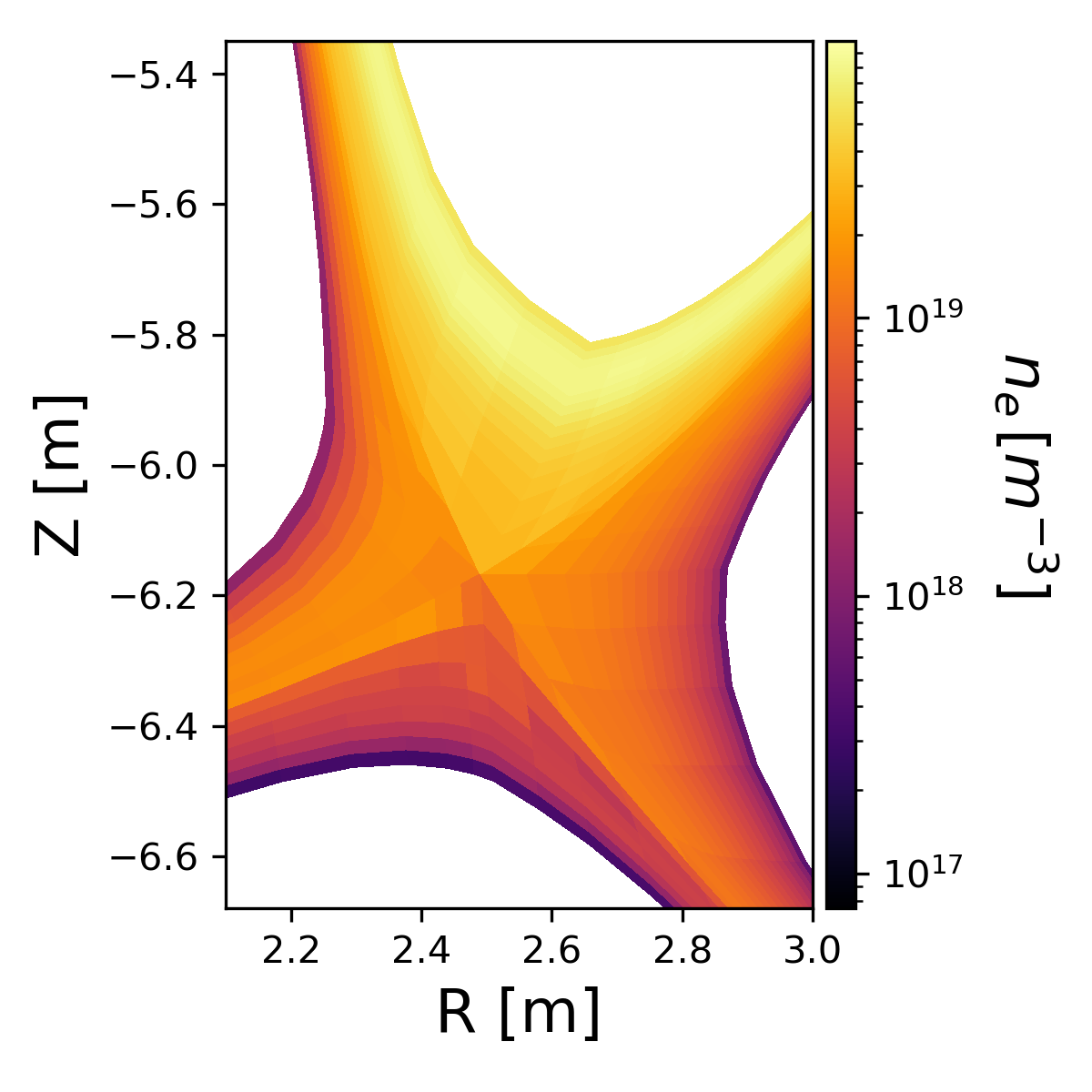}
    }\\
    \subfloat[\label{fig:step_tempzoom}]{
        \includegraphics[width=0.33\textwidth, valign=t]{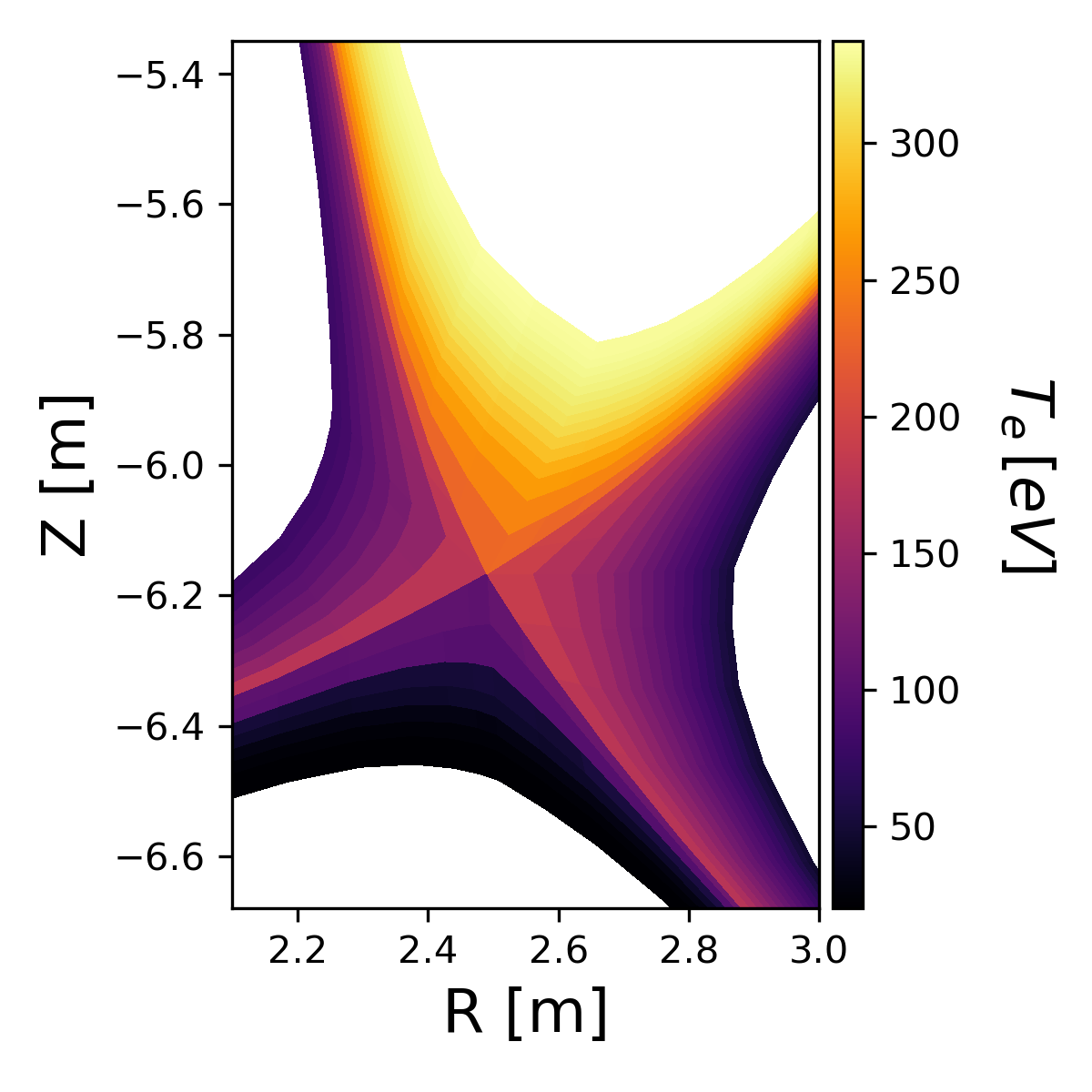}
    }
    \end{flushright}
    \vspace{1cm}
    \caption[Simulation results from a 2D, axisymmetric simulation of the Spherical Tokamak for Energy Production.]{ Simulation results from a 2D, axisymmetric simulation of the Spherical Tokamak for Energy Production. The poloidal projection of the electron density and temperature are shown in (a) and (b) respectively. A close-up of the electron density is shown in (c) and a close-up of the electron temperature is shown in (d).
    \label{fig:step_moments} }
\end{figure}

To demonstrate the effectiveness of our algorithm and geometric consistency at block-boundaries, we conduct a 2-dimensional axisymmetric simulation in the magnetic geometry of STEP with the grid shown in Fig.~\ref{fig:stepgrid}. The simulation consists of a deuterium plasma with 100MW of input power and a particle input of $1.3\times 10^{24}$~m$^{-3}$s$^{-1}$ split evenly between electrons and ions. The particle and heat source is Maxwellian and is present only in the innermost radial cell of the core. Within this first radial cell the particle input rate and temperature of the source is uniform. As is typically done in axisymmetric divertor design codes, an ad-hoc diffusivity is chosen to mimic turbulence which is absent in 2D simulations. Here we choose a particle diffusivity of $D=0.22$~m$^2/$s and a heat diffusivity of $\chi = 0.33$~m$^2/$s to target a heat flux width of $2$~mm. The simulation setup is similar to those in~\cite{Shukla25} where more details on \gke's gyrokinetic model can be found. 
In Fig.~\ref{fig:step_moments} we show the electron density and temperature from the simulation at $t=0.72$~ms. In these figures we can see that the simulation is well-behaved near the X-point; the electron temperature and density do not diverge.

\begin{figure}
    \subfloat[\label{fig:particleerror}]{
    \includegraphics[width=0.45\textwidth, valign=t]{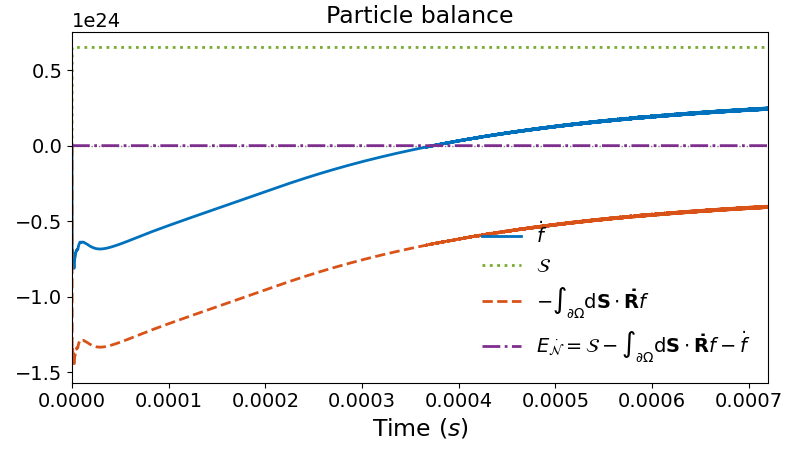}
    }
    \subfloat[\label{fig:particlerelerror}]{
    \includegraphics[width=0.45\textwidth, valign=t]{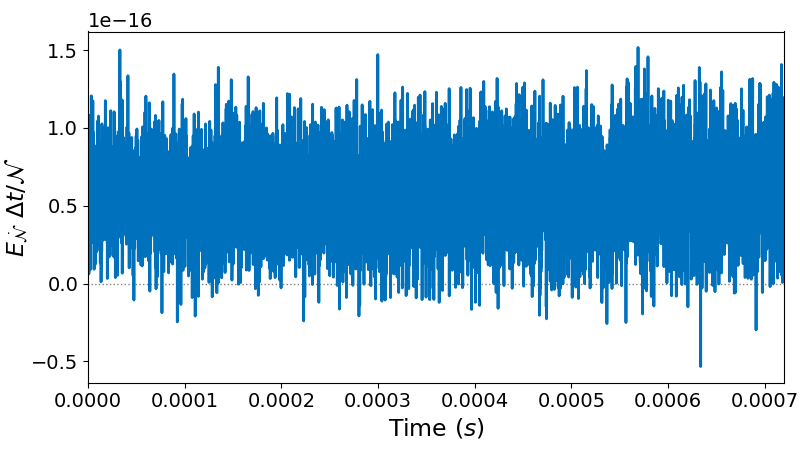}
    }
    \caption[Particle balance (a) and relative error in the number of particles (b) for the STEP simulation]{
    Particle balance (a) and relative error in the number of particles (b) for the STEP simulation. The balance includes the change in a single time step (solid blue) due to fluxes through the boundaries (dashed orange) and sources (dotted green), as well as the error in adding these up (purple dash-dot).
    \label{fig:conservation} }
\end{figure}

As mentioned at the end of section~\ref{sec:discretization}, if the surface normals are not consistent at block boundaries, particle conservation will be broken. In Fig.~\ref{fig:conservation}, we plot the relative error in the number of particles which shows that our algorithm conserves particles to machine precision.


\section{Conclusion}

Field-aligned coordinate systems offer a computational advantage when conducting simulations of tokamaks because they allow for coarse resolution along the field line and larger time steps. However, using field aligned coordinates for simulations that cover both the open and closed field line regions in diverted geometries can be difficult because of the coordinate singularity at the X-point.
Here we have presented a grid generation algorithm along with a phase space discretization scheme that allows for the evolution a gyrokinetic system in X-point tokamak geometries while taking advantage of a field aligned coordinate system.

Our grid generation algorithm described in section~\ref{sec:computation} splits the domain of a tokamak into topologically distinct regions for field line tracing and then further splits the domain at the X-points resulting in a multi-block grid that ensures cell corners lie on the X-point. This grid generation algorithm uses highly accurate integrators for field line tracing and allows for direct calculation of the metric coefficients and other geometric quantities required for evolving the gyrokinetic equation in field aligned coordinates.
In section~\ref{sec:discretization} we describe the key feature of our algorithm that avoids the coordinate singularity at the X-point. Geometric quantities are evaluated at interior and surface quadrature points which do not touch the X-point and thus do not diverge.
In the final section, section~\ref{sec:tests}, we do convergence tests, which show converge, albeit reduced, even in the vicinity of the X-point. We also demonstrate that our algorithm conserves particles to machine precision in an example 2D axisymmetric simulation of a deuterium plasma in the STEP magnetic geometry including the X-point. In the future we hope to use \gke's axisymmetric solver as a complement to fluid divertor design codes and highlight the importance of kinetic effects in divertor design.

Planned improvements to our grid generation methods involve refining our grids and extending these methods for use 3D turbulence simulations. As can be seen in Fig.~\ref{fig:stepgrid} and elsewhere, the grid spacing becomes coarse near the X-point.  There are several ways this could be improved in the future.  One is by using mesh refinement near the X-point.  Another is to use a non-uniform spacing of the $\psi$ grid to give more uniform spacing in real space near the X-point, and merge adjacent DG cells away from X-point if they become more narrow than needed (which would reduce the time step due to the Courant limit).  Another approach could be to switch to a non-aligned grid near the X-point as COGENT does.

We believe the methods described here will also work for 3D turbulence simulations; our method of evaluating geometric quantities and surface fluxes will still avoid the X-point. For 3D simulations, we plan to employ a procedure similar to the shifted metric approach~\citep{Scott2001}, applying a toroidal shift in the coordinate system at parallel block boundaries. We plan to use Gkeyll's previously developed twist-and-shift boundary conditions~\citep{ManaTS} on the distribution function at these parallel block boundaries (for example the boundary between blocks 11 and 12 and the boundary between blocks 2 and 3 in Fig.~\ref{fig:stepgrid}). The extension to 3D will be presented in a future work. 
Detailed physics studies with the grids and algorithms described here, including the effect of neutrals, are presented in chapters~\ref{chap:3} and~\ref{chap:4}.

\chapter{Gyrokinetic Simulations of a Low Recycling Scrape-off Layer without a Lithium Target}
\label{chap:3}
\footnote{All simulations were performed by me. Jonathan Roeltgen and I worked together to write the code required to couple Gkeyll to EIRENE. All analysis was performed by me and the paper~\cite{shukla2025LR} (planned submission to Nuclear Fusion) was written by me.}
Low-recycling regimes are appealing because they entail a high edge temperature and low edge density which are good for core confinement~\citep{Mike23}. 
However, present low recycling scenarios present challenges for heat flux handling and avoiding undue evaporation of lithium. Lithium tends to evaporate quickly at high heat fluxes and materials which handle high heat fluxes with less evaporation are high recycling. 
In addition to physical challenges, there are also modeling challenges associated with the low-recycling regime. Fluid simulations are typically used to study the Scrape-Off Layer (SOL) in tokamaks ~\citep{Hudoba2023, Osawa2023, Rozhansky2021, Zhang2024},  but modeling the collisionless SOL of a low recycling regime in which the fluid assumptions are not valid requires a kinetic treatment. 
Here we explore the feasibility of a low-recycling SOL scenario in the magnetic geometry of STEP. We also compare gyrokinetic and fluid simulations to investigate kinetic effects that can be taken advantage of to address the challenges of sputtering and heat flux handling in the low-recycling regime. We use Gkeyll~\citep{Shukla25,Roeltgen25, Mana25} for gyrokinetic simulations and SOLPS~\citep{Wiesen25,Schneider2006} for fluid simulations. The simulation results presented here indicate that (1) With a sufficiently large pump, a high SOL temperature can be achieved without using a low-recycling material at the target but using it at the side-walls instead, (2) At high SOL temperatures, impurities can be electrostatically confined to the divertor region, and (3) The heat flux width is significantly broadened by mirror trapping. In this chapter we focus on point (1) and in chapter~\ref{chap:4} we focus on points (2) and (3).

\section{Challenges of Maintaining a Low Recycling SOL}
\label{sec:lithium}
Lithium is a commonly used low recycling material but its high vapor pressure makes it unable to go above $\sim$ 400-450 $^\circ$C without evaporating, which tends to counteract the desired physical regime. In addition, at 400-450 $^\circ$C, the chemical gettering of hydrogenic species that gives low recycling is strongly reduced. Keeping the target temperature below 400-450 $^\circ$C is difficult, so it would be difficult to maintain a low temperature, high density SOL using lithium as a target material. The simulations presented in section~\ref{sec:10kev} indicate that, with a sufficiently large pump, a low-recycling SOL can be achieved by coating the side walls of the tokamak with lithium but using a material that can handle high heat fluxes without evaporating, such as refractory solid metals or novel liquid metals~\citep{ARPAE2025, INFUSE2024, SuperXT}, on the target. This addresses a major concern about the feasibility of a low-recycling SOL.

In a low-recycling regime, the high heat fluxes at the target will entail a large amount of sputtering. One concern is that the sputtered impurities will contaminate the core plasma. The kinetic simulations shown in chapter~\ref{chap:4} indicate that, in less collisional regimes, impurities are much better confined to the divertor region than is predicted by fluid simulations. If the SOL temperature is high, the potential drop along the divertor leg can be very large which serves as a barrier preventing impurities from traveling upstream.

Fluid simulations used to model collisional SOLs typically neglect the mirror force. Our gyrokinetic simulation results in chapter~\ref{chap:4} show that in collisionless regimes, the effect of particle drifts in combination with mirror trapping has a large effect on the heat flux width. When drifts are included, kinetic simulations show that the ion heat flux width is broadened to approximately the ion banana width, while the broadening observed in fluid simulations is negligible. In a low-recycling SOL where temperatures can be quite high, the banana width can be quite large which can result in significant increase in the heat flux width and corresponding reduction in the peak heat flux.

\section{Gkeyll-EIRENE Simulations of a Low Recycling Scenario}
\label{sec:10kev}
\begin{figure}[h]
\centering
\includegraphics[width=0.3\textwidth]{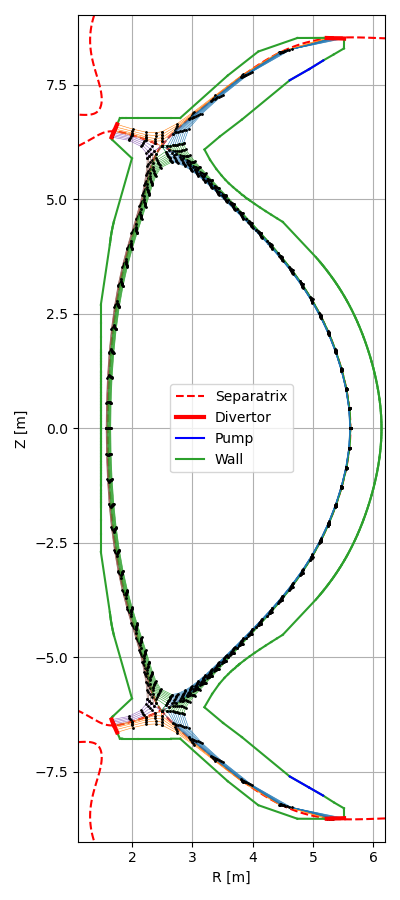}
\caption[Simulation domain used for the coupled Gkeyll-EIRENE simulation.]{Simulation domain used for the coupled Gkeyll-EIRENE simulation. The colored grid lines show the Gkeyll simulation domain and the thick green boundary enclosing the Gkeyll domain indicates the machine boundary which is the EIRENE simulation domain. The separatrix is marked by the dashed red line, the pump on the wall is marked with a blue solid line, and the divertor plates are marked with solid red.
\label{fig:domain}
}
\end{figure}

We conducted a simulation of a low-recycling SOL using the gyrokinetic code Gkeyll's axisymmetric solver~\citep{Shukla25} coupled to the monte-carlo neutral code EIRENE~\citep{Wiesen25}. Gkeyll evolves the plasma and EIRENE evolves the neutral particles. These simulations are similar in concept to SOLPS simulations which couple the fluid solver B2.5 to EIRENE, but B2.5 has been replaced with a gyrokinetic solver to include kinetic effects. The simulation grid is shown in Fig.~\ref{fig:domain}. The simulation grid used in Gkeyll is shown, and the green boundary indicates the machine wall, which is the domain used for EIRENE.

The simulation is sourced by Maxwellian with a temperature of 10keV at the inner radial boundary with an input power of 100 MW. The perpendicular particle and heat diffusivities are 0.5 $m^2/s$ and 0.75 $m^2/s$ respectively, and drifts are turned off. The diffusivities were chosen to target a heat flux width (mapped upstream) of 2 mm. The divertor plate is gallium with a recycling coefficient of 0.99 and the side walls are coated with lithium giving them a recycling coefficient of 0.5 for atomic deuterium and 0.99 for molecular deuterium. The simulation also has a pump marked in blue in Fig.~\ref{fig:domain} which is modeled as a surface with an absorption coefficient of 0.5 for all species. The simulation evolves 5 species: atomic deuterium ions, molecular deuterium ions, electrons, atomic neutral deuterium, and molecular neutral deuterium. The input files containing details of the simulation setup can be found at \url{https://github.com/ammarhakim/gkyl-paper-inp/tree/master/2025_IAEA_STEP}.

\begin{figure}[ht]
\captionsetup[subfloat]{labelformat=empty}
\subfloat[(g)]{
\includegraphics[width=0.48\textwidth]{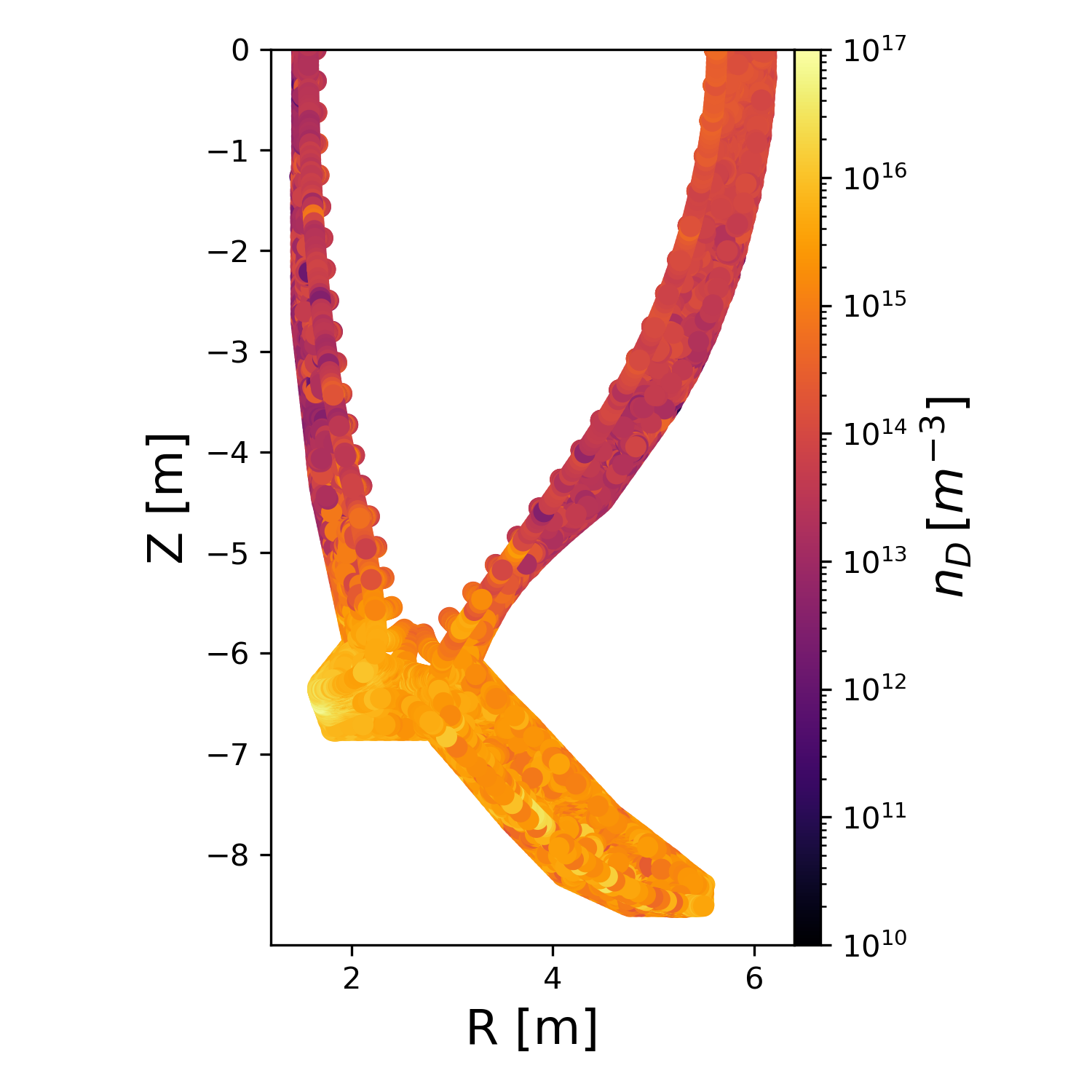}
}
\subfloat[(h)]{
\includegraphics[width=0.48\textwidth]{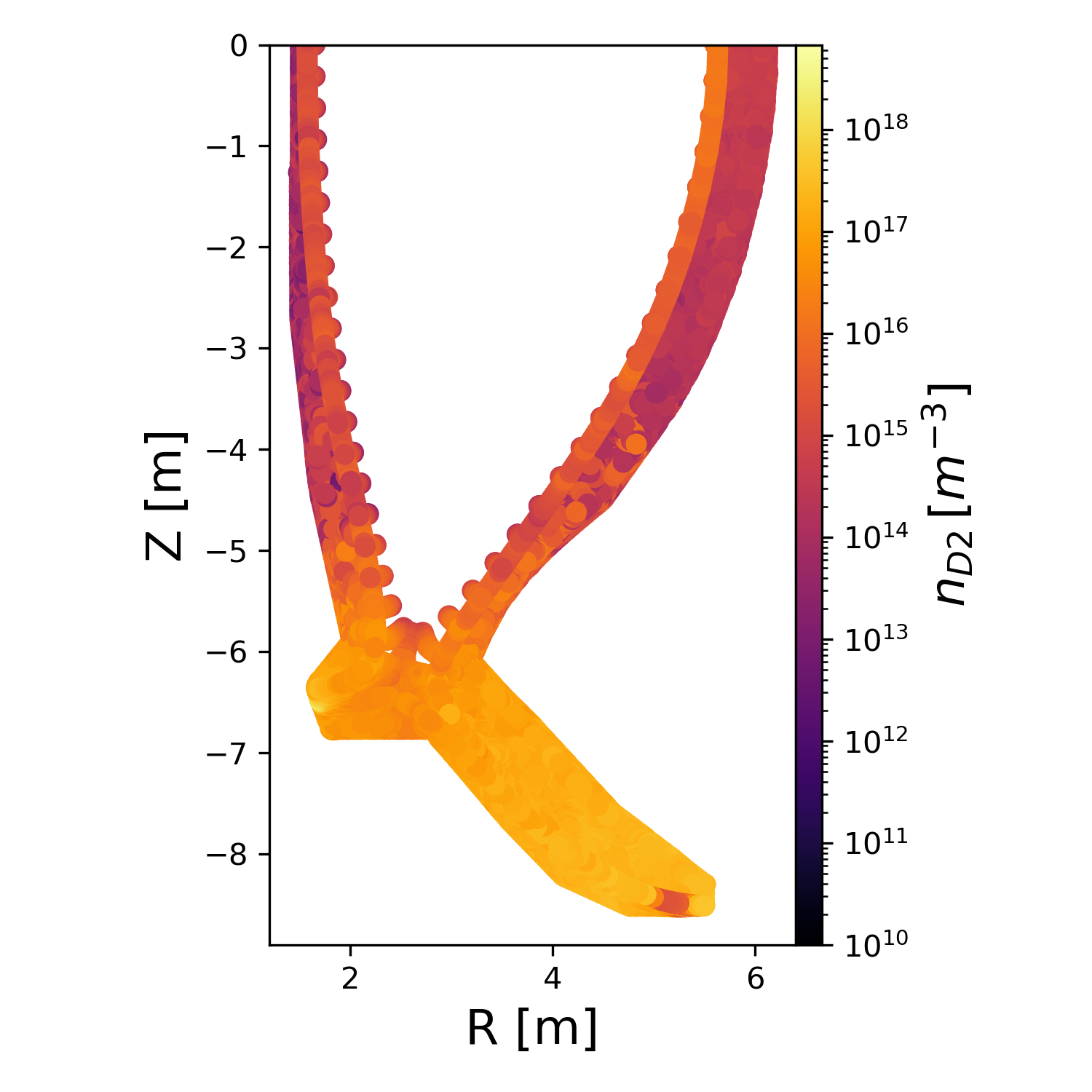}
}
\caption[Simulation results from the coupled Gkeyll-EIRENE simulation of STEP in a low recycling regime.]{Simulation results from the coupled Gkeyll-EIRENE simulation of STEP in a low recycling regime. The electron density (a), atomic deuterium ion density (b), molecular deuterium ion density (c), electron temperature (d), atomic deuterium ion temperature (e), molecular deuterium ion temperature (f), neutral atomic deuterium density (g), and neutral molecular deuterium density (h) are plotted in the poloidal plane. We plot only the lower half of the domain because the simulation is up-down symmetric. The upstream D+ temperature remains at $\sim$8.8 keV, the upstream electron temperature remains at $\sim$5.4 keV, and the downstream density remains low at 5.5$\times$10$^{16}$ $m^{-3}$. The pumping of D2 is effective, so the D2+ density remains at 1.5$\times$10$^{16}$ $m^{-3}$ downstream, which is around one fourth of the D+ density.}
\label{fig:moments}
\end{figure}

\begin{figure}
    \subfloat[]{
    \includegraphics[width=0.45\textwidth]{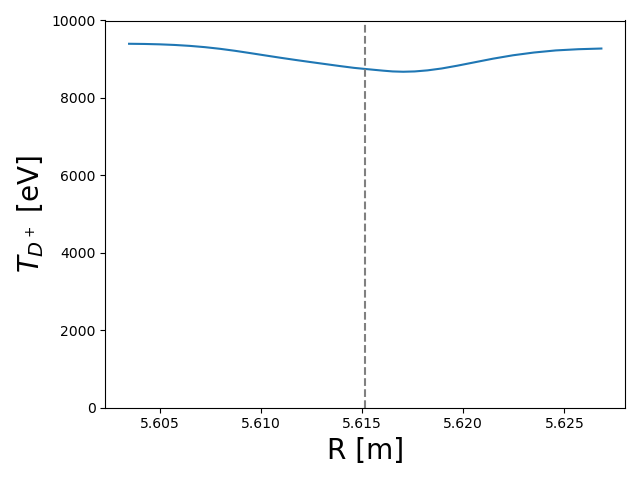}
    }
    \subfloat[]{
    \includegraphics[width=0.45\textwidth]{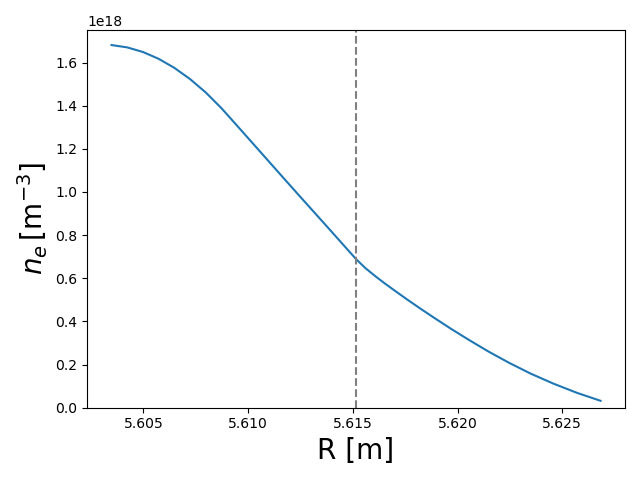}
    }
    \caption[Simulation results from the coupled Gkeyll-EIRENE simulation of STEP in a low recycling regime]{Simulation results from the coupled Gkeyll-EIRENE simulation of STEP in a low recycling regime. The radial profiles of the atomic deuterium ion temperature (a) and electron density (b) are plotted at the OMP. The vertical dashed line indicates the separatrix. The ion temperature at the separatrix is $\sim$ 8.8 keV and the electron density is $\sim$ 6.7 $\times$ 10$^{17}$ m$^{-3}$.\label{fig:OMP}}
\end{figure}

The magnetic geometry features a long outboard divertor leg with tightly packed flux surfaces near the divertor plate. Because the plasma near the divertor plate is narrow and the density is low, most of neutral deuterium ejected from the plate and walls passes through the plasma without undergoing a reaction. The ionization mean free path of the neutral deuterium is more than 10 meters in the  outboard divertor leg, and the SOL-width is about 20 mm, so neutrals pass through the plasma and hit the side walls many times before undergoing a reaction. With a recycling coefficient of 0.5 at the walls, most of the atomic neutral deuterium recycled from the plate is absorbed by the walls before having a chance to react with the plasma.
A large fraction of the deuterium ions incident on the plate are recycled as molecular deuterium rather than atomic deuterium. Since the absorption of molecular deuterium by lithium is much lower than the absorption of atomic deuterium, molecular deuterium must be removed by a different mechanism.
In typical high recycling regimes with a high density, molecular deuterium would quickly dissociate into atomic deuterium, which could be absorbed by lithium. However, in this narrow, low density divertor leg, the dissociation time is long compared to the residence time of molecular deuterium. Thus, the neutral molecular deuterium must be removed with a pump. In our simulation we find that more than 60\% of the molecular deuterium recycled from the plate is removed by the pump. If the neutral molecular deuterium was not pumped effectively, it would result in a large source of cold molecular deuterium ions, which would cool the SOL.
The removal of atomic neutral deuterium by the lithium walls and pump and the removal of molecular neutral deuterium by the pump results in a low neutral and plasma density near the divertor plates and prevents the neutrals from significantly cooling the plasma. The electron density, neutral deuterium density, electron temperature, and ion temperature are shown in Fig.~\ref{fig:moments}. Fig.~\ref{fig:OMP} shows radial profiles of the ion temperature and electron density at the OMP. The upstream density at the OMP is $n_{sep}\sim 6.7\times 10^{17}\, m^{-3}$, $T_{i,sep} \sim 8.8\,keV$, and $T_{e,sep} \sim 5.4\,keV$. The cold D2+ is confined downstream in the divertor region by the potential drop along the divertor leg; downstream, the D2+ density is 20\% of the D+ density, but upstream, the D2+ density is only 4\% of the D+ density.
\vspace{-0.8cm}
\begin{figure}[hb]
\centering
\captionsetup[subfloat]{labelformat=empty}
\subfloat[(a)]{
\includegraphics[width=0.48\textwidth]{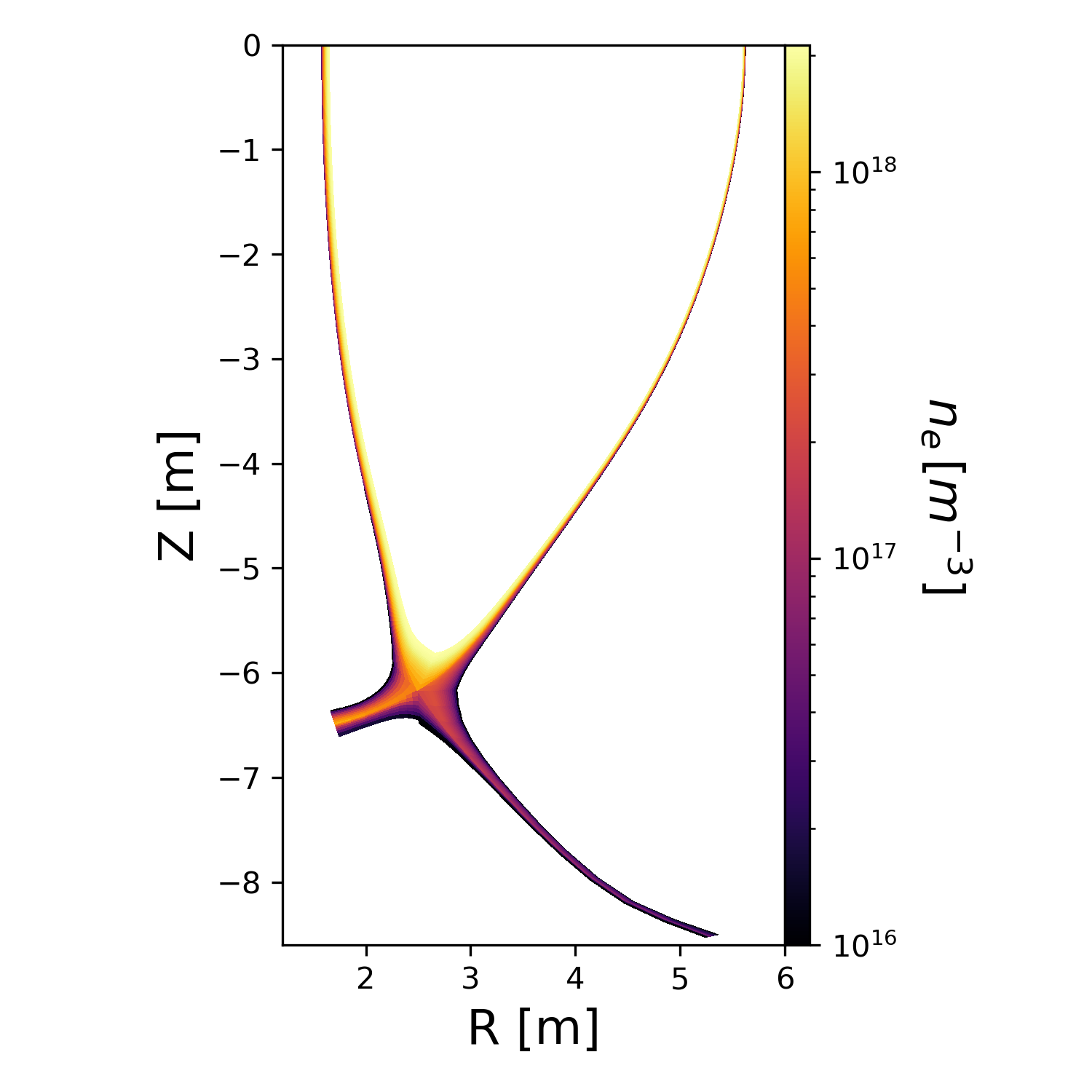}
}
\subfloat[(b)]{
\includegraphics[width=0.48\textwidth]{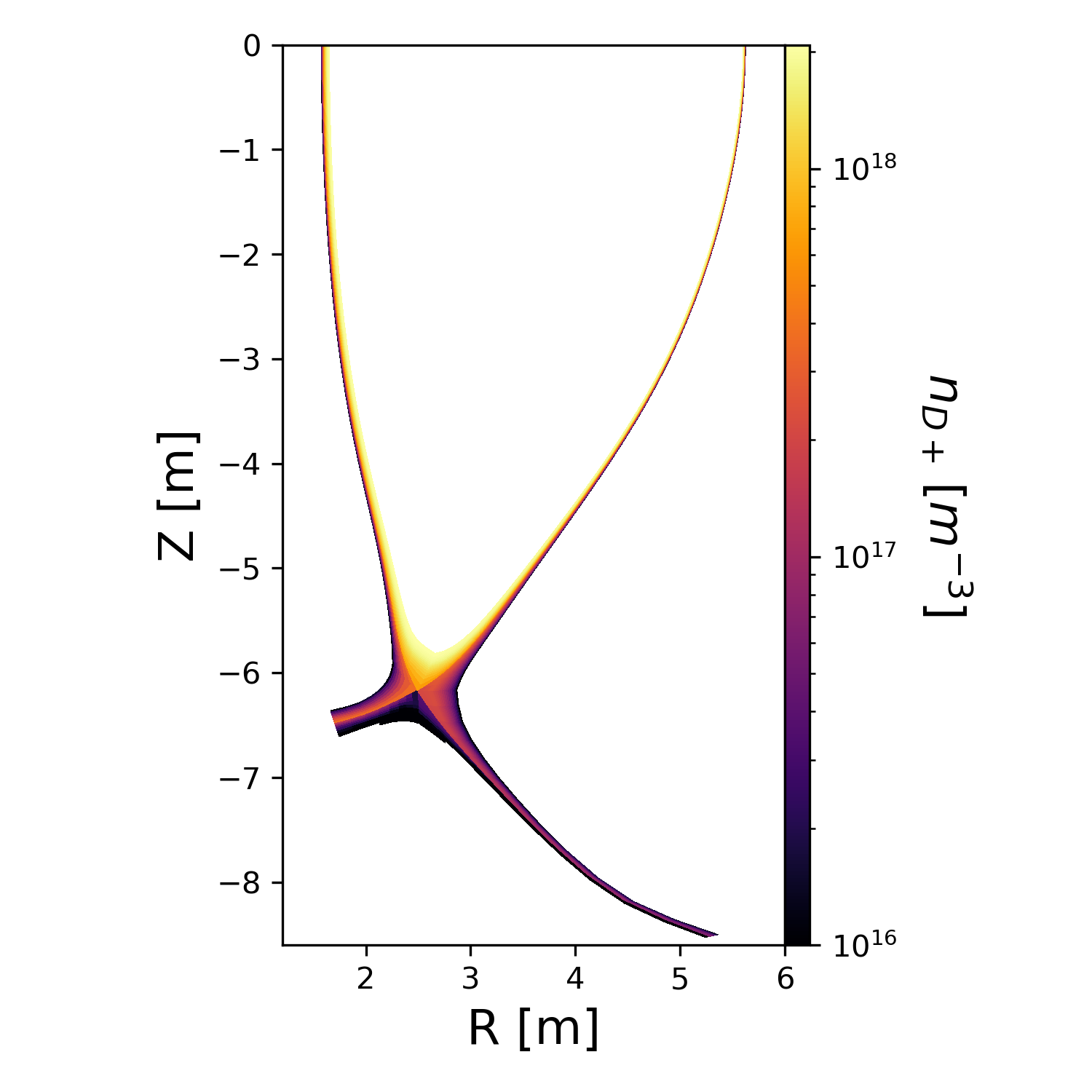}
}
\end{figure}
\clearpage
\begin{figure}[H]
\captionsetup[subfloat]{labelformat=empty}
\subfloat[(c)]{
\includegraphics[width=0.48\textwidth]{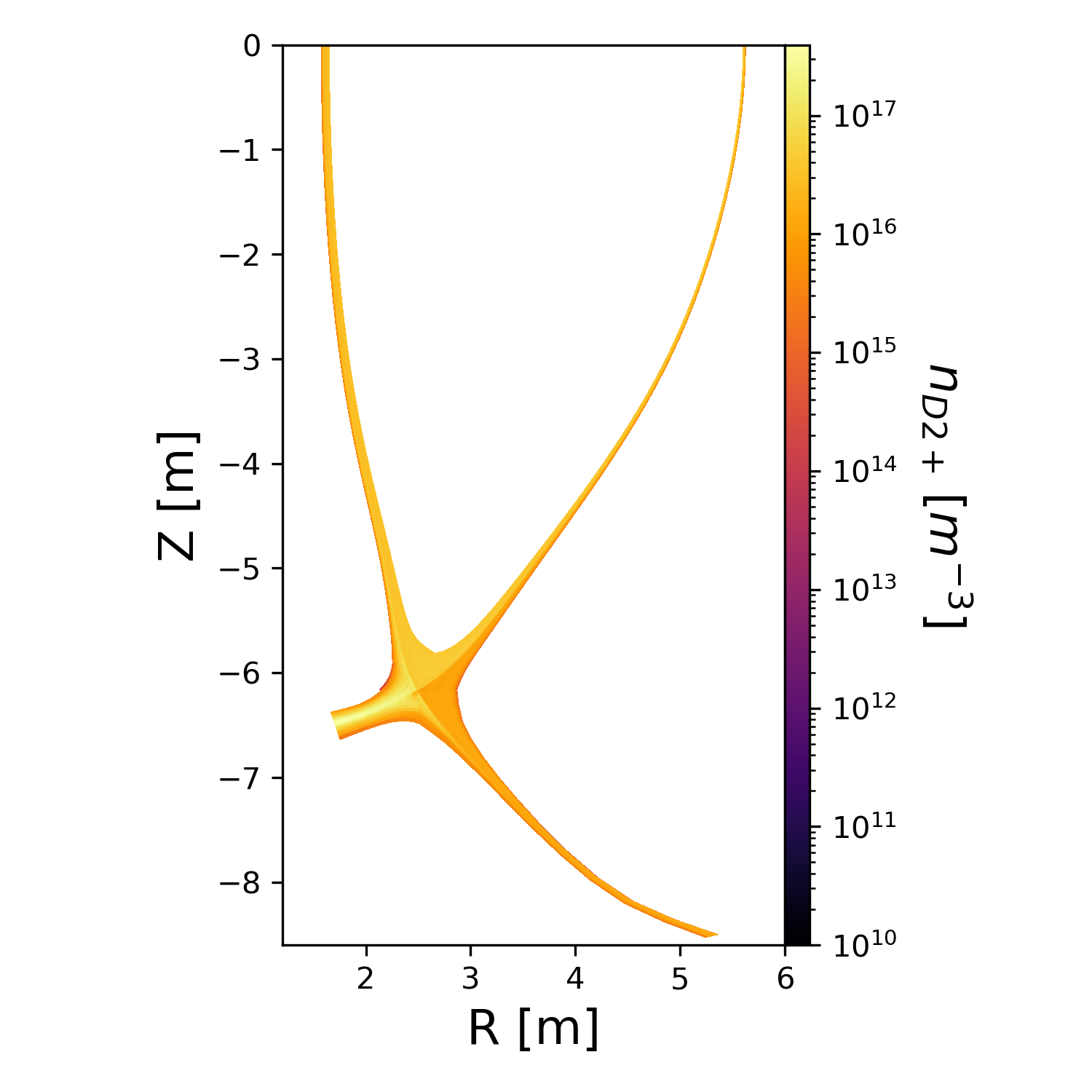}
}
\subfloat[(d)]{
\includegraphics[width=0.48\textwidth]{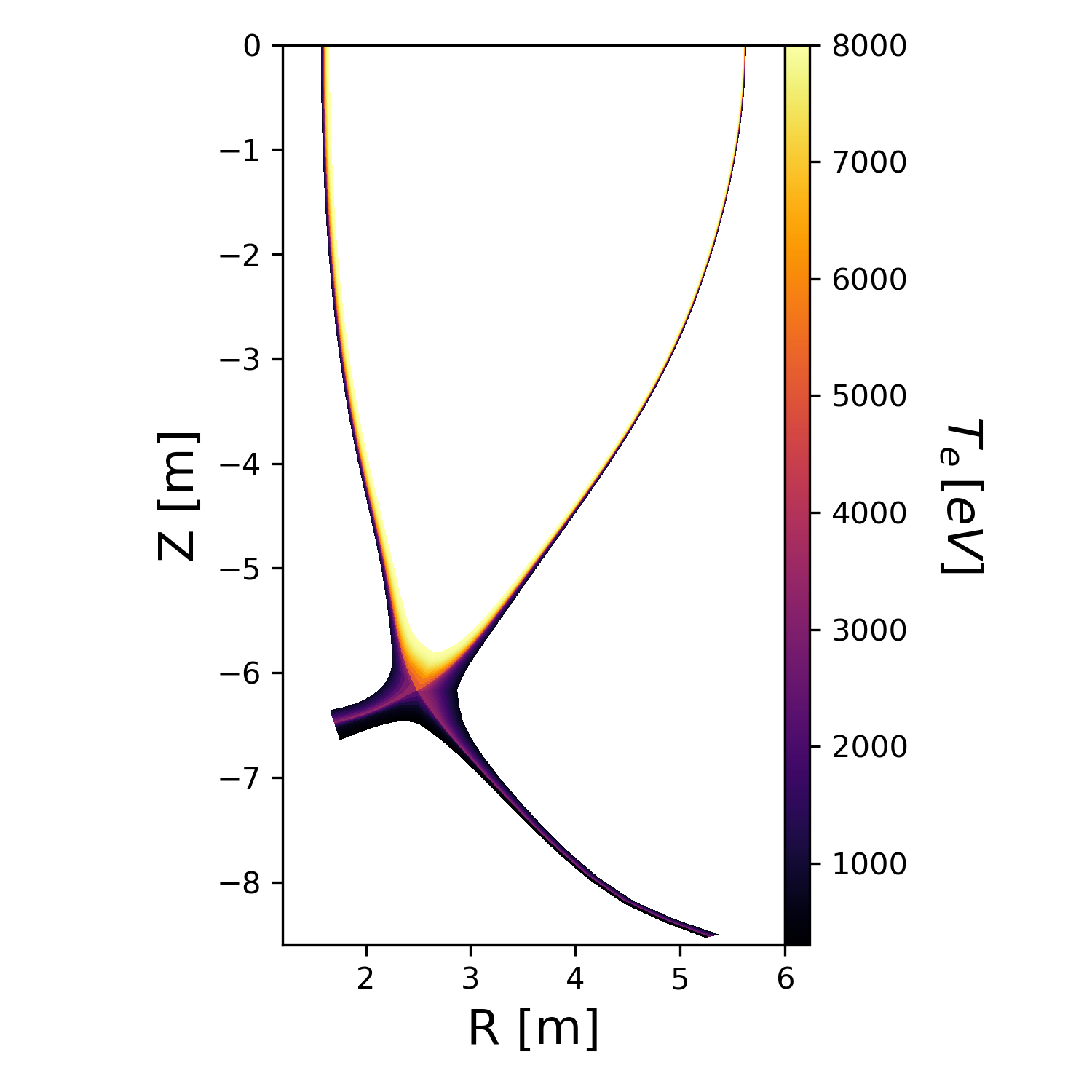}
}\\
\subfloat[(e)]{
\includegraphics[width=0.48\textwidth]{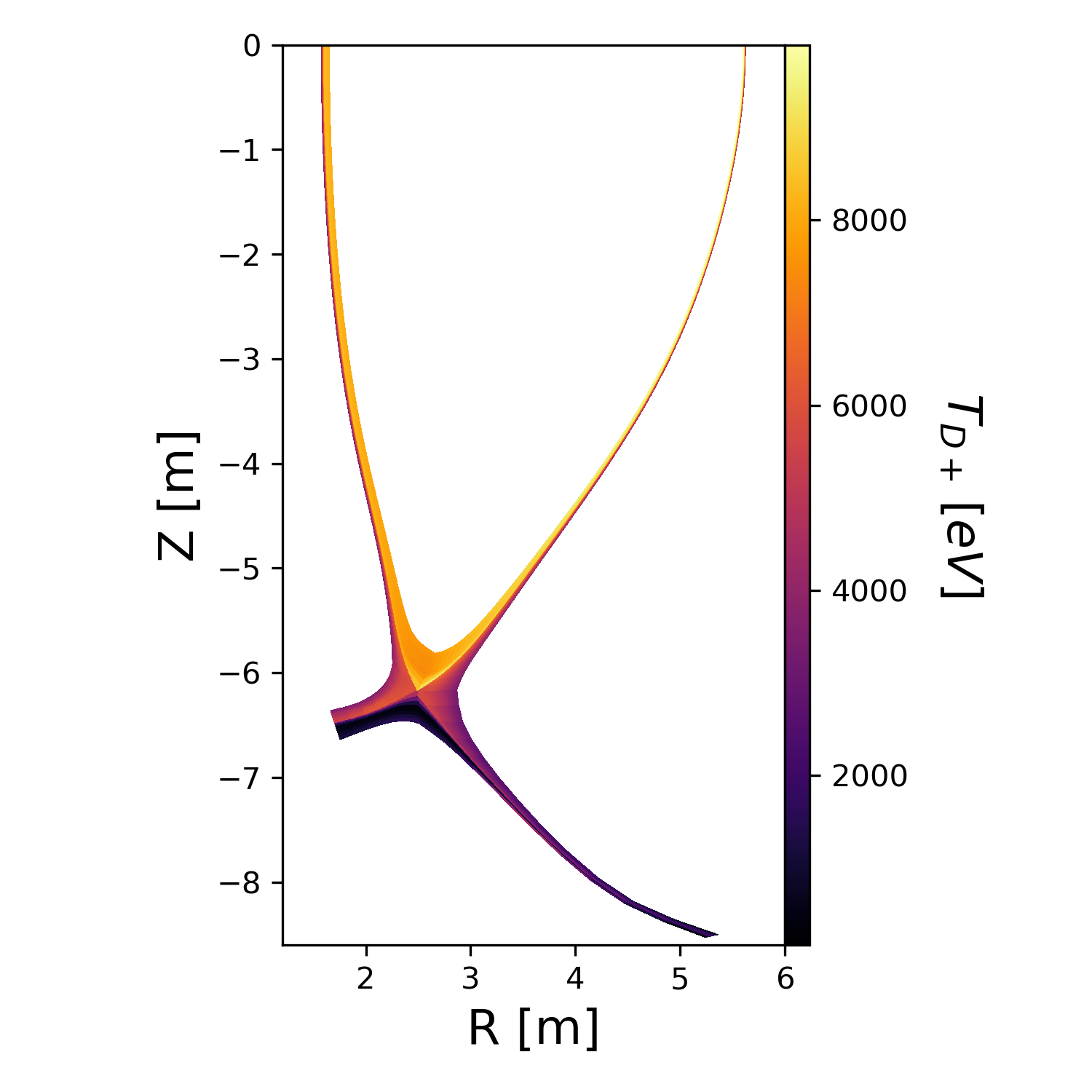}
}
\subfloat[(f)]{
\includegraphics[width=0.48\textwidth]{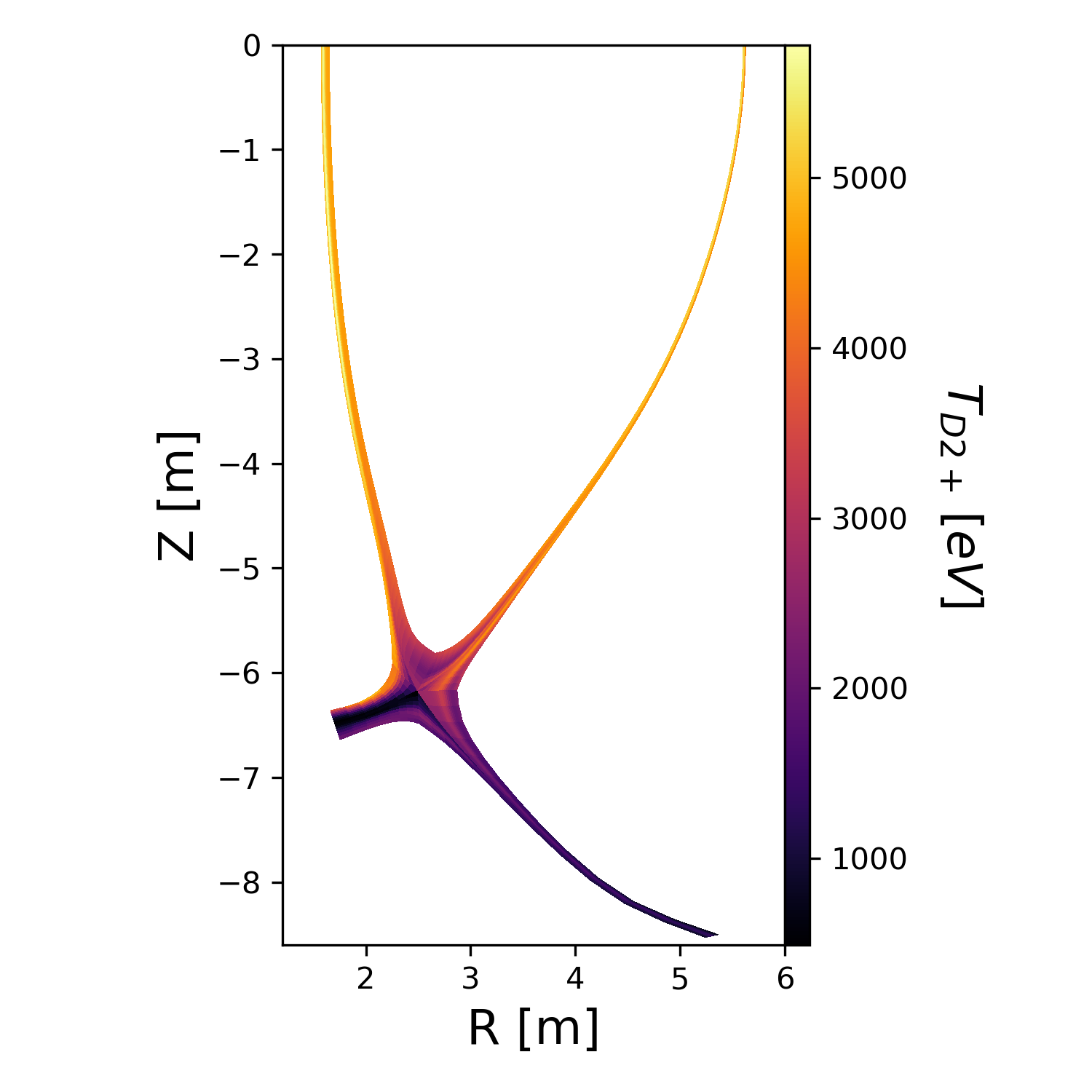}
}
\end{figure}
\clearpage

\begin{figure}[htbp]
\centering
\includegraphics[width=0.45\textwidth]{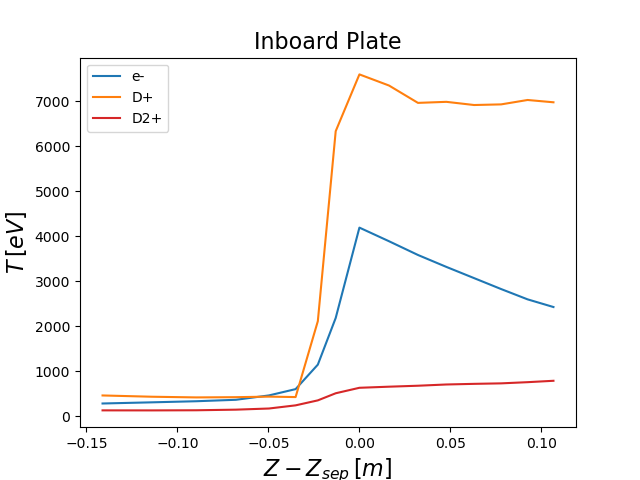}
\includegraphics[width=0.45\textwidth]{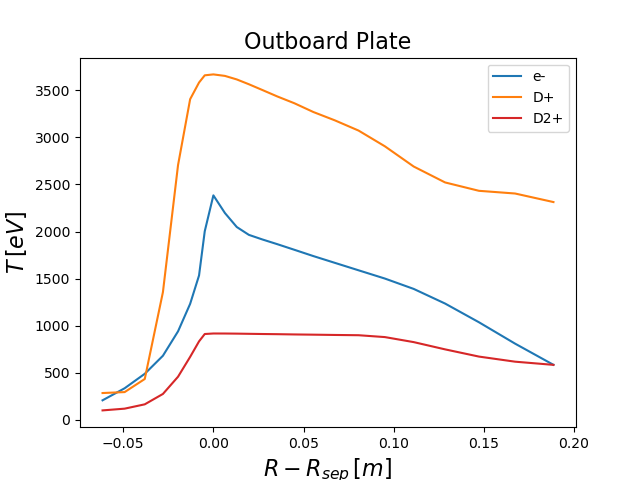}
\includegraphics[width=0.45\textwidth]{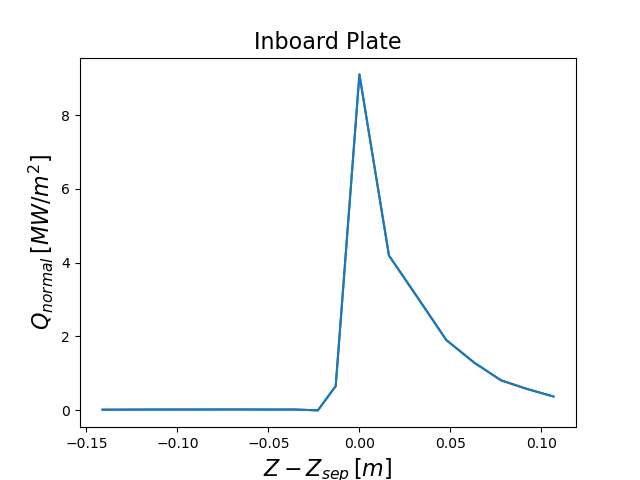}
\includegraphics[width=0.45\textwidth]{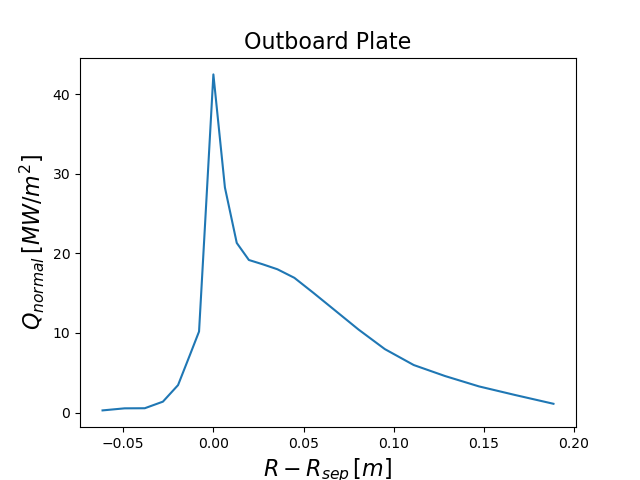}
\caption[Electron temperature, atomic ion temperature, molecular ion temperature, and total heat flux]{
Electron temperature, atomic ion temperature, molecular ion temperature, and total heat flux at the lower inboard and upper outboard divertor plates.
}
\label{fig:plate}
\end{figure}

The total heat flux normal to the plate and the electron and ion temperature at the plate are shown in Fig.~\ref{fig:plate}. The outboard heat flux width mapped upstream is $3.1\,mm$. The peak heat flux on the outboard plate is quite large at 40 $MW/m^2$ and the ions incident at the plate at the peak of the heat flux channel have a temperature of $\sim$3.5 keV. 
If the divertor had a lithium coating, these heat fluxes would quickly cause the target to heat up and the lithium to evaporate. 
Here we make some simple estimates, which indicate that the evaporated lithium would dominate over the recycled deuterium and cool the SOL. This would take the SOL out of the low density, high temperature regime, thus removing the low recycling conditions that are favorable for confinement.

The total number of deuterium ions incident on one of the outboard plates per second is $1\times 10^{22}$. Using the well known lithium vapor pressure and the plasma-wetted area of the outboard target ($\sim 1.7\,m^2$), one can estimate the number of lithium particles evaporated per second based on the target temperature. For temperatures of $450\, ^\circ C$, $550\, ^\circ C$, and $650\, ^\circ C$, the number of lithium particles evaporating from the target per second would be $5\times 10^{21}$, $1.1\times 10^{23}$, and $1.2\times 10^{24}$ respectively. So, the amount of evaporated lithium will be comparable to or much greater than the recycled deuterium, and the amount of evaporated lithium goes up very quickly with target temperature.
At these low electron densities, only a small portion of the evaporated lithium will be ionized before passing through the plasma and sticking to the wall, but this amount would likely be large enough to significantly cool the SOL and raise the density.
An estimate using the width of the plasma near the divertor plate (20 mm), the electron temperature at the divertor plate ($~\sim 2.5\,keV$), and the electron density at the divertor plate ($5.4\times 10^{16}\, m^{-3}$) indicates that around 11\% of the lithium will be ionized before passing through the plasma.

So, if the target reaches a temperature of $550\, ^\circ C$, the amount of lithium ionized would be comparable to the total amount of deuterium hitting the plate. In a low recycling regime, the plate temperature will likely be quite high because of the large heat fluxes. For the parameters of this simulation, which has a peak heat flux of 40 $MW/m^2$, one can estimate the approximate target temperature at 634 $^\circ C$ if the plate was tungsten with a 1 mm thick lithium coating.
Thus, a lithium coating on the divertor plate would likely significantly cool the SOL and increase the density. Fortunately, our simulation indicates that even without a low-recycling target material, a hot, low density SOL could be achieved. This would allow a material with a lower vapor pressure to be used on the target.  

However, there are still challenges associated with heat flux handling to address. Even a material like tungsten would be quickly eroded under the conditions in the simulation. A low-recycling solution would likely have to involve a liquid metal target to avoid detrimental erosion. Additionally, the high ion temperatures will still cause a large degree of sputtering which could contaminate the core. 
Our simulation results presented in chapter~\ref{chap:4} indicate that mirror trapping can significantly broaden the heat flux width and lower the peak heat flux when drifts are included. In chapter~\ref{chap:4}, we also outline kinetic effects that can potentially address the issue of sputtering and core contamination. 

\section{Summary of Results}
\label{sec:IAEA_conclusion}
Low-recycling regimes present advantages for core confinement but the feasibility of a low-recycling SOL is uncertain. 
When using low recycling materials such as lithium as a target material, it is difficult to prevent them from evaporating and cooling the plasma, which would interfere with the confinement advantages low-recycling regimes provide.
Here we show gyrokinetic simulation results that indicate that a low-recycling regime could be achieved by using a low recycling material to coat only the side walls, which receive a much lower heat flux than the target. This is an important step towards demonstrating the feasibility of a low-recycling SOL.

The large target heat fluxes and temperatures present in a low-recycling regime pose challenges such as sputtering and erosion. Here, we note that a low-recycling solution would likely have to involve a liquid metal target to avoid detrimental erosion and also begin to investigate ways to address some of these challenges by taking advantage of kinetic effects in chapter~\ref{chap:4}. By comparing fluid and gyrokinetic simulations in a moderately collisionless SOL, we will find that (1) kinetic effects can help confine impurities downstream, which would help avoid core contamination caused by sputtering and (2) the interaction of mirror trapping and drifts can significantly increase the heat flux width and reduce the peak heat flux.  

\chapter{Direct Comparison of Gyrokinetic and Fluid Scrape-Off Layer Simulations}
\label{chap:4}
\footnote{All Gkeyll simulations were performed by me. I developed the Gkeyll code required to be able to use numerical equilibrium from EFIT in simulations. All analysis was performed by me and the paper~\cite{Shukla25} (published in AIP Advances) was written by me.}
Typically, fluid simulations are used to study the dynamics of the Scrape-Off Layer (SOL) in tokamaks~\citep{Hudoba2023,Osawa2023,Rozhansky2021,Zhang2024}.
These simulations are only valid if the SOL is highly collisional. However, fusion pilot plants will exist in a much less collisional parameter regime, and it is critical to understand the implications.

For detached divertor solutions the downstream SOL is collisional enough to warrant fluid assumptions, but the upstream SOL may be less collisional and warrant a kinetic treatment. Upstream parameters for simulations conducted with SOLPS are sometimes in less collisional parameter regimes with an ion mean free path longer than the parallel length scale~\citep{Subba2021,WANG2025}. For attached divertor solutions, even the downstream SOL may require a kinetic treatment.

This chapter reports on the investigation of SOL scenarios for the proposed Spherical Tokamak for Energy Production (STEP)~\citep{Hendrik24, Waldon23, Baker2024}, systematically comparing fluid and gyrokinetic treatments of the parallel SOL physics in order to identify important kinetic effects. In the scenarios we explore, we emphasize kinetic effects in the upstream SOL. In this work, we uncover potential advantages of low collisionality SOL's for impurity transport and confinement and for heat flux broadening.
We believe that these can likely arise even when the downstream conditions are much more collisional. Even in SOLPS simulations with realistic downstream conditions and recycling, less collisional upstream parameters, in which the effects discussed here should be important, are found.
For example, in~\cite{Subba2021} SOLPS simulations of ITER show upstream conditions with a ratio of $\lambda_{mfp,i}/L_c = 1.14 > 1$ where $\lambda_{mfp,i}$ is the ion mean free path and $L_c$ is the connection length, the parallel length scale. In~\cite{WANG2025} SOLPS simulations of the proposed ENN He-Long 2 (EHL-2) spherical tokamak show much less collisional conditions with $\lambda_{mfp,i}/L_c = 22.4 \gg 1$ upstream. Hence, since SOLPS is often run with upstream parameters similar to those investigated here, which correspond to lower collisionality than the domain of validity of fluid equations, it is important to investigate the importance of kinetic effects in such parameter regimes.
 
Due to the limitations of our gyrokinetic model at the time of these simulations, we conducted SOL-only simulations (rather than including the core \& SOL as done in chapter~\ref{chap:3} and also excluded recycling, focusing on scenarios with relevant upstream parameters. We recognize that the downstream parameters of our simulations are not representative of typical detached divertor conditions such as the ones planned for STEP~\citep{Henderson25}. The extent of the kinetic effects highlighted here may be reduced when more collisional downstream conditions are included in simulations, but we believe they will still be important.
In the future, we will conduct simulations with more relevant downstream conditions to further explore the effects described in this paper. Currently, this work serves as a starting point for axisymmetric gyrokinetic simulations as a new tool for divertor design.
The simplified scenarios explored here serve as a good starting point for understanding differences that will be observed in kinetic and fluid treatments of the SOL in more complex and realistic simulations.

We would also like to note that attached downstream conditions could be relevant in several future experiments. Low recycling and attached divertor conditions will be studied in the National Spherical Torus Experiment-Upgrade (NSTX‑U)~\citep{nstx5year}. Plans for Tokamak Energy's ST40 also include lithium wall conditioning and a low recycling regime~\citep{ST402018}. In low recycling scenarios or scenarios with attached divertor conditions, the kinetic effects found here should be particularly relevant.

In our simulations, we show that the ion mean free path is long compared to the parallel temperature gradient length scale and kinetic effects have a large impact.  The simulations consist of 2D axisymmetric SOL simulations  using a fluid code, the B2.5 portion of the SOLPS-ITER~\citep{Schneider2006,SOLPS} package, and a gyrokinetic code, Gkeyll~\citep{Mana25, Mana23, Noah21, shithesis, Tess22, Gkeyllwebsite}. 
We find that the dynamics of the upstream SOL are kinetic—the velocity distributions are not Maxwellian and there is significant mirror trapping of the ions. This strongly increases the upstream ion temperature (roughly by a factor of three). 
We also find that in a magnetic configuration with a Super-X like divertor, the mirror force accelerates particles along the divertor leg resulting in an enhanced potential drop along the field line~\citep{Mike23}. When drifts are included, we find that the interaction of mirror-trapping and drifts drastically boradens the heat flux width and reduces the peak heat flux in kinetic simulations but has a negligible effect in fluid simulations (since the mirror force is neglected. 
Finally, we demonstrate that the assumption of equal ion and impurity temperatures often made in fluid codes is violated. The combination of these effects results in superior confinement of impurities to the divertor region in kinetic simulations.
 
These findings have important implications for both high and low recycling regimes. 
Higher SOL ion temperature might lead to higher pedestal ion temperature. Via stiff transport, this can result in higher core fusion power. This could also increase the energy of neutrals hitting the main chamber wall, due to charge exchange of the hot plasma with cold recycled neutrals. This could increase wall erosion.
The heat-flux broadening effect cause by the interaction of mirror trapping and drifts could also be important for reducing the large heat fluxes characteristic of the low recycling regime mentioned in chapter~\ref{chap:3}. 
Also, the kinetic regimes characteristic of future devices may be able to support
larger downstream impurity densities (and hence more radiated power) than would be suggested by SOLPS. Higher confinement of impurities to the divertor region would entail at least two benefits: (1) avoidance of impurity contamination of the core plasma, and (2) avoidance of high upstream densities, which can degrade confinement according to~\cite{Mike23}.

We explore the differences in the steady state profiles and impurity distribution produced by both codes in several scenarios. We start with a base case comparison of highly collisional simulations in a slab to establish agreement between the two codes. We then investigate cases in the STEP outboard SOL geometry without and with drifts and then with impurities but without drifts.

This chapter is outlined as follows: In section~\ref{sec:setup}, we describe the magnetic geometry, the simulation models used, and the simulations' setup. In section~\ref{sec:slab_comparison}, we establish baseline agreement between SOLPS and Gkeyll in simulations with a slab geometry. In section~\ref{sec:plasma_only}, we compare Gkeyll and SOLPS simulations in the outboard STEP SOL with no impurities or drifts. In section~\ref{sec:plasma_only_drifts}, we compare Gkeyll and SOLPS simulations in the outboard STEP SOL with drifts but without impurities. In section~\ref{sec:impurities}, we compare Gkeyll and SOLPS simulations in the outboard STEP SOL without drifts including argon impurities. Finally, we conclude in section~\ref{sec:conclusion}.

\section{Simulation Setup}\label{sec:setup}
Three types of simulations were employed in this work. First we ran simulations in a slab geometry with only electrons and deuterium ions to establish a baseline agreement between SOLPS and Gkeyll. Second, we ran simulations in the STEP geometry described in sec~\ref{sec:magnetic_geometry} with only electrons and deuterium ions. Finally, we added argon impurities to the STEP simulation and conducted simulations including charge states up to Ar$^{4+}$ and up to Ar$^{8+}$ with various neutral argon densities and temperatures. In this section we will describe the models used by both SOLPS and Gkeyll as well as the magnetic geometry, sources, and boundary conditions used in the simulations. Further details of the resolution and cost of each simulation as well as a link to the input files can be found in appendices~\ref{sec:resolution} and~\ref{sec:cost}.

\subsection{Model Descriptions}\label{sec:model_descriptions}
\subsubsection{Gkeyll}
\label{sec:Gkeyll_model}
Gkeyll is a full-f, long-wavelength gyrokinetic code using a Discontinuous Galerkin (DG) method for spatial discretization and Runge-Kutta for discretization in time. We use the electrostatic version of the code which solves the gyrokinetic equation
\begin{eqnarray}
\frac{\partial f_s}{\partial t}+\dot{\boldsymbol{R}} \cdot \nabla f_s+\dot{v}_{\|} \frac{\partial f_s}{\partial v_{\|}} - \nabla \cdot (\mathbf D \cdot \nabla f_s) = \nonumber \\
C\left[f_s\right]+S_s+C^{iz}_s+C^{rec}_s +  C^{rad}_s,
\label{eq:gkeq}
\end{eqnarray}
with $\dot{\mathbf{R}}=\{\mathbf{R}, H\}, \dot{v}_{\|}=\left\{v_{\|}, H\right\}, \textrm{ and } H_s=\frac{1}{2} m_s v_{\|}^2+\mu B+q_s \phi$
along with the gyrokinetic Poisson equation
\begin{equation}
-\nabla_{\perp} \cdot\left(\sum_s\frac{m_sn_{0s}}{B_0^2} \nabla_{\perp} \phi\right) = \sum_sq_s n_s(\boldsymbol{R}),
\label{eq:poisson}
\end{equation}
where  $f_s = f_s(\mathbf{R}, v_\parallel,\mu.t)$ is the gyrocenter distribution function for species $s$, $\mathbf R$ is the guiding center position, $\mathbf D$ is the particle diffusivity, $v_\parallel$ is the velocity parallel to the magnetic field, $\mu = \frac{mv_\perp^2}{2B}$ is the magnetic moment, $v_\perp$ is the velocity perpendicular to the magnetic field, $B$ is the magnetic field magnitude, $B_0$ is the magnetic field magnitude at the center of the simulation domain, $H_s$ is the gyrocenter center Hamiltonian of species s, $\phi$ is the electrostatic potential, $n_s$ is the guiding center density of species s, $n_{0s}$ is a reference density for species s, $q_i$ is the ion charge, and $e$ is the elementary charge.

The gyrokinetic Poisson bracket is given by~\citep{Noah21}
\begin{equation}
\{F, G\}=\frac{\boldsymbol{B}^*}{m B_{\|}^*} \cdot\left(\nabla F \frac{\partial G}{\partial v_{\|}}-\frac{\partial F}{\partial v_{\|}} \nabla G\right)-\frac{\mathbf{b}}{q B_{\|}^*} \times \nabla F \cdot \nabla G,
\end{equation}
with $\mathbf{B^*} = \mathbf{B} + (mv_\parallel/q)\nabla\times\mathbf{b}$ and $B_\parallel^* = \mathbf{b} \cdot \mathbf{B^*} \approx B$ where $\mathbf{b} = \mathbf{B}/B$ is the unit vector along the magnetic field.

The right-hand side of Eq.~\ref{eq:gkeq} contains the collision term~\citep{Mana22} $C[f_s]$, volumetric source terms $S_s$, ionization and recombination terms $C^{iz}_s$ and $C^{rec}_s$~\citep{Tess22}, and the radiation term $C^{rad}_s$~\citep{radiation}. 
The radiation term removes energy from the electrons based on the electron distribution function and the densities of the electrons and radiating impurities~\citep{Roeltgen25}.
Gkeyll uses the Dougherty collision operator~\citep{Dougherty1964, Mana22} which lacks the full velocity dependence of the full Fokker-Planck collision operator and thus does not match the Spitzer parallel heat conductivity in the collisional limit. Since we are comparing Gkeyll with SOLPS, we have adjusted the collision frequency in Gkeyll such that the parallel heat conductivity matches the Spitzer conductivity in the collisional limit. This amounts to a factor of 4 reduction in the collision frequency in Gkeyll. 

Note that Gkeyll implements the drift-kinetic limit of the gyrokinetic equation, neglecting all gyroaveraging operations. However, first order finite Larmor radius (FLR) effects are present in the ion polarization term in the gyrokinetic poisson equation, which distinguishes the long-wavelength gyrokinetic model from a drift-kinetic model~\citep{Noah21}.

In this paper, we conduct 2D axisymmetric simulations with field aligned coordinates, radial coordinate $\psi$ and poloidal coordinate $\theta$, as described in section~\ref{sec:magnetic_geometry}; our simulation grid does not include the binormal direction. So, like SOLPS, we are not directly simulating the turbulence and instead account for it by including radial diffusion as described in section~\ref{sec:sources}. 
We have also turned off drifts (the $\mathbf{E}\times \mathbf{B}$ drift, the $\nabla \mathbf{B}$ drift, and curvature drift) in both codes to simplify our comparison. In Gkeyll's axisymmetric model, turning off drifts amounts to setting the binormal component of the magnetic field unit vector $\mathbf b$ to zero.
Gkeyll's axisymmetric model does not yet support dynamic neutral species. So, for simulations with impurities, we use a static background neutral species as described in section~\ref{sec:impurity_sources}.


\subsubsection{SOLPS}
The SOLPS-ITER package consists of a 2D multi-fluid transport code, B2.5 and a 3D kinetic Monte Carlo neutral transport code EIRENE. We use only the fluid plasma edge solver B2.5 part of the SOLPS-ITER package described in~\cite{Schneider2006} and~\cite{SOLPS} in our simulations, since we are not evolving the neutral species. 
B2.5 evolves a continuity and momentum equation for each species, a total ion energy equation (one energy equation for all the ion species), and an electron energy equation. 
Section~\ref{sec:impurity_sources} describes how we freeze the neutral species in B2.5. For the artificial parallel flux limiters in SOLPS, we used the default coefficients as we did not have good physical motivation for modifying them. Details of the SOLPS model equations can be found in~\cite{Schneider2006}.

\subsubsection{Differences between Models}
There are 3 main differences between the two models. First, SOLPS makes the fluid assumption of lowest order Maxwellian velocity distributions for all species while Gkeyll does not. Second, SOLPS has just one energy equation for all of the ion species, which means that all ion species will have a common temperature; Gkeyll does not group the ions together so they are free to evolve to have different temperatures if the physics so dictates. Third, the mirror force is included via the parallel acceleration term in Gkeyll but not in SOLPS.
Note that while SOLPS's form of the fluid equations does not include the mirror force, it is possible to include some effects of the mirror force in a fluid code by including temperature anisotropy as is done in the 2D transport code UEDGE~\citep{Zhao2021, Zhao20212, Zhao2022}. 

We recognize that the parameters regimes explored here are outside the applicability of SOLPS's fluid equations and we do not expect the two codes to agree entirely. However, SOLPS is often run with upstream parameters similar to those shown here, so it is important to investigate the importance of kinetic effects in these parameter regimes.

\subsection{Magnetic Geometry}\label{sec:magnetic_geometry}
\begin{figure}
    \centering
    \subfloat[Simulation domain used for STEP SOL simulations. The solid black lines indicate two flux surfaces, the separatrix and the outermost flux surface of the domain. The blue lines indicate the inner and outer radial boundaries of the simulation domain and the orange lines indicate the divertor plates which are the parallel boundaries. The simulation domain shown is 6cm wide at the OMP and the connection length from midplane to divertor plate is approximately 60m.]{
    \includegraphics[width=0.45\textwidth]{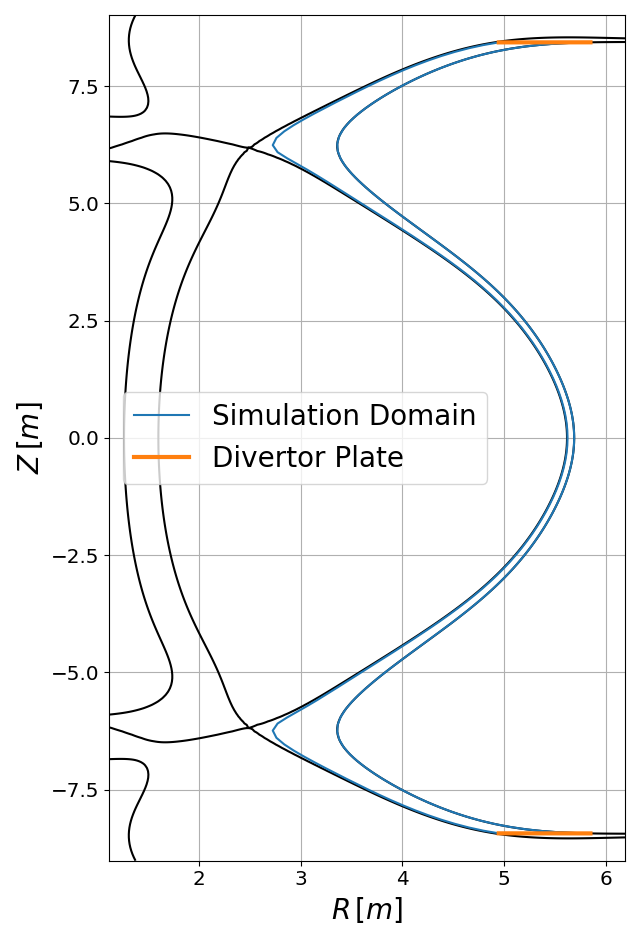}
    }\hspace{0.5cm}
    \subfloat[Simulation domain used for slab simulations. This is a straight SOL which is 6cm wide and 120m long to match the width and connection length of the STEP simulation domain.]{
    \includegraphics[width=0.45\textwidth]{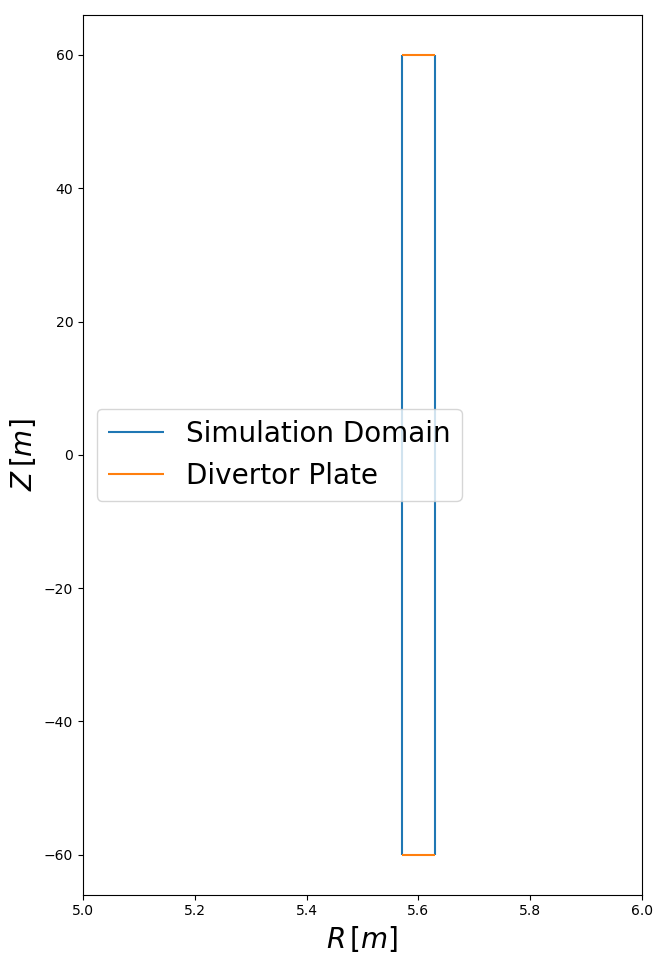}
    }
    \caption{
    STEP (a) and slab (b) simulation domains.
    \label{fig:geometry}
    }
\end{figure}

For simulations in the STEP geometry we use the equilibrium depicted in Fig.~\ref{fig:geometry}. This is a prospective equilibrium for STEP similar to the ones described in~\cite{Hudoba2023}. The parameters of the equilibrium are listed in Table~\ref{tab:params}. The domain of these simulations is just the SOL rather than the core \& SOL like those shown in chapter~\ref{chap:3} because at the time these simulations were run, \gke did not yet have the capability of simultaneously simulating the core \& SOL in X-point geometry.
\begin{table}
  \begin{center}
\def~{\hphantom{0}}
  \begin{tabular}{|l|c|}
      \hline
      $R$        &  4 m    \\
      \hline
      $R_{axis}$ & 4.24 m \\
      \hline
      $a$        &  2 m    \\
      \hline
      $\kappa$   &  3.1    \\
      \hline
      $A$        &  2      \\
      \hline
      $I_p$      &  20.37 MA  \\
      \hline
      $B_{0}$ & 2.88 T     \\
      \hline
  \end{tabular}
  \caption[Magnetic Equilibirum Parameters]{Magnetic Equilibirum Parameters: Major radius $R$, magnetic axis $R_{axis}$, minor radius $a$, elongation $\kappa$, aspect ratio $A$, plasma current $I_p$, toroidal magnetic field at Major radius, $B_0$.
  \label{tab:params}
  }
  \end{center}
\end{table}

The connection length from midplane to divertor plate is approximately 60m. We are targeting an SOL width of 2mm at the outboard midplane (OMP), so we chose to make the simulation domain much wider than that--3cm in Gkeyll and 6cm in SOLPS. We found that this width was enough to decouple the effect of the radial boundary conditions described in section~\ref{sec:BCs} from the dynamics in the main heat channel in both codes.

Gkeyll uses a field aligned coordinate system with poloidal flux ($\psi$) as the radial coordinate and normalized poloidal arc length ($\theta$) as the parallel coordinate. The parallel coordinate is normalized such that the upper and lower divertors are located at $\theta= \pm \pi$ respectively. The poloidal flux at the separatrix is $\psi_{sep} = 1.50982$ and $\psi$ increases toward the magnetic axis. Throughout this paper, we plot profiles using Gkeyll's coordinate system ($\psi, \theta$).

Gkeyll and SOLPS STEP simulations are both radially centered at the flux surface $\psi=1.2014$ which is 3.7cm away from the separatrix at the outboard midplane. The SOLPS simulation domain extends from $\psi = 0.934$ to $\psi = 1.469$ and the Gkeyll simulation domain extends from $\psi = 1.068$ to $\psi = 1.335$.

For the simulation in slab geometry, we use a box which is 6cm wide in the radial direction and 120m long in the direction parallel to the magnetic field to mimic the STEP SOL simulations. In Gkeyll, constructing the slab geometry shown in Fig.~\ref{fig:geometry}b is straightforward. In SOLPS the slab geometry is created using the same method (and scripts) as the isolated divertor in~\cite{Moulton2017}, but the radius is set to 5500m so the toroidal B field is essentially constant. The field ratio is then set based on the radius to get the desired poloidal magnetic field.

\subsection{Sources \& Diffusion}\label{sec:sources}
For simulations in both the slab and STEP geometry, we use a particle source which is a Gausian centered at the radial center of the domain with a 0.76mm half-width at the OMP. The particle source is also a Gaussian along the field line constructed such that its value at the X-point is $10^2$ times smaller than its peak value at the OMP. Since the slab simulation does not have X-points, we construct the source so that it decays by a factor of $10^5$ 40m from the OMP. This decay length is approximately the same as in the STEP geometry.

\begin{figure}
    \subfloat[]{
    \includegraphics[width=0.45\textwidth]{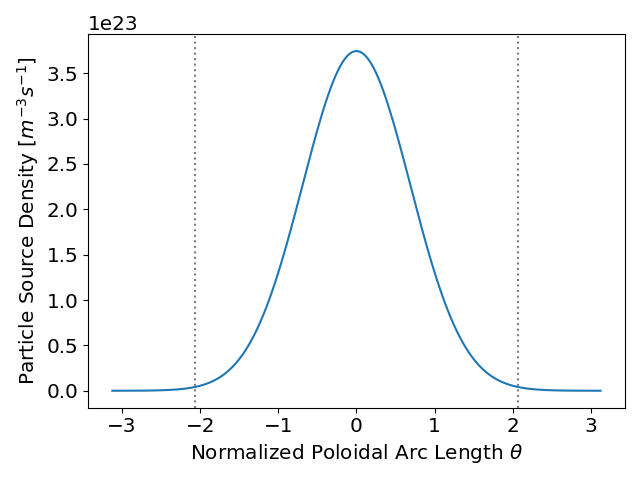}
    }
    \subfloat[]{
    \includegraphics[width=0.45\textwidth]{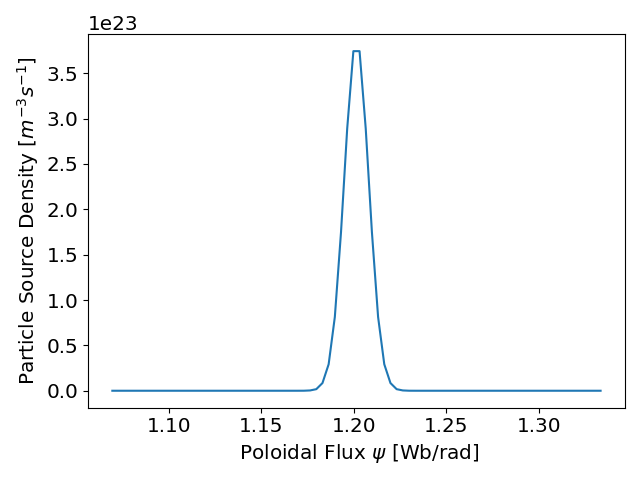}
    }
    \caption[Particle source density]{
    Particle source density plotted along the field line at the radial center of the domain (a), and radially at the midplane (b). The source is constructed such that the value at the X-points, located at  $\theta = \pm$ 2.07 (marked by the gray dotted vertical lines), is 10$^2$ times smaller than the peak value. The radial half-width of the source at the OMP is 0.76mm.
    \label{fig:sourcedensity_plasma}
    }
\end{figure}

The source is localized between the X-points to mimic leakage of particles from the core into the SOL. We chose to center it at the radial center of the domain rather than the inner radial edge to decouple the effect of the boundary conditions from the dynamics in the main heat channel. In the radial direction, we chose the minimum width that could be represented without introducing numerical issues. We chose the minimum width because, in reality, a particle source in the SOL has no width; the particle source is radial diffusion of particles from the core.
The source density profiles, which are common to SOLPS and Gkeyll, are plotted radially and along the field line in Fig.~\ref{fig:sourcedensity_plasma} and are given by
\begin{equation}
n_{source}(R,Z) = \dot{n}_{peak}exp\{ -(R-R_{center})^2/2c_R^2 \} exp\{ -Z^2/2c_Z^2 \}
\end{equation}
with $\dot{n}_{peak} = 6.1 \times 10^{23} m^{-3}s^{-1}$, $c_R = 0.00074$ m, $c_Z = 7.22$ m, and $R_{center} = 0.003$ m for the slab simulations and
\begin{equation}
n_{source}(\psi,\theta) = \dot{n}_{peak}exp\{ -(\psi-\psi_{center}^2/2c_\psi^2 \}exp\{ -\theta^2/2c_\theta^2 \}
\end{equation}
with $\dot{n}_{peak} = 3.9 \times 10^{23} m^{-3}s^{-1}$, $c_\psi = 0.0065$ Wb/rad, $c_\theta = 0.688$, and $\psi_{center} = 1.2014$ Wb/rad for the STEP simulations.

We were targeting upstream parameters typical of a conventional high recycling regime, an upstream density of $3 \times 10^{19} m^{-3}$ and 80MW of input power, so we adjusted the particle source amplitude and temperature accordingly. The temperature of the source is 1.037 keV for both electrons and deuterium. The choice of 80 MW of input power is based on 100 MW of input power crossing the separatrix~\citep{Hudoba2023} with 20 MW going to the inboard SOL and 80 MW going to the outboard.

We ran both Gkeyll and SOLPS without drifts and with constant radial diffusion coefficients. The particle diffusivity was $D_\perp = 0.03\, m^2/s$ and heat diffusivity was $\chi_\perp = 0.045 \,m^2/s$. These diffusivities were chosen in combination with the particle source width so that the half width of heat flux channel was approximately 2mm in the steady state. The proportion $\chi_\perp/D_\perp = 3/2$ is fixed by Gkeyll's implementation of diffusion; Gkeyll's diffusion coefficient, $\mathbf D$, introduced in section~\ref{sec:Gkeyll_model} is applied to the distribution function and does not depend on velocity. In the future we plan to improve Gkeyll's diffusion operator by adding velocity dependence to $\mathbf D$ to allow for decoupling of the particle and heat diffusivities.

\subsection{Sources of Impurities}
\label{sec:impurity_sources}
\begin{figure}
    \includegraphics[width=\columnwidth]{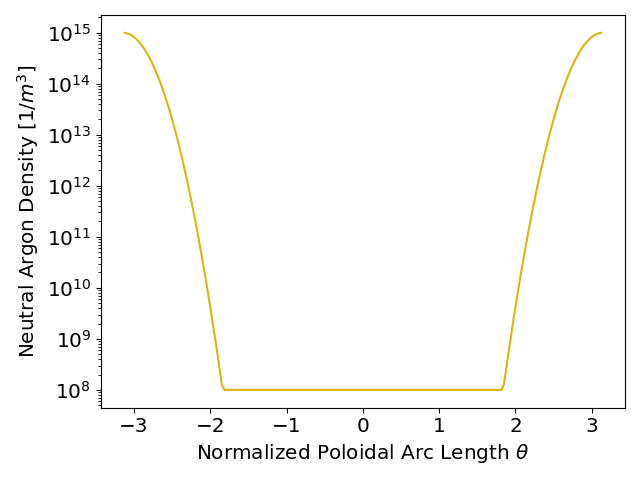}
    \caption[Neutral argon profile along the field line]{Neutral argon profile along the field line. The profile is uniform in $\psi$. The shape of the profile, peaked near the divertor plates, mimics gas puffing and recycling in the target region. In all simulations, we use the same shape for the profile and scale it to have a specified density at the divertor plate. The minimum density for the profile is always $n = 10^8 \textrm{ m}^{-3}$.
    \label{fig:Ar0}
    }
\end{figure}

In simulations with impurities we use a static background neutral argon profile concentrated downstream near the divertor plates as shown in Fig.~\ref{fig:Ar0}. The neutral argon profile is uniform in the radial direction ($\psi$) and is given by
\begin{equation}
n(\theta) =
\begin{cases}
    max(n_{peak} exp\{ -(\theta-\pi)^2/2c_\theta^2 \}, 10^8), & \theta \geq 0 \\
    max(n_{peak} exp\{ -(\theta+\pi)^2/2c_\theta^2 \}, 10^8), & \theta < 0 \\
\end{cases}
\end{equation}
with $n_{peak}$ varying between simulations and $c_{\theta} = 0.25$.
We include ionization and recombination for the neutral and charged argon species. The shape of the argon neutral profile is meant to mimic puffing and recycling near the divertor plate.

Gkeyll explicitly supports static neutral species while SOLPS does not. In SOLPS, we approximate a static neutral species by turning off the continuity and momentum equations for the fluid neutrals. The fluid neutrals still enter into the ion energy equation, but we expect the effect on the ion energy equation to be minimal due to the comparatively low neutral density.

We acknowledge that including dynamic neutrals in Gkeyll and using the full SOLPS-ITER package (B2.5 coupled to EIRENE) would be preferable to running these codes with static neutrals. However, Gkeyll's axisymmetric model does not currently support dynamic neutrals. We believe that, qualitatively, the major results of this work will hold even when dynamic  neutrals are included. In a future work, we will conduct a comparison including dynamic neutrals.

It is also important to note that, in a realistic scenario, neutral recycling would provide a large ionization source of deuterium downstream near the divertor plates. As noted in the introduction, the current limitations of Gkeyll prevent us from exploring those scenarios. In the future we plan to carry out simulations including a recycling source.  

\subsection{Boundary Conditions}
\label{sec:BCs}
We adjusted the boundary conditions in the two codes to be as similar as possible. In the parallel (poloidal) direction, we use sheath boundary conditions on the particles. 
In Gkeyll, there is no boundary condition on the potential in the parallel direction~\citep{shithesis}. In Gkeyll's conducting sheath boundary condition, the potential at the sheath entrance, $\phi_{sh}$, is obtained from the gyrokinetic Poisson equation, Eq.~\ref{eq:poisson}, and electrons with energy less than $e\phi_{sh}$ are reflected back into the simulation domain~\citep{shithesis, Shi17}. 
In SOLPS STEP simulations, we use the standard sheath boundary condition, option 15 in the SOLPS manual~\citep{Schneider2006}, which enforces the Bohm sheath condition at the sheath entrance~\citep{Schneider2006}.
In SOLPS slab simulations, we used an alternative sheath boundary condition for the energy equation that allows one to specify the sheath heat transmission coefficient for each species.
This is option 12 in the SOLPS manual~\citep{Schneider2006} and enforces a heat flux to the sheath entrance of $Q_s = \delta_s \Gamma_s T_s$ where $Q_s$ is the heat flux of species s, $\delta_s$ is the transmission coefficient for  species s, $T_s$ is the temperature of species s, and $\Gamma_s$ is the particle flux of species s. We used this boundary condition to match the electron and ion transmission coefficients in SOLPS to the radially averaged transmission coefficients observed in Gkeyll which were 4.62 and 2.88 respectively.

In Gkeyll, we apply absorbing boundary conditions to the particles in the radial direction--all particle species leave freely at the radial boundaries. We apply Dirichlet boundary conditions, $\phi=0$, to the potential at both radial boundaries for both Gkeyll and SOLPS.
For SOLPS, we use leakage boundary conditions in the radial direction with a large leakage coefficient of -1000 for both the continuity and energy equations.
The leakage boundary condition allows one to specify the radial velocity at the boundary as the leakage coefficient times the species' thermal velocity and the radial particle flux to the boundary as a leakage coefficient times the density times the sound speed. This leakage boundary condition with a large coefficient mimics the absorbing radial boundary conditions used in Gkeyll.
For the radial boundary condition of the momentum equation, we set the radial derivative of the momentum to zero to minimize the effect of the boundary condition on the momentum. Using a leakage boundary condition in the momentum equation would lead to an unphysically large velocity.

\section{Slab Comparison}\label{sec:slab_comparison}
As a base case comparison to establish reasonable agreement, we set up a simulation in a 2D (radial, poloidal) slab geometry without drifts. For this simulation we reduced the input power from 80MW to 50MW to target a more collisional regime where the codes are likely to agree. 
These simulations serve to verify that we are in fact using equivalent boundary conditions, sourcing, and diffusion in the two codes.

\begin{figure}
  \subfloat[]{\label{fig:slab_density_x}
    \includegraphics[width=0.32\textwidth]{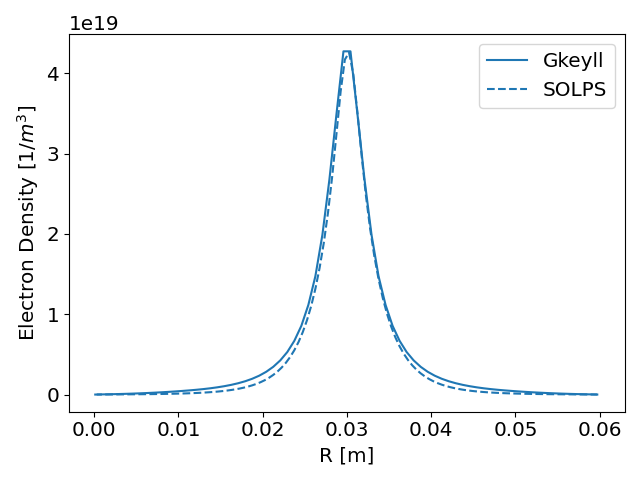}
  }
  \subfloat[]{\label{fig:slab_density_z}
    \includegraphics[width=0.32\textwidth]{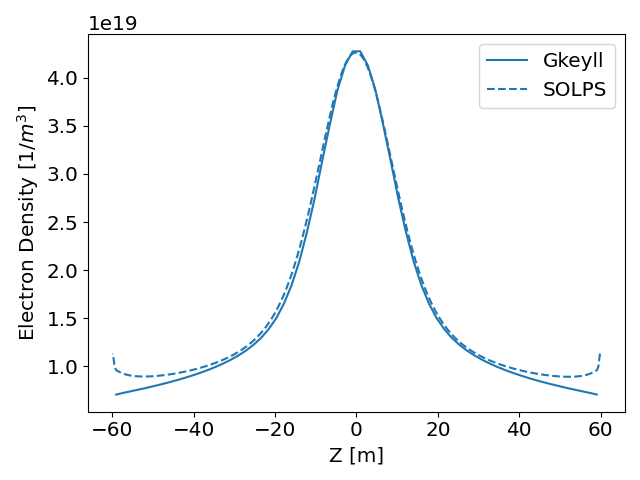}
  }
  \subfloat[]{\label{fig:slab_temp_z}
    \includegraphics[width=0.32\textwidth]{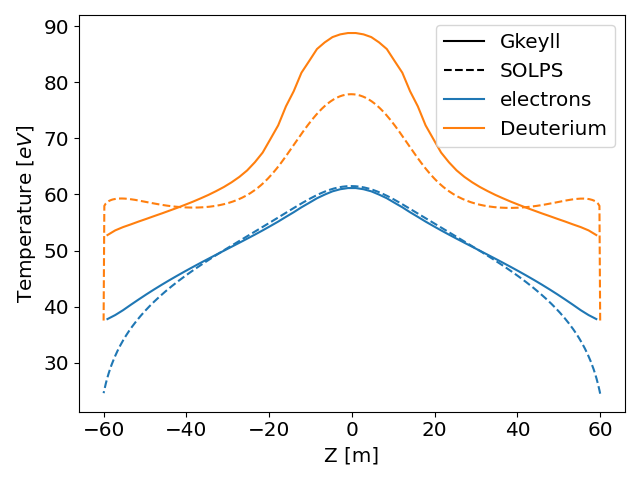}
  }
  \caption[Electron density plotted (a) radially (vs. R) and (b) along the field line (vs. Z)]{
  Electron density plotted (a) radially (vs. R) and (b) along the field line (vs. Z)  and electron and ion temperature plotted 
  and (c) along the field line
  for both SOLPS and Gkeyll from the simulations with slab geometry. The density profiles are similar, but kinetic effects influence the parallel heat conduction and raise the ion temperature in Gkeyll relative to SOLPS.
    \label{fig:slab_moments}
    }
\end{figure}

In Fig.~\ref{fig:slab_moments}, we plot the steady state temperature and density profiles along the field line and radially. The density and temperature profiles in SOLPS and Gkeyll agree quite well. 
The peak ion temperature is 12\% larger in Gkeyll.
This magnitude of difference could possibly be due to long mean free path effects upon the thermal conduction, since this is known to be rather sensitive to such effects. Another explanation could be due to a difference in the electron-ion energy equilibration term caused by our reduction in the collision frequency. However, we conducted tests with the original collision frequency restored in Gkeyll, and these tests did not remove the temperature difference between SOLPS and Gkeyll, so we find the first explanation more likely.
Although the Dougherty operator produces the correct ion-electron equilibration behavior~\citep{Mana22}, the reduction in the collision frequency affects this behavior. We chose to reduce the collision frequency as described in section~\ref{sec:Gkeyll_model} to match the parallel heat conduction which will be more important than the electron-ion energy transfer in the less collisional simulations conducted in the STEP geometry. However in this collisional slab case, the electron-ion energy transfer term is dominant. The ratio of the divergence of the ion heat conductivity and the electron-ion energy transfer term can be estimated as~\citep{Braginskii1965}

\begin{equation}
    \frac{\nabla_\parallel \mathbf q_i}{Q_\Delta} \sim \sqrt{\frac{m_e}{m_i}}\Big(\frac{T_e}{T_i}\Big)^{3/2}\frac{\Delta T_i}{T_e-T_i} \Big(\frac{\lambda_{mfp,i}}{L_\parallel}\Big)^2.
\end{equation}

\begin{figure}
        \includegraphics[width=\textwidth]{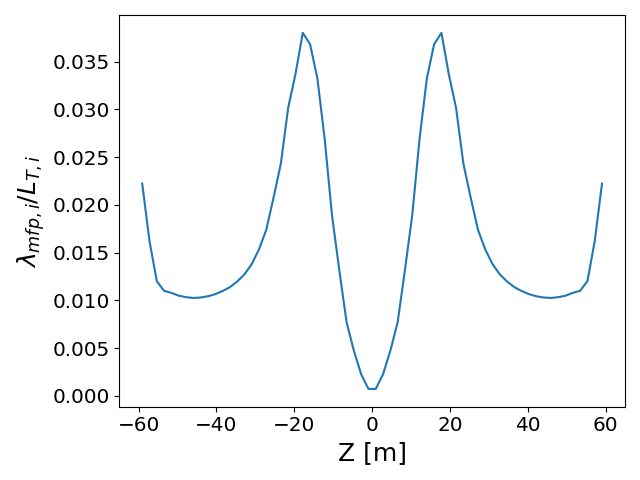}
    \caption[Ion mean free  path normalized to $L_T$ from the Gkeyll simulation with slab geometry]{
    Ion mean free  path normalized to $L_T$ from the Gkeyll simulation with slab geometry plotted along the field line at the radial center.
    \label{fig:mfp_slab}
    }
\end{figure}

In this case, this ratio is $\sim 10^{-2}$ so the electron-ion energy transfer can be comparable to or even dominate over ion heat conduction. However in the simulations in STEP geometry in the following sections, this ratio is $\sim 10^3$ so the ion heat conduction dominates. Fig.~\ref{fig:mfp_slab} shows the ion mean free path normalized to the parallel length scale, $L_{\parallel} = L_T$, where $L_T$ is the parallel temperature gradient scale length.

Having established baseline agreement in this simple and more collisional case, we will now move on to less collisional simulations in the STEP magnetic geometry. Note that the difference in the electron-ion energy equilibration term described in this section should not be important in the following simulations because this term is very small compared to the ion heat conduction.

\section{Plasma Only Simulations}
\subsection{Without Drifts}
\label{sec:plasma_only}
As a next step, we ran simulations in the STEP SOL geometry. 
Since we kept drifts turned off for both codes and matched the diffusivities, differences between the two simulations beyond what was seen in section~\ref{sec:slab_comparison} can be attributed to the mirror force and non-Maxwellian velocity distributions.

We note that it would be preferable to start with a lower power higher collisionality case with 50MW of input power like the slab case in Sec.~\ref{sec:slab_comparison} to establish agreement in the STEP geometry. However, at high collisionality, Gkeyll's explicit collision operator causes a severe limitation on the timestep and it was not feasible to run a higher collisionality case in the STEP geometry. So, we instead start with 80MW of input power. 

\begin{figure}
    \subfloat[\label{fig:density_plasma_z}]{
        \includegraphics[width=0.45\textwidth]{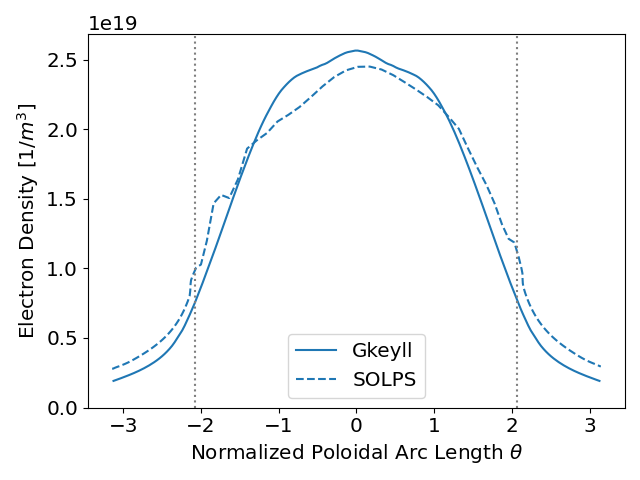}
    }
    \subfloat[]{
        \includegraphics[width=0.45\textwidth]{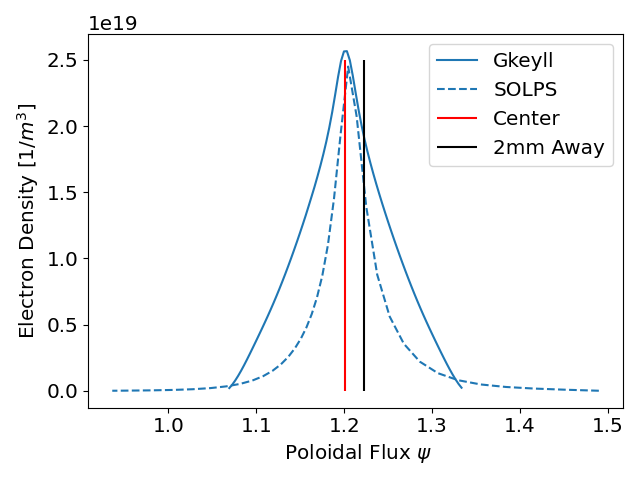}
    }
    \caption[Electron density plotted along the field line at the radial center of the domain]{
    Electron density plotted along the field line at the radial center of the domain (a), and radially at the midplane (b) for Gkeyll (solid) and SOLPS (dashed) simulations in STEP geometry. The vertical red line marks the flux surface at the radial center of the simulation at the OMP and the black line marks a surface 2mm away from the radial center. The density profiles given by Gkeyll and SOLPS are similar along  the field, but Gkeyll's radial density profile is slightly wider due to mirror trapping of the ions which broadens the electron density to maintain quasineutrality.
    \label{fig:density_plasma}
    }
\end{figure}

First we compare the steady state density profiles both along the field line at the radial center and radially at the midplane shown in Fig.~\ref{fig:density_plasma}. The profiles are similar for SOLPS and Gkeyll.
However, the radial density profile is wider in Gkeyll due to mirror trapping of the ions. The electron density broadens along with the ion profile to maintain quasineutrality. 
One can estimate the broadening of the density due to mirror trapping based on the ion collision time, $\tau_i$ and the particle diffusivity, $D_\perp$. The ion collision time is $\tau_i = 2.7\times 10^{-3}\, s$ and the particle diffusivity is $D_\perp = 0.03\, m^2/s$. If trapped ions take an ion collision time to de-trap, they will diffuse a radial distance of $l_{D,coll} = \sqrt{D _\perp\tau_i} = 9\, mm$ before de-trapping. At the radial center of the domain, the maximum magnetic field is $B_{max} = 3.69 \,T$ and the minimum is $B_{min} = 2.36\,T$ which gives an approximate trapped fraction of $f_{trap} = \sqrt{1-B_{min}/B_{max}} = 0.6$. The ion transit time is $\tau_{transit} = 2.5\times 10^{-4} s$, so passing ions will diffuse $l_{D,transit} = \sqrt{D_\perp \tau_{transit}} = 2.7 \, mm$ before being lost. A simple estimate using a weighted average would give an expected ion density half width of $w = f_{trap}l_{D,coll} + (1-f_{trap})l_{D,transit} = 6.5\,mm$. In Gkeyll we observe approximately this amount of broadening; in Fig.~\ref{fig:density_plasma} the radial half width of the Gkeyll density is $6\, mm$. 
The radial half width of the SOLPS electron density is $2.7\, mm$  which is equal to $l_{D,transit}$, consistent with an assumption of no trapped particles. Looked at from another perspective, the kinetic parallel dynamics (mirror trapping) effectively increase the perpendicular particle diffusivity. A simple estimate of the effective diffusivity in Gkeyll can be calculated from the Gkeyll density half width and the ion transit time as $D_{\perp,effective} = w^2/\tau_{transit} = 0.135\, m^2/s$. One could try to incorporate this effect into a fluid code by estimating and using an effective perpendicular diffusivity caused by mirror trapping.

\begin{figure}
    \subfloat[\label{fig:temperature_plasma_z}]{
        \includegraphics[width=0.45\textwidth]{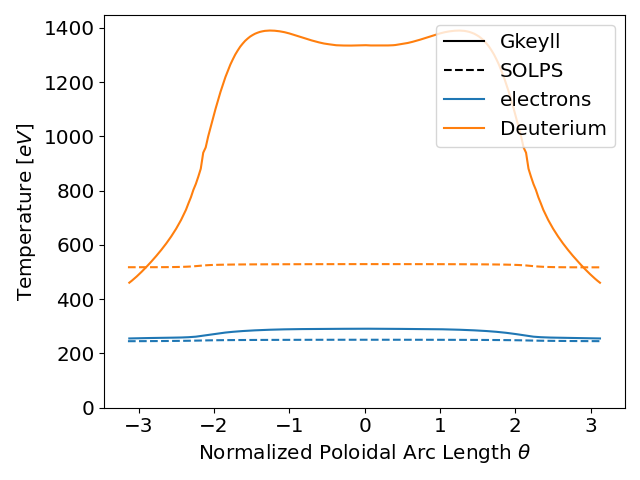}
    }
    \subfloat[]{
        \includegraphics[width=0.45\textwidth]{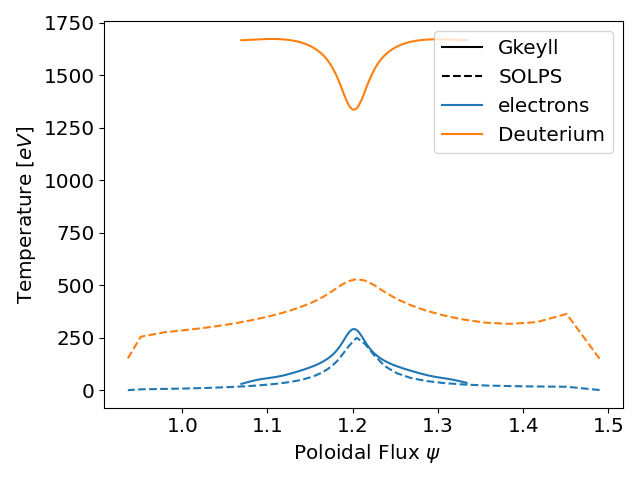}
    }
    \caption[Electron and ion temperature plotted along the field line at the radial center]{
    Electron and ion temperature plotted along the field line at the radial center of the domain (a) and radially at the midplane (b) for Gkeyll (solid) and SOLPS (dashed) simulations in STEP geometry. The upstream ion temperature is much higher in Gkeyll because the distribution function is depleted at large $v_\parallel$ and heat leaves more slowly along the field line.
    \label{fig:temperature_plasma}
    }
\end{figure}

\begin{figure}
    \includegraphics[width=\textwidth]{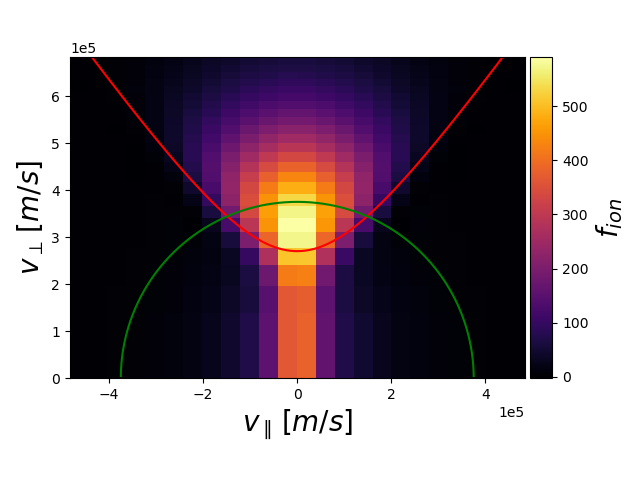}
    \caption[Ion distribution function 2mm away from the radial center at the midplane.]{Ion distribution function 2mm away from the radial center at the midplane. The red line indicates the trapped-passing boundary including the effect of the potential and the green line is a contour of a Maxwellian with the same temperature as the distribution function. The ions are clearly not Maxwellian; they consist primarily of trapped particles. The depletion of the distribution function at large $v_\parallel$ is responsible for the difference in ion temperature between Gkeyll and SOLPS.
    \label{fig:ion_upstream}
    }
\end{figure}

Next we examine the temperature profiles in Fig.~\ref{fig:temperature_plasma} and see an obvious difference in the ion temperatures - the Gkeyll ion temperature is much higher.
The effect responsible for this difference is mirror trapping of the ions. The mean free path is long relative to the the connection length ($\lambda_{mfp,i}/L_\parallel = 12.2$ in Gkeyll and 2.3 in SOLPS upstream), so we should see mirror trapping of the ions upstream. The trapping will result in an ion distribution function that is depleted at large $v_\parallel$. This depletion will cause heat to move more slowly along the field line. Since heat leaves more slowly, for the same heating power, the result is a higher ion temperature in the Gkeyll simulations.
If we inspect the distribution function in Fig.~\ref{fig:ion_upstream}, we can see that the ions are clearly not Maxwellian and are instead trapped.
In Fig.~\ref{fig:ion_upstream}, the red line indicates the trapped-passing boundary
including the effect of the potential and the green line is a contour
of a Maxwellian with the same temperature as the distribution function.


An enhancement of the potential drop along the divertor leg due to the mirror force in a Super-X like divertor configuration is predicted in~\cite{Mike23}. The long outer leg results in a drop in $B$ along the field line. Thus, the mirror force accelerates particles toward the divertor plate. This acceleration results in an increased average velocity parallel to the field line, $u_\parallel$, and is known as the "magnetic Laval nozzle" effect in magnetic mirrors~\citep{Xianzhu14,Andersen2003}. Particle conservation requires that $nu_\parallel/B$ remains constant, so the density must drop along the field line. If electrons have approximately a Boltzmann response, this will result in a potential drop along the field line.

\begin{figure}
    \centering
    \includegraphics[width=\textwidth]{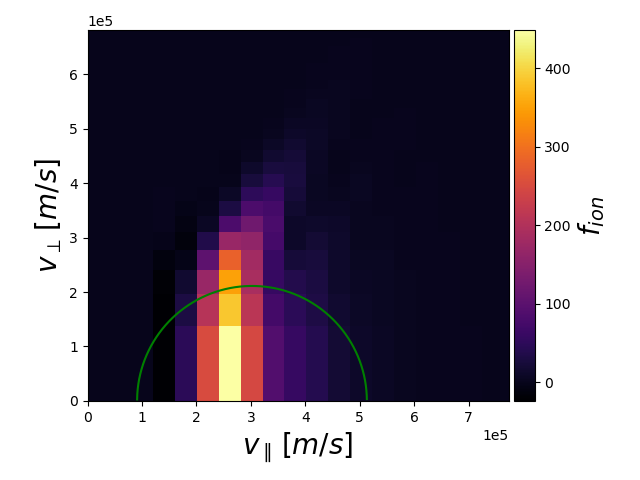}
    \caption[Ion distribution function downstream near the divertor plate]{
    Ion distribution function downstream near the divertor plate 2mm away from the radial center. The green line is a contour of a Maxwellian with the same temperature as the distribution function. Ions with large $v_\perp$ are pushed towards higher $v_\parallel$ by the mirror force as can be seen by the way the distribution function is skewed to the right at large $v_\perp$. The mirror force is responsible for raising the ion exit velocity in Gkeyll relative to SOLPS.
    \label{fig:ion_downstream}
    }
\end{figure}

\begin{figure}
    \centering
    \includegraphics[width=\textwidth]{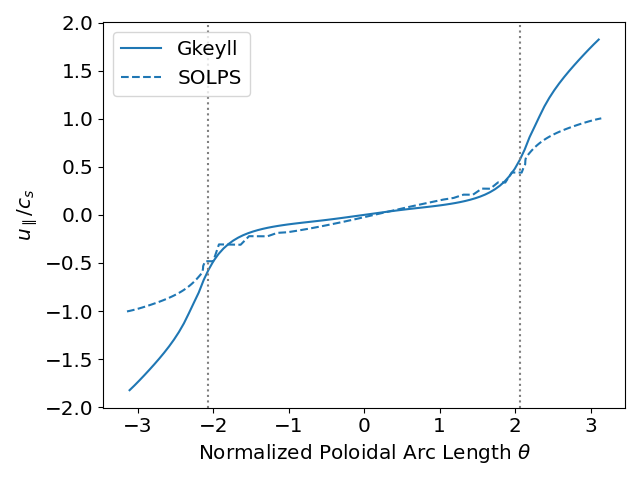}
    \caption[Average ion parallel velocity, $u_\parallel$, normalized to the sound speed]{
    Average ion parallel velocity, $u_\parallel$, normalized to the sound speed, $c_s = \sqrt{(T_e + T_i)/m_i}$, plotted along the field line for SOLPS and Gkeyll simulations in STEP geometry. The X-points are located at $\theta  =\pm$ 2.07 (marked by the vertical gray dotted lines) and the divertor plates are located at $\theta = \pm\pi$. The ion velocity in Gkeyll exceeds that of SOLPS at the divertor plates due to acceleration by the mirror force along the divertor legs.
    \label{fig:upar_plasma}
    }
\end{figure}

\begin{figure}
    \includegraphics[width=\textwidth]{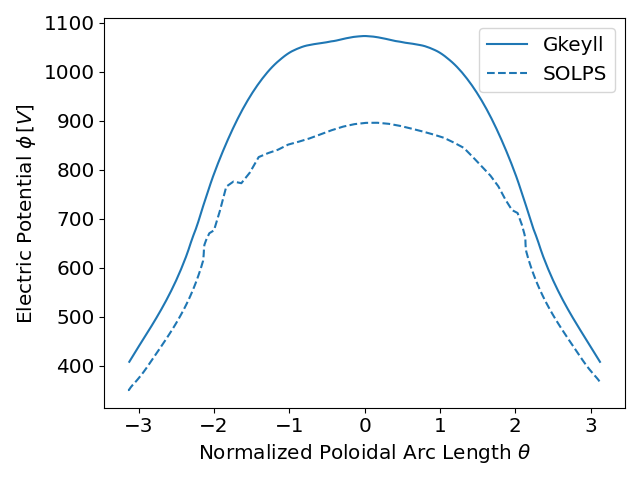}
    \caption[Electrostatic potential, $\phi$, plotted along the field line]{
    Electrostatic potential, $\phi$, plotted along the field line at the radial center for SOLPS and Gkeyll simulations in STEP geometry. The potential drop from the midplane ($\theta=0$) to the divertor plate ($\theta = \pm \pi$) is 26\% larger in Gkeyll. The increased ion parallel velocity in Gkeyll results in this enhanced potential drop.  
    \label{fig:phi_plasma}
    }
\end{figure}

This effect can be seen clearly in figures~\ref{fig:ion_downstream},~\ref{fig:upar_plasma}, ~\ref{fig:density_plasma_z}, and~\ref{fig:phi_plasma}. In Fig~\ref{fig:ion_downstream} we see that high $\mu$ particles are pushed towards high $v_\parallel$. This is due to the mirror force accelerating particles towards the divertor plate. In Fig.~\ref{fig:upar_plasma} we see that in the region beyond the X-point, the Gkeyll $u_\parallel$ exceeds that of SOLPS as predicted. In SOLPS, the Bohm condition, $u_\parallel = c_s$ where $c_s = \sqrt{(T_e + T_i)/m_i}$ is the local sound speed, is enforced at the divertor plate, but in Gkeyll the ion outflow is supersonic.
This effect has also been explored in simplified magnetic geometries in~\cite{Xianzhu14} and~\cite{Sabo2022}. 
The effect on the density drop along the divertor leg is not dramatic but can be seen in Fig.~\ref{fig:density_plasma_z}. We can calculate the expected ratio of the density near the X-point to the density at the plate based on the requirement that $nu_\parallel/B$ is equal at the X-point and divertor plate, $u_\parallel$ from the codes, and the magnetic field. This calculation gives an expected ratio of density near the X-point to density at the plate of 3.1 in SOLPS and 4.1 in Gkeyll. In Fig.~\ref{fig:density_plasma_z} we observe a ratio of 3.4 in SOLPS and 4.1 in Gkeyll.
Finally, we see the effect of the increased ion outflow velocity and resultant density drop on the potential in Fig.~\ref{fig:phi_plasma}. The effect is moderate with a 26\% larger potential drop from midplane to divertor plate in Gkeyll.


\subsection{With Drifts}
\label{sec:plasma_only_drifts}
We also conducted a set of SOL-only simulations using both SOLPS and Gkeyll with and without drifts to investigate the effect of drifts on the peak heat flux and heat flux width. These simulations are identical to those described in section~\ref{sec:plasma_only} except that drifts are included and the magnetic field is halved to emphasize the effect of drifts. We found that while drifts had a negligible effect on the heat flux width in fluid simulations, there was signifcant broadening in kinetic simulations. In kinetic simulations, drifts caused the heat flux channel to spread out to approximately the ion banana orbit width (or poloidal ion gyroradius). This is due to mirror trapping: in a collisionless regime, an ST has a significant trapped fraction on the outboard, and ions spread out radially during banana orbits before being de-trapped and lost to the plate. Because the mirror force is neglected by SOLPS's fluid equations, this effect is only observed in the kinetic simulations. As seen in the previous section~\ref{sec:plasma_only}, Fig.~\ref{fig:ion_upstream} clearly shows that the ions are trapped upstream.

\begin{figure}[htbp]
\centering
\includegraphics[width=0.45\textwidth]{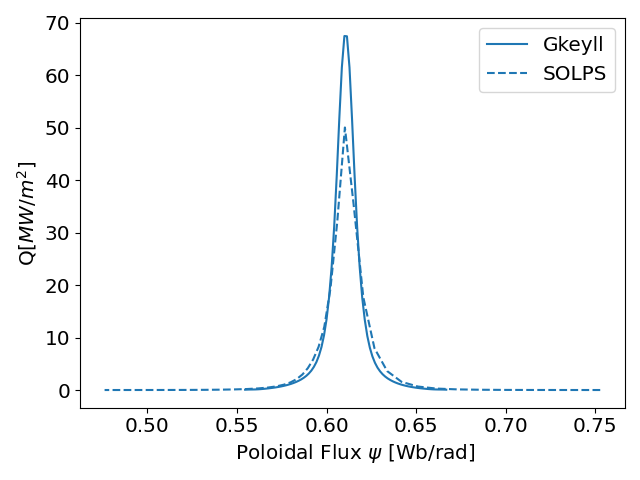}
\includegraphics[width=0.45\textwidth]{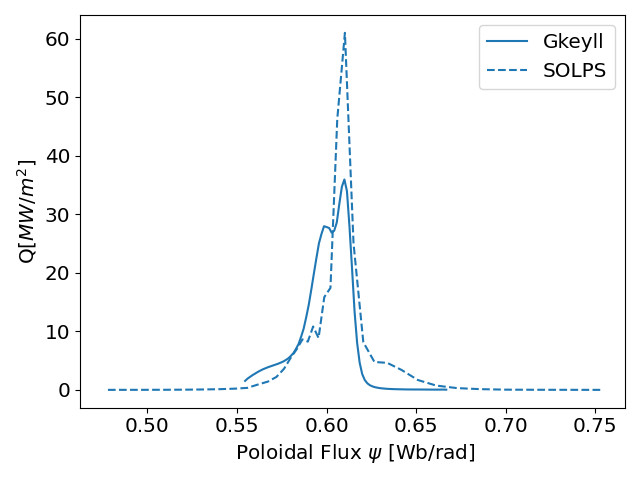}
\caption[Heat Flux at upper outboard plate from SOLPS and Gkeyll simulations without drifts (left) and with drifts (right).]{Heat Flux at upper outboard plate from SOLPS and Gkeyll simulations without drifts (left) and with drifts (right). Without drifts, both codes have a heat flux width (mapped upstream) of 3mm. With drifts, the SOLPS width remains at 3mm, but the Gkeyll width is increased to the 6.4mm, which is the poloidal ion gyroradius, and the peak heat flux is reduced by a factor of 2.
}
\label{fig:drifts}
\end{figure}

In Fig~\ref{fig:drifts} we plot the heat flux incident at the outboard plate from Gkeyll and SOLPS simulations with and without drifts. The heat flux width in SOLPS is 3 mm with and without drifts; the width is barely affected by the inclusion of drifts. In Gkeyll, the heat flux width is 3 mm without drifts but 6.4 mm (the ion banana width) with drifts and the peak heat flux is reduced by a factor of 2. This result indicates that in a collisionless low-recycling SOL, we can take advantage of the kinetic parallel dynamics to broaden the heat flux channel and reduce peak heat loads. In an even less collisional, hotter SOL, like the 10keV case simulated in section~\ref{sec:10kev}, the ion banana width is 1 cm, so the broadening effect would likely be larger as well.

\section{Simulations with Argon Impurities}\label{sec:impurities}

\subsection{500 eV Neutral Argon}
\label{sec:hotar}
In order to establish a baseline comparison, we added a static 500 eV neutral argon background localized near the divertor plate. The neutral argon profile is shown in Fig.~\ref{fig:Ar0}.
In the initial conditions of the simulation, the density of all charged argon states is zero. The only sources of Ar$^{+1}$ are ionization of neutral argon and recombination of Ar$^{+2}$, the only sources of Ar$^{2+}$ are ionization of Ar$^{+1}$ and recombination of Ar$^{+3}$, and so on.
We chose 500 eV because the deuterium ion temperature (and thus the argon ion temperature) in SOLPS is approximately 500 eV. The argon density in this simulation is sufficiently low that the inclusion of argon does not significantly affect the deuterium or electron profiles. Due to the computational cost of including more charge states in Gkeyll, especially for the higher radiation scenario described in Sec.~\ref{sec:high_frac}, we included only up to argon charge state 4. In appendix~\ref{sec:ar8} we show one case with up to charge state 8 and show that the change in upstream and downstream argon density is negligible compared to the case which includes only up to argon charge state 4.   

In this case, we expect the downstream and upstream argon temperatures in Gkeyll and SOLPS to aree reasonably well.
The neutral argon profiles in SOLPS and Gkeyll are identical and the electron density and temperature profiles are very similar as seen in Sec.~\ref{sec:plasma_only}. 

Thus, the differences in the steady state profiles for argon charge states 1 through 4 can be mostly attributed to the difference in the potential drop described in the previous section as well as the difference in argon and ion temperatures included in Gkeyll. In SOLPS, the argon is assumed to be the same temperature as the main ion species, deuterium. This assumption is not a choice made by the user; SOLPS has only one ion energy equation, so there is only one ion temperature.

\begin{figure}
    \includegraphics[width=\columnwidth]{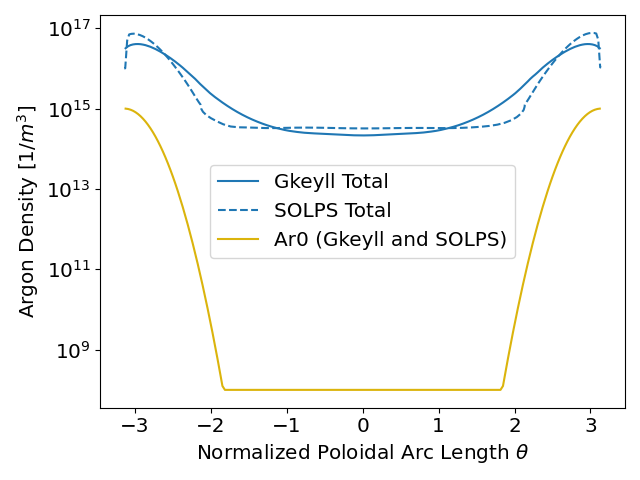}
    \caption[Neutral and charged argon density]{
    Neutral and charged argon density (summed over charge states 1 through 4) plotted along the field line in SOLPS and Gkeyll for the Gkeyll simulation with 500 eV neutral argon and charge states up to Ar$^{4+}$. In this case the upstream argon densities in Gkeyll and SOLPS are only a factor of 2 different. This is expected and indicates that the low impurity temperature is primarily responsible for the huge differences in upstream density we saw with 10eV neutral argon.
    \label{fig:density_arhot_total}
    }
\end{figure}

As seen in Fig.~\ref{fig:density_arhot_total}, the upstream and downstream argon densities in Gkeyll are approximately a factor of 2 lower in Gkeyll. We cannot expect complete agreement in this case due to the difference in the potential drop between Gkeyll and SOLPS. Instead, this case serves as a baseline for the comparison in the next section where we will reduce the temperature of the neutral argon to 10eV. In this next case we will see much larger differences in the upstream argon density.



\subsection{10 eV Neutral Argon up to Ar$^{4+}$}
\label{sec:ar4}
In order to see how the low impurity temperature along with the increased potential drop produced by the mirror force affects impurity confinement, we conducted a simulation identical to the one in the previous section, Sec.~\ref{sec:hotar} but changed the neutral argon temperature in Gkeyll from 500 eV to 10 eV. So, differences in the argon density beyond those seen in the previous section can be attributed to the low impurity temperature relative to the deuterium temperature.

In~\cite{Mike23}, it is argued that the assumption of equal main ion and impurity temperature will not be valid if the mean free path is long. The heating rate for a cold impurity species by deuterium is given by 
\begin{equation}
\frac{d T_Z}{d t} \sim v_D \frac{m_D}{m_Z} Z^2 T_D,
\end{equation}
where $\nu_D$ is the deuterium-deuterium collision frequency.
A simple estimate for the time, $\tau$, for an impurity to be expelled by the potential $\phi$ over a length $L_\parallel$ is
\begin{equation}
\tau \sim L_{\|}\left(Z e \phi / m_Z\right)^{-1 / 2}
\end{equation}

Heating over this time will give an impurity temperature $T_Z \sim \tau \dv{T_z}{t}$ giving a temperature ratio of
\begin{equation}
\frac{T_Z}{T_D} \sim L_{\|} Z^2 v_D \frac{m_D}{m_Z}\left(Z e \phi / m_Z\right)^{-1 / 2}
\label{eq:heating_1}
\end{equation}

From the simulation results, we see that $e\phi\sim T_D$. Substituting this and $\nu_D = \sqrt{T_D/m_D}/\lambda_{mfp,D}$ into Eq.~\ref{eq:heating_1}, we get
\begin{equation}
\frac{T_Z}{T_D} \sim\left(\frac{L_{\|}}{\lambda_{m f p, D}}\right)\left(\frac{m_D}{m_Z}\right)^{1 / 2} Z^{3 / 2}
\label{eq:temp_ratio}
\end{equation}

In the regime we are simulating, the first term, $L_\parallel/\lambda_{mfp,D}$ is small; it is approximately 0.02 near the divertor plate at the radial center where the impurities are concentrated. The mass ratio will be small for impurities with a large atomic number like argon. The heating rate also scales with the charge as $Z^{3/2}$, so higher charge state impurities will be heated more quickly.

If impurities are much colder than the deuterium and $e\phi \sim T_D$, the amount of impurities able to overcome the potential barrier and make it upstream will be small. This shielding effect will be stronger for larger mass and lower charge impurities.

\begin{figure}
    \includegraphics[width=\columnwidth]{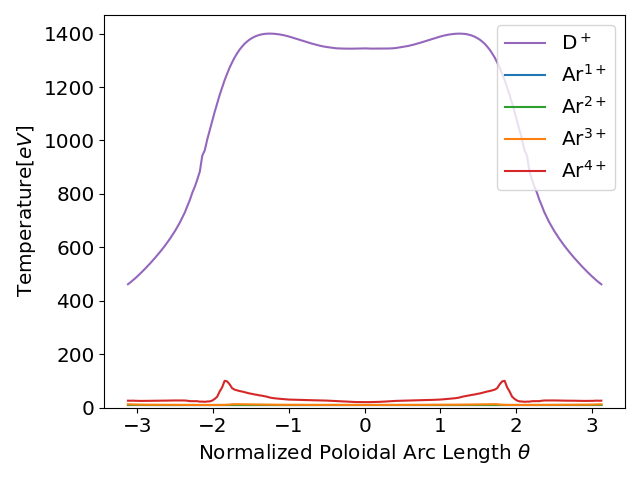}
    \caption[Deuterium and argon charge state temperatures]{
    Deuterium and argon charge state temperatures plotted along the field line at the radial center for the Gkeyll simulation with 10eV neutral argon and charge states up to Ar$^{4+}$. All of the argon charge states are much colder than the deuterium, which invalidates the assumption of equal ion and impurity temperature made by SOLPS. The argon is expelled by the potential before it has time to thermally equilibrate with the deuterium.
    \label{fig:temperature_ar}
    }
\end{figure}

In Fig.~\ref{fig:temperature_ar}, we can see that the assumption of equal deuterium and impurity temperatures is invalidated by Gkeyll; the assumption is inaccurate this kinetic regime. Only the Ar$^{4+}$ is significantly heated, but even the Ar$^{4+}$ temperature is drastically lower than the deuterium temperature. We also see that the downstream Ar$^{4+}$ temperature is lower than the upstream temperature; the downstream temperature is what will determine how much Ar$^{4+}$ can make it upstream.

\begin{figure}
    \includegraphics[width=\columnwidth]{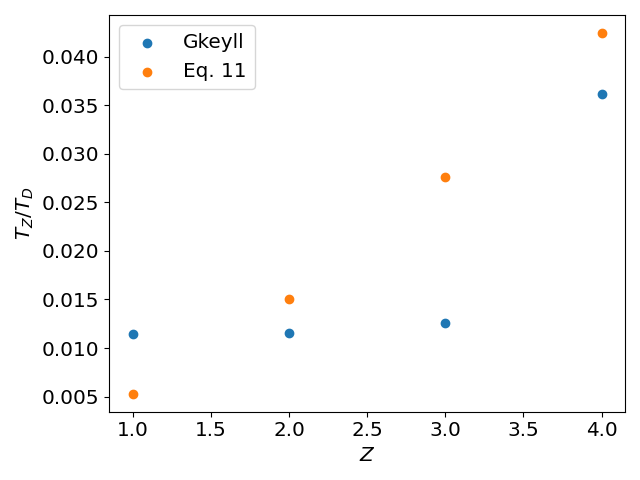}
    \caption[Poloidally averaged ratio of argon charge state temperature to deuterium temperature]{
        Poloidally averaged ratio of argon charge state temperature to deuterium temperature predicted by Eq.~\ref{eq:temp_ratio} and observed in the simulation plotted at the radial center of the simulation domain vs. charge state (Z) for the Gkeyll simulation with 10eV neutral argon and charge states up to Ar$^{4+}$. The predicted ratio agrees quite well with the observed ratio for the highest charge state, Ar$^{4+}$ but, as expected, overestimates the ratio for several of the lower charge states. Lower charge states are ionized quickly and do not have as much time to be heated as Eq.~\ref{eq:temp_ratio} assumes.
    \label{fig:temp_ratio}
    }
\end{figure}

In Fig.~\ref{fig:temp_ratio}, we compare the argon charge state temperatures observed in Gkeyll with those predicted by Eq.~\ref{eq:temp_ratio}. It is important to note that Eq.~\ref{eq:temp_ratio} takes into account losses due to expulsion by the potential but not due to ionization. 
Lower charge states are ionized very quickly at the electron temperatures observed in these simulations, so they have less time to be heated than assumed by Eq.~\ref{eq:temp_ratio}. Thus, Eq.~\ref{eq:temp_ratio} will overestimate the temperature of low charge states. In Fig.~\ref{fig:temp_ratio} we plot the ratio $T_Z/T_D$ for each argon charge state at the radial center of the simulation domain averaged in the parallel direction along with the average temperature predicted by Eq~\ref{eq:temp_ratio}. The predicted ratio agrees quite well with the observed ratio for the highest charge state, Ar$^{4+}$ but, as expected, overestimates the ratio for several of the lower charge states. In the next section, we will compare the predicted and observed temperature ratio for more charge states and the trend will become clearer.

\begin{figure}
    \subfloat[\label{fig:density_ar_total}Neutral and charged argon density (summed over charge states 1 through 4) plotted along the field line in Gkeyll and SOLPS.]{
        \includegraphics[width=0.45\textwidth]{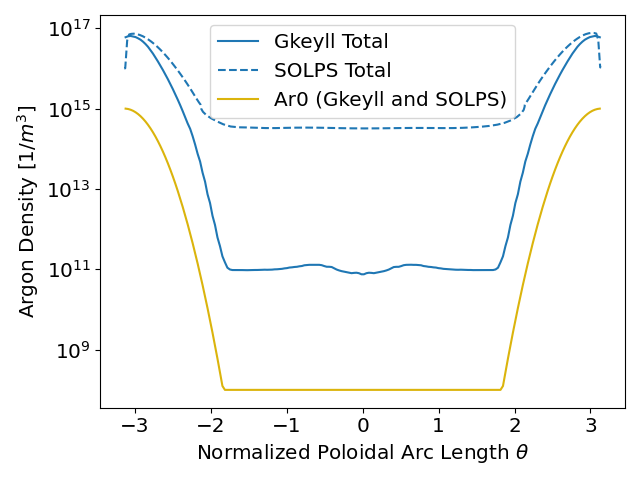}
    }\hspace{0.4cm}
    \subfloat[\label{fig:density_ar_individual}Neutral, Ar$^{+1}$, and Ar$^{+4}$ density plotted along the field line in Gkeyll and SOLPS.]{
        \includegraphics[width=0.45\textwidth]{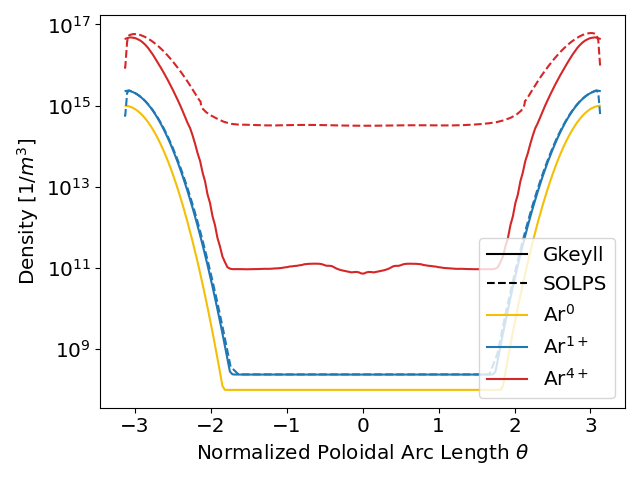}
    }
    \caption[Argon density plotted along the field line]{Argon density plotted along the field line at the radial center in SOLPS and Gkeyll for the Gkeyll simulation with 10 eV neutral argon and charge states up to Ar$^{4+}$. The total argon density is shown in (a) and individual charge state densities are shown in (b). The downstream density of charged argon is similar, but Gkeyll's upstream density is orders of magnitude lower.}
\end{figure}

Next we can see the effect of this temperature difference as well as the enhanced potential drop on the argon density profiles. In Fig.~\ref{fig:density_ar_total}, we plot the neutral argon density along with the total charged argon density, summed over charge states 1 through 4, along the field line for both SOLPS and Gkeyll. The shielding effect is quite clear; the total charged argon density in Gkeyll is much lower upstream. The charged argon density is similar downstream near the divertor plate, but Gkeyll's upstream argon density is four orders of manitude lower. This drastic difference can be attributed to the shielding effect created by the enhanced potential drop and the fact that the argon is much colder than the deuterium.

In Fig.~\ref{fig:density_ar_individual}, we look at some of the individual charge state densities for both Gkeyll and SOLPS. We can see here that Ar$^{4+}$ dominates the density both upstream and downstream. In light of Fig.~\ref{fig:temperature_ar} and Eq.~\ref{eq:temp_ratio}, it makes sense that higher charge states should dominate upstream.
Only high charge states are heated enough overcome the potential barrier and travel upstream. Furthermore, at these electron temperatures, any low charge states that do make it upstream will be quickly ionized. The Ar$^{+1}$ profiles in Gkeyll and SOLPS agree quite well both downstream and upstream because, at these electron temperatures, the steady state Ar$^{+1}$ density is set almost entirely by the ionization rate of neutral argon and Ar$^{+1}$.

We acknowledge that for more collisional downstream conditions with a lower downstream deuterium temperature, the difference between the deuterium and impurity temperature would be reduced and thus the shielding effect would be reduced. We will study this in more detail in future work when we are capable of simulating more collisional downstream conditions in Gkeyll.


\subsection{Higher Radiation Fractions}
\label{sec:high_frac}

The simulations we have shown so far have low enough argon density that radiation is negligible ($<1\textrm{ MW}$) and the deuterium and electron profiles are unaffected. In a practical scenario, most of the electron power will need to be radiated. In order to verify that the superior impurity confinement seen in kinetic simulations holds in regimes with higher radiation fraction, we ran one final set of simulations.

\begin{figure}
    \subfloat[ \label{fig:density_arhigh_total} Charged argon density (summed over charge states 1 through 4) plotted along the field line at the radial center in Gkeyll and SOLPS.
    ]{
        \includegraphics[width=0.45\textwidth]{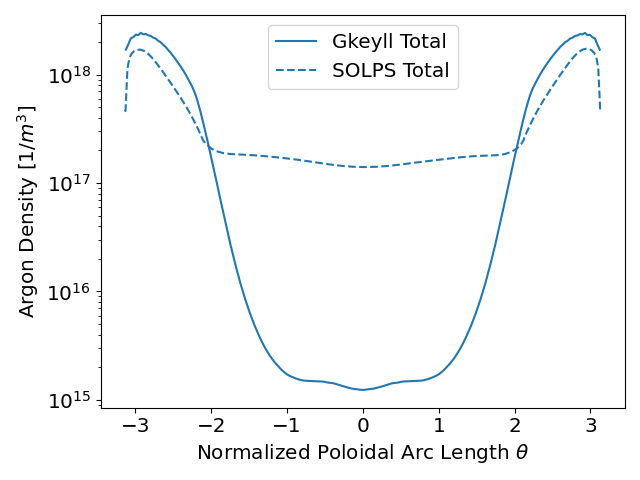}
    }\hspace{0.4cm}
    \subfloat[\label{fig:high_frac_artemp}
    Deuterium and argon temperature plotted along the field line at the radial center in Gkeyll.]{
        \includegraphics[width=0.45\textwidth]{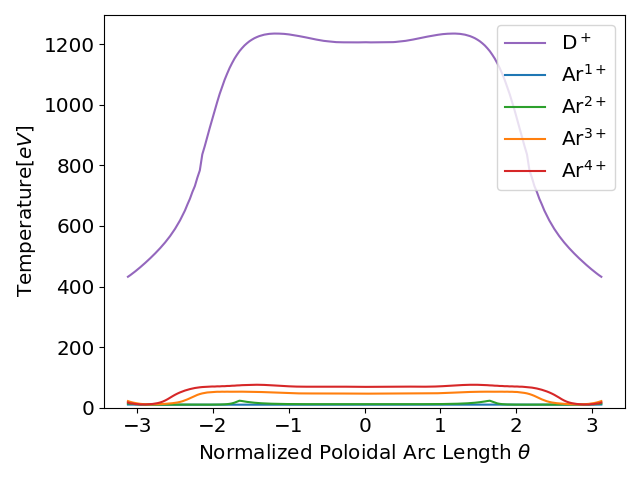}
    }
    \caption[Argon densities (a) and deuterium and argon temperatures (b) from the high radiation simulations]{
    Argon densities (a) and deuterium and argon temperatures (b) from the high radiation simulations with argon charge states up to Ar$^{4+}$. The neutral argon density at the divertor plate is 1.5$\times$10$^{16}$ in Gkeyll and  1$\times$10$^{16}$ in SOLPS. This results in 31MW of radiation in Gkeyll and 27 MW in SOLPS. The upstream argon density in Gkeyll is a factor of 7 lower.
    }
\end{figure}

This set of simulations is similar to the set described in subsection~\ref{sec:ar4}, but instead of matching the neutral argon densities in SOLPS and Gkeyll, we scaled the neutral argon profiles until we achieved a similar amount of total radiated power in the two codes. 
We used a peak neutral argon density of 1$\times$ 10$^{16}$ in SOLPS and 8.1$\times$ 10$^{15}$ in Gkeyll. The neutral argon temperature is 10 eV. This resulted in 27 MW of radiated power in SOLPS and 25 MW in Gkeyll. The total upstream argon density in Gkeyll for this case is two orders of magnitude lower than in SOLPS as seen in Fig.~\ref{fig:density_arhigh_total}. In total, 40MW of power are put into the electrons, so 25MW of radiation constitutes a significant radiation fraction.

In this high radiation case we see that, while still very large, the difference in upstream impurity density between SOLPS and Gkeyll is reduced relative to the low radiation case; the difference was a factor of 10$^4$ in the low radiation case and is a factor of  10$^2$ here.
However, this case shows that the shielding effect is still very strong even at higher radiation fractions; there is still drastically superior confinement of impurities to the divertor region in Gkeyll.

\begin{figure}
    \subfloat[\label{fig:high_frac_temp} Electron and deuterium temperatures from SOLPS and Gkeyll plotted along the field line at the radial center.]{
    \includegraphics[width=0.3\textwidth]{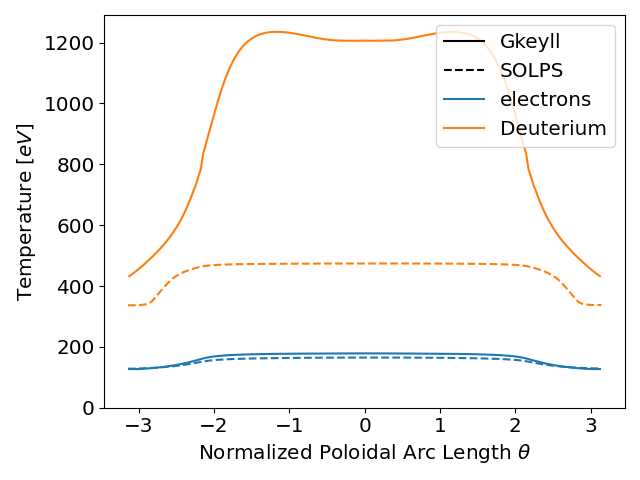}
    }\hspace{0.1cm}
    \subfloat[ \label{fig:high_frac_phi} Electrostatic potential from SOLPS and Gkeyll plotted along the field line at the radial center. ]{
    \includegraphics[width=0.3\textwidth]{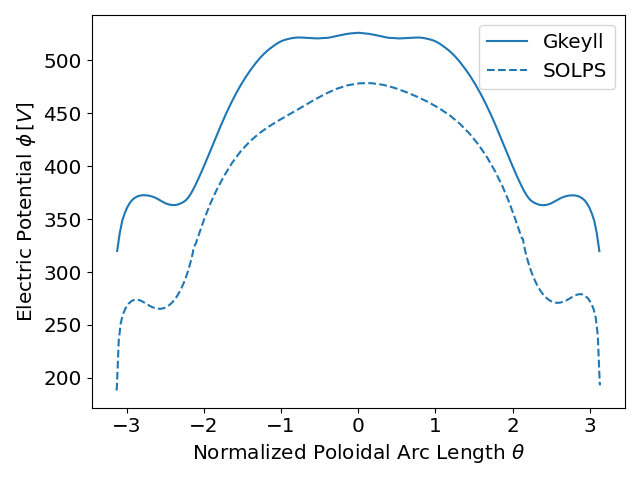}
    }\hspace{0.1cm}
    \subfloat[ \label{fig:high_frac_density} Electron density from SOLPS and Gkeyll plotted along the field line at the radial center. ]{
    \includegraphics[width=0.3\textwidth]{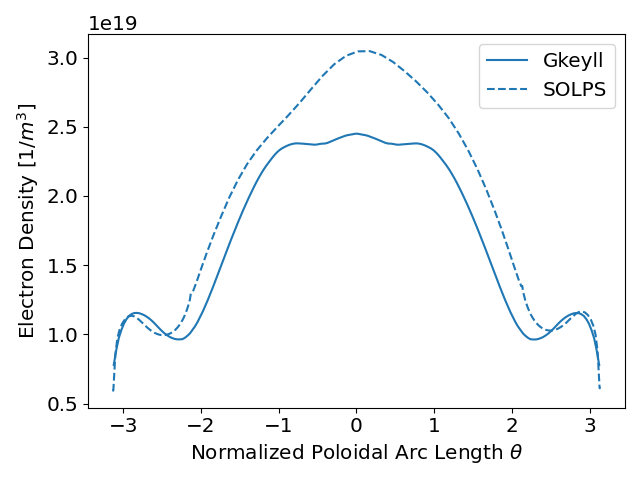}
    }
    \caption[Electron and deuterium temperatures (a), electrostatic potential (b), and electron density (c)]{
    Electron and deuterium temperatures (a), electrostatic potential (b), and electron density (c) from the high radiation simulations with argon charge states up to Ar$^{4+}$. The electron temperature is greatly reduced near the divertor plate relative to the simulations without impurities. The reduction in electron temperature results in a reduction in the potential drop from midplane to divertor plate relative to simulations without impurities. The upstream electron  density in Gkeyll is lower than in SOLPS as a result of the lower upstream impurity density.
    }
\end{figure}

We also see that the electron temperature is significantly reduced by the radiation; it is reduced by a factor of 2 downstream relative to the simulation without impurities as can be seen by comparing Fig.~\ref{fig:high_frac_temp} to Fig.~\ref{fig:temperature_plasma_z}. The electrostatic potential drop from midplane to divertor plate is proportional to the electron temperature, so the drop in electron temperature results in a reduction in the potential drop as can be seen by Fig.~\ref{fig:high_frac_phi} to Fig.~\ref{fig:phi_plasma} (a).

In this simulation, the upstream electron density in Gkeyll is significantly lower than in SOLPS as shown in Fig.~\ref{fig:high_frac_density}. Lower upstream impurity densities help maintain lower upstream densities for the main plasma which can be beneficial for confinement~\citep{Mike23}.
These results in this section indicate that kinetic effects may enable more effective strategies for reactor scenarios to radiate power while avoiding core contamination and maintaining good confinement.



The high radiation Gkeyll simulation is not completely converged as we did not have enough time to run it to convergence. However, most quantities from the simulation have converged, and we believe the current results are resonable. Time traces of various quantities from this simulation are included in Appendix~\ref{sec:convergence}.

\section{Conclusion}\label{sec:conclusion}
We have conducted a comparison of axisymmetric fluid and gyrokinetic simulations of a proposed STEP SOL using SOLPS and Gkeyll. We first established baseline agreement between the codes in a slab geometry in section~\ref{sec:slab_comparison}. We then demonstrated in section~\ref{sec:plasma_only} that, with upstream parameters typical for a pilot plant, kinetic effects play an important role. Mirror trapping of the ions results in a much higher ion temperature in kinetic simulations. We also found that a super-X like divertor with a long outer leg results in an enhanced ion outflow speed and electrostatic potential drop which can be seen in Gkeyll but not SOLPS because SOLPS excludes the mirror force.

In section~\ref{sec:impurities} we included argon impurities in both codes and found that the assumption of equal main ion and impurity temperature made in SOLPS is violated. In agreement with predictions made in~\cite{Mike23}, the low impurity temperature prevents impurities from travelling upstream in kinetic simulations. With identical neutral density profiles, kinetic simulations achieved a similar downstream impurity density to fluid simulations but maintained a much lower upstream impurity. At high radiation fractions, kinetic simulations still demonstrated much lower upstream impurity densities.


These findings have important implications for reactor relevant regimes. 
Higher SOL ion temperatures observed in kinetic simulations might lead to higher pedestal ion temperature which can result in higher core fusion power. Higher ion temperatures can also increase wall erosion by increasing the energy of neutrals hitting the main chamber wall, due to charge exchange of the hot plasma with cold recycled neutrals.
Our results also indicate that we may be able to support much larger downstream impurity densities (and hence more radiated power) while avoiding unacceptable upstream densities than previously thought based on SOLPS simulations. Better confinement of impurities to the divertor region entails at least two benefits: (1) avoidance of impurity contamination of the core plasma, and (2) avoidance of high upstream densities, which can degrade confinement.

While the results presented here are promising, we note that simulations with more realistic downstream parameters will need to be conducted in order to see whether we can take advantage of the effects described here in a reactor. Nonetheless, this comparison highlights the potential importance of kinetic effects in reactor SOL's. These effects could be significant in the upstream SOL of detached regimes and even downstream for low recycling or attached regimes.

We would like to mention that, while much more expensive than a fluid model, Gkeyll's axisymmetric kinetic model is not prohibitively expensive. Most of the simulations conducted in this work took 1-3 days on a modest number of GPUs. Once we have implemented an implicit collision operator in Gkeyll to alleviate restrictions on the timestep and coupled Gkeyll to a more sophisticated neutral model, we hope to use it for divertor studies and to conduct parameter scans. Details of the runtime and cost for each simulation can be found in appendix~\ref{sec:cost}.

In future work, we will explore more realistic scenarios by including dynamic neutrals, recycling and charge exchange in the scenario described in section~\ref{sec:high_frac}. Charge exchange will be essential for removing power from the ions and reducing the heat load to the divertor plates to acceptable levels. We will also explore the effects of drifts in axisymmetric simulations.


\chapter{Conclusion}
\label{chap:conclusion}
\section{Summary}
The main technical advance in this thesis is the extension of \gke's gyrokinetic solver to X-point geometry for 4D simulations described in~\ref{chap:2}. This is an important step because this is the first time a gyrokinetic simulation has been able to take advantage of a field aligned coordinate system over the open and closed field lines regieons of a diverted tokamak.
Another significant technical advance is the ability to couple \gke\ to EIRENE for plasma wall interactions and neutral evolution shown in ~\ref{chap:3}.

The main physics results presented in this thesis are centered around low recycling regimes. Here we have presented simulation results in chapters~\ref{chap:3} and~\ref{chap:4} that take initial steps towards demonstrating the feasibility of low recycling SOLs. The results indicate that a high temperature, low density SOL could be achieved and that, with the help of developing liquid metal technologies, the heat flux on the divertor target can be made manageable. 

Low-recycling regimes present advantages for core confinement but the feasibility of a low-recycling SOL is uncertain. 
When using low recycling materials such as lithium as a target material, it is difficult to prevent them from evaporating and cooling the plasma, which would interfere with the confinement advantages low-recycling regimes provide.
Here we show gyrokinetic simulation results that indicate that a low-recycling regime could be achieved by using a low recycling material to coat only the side walls, which receive a much lower heat flux than the target. This is an important step towards demonstrating the feasibility of a low-recycling SOL.

The large target heat fluxes and temperatures present in a low-recycling regime pose challenges such as sputtering and erosion. Here, we note that a low-recycling solution would likely have to involve a liquid metal target to avoid detrimental erosion and also begin to investigate ways to address some of these challenges by taking advantage of kinetic effects. By comparing fluid and gyrokinetic simulations in a moderately collisionless SOL, we find that (1) kinetic effects can help confine impurities downstream, which would help avoid core contamination caused by sputtering and (2) the interaction of mirror trapping and drifts can significantly increase the heat flux width and reduce the peak heat flux.

\section{Future Work}
Although we have developed the capability to model X-points in 4D simulations, we have not yet extended the capabilities to 5D gyrokinetic simulations with turbulence. Although 4D simulations are computationally cheaper and remain useful, we would like to develop the capability to simulate turbulence in diverted geometry. Turbulent 5D simulations can either be used independently or in combination with 4D simulations - turbulent simulations can be used to inform the diffusivities used in 4D simulations. We are already working towards this capability.

One important design issue we have not yet explored is helium exhaust in a low recycling SOL. If the neutral mean free path is long relative to the pump opening, it may be difficult to pump out helium. However, with a long-legged divertor like the one featured in the STEP magnetic geometry, there could be room to make a very large pump opening to ease the pumping of neutral helium. Additionally, the low electron density in the divertor leg will reduce the helium ionization rate and could make it easier to pump out helium before it reacts with and cools the plasma. In future work, we plan to conduct simulations including helium to address the issue of helium exhaust in a low recycling SOL.

%
\appendices

\chapter{SOL-Only Simulation Details}
\section{10 eV Neutral Argon up to Ar$^{8+}$}
\label{sec:ar8}
One may think that higher charge states will acquire a significant density upstream because they will be hotter and more capable of overcoming the potential barrier, but this does not happen. 
The reason is that higher charge states must come from ionization of lower charge states, which are strongly shielded and confined to the divertor region.

High Z impurities are expelled very strongly by the potential since the electrostatic force is proportional to Z. Low Z impurities are strongly shielded since they do not have time to heat up and will not make it upstream to be ionized to a higher charge state. Higher charge states originate from ionization of lower chage states, so the shielding of low charge states should result in a lower concentration of high charge states even if the higher charge states are thermally equilibrated. In this section we provide evidence to support this argument; we ran simulations identical to those described in the previous subsection~\ref{sec:ar4} but included up to  Ar$^{8+}$.

\begin{figure}
    \centering
    \includegraphics[width=\columnwidth]{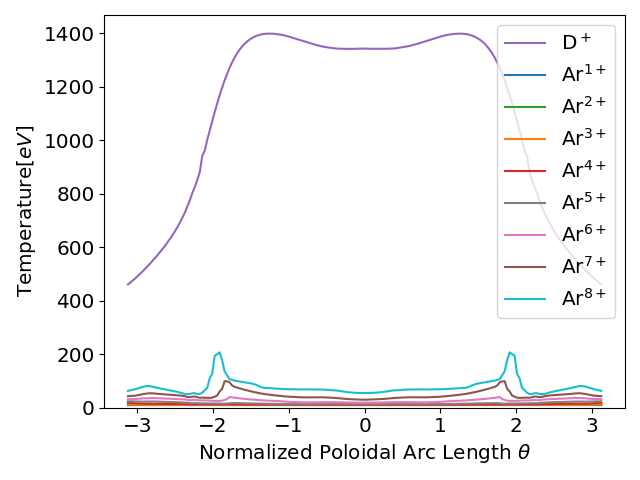}
    \caption[Deuterium and argon charge state temperatures]{Deuterium and argon charge state temperatures plotted along the field line at the radial center for the Gkeyll simulation with 10eV neutral argon and charge states up to Ar$^{8+}$. Higher argon charge states have higher temperatures, but their temperature is still much lower than the deuterium temperature.}
    \label{fig:temperature_ar8}
\end{figure}
\begin{figure}
\centering
    \includegraphics[width=\columnwidth]{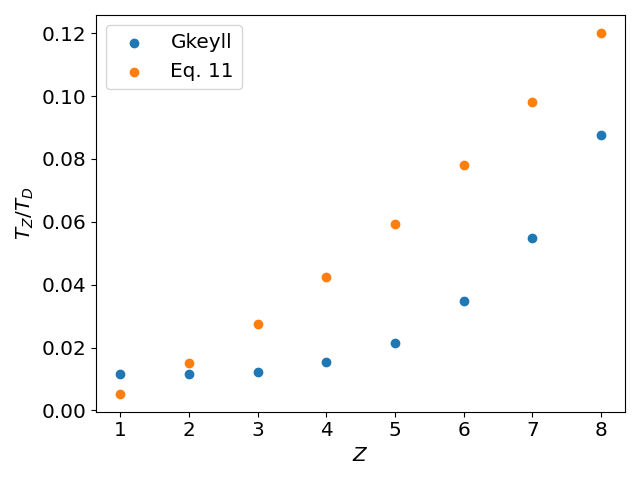}
    \caption[Poloidally averaged ratio of argon charge state temperature to deuterium temperature]{
        Poloidally averaged ratio of argon charge state temperature to deuterium temperature predicted by Eq.~\ref{eq:temp_ratio} and observed in the simulation plotted at the radial center of the simulation domain vs. charge state (Z) for the Gkeyll simulation with 10eV neutral argon and charge states up to Ar$^{8+}$. The predicted ratio agrees quite well with the observed ratio for the highest charge state, Ar$^{8+}$ but, as expected, overestimates the ratio for most of the lower charge states. Lower charge states are ionized quickly and do not have as much time to be heated as Eq.~\ref{eq:temp_ratio} assumes.
    \label{fig:temp_ratio_ar8}
    }
\end{figure}

In Fig.~\ref{fig:temperature_ar8}, we plot the argon charge state temperatures along the field line. Ar$^{4+}$ is not as hot as it was in the simulation which included only up to charge state 4 because ionization of Ar$^{4+}$ reduces the amount of time available for it to be heated. Higher charge states are indeed hotter than lower charge states as predicted by Eq.~\ref{eq:temp_ratio}. As observed in the previous section, the downstream temperature is significantly lower than the upstream temperature.
In Fig.~\ref{fig:temp_ratio_ar8} we plot the ratio $T_Z/T_D$ for each argon charge state at the radial center of the simulation domain averaged in the parallel direction along with the average temperature predicted by Eq~\ref{eq:temp_ratio}. The predicted ratio agrees quite well with the observed ratio for the highest charge state, Ar$^{8+}$ but again overestimates the ratio for several of the lower charge states.

\begin{figure}
\centering
    \subfloat[\label{fig:density_ar8_total}Neutral and charged argon density (summed over charge states 1 through 8) plotted along the field line in Gkeyll and SOLPS.]{
        \includegraphics[width=0.45\textwidth]{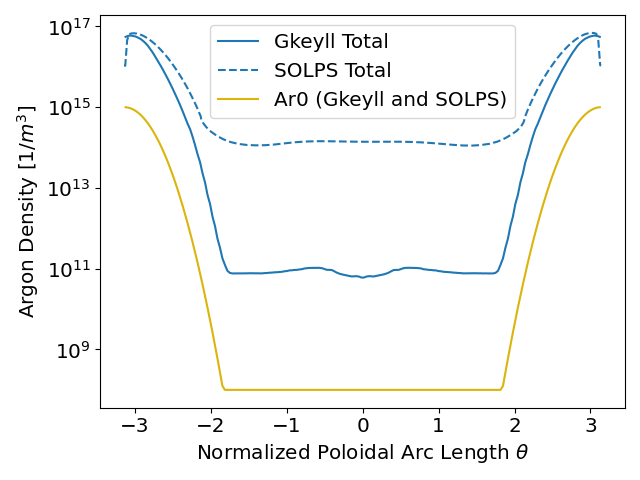}
    }\hspace{0.5cm}
    \subfloat[\label{fig:density_ar8_individual} Neutral, Ar$^{+1}$, Ar$^{+5}$, and Ar$^{+8}$ density plotted along the field line in Gkeyll and SOLPS.]{
        \includegraphics[width=0.45\textwidth]{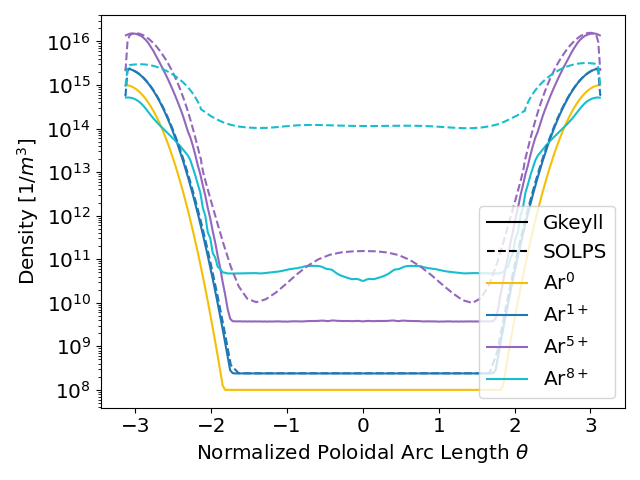}
    }
    \caption[Argon density plotted along the field line in SOLPS and Gkeyll]{Argon density plotted along the field line in SOLPS and Gkeyll for the Gkeyll simulation with 10eV neutral argon and charge states up to Ar$^{8+}$. The total argon density is shown in (a) and individual charge state densities are shown in (b). Charge state 5 dominates near the divertor plate because charge states higher than 5 are pushed out more quickly by  the potential. The total argon density is nearly identical to that of the simulation which included only up to Ar$^{4+}$.
    }
\end{figure}

In Fig.~\ref{fig:density_ar8_individual} we plot some of the individual charge state densities. In both codes, Ar$^{5+}$ dominates near the divertor plate and higher charge states such as Ar$^{8+}$ have a lower downstream density. This can be attributed to the fact that higher charge states are expelled more quickly by the potential. Both codes also show that Ar$^{8+}$ dominates upstream which also makes sense; in Gkeyll only higher charge states should be able to travel upstream and in both codes any low charge states that make it upstream will be quickly ionized.

In Fig.\ref{fig:density_ar8_total}, we plot the total argon density summed over charge states 1 through 8 along the field line for both SOLPS and Gkeyll. This figure supports the argument that the strong shielding effect for low charge states does in fact reduce the upstream density of high charge states. Comparing Fig.\ref{fig:density_ar8_total} and Fig.~\ref{fig:density_ar_total} we see that the upstream and downstream total charged Argon density is nearly identical when we include up to Ar$^{4+}$ or Ar$^{8+}$. So, the inclusion of higher charge states, which heat faster, did not have significant effect on the total upstream or downstream impurity density.

In Gkeyll, including charge states beyond Ar$^{8+}$ is computationally expensive. However, in SOLPS, we ran another simulation including up to Ar$^{12+}$, not shown here, which confirmed that the inclusion of higher charge states did not have a significant effect on the total upstream or downstream argon density.

\section{Simulation Details}
\label{sec:resolution}
In this section we detail the resolution of each Gkeyll simulation we conducted. The simulation with slab geometry described in section~\ref{sec:slab_comparison} used 72 cells in the radial direction and 64 cells in the parallel direction ($N_R\times N_Z = 72x64$). All simulations in the ST geometry had the same resolution in configuration space: 40 cells in the radial direction ($N_\psi = 40$) and 96 cells in the parallel direction ($N_\theta = 96$). The details of the velocity space resolution can be found in tables~\ref{tab:resolution} and~\ref{tab:extents}.

The version of the Gkeyll code used was commit 5d78312 on the gk-g0-app branch of \url{https://github.com/ammarhakim/gkylzero}. In order to make the simulation results reproducible, we have made the input files, which contain all of the details of the simulation setup, for the simulations conducted in this work available at \url{https://github.com/ammarhakim/gkyl-paper-inp/tree/master/2024_PoP_GKST}. 

\begin{table}
\centering
\resizebox{\columnwidth}{!}{
\begin{tabular}{|c|c|c|c|c|c|c|c|c|c|c|}
 Section&electrons&$D^+$&$Ar^{+1}$&$Ar^{+2}$&$Ar^{+3}$&$Ar^{+4}$&$Ar^{+5}$&$Ar^{+6}$&$Ar^{+7}$&$Ar^{+8}$\\ \hline
 ~\ref{sec:slab_comparison}&16x12&16x12&--&--&--&--&--&--&--&--\\
 ~\ref{sec:plasma_only}&16x12&16x12&--&--&--&--&--&--&--&--\\
 ~\ref{sec:ar4}&16x12&16x12&16x12&16x12&16x12&16x12&--&--&--&--\\
 ~\ref{sec:ar8}&16x12&16x12&16x12&16x12&16x12&16x12&16x12&16x12&16x12&16x12\\
 ~\ref{sec:hotar}&16x12&16x12&16x12&16x12&16x12&16x12&--&--&--&--\\
 ~\ref{sec:high_frac}&16x12&16x12&16x12&32x12&48x12&96x12&--&--&--&--\\
\end{tabular}
}
\caption[Velocity space resolution for each species]{\label{tab:resolution} Velocity space resolution for each species in each simulation. Each entry in the table gives the velocity space resolution ($N_{v_\parallel} \times N_\mu$) for the species in the corresponding column in the simulation in the corresponding row.}
\end{table}

\begin{table}
\centering
\resizebox{\columnwidth}{!}{
\begin{tabular}{|c|c|c|c|c|c|c|c|c|c|c|}
 Section&electrons&$D^+$&$Ar^{+1}$&$Ar^{+2}$&$Ar^{+3}$&$Ar^{+4}$&$Ar^{+5}$&$Ar^{+6}$&$Ar^{+7}$&$Ar^{+8}$\\ \hline
 ~\ref{sec:slab_comparison}&63,6,12&94,6,12&--&--&--&--&--&--&--&--\\
 ~\ref{sec:plasma_only}&364, 6, 18&546, 6, 18&--&--&--&--&--&--&--&--\\
 ~\ref{sec:ar4}&364, 6, 18&546, 6, 18&10, 4, 18&10, 8, 18&10, 12, 36&10, 22, 72&--&--&--&--\\
 ~\ref{sec:ar8}&364, 6, 18&546, 6, 18&10, 4, 18&10, 8, 18&10, 12, 36&10, 22, 72&10, 6, 72&110,8.5,144&10, 9, 144&10,9,144\\
 ~\ref{sec:hotar}&364, 6, 18&546, 6, 18&500, 6, 18&500, 6, 18&500, 6, 18&500, 6, 18&--&--&--&--\\
 ~\ref{sec:high_frac}&364, 6, 18&546, 6, 18&10, 4, 18 &10, 8, 18&10, 12, 18&10, 24, 18&--&--&--&--\\
\end{tabular}
}
\caption[Parallel velocity extents for each species]{\label{tab:extents} Parallel velocity extents for each species in each simulation. Each entry lists the reference temperature, $T_{ref}$, in eV, the maximum parallel velocity, $v_{\parallel,max}$, normalized to $v_t = \sqrt{T_{ref}/m}$, and the maximum magnetic moment, $\mu_{max}$, normalized to $mv_t^2/2B_0$. $B_0$=2.51 for all simulations.}
\end{table}

\section{Simulation Cost}
\label{sec:cost}
The Gkeyll simulation section~\ref{sec:slab_comparison} was run on 2 nodes (4 GPUs per node) and all other Gkeyll simulations were run on 8 nodes. Table~\ref{tab:runtime} lists the number of node hours used and the final simulation time for each Gkeyll simulation. Simulations with impurities are started with initial conditions for the deuterium and electrons given by the steady of the plasma only simulation in section~\ref{sec:plasma_only}.  

\begin{table}
\centering
\begin{tabular}{|c|c|c|c|c|c|c|c|c|c|c|}
 Section&Node Hours& End time  (ms)\\ \hline
 ~\ref{sec:slab_comparison}&96&4\\
 ~\ref{sec:plasma_only}&160&8\\
 ~\ref{sec:ar4}&256&1.83\\
 ~\ref{sec:ar8}&576&2.16\\
 ~\ref{sec:hotar}&384&2.75\\
 ~\ref{sec:high_frac}&3648&25.7\\
\end{tabular}
\caption{\label{tab:runtime} Simulation cost and end time.}
\end{table}

\section{Convergence of  the High Radiation Case}
\label{sec:convergence}
\begin{figure}
    \subfloat[\label{fig:trace_density} ]{
    \includegraphics[width=0.45\textwidth]{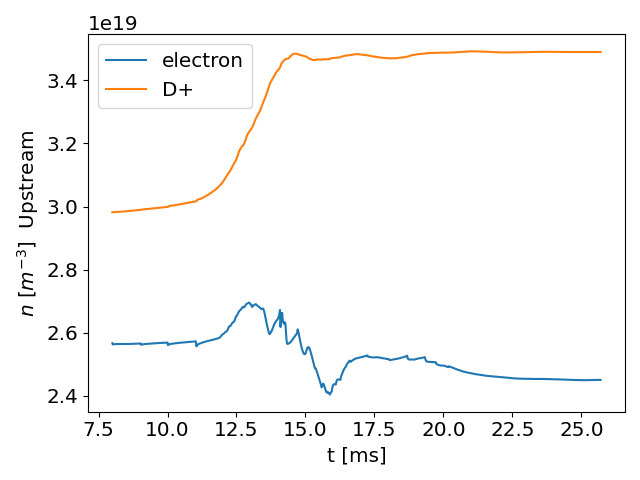}
    }
    \subfloat[\label{fig:trace_density_down} ]{
    \includegraphics[width=0.45\textwidth]{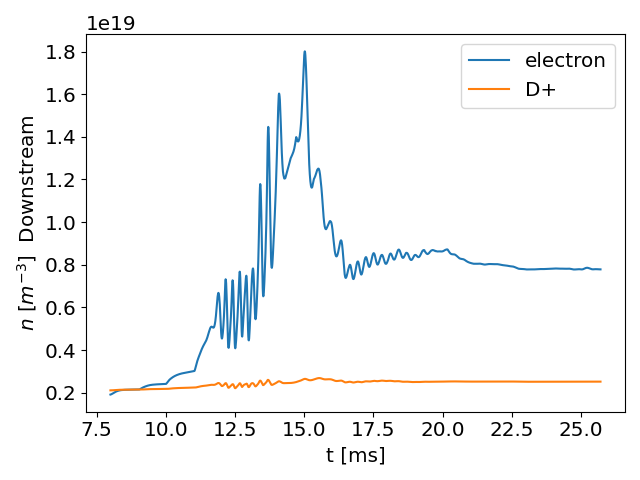}
    }\\
    \subfloat[\label{fig:trace_ardensity} ]{
    \includegraphics[width=0.45\textwidth]{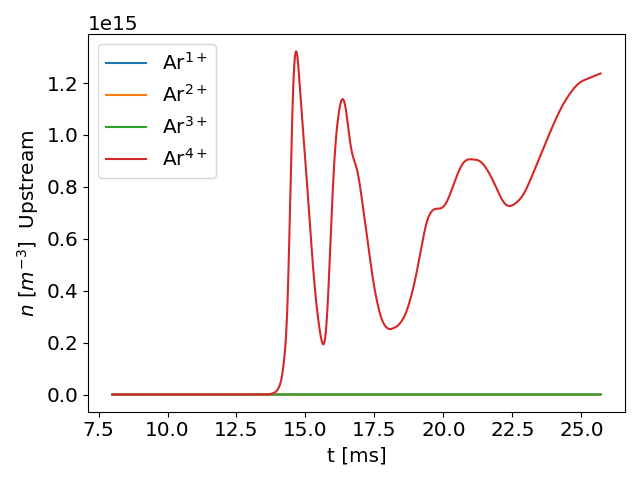}
    }
    \subfloat[\label{fig:trace_ardensity_down} ]{
    \includegraphics[width=0.45\textwidth]{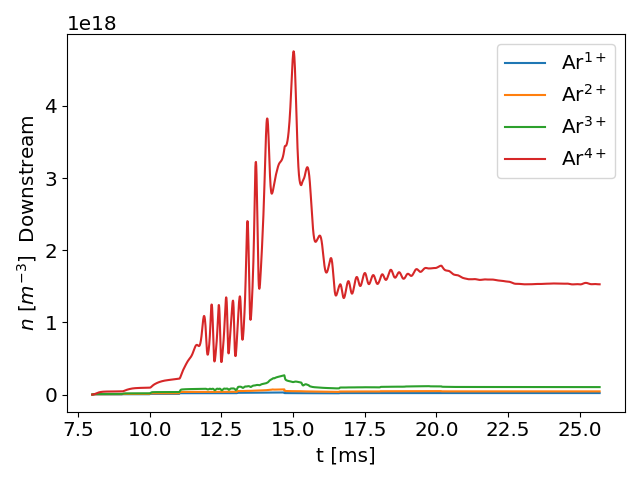}
    }
    \caption[Time traces of the electron and deuterium density upstream]{
    \label{fig:density_trace}
    Time traces of the electron and deuterium density upstream (a) and downstream (b) and of the argon density upstream (c) and downstream (d) at the radial center of the domain from the Gkeyll simulation  in Sec.~\ref{sec:high_frac}.
    }
\end{figure}

\begin{figure}
    \subfloat[\label{fig:trace_temp} ]{
    \includegraphics[width=0.45\textwidth]{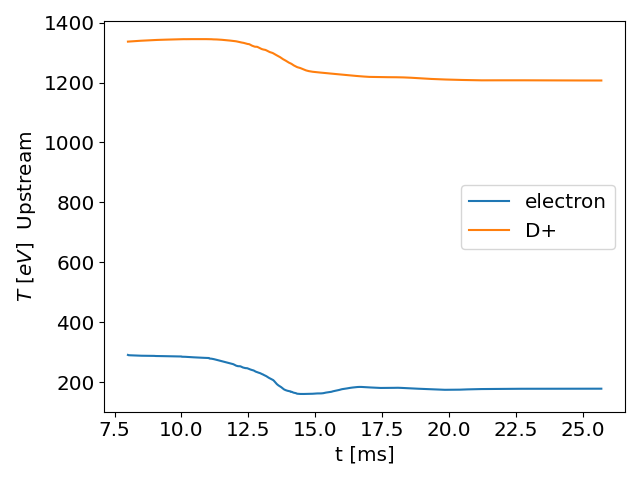}
    }
    \subfloat[\label{fig:trace_temp_down} ]{
    \includegraphics[width=0.45\textwidth]{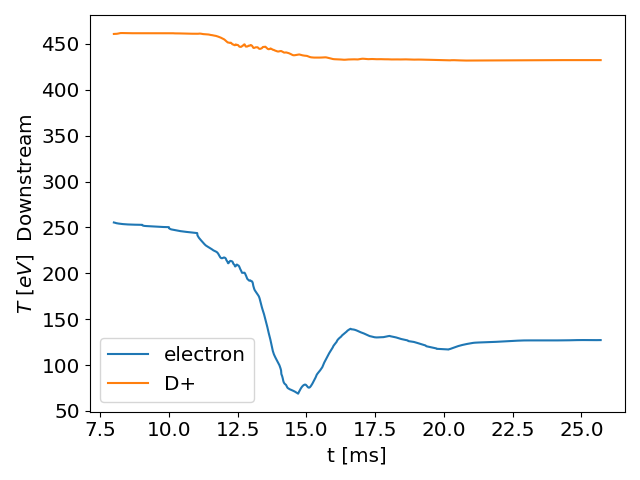}
    }\\
    \subfloat[\label{fig:trace_artemp} ]{
    \includegraphics[width=0.45\textwidth]{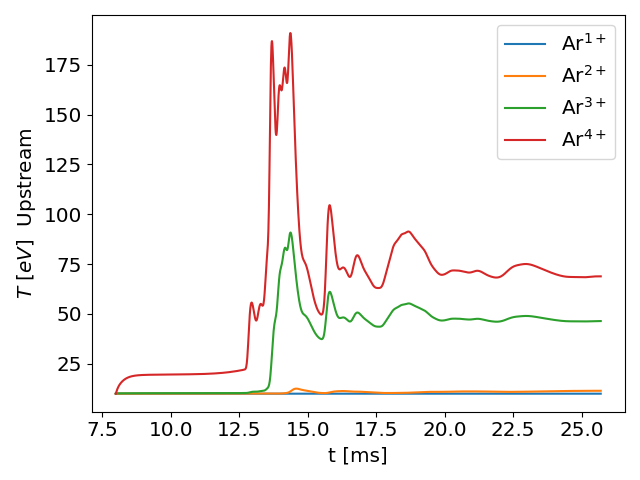}
    }
    \subfloat[\label{fig:trace_artemp_down} ]{
    \includegraphics[width=0.45\textwidth]{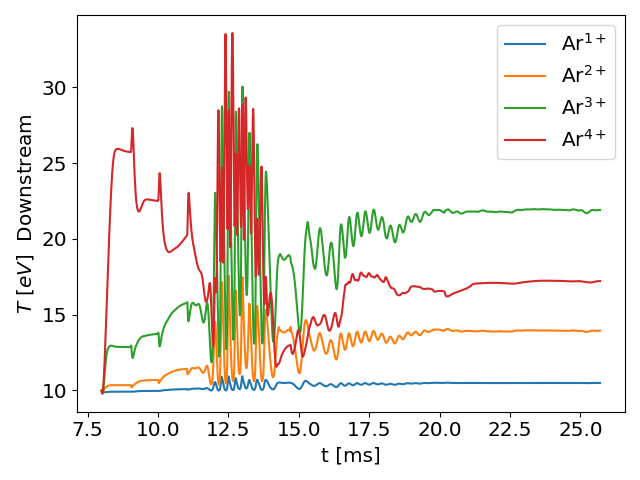}
    }
    \caption[Time traces of the electron and deuterium temperature upstream]{
    \label{fig:temp_trace}
    Time traces of the electron and deuterium temperature upstream (a) and downstream (b) and of the argon temperature upstream (c) and downstream (d) at the radial center of the domain from the Gkeyll simulation  in Sec.~\ref{sec:high_frac}.
    }
\end{figure}

As noted in the main text, the high radiation case is not completely converged. In this appendix we include time traces from the Gkeyll simulation in Sec.~\ref{sec:high_frac} of the temperature and density of all species at the radial both upstream (at the midplane) and downstream (at the divertor plate). Densities are plotted in Fig.~\ref{fig:density_trace} and temperatures are plotted in Fig.~\ref{fig:temp_trace}.

All quantities except the upstream density of Ar$^{4+}$ seem to have converged both upstream and downstream. The upstream density of Ar$^{4+}$ has not completely flattened out, but it is more than 100 times lower than the upstream SOLPS density and does not seem to be increasing very fast.

As noted in Appendix~\ref{sec:cost}, this simulation  was restarted from the steady state of the  simulation in Sec.~\ref{sec:plasma_only}. This is why all the time traces start from 8 ms.

\bibliographystyle{plainnat}  
\bibliography{references}        

\begin{thebibliography}{130}
\providecommand{\natexlab}[1]{#1}
\providecommand{\url}[1]{\texttt{#1}}
\expandafter\ifx\csname urlstyle\endcsname\relax
  \providecommand{\doi}[1]{doi: #1}\else
  \providecommand{\doi}{doi: \begingroup \urlstyle{rm}\Url}\fi

\bibitem[Abu-Shawareb et~al.(2024)Abu-Shawareb, Acree, Adams, Adams, and et~al.]{NIF2024}
H.~Abu-Shawareb, R.~Acree, P.~Adams, J.~Adams, and et~al.
\newblock Achievement of target gain larger than unity in an inertial fusion experiment.
\newblock \emph{Phys. Rev. Lett.}, 132:\penalty0 065102, Feb 2024.
\newblock \doi{10.1103/PhysRevLett.132.065102}.

\bibitem[Amorisco et~al.(2024)Amorisco, Agnello, Holt, Mars, Buchanan, and Pamela]{amorisco2024}
Nicola~C. Amorisco, A.~Agnello, G.~Holt, M.~Mars, J.~Buchanan, and S.~Pamela.
\newblock {FreeGSNKE}: A python-based dynamic free-boundary toroidal plasma equilibrium solver.
\newblock \emph{Physics of Plasmas}, 31\penalty0 (4):\penalty0 042517, 2024.
\newblock \doi{10.1063/5.0188467}.

\bibitem[Andersen et~al.(1969)Andersen, Jensen, Nielsen, and D'Angelo]{Andersen2003}
S.~A. Andersen, V.~O. Jensen, P.~Nielsen, and N.~D'Angelo.
\newblock {Continuous Supersonic Plasma Wind Tunnel}.
\newblock \emph{The Physics of Fluids}, 12\penalty0 (3):\penalty0 557--560, 03 1969.
\newblock ISSN 0031-9171.
\newblock \doi{10.1063/1.1692519}.
\newblock URL \url{https://doi.org/10.1063/1.1692519}.

\bibitem[Bailey and Borwein(2011)]{Bailey}
D.H. Bailey and J.M. Borwein.
\newblock {High-precision numerical integration: Progress and challenges}.
\newblock \emph{Journal of Symbolic Computation}, 46\penalty0 (7):\penalty0 741--754, 2011.
\newblock ISSN 0747-7171.
\newblock \doi{https://doi.org/10.1016/j.jsc.2010.08.010}.
\newblock URL \url{https://www.sciencedirect.com/science/article/pii/S0747717110001409}.
\newblock Special Issue in Honour of Keith Geddes on his 60th Birthday.

\bibitem[Baker(2024)]{Baker2024}
Adam Baker.
\newblock The spherical tokamak for energy production (step) in context: Uk public sector approach to fusion energy.
\newblock \emph{Philosophical Transactions of the Royal Society A: Mathematical, Physical and Engineering Sciences}, 382\penalty0 (2280):\penalty0 20230401, 2024.
\newblock \doi{10.1098/rsta.2023.0401}.
\newblock URL \url{https://royalsocietypublishing.org/doi/abs/10.1098/rsta.2023.0401}.

\bibitem[Barnes et~al.(2024)Barnes, Dickinson, Dorland, Hill, Parker, Roach, Giacomin, Mandell, Numata, Schuett, Biggs-Fox, Christen, Parisi, Wilkie, Anton, Ball, Baumgaertel, Colyer, Hardman, Hein, Highcock, Howes, Jackson, Kotschenreuther, Lee, Leggate, Mauriya, Patel, Tatsuno, and Van~Wyk]{GS2-zenodo}
Michael Barnes, David Dickinson, William Dorland, Peter~Alec Hill, Joseph~Thomas Parker, Colin~Malcolm Roach, Maurizio Giacomin, Noah Mandell, Ryusuke Numata, Tobias Schuett, Stephen Biggs-Fox, Nicolas Christen, Jason Parisi, George Wilkie, Lucian Anton, Justin Ball, Jessica Baumgaertel, Greg Colyer, Michael Hardman, Joachim Hein, Edmund Highcock, Gregory Howes, Adrian Jackson, Michael~T. Kotschenreuther, Jungpyo Lee, Huw Leggate, Adwiteey Mauriya, Bhavin Patel, Tomo Tatsuno, and Ferdinand Van~Wyk.
\newblock Gs2 v8.2.1, 11 2024.
\newblock URL \url{https://doi.org/10.5281/zenodo.14022029}.

\bibitem[Becker(2004)]{Becker04}
G.~Becker.
\newblock Scaling law for effective heat diffusivity in elmy h-mode plasmas.
\newblock \emph{Nuclear Fusion}, 44\penalty0 (11):\penalty0 L26, 11 2004.
\newblock \doi{10.1088/0029-5515/44/11/L02}.
\newblock URL \url{https://dx.doi.org/10.1088/0029-5515/44/11/L02}.

\bibitem[Beer et~al.(1995)Beer, Cowley, and Hammett]{beer95}
M~A Beer, S~C Cowley, and G~W Hammett.
\newblock Field-aligned coordinates for nonlinear simulations of tokamak turbulence.
\newblock \emph{Physics of Plasmas}, 2\penalty0 (7), 07 1995.
\newblock \doi{10.1063/1.871232}.
\newblock URL \url{https://www.osti.gov/biblio/71486}.

\bibitem[Bernard et~al.(2022)Bernard, Halpern, Francisquez, Mandell, Juno, Hammett, Hakim, Wilkie, and Guterl]{Tess22}
T.~N. Bernard, F.~D. Halpern, M.~Francisquez, N.~R. Mandell, J.~Juno, G.~W. Hammett, A.~Hakim, G.~J. Wilkie, and J.~Guterl.
\newblock {Kinetic modeling of neutral transport for a continuum gyrokinetic code}.
\newblock \emph{Physics of Plasmas}, 29\penalty0 (5):\penalty0 052501, 05 2022.
\newblock ISSN 1070-664X.
\newblock \doi{10.1063/5.0087131}.
\newblock URL \url{https://doi.org/10.1063/5.0087131}.

\bibitem[Betti(2025)]{Betti2025}
Riccardo Betti.
\newblock Status and prospects for inertial fusion energy via lasers.
\newblock volume 13358. University of Rochester, NY (United States), 03 2025.
\newblock \doi{10.1117/12.3049450}.
\newblock URL \url{https://www.osti.gov/biblio/2561299}.

\bibitem[{Braginskii}(1965)]{Braginskii1965}
S.~I. {Braginskii}.
\newblock {Transport Processes in a Plasma}.
\newblock \emph{Reviews of Plasma Physics}, 1:\penalty0 205, 01 1965.

\bibitem[Bufferand et~al.(2024)Bufferand, Ciraolo, Düll, Falchetto, Fedorczak, Marandet, Quadri, Raghunathan, Rivals, Schwander, Serre, Sureshkumar, Tamain, and Varadarajan]{BUFFERAND2024}
H.~Bufferand, G.~Ciraolo, R.~Düll, G.~Falchetto, N.~Fedorczak, Y.~Marandet, V.~Quadri, M.~Raghunathan, N.~Rivals, F.~Schwander, E.~Serre, S.~Sureshkumar, P.~Tamain, and N.~Varadarajan.
\newblock Global 3d full-scale turbulence simulations of tcv-x21 experiments with soledge3x.
\newblock \emph{Nuclear Materials and Energy}, 41:\penalty0 101824, 2024.
\newblock ISSN 2352-1791.
\newblock \doi{https://doi.org/10.1016/j.nme.2024.101824}.

\bibitem[Candy(2009)]{Candy2009}
J~Candy.
\newblock {A unified method for operator evaluation in local Grad–Shafranov plasma equilibria}.
\newblock \emph{Plasma Physics and Controlled Fusion}, 51\penalty0 (10):\penalty0 105009, 9 2009.
\newblock \doi{10.1088/0741-3335/51/10/105009}.
\newblock URL \url{https://dx.doi.org/10.1088/0741-3335/51/10/105009}.

\bibitem[Candy and Belli(2010)]{Candy2010}
J~Candy and E~Belli.
\newblock Gyro technical guide.
\newblock \emph{General Atomics, PO Box}, 85608:\penalty0 92186--5608, 2010.

\bibitem[Cerfon and Freidberg(2010)]{Cerfon2010}
Antoine~J. Cerfon and Jeffrey~P. Freidberg.
\newblock One size fits all analytic solutions to the grad-shafranov equation.
\newblock \emph{Physics of Plasmas}, 17\penalty0 (3):\penalty0 032502, 03 2010.
\newblock ISSN 1070-664X.
\newblock \doi{10.1063/1.3328818}.
\newblock URL \url{https://doi.org/10.1063/1.3328818}.

\bibitem[Cowley(2016)]{Cowley2016}
Steven~C. Cowley.
\newblock The quest for fusion power.
\newblock \emph{Nature Physics}, 12\penalty0 (5):\penalty0 384--386, 2016.
\newblock \doi{10.1038/nphys3719}.

\bibitem[Creely et~al.(2020)Creely, Greenwald, Ballinger, Brunner, Canik, Doody, Fülöp, Garnier, Granetz, Gray, and et~al.]{SPARCGain2}
A.~J. Creely, M.~J. Greenwald, S.~B. Ballinger, D.~Brunner, J.~Canik, J.~Doody, T.~Fülöp, D.~T. Garnier, R.~Granetz, T.~K. Gray, and et~al.
\newblock Overview of the sparc tokamak.
\newblock \emph{Journal of Plasma Physics}, 86\penalty0 (5):\penalty0 865860502, 2020.
\newblock \doi{10.1017/S0022377820001257}.

\bibitem[Crownhart(2024)]{Crownhart2024}
Casey Crownhart.
\newblock Inside a fusion energy facility.
\newblock \emph{MIT Technology Review}, 10 2024.
\newblock URL \url{https://www.technologyreview.com/2024/10/31/1106384/inside-a-fusion-energy-facility/}.

\bibitem[Dhaeseleer et~al.(1991)Dhaeseleer, Hitchon, Shohet, Callen, and Kerst]{DHaessler}
W.~D. Dhaeseleer, W.~N.G. Hitchon, J.~L. Shohet, J.~D. Callen, and D.~W. Kerst.
\newblock Flux coordinates and magnetic field structure. a guide to a fundamental tool of plasma theory, Jan 1991.

\bibitem[Dimits(1993)]{Dimits1993}
A.~M. Dimits.
\newblock Fluid simulations of tokamak turbulence in quasiballooning coordinates.
\newblock \emph{Phys. Rev. E}, 48:\penalty0 4070--4079, Nov 1993.
\newblock \doi{10.1103/PhysRevE.48.4070}.
\newblock URL \url{https://link.aps.org/doi/10.1103/PhysRevE.48.4070}.

\bibitem[Dominski et~al.(2024)Dominski, Maget, Manas, Morales, Ku, Scheinberg, Chang, Hager, O’Mullane, and the WEST~team]{Dominski2025}
J.~Dominski, P.~Maget, P.~Manas, J.~Morales, S.~Ku, A.~Scheinberg, C.S. Chang, R.~Hager, M.~O’Mullane, and the WEST~team.
\newblock Gyrokinetic prediction of core tungsten peaking in a west plasma with nitrogen impurities.
\newblock \emph{Nuclear Fusion}, 65\penalty0 (1):\penalty0 016003, nov 2024.
\newblock \doi{10.1088/1741-4326/ad8c63}.
\newblock URL \url{https://doi.org/10.1088/1741-4326/ad8c63}.

\bibitem[Dorf et~al.(2016)Dorf, Dorr, Hittinger, Cohen, and Rognlien]{Dorf16}
M.~A. Dorf, M.~R. Dorr, J.~A. Hittinger, R.~H. Cohen, and T.~D. Rognlien.
\newblock Continuum kinetic modeling of the tokamak plasma edge.
\newblock \emph{Physics of Plasmas}, 23\penalty0 (5):\penalty0 056102, 03 2016.
\newblock ISSN 1070-664X.
\newblock \doi{10.1063/1.4943106}.
\newblock URL \url{https://doi.org/10.1063/1.4943106}.

\bibitem[Dorland et~al.(2000)Dorland, Jenko, Kotschenreuther, and Rogers]{Dorland2000}
W.~Dorland, F.~Jenko, M.~Kotschenreuther, and B.~N. Rogers.
\newblock Electron temperature gradient turbulence.
\newblock \emph{Phys. Rev. Lett.}, 85:\penalty0 5579--5582, Dec 2000.
\newblock \doi{10.1103/PhysRevLett.85.5579}.
\newblock URL \url{https://link.aps.org/doi/10.1103/PhysRevLett.85.5579}.

\bibitem[Dougherty(1964)]{Dougherty1964}
J.~P. Dougherty.
\newblock Model fokker-planck equation for a plasma and its solution.
\newblock \emph{The Physics of Fluids}, 7\penalty0 (11):\penalty0 1788--1799, 11 1964.
\newblock ISSN 0031-9171.
\newblock \doi{10.1063/1.2746779}.
\newblock URL \url{https://doi.org/10.1063/1.2746779}.

\bibitem[Dudson and developers(2025)]{freegs-docs}
Ben Dudson and FreeGS developers.
\newblock Freegs documentation.
\newblock \url{https://freegs.readthedocs.io/}, 2025.
\newblock Version 0.8.3.dev18+ (accessed 2025-10-13).

\bibitem[Dudson et~al.(2024)Dudson, Kryjak, Muhammed, Hill, and Omotani]{Dudson2024}
Ben Dudson, Mike Kryjak, Hasan Muhammed, Peter Hill, and John Omotani.
\newblock Hermes-3: Multi-component plasma simulations with bout++.
\newblock \emph{Computer Physics Communications}, 296:\penalty0 108991, 2024.
\newblock ISSN 0010-4655.
\newblock \doi{https://doi.org/10.1016/j.cpc.2023.108991}.
\newblock URL \url{https://www.sciencedirect.com/science/article/pii/S0010465523003363}.

\bibitem[Dunlap(2021)]{Dunlap21}
Richard~A Dunlap.
\newblock \emph{Energy from Nuclear Fusion}.
\newblock 2053-2563. IOP Publishing, 2021.
\newblock ISBN 978-0-7503-3307-8.
\newblock \doi{10.1088/978-0-7503-3307-8}.
\newblock URL \url{https://doi.org/10.1088/978-0-7503-3307-8}.

\bibitem[Editors et~al.(1999)Editors, Chairs, Co-Chairs, Team, and Unit]{ITER}
ITER Physics~Basis Editors, ITER Physics Expert~Group Chairs, Co-Chairs, ITER Joint~Central Team, and Physics~Integration Unit.
\newblock Chapter 1: Overview and summary.
\newblock \emph{Nuclear Fusion}, 39\penalty0 (12):\penalty0 2137, 12 1999.
\newblock \doi{10.1088/0029-5515/39/12/301}.
\newblock URL \url{https://dx.doi.org/10.1088/0029-5515/39/12/301}.

\bibitem[Eich et~al.(2011)Eich, Sieglin, Scarabosio, Fundamenski, Goldston, and Herrmann]{Eich2011}
T.~Eich, B.~Sieglin, A.~Scarabosio, W.~Fundamenski, R.~J. Goldston, and A.~Herrmann.
\newblock Inter-elm power decay length for jet and asdex upgrade: Measurement and comparison with heuristic drift-based model.
\newblock \emph{Phys. Rev. Lett.}, 107:\penalty0 215001, 11 2011.
\newblock \doi{10.1103/PhysRevLett.107.215001}.
\newblock URL \url{https://link.aps.org/doi/10.1103/PhysRevLett.107.215001}.

\bibitem[Eich et~al.(2013)Eich, Leonard, Pitts, Fundamenski, Goldston, Gray, Herrmann, Kirk, Kallenbach, Kardaun, Kukushkin, LaBombard, Maingi, Makowski, Scarabosio, Sieglin, Terry, Thornton, Team, and Contributors]{Eich2013}
T.~Eich, A.W. Leonard, R.A. Pitts, W.~Fundamenski, R.J. Goldston, T.K. Gray, A.~Herrmann, A.~Kirk, A.~Kallenbach, O.~Kardaun, A.S. Kukushkin, B.~LaBombard, R.~Maingi, M.A. Makowski, A.~Scarabosio, B.~Sieglin, J.~Terry, A.~Thornton, ASDEX~Upgrade Team, and JET~EFDA Contributors.
\newblock Scaling of the tokamak near the scrape-off layer h-mode power width and implications for iter.
\newblock \emph{Nuclear Fusion}, 53\penalty0 (9):\penalty0 093031, 08 2013.
\newblock \doi{10.1088/0029-5515/53/9/093031}.
\newblock URL \url{https://doi.org/10.1088/0029-5515/53/9/093031}.

\bibitem[{ExoFusion (ARPA-E)}(2025)]{ARPAE2025}
{ExoFusion (ARPA-E)}.
\newblock Novel liquid metal plasma facing component alloys.
\newblock ARPA-E project description, “Novel Liquid Metal Plasma Facing Component Alloys”, 4 2025.
\newblock URL \url{https://arpa-e.energy.gov/programs-and-initiatives/search-all-projects/novel-liquid-metal-plasma-facing-component-alloys}.
\newblock Developed by ExoFusion under the CHADWICK program; involves designing low-vapor-pressure, low-melting-point liquid metals for continuously replenished fusion first-wall components.

\bibitem[{ExoFusion (INFUSE)}(2024)]{INFUSE2024}
{ExoFusion (INFUSE)}.
\newblock Testing novel liquid metal pfc compositions.
\newblock INFUSE project description, “Testing Novel Liquid Metal PFC compositions”, 8 2024.
\newblock URL \url{https://infuse.ornl.gov/awards/testing-novel-liquid-metal-pfc-compositions/}.
\newblock Experimental testing of novel liquid metal plasma-facing component alloys at Penn State University’s Radiation Surface Science and Engineering Laboratory.

\bibitem[Fitzpatrick(n.d.{\natexlab{a}})]{FitzLawson}
Richard Fitzpatrick.
\newblock The lawson criterion, n.d.{\natexlab{a}}.
\newblock URL \url{https://farside.ph.utexas.edu/teaching/plasma1/Fusionhtml/node7.html}.
\newblock Online lecture notes, University of Texas at Austin.

\bibitem[Fitzpatrick(n.d.{\natexlab{b}})]{FitzReact}
Richard Fitzpatrick.
\newblock Nuclear fusion reactions, n.d.{\natexlab{b}}.
\newblock URL \url{https://farside.ph.utexas.edu/teaching/plasma1/Fusionhtml/node6.html}.
\newblock Online educational module, University of Texas at Austin.

\bibitem[Francisquez et~al.(2022)Francisquez, Juno, Hakim, Hammett, and Ernst]{Mana22}
Manaure Francisquez, James Juno, Ammar Hakim, Gregory~W. Hammett, and Darin~R. Ernst.
\newblock Improved multispecies dougherty collisions.
\newblock \emph{Journal of Plasma Physics}, 88\penalty0 (3):\penalty0 905880303, 2022.
\newblock \doi{10.1017/S0022377822000289}.

\bibitem[Francisquez et~al.(2023)Francisquez, Rosen, Mandell, Hakim, Forest, and Hammett]{Mana23}
Manaure Francisquez, Maxwell~H. Rosen, Noah~R. Mandell, Ammar Hakim, Cary~B. Forest, and Gregory~W. Hammett.
\newblock Toward continuum gyrokinetic study of high-field mirrors.
\newblock \emph{Physics of Plasmas}, 30\penalty0 (10), 10 2023.
\newblock ISSN 1070-664X.
\newblock \doi{10.1063/5.0152440}.
\newblock URL \url{https://www.osti.gov/biblio/2217298}.

\bibitem[Francisquez et~al.(2024)Francisquez, Mandell, Hakim, and Hammett]{ManaTS}
Manaure Francisquez, Noah~R. Mandell, Ammar Hakim, and Gregory~W. Hammett.
\newblock Conservative discontinuous galerkin interpolation: Sheared boundary conditions.
\newblock \emph{Computer Physics Communications}, 298:\penalty0 109109, 2024.
\newblock ISSN 0010-4655.
\newblock \doi{https://doi.org/10.1016/j.cpc.2024.109109}.
\newblock URL \url{https://www.sciencedirect.com/science/article/pii/S0010465524000328}.

\bibitem[Francisquez et~al.(2025)Francisquez, Cagas, Shukla, Juno, and Hammett]{Mana25}
Manaure Francisquez, Petr Cagas, Akash Shukla, James Juno, and Gregory~W. Hammett.
\newblock {Conservative velocity mappings for discontinuous Galerkin kinetics}, 2025.
\newblock URL \url{https://arxiv.org/abs/2505.10754}.

\bibitem[Giacomin et~al.(2022)Giacomin, Ricci, Coroado, Fourestey, Galassi, Lanti, Mancini, Richart, Stenger, and Varini]{GIACOMIN2022}
M.~Giacomin, P.~Ricci, A.~Coroado, G.~Fourestey, D.~Galassi, E.~Lanti, D.~Mancini, N.~Richart, L.N. Stenger, and N.~Varini.
\newblock The gbs code for the self-consistent simulation of plasma turbulence and kinetic neutral dynamics in the tokamak boundary.
\newblock \emph{Journal of Computational Physics}, 463:\penalty0 111294, 2022.
\newblock ISSN 0021-9991.
\newblock \doi{https://doi.org/10.1016/j.jcp.2022.111294}.

\bibitem[Goldston(2011)]{Goldston2011}
R.J. Goldston.
\newblock Heuristic drift-based model of the power scrape-off width in low-gas-puff h-mode tokamaks.
\newblock \emph{Nuclear Fusion}, 52\penalty0 (1):\penalty0 013009, 12 2011.
\newblock \doi{10.1088/0029-5515/52/1/013009}.
\newblock URL \url{https://doi.org/10.1088/0029-5515/52/1/013009}.

\bibitem[Goodin et~al.(2006)Goodin, Alexander, Besenbruch, Bozek, Brown, Flint, Kilkenny, McQuillan, Nikroo, Paguio, et~al.]{Goodin2006}
D~T Goodin, N~B Alexander, G~E Besenbruch, A~S Bozek, L~C Brown, G~W Flint, J~D Kilkenny, B~W McQuillan, A~Nikroo, R~R Paguio, et~al.
\newblock Developing a commercial production process for 500,000 targets per day: A key challenge for inertial fusion energy.
\newblock \emph{Physics of Plasmas}, 13\penalty0 (5), 05 2006.
\newblock ISSN ISSN 1070-664X.
\newblock \doi{10.1063/1.2177129}.
\newblock URL \url{https://www.osti.gov/biblio/20783153}.

\bibitem[Goodin et~al.(2004)Goodin, Alexander, Brown, Frey, Gallix, Gibson, Maxwell, Nobile, Olson, Petzoldt, Raffray, Rochau, Schroen, Tillack, Rickman, and Vermillion]{Goodin2004}
D.T. Goodin, N.B. Alexander, L.C. Brown, D.T. Frey, R.~Gallix, C.R. Gibson, J.L. Maxwell, A.~Nobile, C.~Olson, R.W. Petzoldt, R.~Raffray, G.~Rochau, D.G. Schroen, M.~Tillack, W.S. Rickman, and B.~Vermillion.
\newblock A cost-effective target supply for inertial fusion energy.
\newblock \emph{Nuclear Fusion}, 44\penalty0 (12):\penalty0 S254, nov 2004.
\newblock \doi{10.1088/0029-5515/44/12/S17}.
\newblock URL \url{https://doi.org/10.1088/0029-5515/44/12/S17}.

\bibitem[Greenwald et~al.(1988)Greenwald, Terry, Wolfe, Ejima, Bell, Kaye, and Neilson]{Greenwald1988}
M.~Greenwald, J.L. Terry, S.M. Wolfe, S.~Ejima, M.G. Bell, S.M. Kaye, and G.H. Neilson.
\newblock A new look at density limits in tokamaks.
\newblock \emph{Nuclear Fusion}, 28\penalty0 (12):\penalty0 2199, 12 1988.
\newblock \doi{10.1088/0029-5515/28/12/009}.
\newblock URL \url{https://doi.org/10.1088/0029-5515/28/12/009}.

\bibitem[Gryaznevich et~al.(2028)Gryaznevich, Sykes, Noonan, Kingham, Buxton, Nicolai, Lister, Costley, Shevchenko, McFarland, and et~al.]{ST402018}
M.~Gryaznevich, A.~Sykes, P.~Noonan, D.~Kingham, P.~Buxton, A.~Nicolai, J.B. Lister, A.~Costley, V.~Shevchenko, A.~McFarland, and et~al.
\newblock Overview and status of construction of st40.
\newblock In \emph{26th IAEA Fusion Energy Conference (FEC 2016)}, Kyoto, Japan, 05 2028. IAEA.

\bibitem[Guo et~al.(2014)Guo, Tang, and McDevitt]{Xianzhu14}
Zehua Guo, Xian-Zhu Tang, and Chris McDevitt.
\newblock {Parallel heat flux and flow acceleration in open field line plasmas with magnetic trapping}.
\newblock \emph{Physics of Plasmas}, 21\penalty0 (10):\penalty0 102512, 10 2014.
\newblock ISSN 1070-664X.
\newblock \doi{10.1063/1.4900407}.
\newblock URL \url{https://doi.org/10.1063/1.4900407}.

\bibitem[Görler et~al.(2011)Görler, Lapillonne, Brunner, Dannert, Jenko, Merz, and Told]{Gorler2011}
T.~Görler, X.~Lapillonne, S.~Brunner, T.~Dannert, F.~Jenko, F.~Merz, and D.~Told.
\newblock {The global version of the gyrokinetic turbulence code GENE}.
\newblock \emph{Journal of Computational Physics}, 230\penalty0 (18):\penalty0 7053--7071, 2011.
\newblock ISSN 0021-9991.
\newblock \doi{https://doi.org/10.1016/j.jcp.2011.05.034}.
\newblock URL \url{https://www.sciencedirect.com/science/article/pii/S0021999111003457}.

\bibitem[Haasz and Davis(2005)]{Haasz2005}
A.A. Haasz and J.W. Davis.
\newblock \emph{Hydrogen Retention in and Release from Carbon Materials}, pages 225--248.
\newblock Springer Berlin Heidelberg, Berlin, Heidelberg, 2005.
\newblock ISBN 978-3-540-27362-2.
\newblock \doi{10.1007/3-540-27362-X_10}.
\newblock URL \url{https://doi.org/10.1007/3-540-27362-X_10}.

\bibitem[Hakim et~al.(2019)Hakim, Hammett, Shi, and Mandell]{Ammar2019}
A.~Hakim, G.~Hammett, E.~Shi, and N.~Mandell.
\newblock {Discontinuous Galerkin schemes for a class of Hamiltonian evolution equations with applications to plasma fluid and kinetic problems}, 2019.
\newblock URL \url{https://arxiv.org/abs/1908.01814}.

\bibitem[Hakim and Juno(2020)]{HakimandJuno2020}
Ammar Hakim and James Juno.
\newblock Alias-free, matrix-free, and quadrature-free discontinuous galerkin algorithms for (plasma) kinetic equations.
\newblock In \emph{Proceedings of the International Conference for High Performance Computing, Networking, Storage and Analysis}, SC '20. IEEE Press, 2020.
\newblock ISBN 9781728199986.

\bibitem[Hariri and Ottaviani(2013)]{HARIRI20132419}
F.~Hariri and M.~Ottaviani.
\newblock A flux-coordinate independent field-aligned approach to plasma turbulence simulations.
\newblock \emph{Computer Physics Communications}, 184\penalty0 (11):\penalty0 2419--2429, 2013.
\newblock ISSN 0010-4655.
\newblock \doi{https://doi.org/10.1016/j.cpc.2013.06.005}.
\newblock URL \url{https://www.sciencedirect.com/science/article/pii/S0010465513001999}.

\bibitem[Henderson et~al.(2024)Henderson, Osawa, Newton, Moulton, Xiang, Futtersack, Kryjak, Ridgers, Karhunen, Jarvinen, Hudoba, Bakes, Eriksson, Meyer, Lord, Tarazona, Cureton, Barth, Chuilon, Hebrard, Wang, Vizvary, Vaccaro, Perez~Smith, Farrington, Harrison, Dudson, and Lipschultz]{Henderson25}
S.S. Henderson, R.T. Osawa, S.L. Newton, D.~Moulton, L.~Xiang, R.~Futtersack, M.~Kryjak, C.~Ridgers, J.~Karhunen, A.~Jarvinen, A.~Hudoba, S.~Bakes, F.~Eriksson, H.~Meyer, M.~Lord, A.~Tarazona, A.~Cureton, A.~Barth, B.~Chuilon, T.~Hebrard, S.~Wang, Z.~Vizvary, D.~Vaccaro, F.~Perez~Smith, J.~Farrington, J.~Harrison, B.~Dudson, and B.~Lipschultz.
\newblock An overview of the step divertor design and the simple models driving the plasma exhaust scenario.
\newblock \emph{Nuclear Fusion}, 65\penalty0 (1):\penalty0 016033, 11 2024.
\newblock \doi{10.1088/1741-4326/ad93e7}.
\newblock URL \url{https://dx.doi.org/10.1088/1741-4326/ad93e7}.

\bibitem[Hudoba et~al.(2023)Hudoba, Newton, Voss, Cunningham, and Henderson]{Hudoba2023}
A.~Hudoba, S.~Newton, G.~Voss, G.~Cunningham, and S.~Henderson.
\newblock Divertor optimisation and power handling in spherical tokamak reactors.
\newblock \emph{Nuclear Materials and Energy}, 35:\penalty0 101410, 2023.
\newblock ISSN 2352-1791.
\newblock \doi{https://doi.org/10.1016/j.nme.2023.101410}.
\newblock URL \url{https://www.sciencedirect.com/science/article/pii/S2352179123000492}.

\bibitem[Jardin(2010)]{Jardin}
Stephen Jardin.
\newblock \emph{{Computational Methods in Plasma Physics}}.
\newblock CRC Press, Inc., USA, 1st edition, 2010.
\newblock ISBN 1439810214.

\bibitem[Jenko et~al.(2000)Jenko, Dorland, Kotschenreuther, and Rogers]{Jenko2000}
F.~Jenko, W.~Dorland, M.~Kotschenreuther, and B.~N. Rogers.
\newblock Electron temperature gradient driven turbulence.
\newblock \emph{Physics of Plasmas}, 7\penalty0 (5):\penalty0 1904--1910, 05 2000.
\newblock ISSN 1070-664X.
\newblock \doi{10.1063/1.874014}.
\newblock URL \url{https://doi.org/10.1063/1.874014}.

\bibitem[Juno et~al.(2018)Juno, Hakim, TenBarge, Shi, and Dorland]{Juno18}
James Juno, Ammar Hakim, Jason TenBarge, Eric Shi, and William Dorland.
\newblock Discontinuous galerkin algorithms for fully kinetic plasmas.
\newblock \emph{Journal of Computational Physics}, 353:\penalty0 110--147, 2018.

\bibitem[Kaita et~al.(2007)Kaita, Majeski, Doerner, Gray, Kugel, Lynch, Maingi, Mansfield, Soukhanovskii, Spaleta, Timberlake, and Zakharov]{Kaita2007}
R.~Kaita, R.~Majeski, R.~Doerner, T.~Gray, H.~Kugel, T.~Lynch, R.~Maingi, D.~Mansfield, V.~Soukhanovskii, J.~Spaleta, J.~Timberlake, and L.~Zakharov.
\newblock Extremely low recycling and high power density handling in cdx-u lithium experiments.
\newblock \emph{Journal of Nuclear Materials}, 363-365:\penalty0 1231--1235, 2007.
\newblock ISSN 0022-3115.
\newblock \doi{https://doi.org/10.1016/j.jnucmat.2007.01.229}.
\newblock URL \url{https://www.sciencedirect.com/science/article/pii/S0022311507002504}.
\newblock Plasma-Surface Interactions-17.

\bibitem[Karhunen et~al.(2024)Karhunen, Henderson, Järvinen, Moulton, Newton, and Osawa]{Karhunen2024}
J.~Karhunen, S.S. Henderson, A.~Järvinen, D.~Moulton, S.~Newton, and R.T. Osawa.
\newblock First solps-iter predictions of the impact of cross-field drifts on divertor and scrape-off layer conditions in double-null configuration of step.
\newblock \emph{Nuclear Fusion}, 64\penalty0 (9):\penalty0 096021, jul 2024.
\newblock \doi{10.1088/1741-4326/ad63ba}.
\newblock URL \url{https://doi.org/10.1088/1741-4326/ad63ba}.

\bibitem[Kim et~al.(2014)Kim, Kim, and Kim]{Kim2014}
Younghwan Kim, Wonjoon Kim, and Minki Kim.
\newblock An international comparative analysis of public acceptance of nuclear energy.
\newblock \emph{Energy Policy}, 66:\penalty0 475--483, 2014.
\newblock ISSN 0301-4215.
\newblock \doi{https://doi.org/10.1016/j.enpol.2013.11.039}.
\newblock URL \url{https://www.sciencedirect.com/science/article/pii/S0301421513011464}.

\bibitem[Kotschenreuther et~al.(2023{\natexlab{a}})Kotschenreuther, Hatch, Mahajan, Merlo, Liu, and Shukla]{Mike}
M.~Kotschenreuther, D.R. Hatch, S.~Mahajan, G.~Merlo, X.~Liu, and A.~Shukla.
\newblock Novel methods to induce transport barriers and improve confinement, 2023{\natexlab{a}}.

\bibitem[Kotschenreuther et~al.(2023{\natexlab{b}})Kotschenreuther, Hatch, Mahajan, Merlo, Liu, Shukla, Garofalo, Ding, Park, Hassan, Wang, Hu, and Gong]{Mike23}
M.~Kotschenreuther, D.R. Hatch, S.~Mahajan, G.~Merlo, X.~Liu, A.~Shukla, A.~Garofalo, S.~Ding, J.~M. Park, E.~Hassan, Z.~Wang, Y.~Hu, and X.~Gong.
\newblock Physics based routes to increased confinement.
\newblock In \emph{29th IAEA 29 Fusion Energy Conference}, London, UK, 10 2023{\natexlab{b}}. IAEA.

\bibitem[Kotschenreuther et~al.(2024)Kotschenreuther, Liu, Hatch, and Mahajan]{MikeArxiv2024}
M.~Kotschenreuther, X.~Liu, D.~R. Hatch, and S.~M. Mahajan.
\newblock The effect of separatrix density and pfc material on h-mode confinement in the itpa global h-mode database, 2024.
\newblock URL \url{https://arxiv.org/abs/2406.15693}.

\bibitem[Kotschenreuther(2023)]{SuperXT}
Michael Kotschenreuther.
\newblock Increasing energy gain in magnetically confined fusion plasmas by increasing the edge temperature: The super-xt divertor, 8 2023.

\bibitem[Krasheninnikov et~al.(2003)Krasheninnikov, Zakharov, and Pereverzev]{Krasheninnikov2003}
S.~I. Krasheninnikov, L.~E. Zakharov, and G.~V. Pereverzev.
\newblock On lithium walls and the performance of magnetic fusion devices.
\newblock \emph{Physics of Plasmas}, 10\penalty0 (5):\penalty0 1678--1682, 05 2003.
\newblock ISSN 1070-664X.
\newblock \doi{10.1063/1.1558293}.
\newblock URL \url{https://doi.org/10.1063/1.1558293}.

\bibitem[Lao(1997)]{lao1997geqdsk}
L.~L. Lao.
\newblock G\_eqdsk format: Specification of the eqdsk data file format.
\newblock \url{https://w3.pppl.gov/ntcc/TORAY/G\_EQDSK.pdf}, February 1997.
\newblock Accessed: 2025-10-13.

\bibitem[Leddy et~al.(2017)Leddy, Dudson, Romanelli, Shanahan, and Walkden]{leddy2017}
Jarrod Leddy, Ben Dudson, Michele Romanelli, Brian Shanahan, and Nick Walkden.
\newblock A novel flexible field-aligned coordinate system for tokamak edge plasma simulation.
\newblock \emph{Computer Physics Communications}, 212:\penalty0 59--68, 2017.

\bibitem[Lomanowski et~al.(2022)Lomanowski, Dunne, Vianello, Aleiferis, Brix, Canik, Carvalho, Frassinetti, Frigione, Garzotti, Groth, Meigs, Menmuir, Maslov, Pereira, von Thun, Reinke, Refy, Rimini, Rubino, Schneider, Sergienko, Uccello, Eester, and Contributors]{Lomanowski22}
B.~Lomanowski, M.~Dunne, N.~Vianello, S.~Aleiferis, M.~Brix, J.~Canik, I.S. Carvalho, L.~Frassinetti, D.~Frigione, L.~Garzotti, M.~Groth, A.~Meigs, S.~Menmuir, M.~Maslov, T.~Pereira, C.~Perez von Thun, M.~Reinke, D.~Refy, F.~Rimini, G.~Rubino, P.A. Schneider, G.~Sergienko, A.~Uccello, D.~Van Eester, and JET Contributors.
\newblock Experimental study on the role of the target electron temperature as a key parameter linking recycling to plasma performance in jet-ilw*.
\newblock \emph{Nuclear Fusion}, 62\penalty0 (6):\penalty0 066030, 4 2022.
\newblock \doi{10.1088/1741-4326/ac5668}.
\newblock URL \url{https://dx.doi.org/10.1088/1741-4326/ac5668}.

\bibitem[Lore et~al.(2024)Lore, Park, Eich, Kuang, Reinke, De~Pascuale, Lomanowski, Creely, and Canik]{Lore2024}
Jeremy~D. Lore, Jae-Sun Park, Thomas Eich, Adam~Q. Kuang, Matthew~L. Reinke, Sebastian De~Pascuale, Bart Lomanowski, Alex Creely, and John~M. Canik.
\newblock Evaluation of sparc divertor conditions in h-mode operation using solps-iter.
\newblock \emph{Nuclear Fusion}, 64\penalty0 (12):\penalty0 126054, 10 2024.
\newblock \doi{10.1088/1741-4326/ad85f3}.
\newblock URL \url{https://doi.org/10.1088/1741-4326/ad85f3}.

\bibitem[Maingi et~al.(2012)Maingi, Boyle, Canik, Kaye, Skinner, Allain, Bell, Bell, Gerhardt, Gray, Jaworski, Kaita, Kugel, LeBlanc, Manickam, Mansfield, Menard, Osborne, Raman, Roquemore, Sabbagh, Snyder, and Soukhanovskii]{Maingi2012}
R.~Maingi, D.P. Boyle, J.M. Canik, S.M. Kaye, C.H. Skinner, J.P. Allain, M.G. Bell, R.E. Bell, S.P. Gerhardt, T.K. Gray, M.A. Jaworski, R.~Kaita, H.W. Kugel, B.P. LeBlanc, J.~Manickam, D.K. Mansfield, J.E. Menard, T.H. Osborne, R.~Raman, A.L. Roquemore, S.A. Sabbagh, P.B. Snyder, and V.A. Soukhanovskii.
\newblock The effect of progressively increasing lithium coatings on plasma discharge characteristics, transport, edge profiles and elm stability in the national spherical torus experiment.
\newblock \emph{Nuclear Fusion}, 52\penalty0 (8):\penalty0 083001, 06 2012.
\newblock \doi{10.1088/0029-5515/52/8/083001}.
\newblock URL \url{https://doi.org/10.1088/0029-5515/52/8/083001}.

\bibitem[Mandell et~al.(2020)Mandell, Hakim, Hammett, and Francisquez]{Mandell2020}
N.~R. Mandell, A.~Hakim, G.~W. Hammett, and M.~Francisquez.
\newblock {Electromagnetic full-$f$ gyrokinetics in the tokamak edge with discontinuous Galerkin methods}.
\newblock \emph{Journal of Plasma Physics}, 86\penalty0 (1):\penalty0 905860109, 2020.
\newblock \doi{10.1017/S0022377820000070}.

\bibitem[Mandell(2021{\natexlab{a}})]{MandellThesis}
Noah Mandell.
\newblock \emph{{Magnetic Fluctuations in Gyrokinetic Simulations of Tokamak Scrape-Off Layer Turbulence}}.
\newblock PhD thesis, Princeton University, 2021{\natexlab{a}}.
\newblock URL \url{https://arxiv.org/abs/2103.16062}.

\bibitem[Mandell(2021{\natexlab{b}})]{Noah21}
Noah~Roth Mandell.
\newblock \emph{Magnetic fluctuations in gyrokinetic simulations of tokamak scrape-off layer turbulence}.
\newblock PhD thesis, Princeton University, 2021{\natexlab{b}}.

\bibitem[Mattor(1995)]{Mattor1995594}
Nathan Mattor.
\newblock Coordinate system for use around an x point.
\newblock \emph{Physics of Plasmas}, 2\penalty0 (3):\penalty0 594 – 598, 1995.
\newblock \doi{10.1063/1.871411}.

\bibitem[McCorquodale et~al.(2015)McCorquodale, Dorr, Hittinger, and Colella]{MCCORQUODALE2015181}
P.~McCorquodale, M.R. Dorr, J.A.F. Hittinger, and P.~Colella.
\newblock High-order finite-volume methods for hyperbolic conservation laws on mapped multiblock grids.
\newblock \emph{Journal of Computational Physics}, 288:\penalty0 181--195, 2015.
\newblock ISSN 0021-9991.
\newblock \doi{https://doi.org/10.1016/j.jcp.2015.01.006}.
\newblock URL \url{https://www.sciencedirect.com/science/article/pii/S002199911500008X}.

\bibitem[Meschini et~al.(2023)Meschini, Laviano, Ledda, Pettinari, Testoni, Torsello, and Panella]{Meschini2023}
Samuele Meschini, Francesco Laviano, Federico Ledda, Davide Pettinari, Raffella Testoni, Daniele Torsello, and Bruno Panella.
\newblock Review of commercial nuclear fusion projects.
\newblock \emph{Frontiers in Energy Research}, 11:\penalty0 1157394, 2023.
\newblock \doi{10.3389/fenrg.2023.1157394}.
\newblock URL \url{https://doi.org/10.3389/fenrg.2023.1157394}.

\bibitem[Meyer and null(2024)]{Hendrik24}
Hendrik Meyer and null null.
\newblock Plasma burn—mind the gap.
\newblock \emph{Philosophical Transactions of the Royal Society A: Mathematical, Physical and Engineering Sciences}, 382\penalty0 (2280):\penalty0 20230406, 2024.
\newblock \doi{10.1098/rsta.2023.0406}.
\newblock URL \url{https://royalsocietypublishing.org/doi/abs/10.1098/rsta.2023.0406}.

\bibitem[Michels et~al.(2021)Michels, Stegmeir, Ulbl, Jarema, and Jenko]{Michels21}
Dominik Michels, Andreas Stegmeir, Philipp Ulbl, Denis Jarema, and Frank Jenko.
\newblock {GENE-X: A full-f gyrokinetic turbulence code based on the flux-coordinate independent approach}.
\newblock \emph{Computer Physics Communications}, 264:\penalty0 107986, 2021.
\newblock ISSN 0010-4655.
\newblock \doi{https://doi.org/10.1016/j.cpc.2021.107986}.
\newblock URL \url{https://www.sciencedirect.com/science/article/pii/S0010465521000989}.

\bibitem[Moulton et~al.(2017)Moulton, Harrison, Lipschultz, and Coster]{Moulton2017}
D~Moulton, J~Harrison, B~Lipschultz, and D~Coster.
\newblock Using solps to confirm the importance of total flux expansion in super-x divertors.
\newblock \emph{Plasma Physics and Controlled Fusion}, 59\penalty0 (6):\penalty0 065011, 05 2017.
\newblock \doi{10.1088/1361-6587/aa6b13}.
\newblock URL \url{https://dx.doi.org/10.1088/1361-6587/aa6b13}.

\bibitem[{NSTX-U Program Team}(2013)]{nstx5year}
{NSTX-U Program Team}.
\newblock {NSTX-U 5 Year Plan for FY2014–2018}.
\newblock Technical report, Princeton Plasma Physics Laboratory, 2013.
\newblock URL \url{https://nstx.pppl.gov/DragNDrop/Five_Year_Plans/2014_2018/chapter_text/full_text/NSTXU_5YearPlan_text.pdf}.
\newblock Research plan for the National Spherical Torus eXperiment - Upgrade (NSTX-U) covering FY2014–2018, outlining mission elements and major research goals.

\bibitem[Osawa et~al.(2023)Osawa, Moulton, Newton, Henderson, Lipschultz, and Hudoba]{Osawa2023}
R.T. Osawa, D.~Moulton, S.L. Newton, S.S. Henderson, B.~Lipschultz, and A.~Hudoba.
\newblock Solps-iter analysis of a proposed step double null geometry: impact of the degree of disconnection on power-sharing.
\newblock \emph{Nuclear Fusion}, 63\penalty0 (7):\penalty0 076032, 06 2023.
\newblock \doi{10.1088/1741-4326/acd863}.
\newblock URL \url{https://dx.doi.org/10.1088/1741-4326/acd863}.

\bibitem[Pitts et~al.(2011)Pitts, Carpentier, Escourbiac, Hirai, Komarov, Kukushkin, Lisgo, Loarte, Merola, Mitteau, Raffray, Shimada, and Stangeby]{Pitts2011}
R.A. Pitts, S.~Carpentier, F.~Escourbiac, T.~Hirai, V.~Komarov, A.S. Kukushkin, S.~Lisgo, A.~Loarte, M.~Merola, R.~Mitteau, A.R. Raffray, M.~Shimada, and P.C. Stangeby.
\newblock Physics basis and design of the iter plasma-facing components.
\newblock \emph{Journal of Nuclear Materials}, 415\penalty0 (1, Supplement):\penalty0 S957--S964, 2011.
\newblock ISSN 0022-3115.
\newblock \doi{https://doi.org/10.1016/j.jnucmat.2011.01.114}.
\newblock URL \url{https://www.sciencedirect.com/science/article/pii/S0022311511001620}.
\newblock Proceedings of the 19th International Conference on Plasma-Surface Interactions in Controlled Fusion.

\bibitem[Pitts et~al.(2019)Pitts, Bonnin, Escourbiac, Frerichs, Gunn, Hirai, Kukushkin, Kaveeva, Miller, Moulton, Rozhansky, Senichenkov, Sytova, Schmitz, Stangeby, {De Temmerman}, Veselova, and Wiesen]{Pitts2019}
R.A. Pitts, X.~Bonnin, F.~Escourbiac, H.~Frerichs, J.P. Gunn, T.~Hirai, A.S. Kukushkin, E.~Kaveeva, M.A. Miller, D.~Moulton, V.~Rozhansky, I.~Senichenkov, E.~Sytova, O.~Schmitz, P.C. Stangeby, G.~{De Temmerman}, I.~Veselova, and S.~Wiesen.
\newblock Physics basis for the first iter tungsten divertor.
\newblock \emph{Nuclear Materials and Energy}, 20:\penalty0 100696, 2019.
\newblock ISSN 2352-1791.
\newblock \doi{https://doi.org/10.1016/j.nme.2019.100696}.
\newblock URL \url{https://www.sciencedirect.com/science/article/pii/S2352179119300237}.

\bibitem[Reinke et~al.(2024)Reinke, Abramovic, Albert, Asai, Ball, Batko, Brettingen, Brunner, Cario, Carmichael, Chrobak, Creely, Cykman, Dalla~Rosa, Dubas, Downey, Ferrera, Frenje, Fox-Widdows, Gocht, Gorini, Granetz, Greenwald, Grieve, Hanson, Hawke, Henderson, Hicks, Hillesheim, Hoffmann, Holmes, Howard, Hubbard, Hughes, Ilagan, Irby, Jean, Kaur, Kennedy, Kowalski, Kuang, Kulchy, LaCapra, Lafleur, Lagieski, Li, Lin, Looby, Zubieta~Lupo, Mackie, Marmar, McKanas, Moncada, Mumgaard, Myers, Nikolaeva, Nocente, Normile, Novoa, Ouellet, Panontin, Paz-Soldan, Pentecost, Perks, Petruzzo, Quinn, Raimond, Raj, Rebai, Riccardo, Rigamonti, Rice, Rosenthal, Safabakhsh, Saltos, Shanahan, Silva~Sa, Song, Souza, Stein-Lubrano, Stewart, Sweeney, Tardocchi, Tinguely, Vezinet, Wang, and Witham]{SPARCDutyCycle}
M.~L. Reinke, I.~Abramovic, A.~Albert, K.~Asai, J.~Ball, J.~Batko, J.~Brettingen, D.~Brunner, M.~Cario, J.~Carmichael, C.~Chrobak, A.~Creely, D.~Cykman, M.~Dalla~Rosa, E.~Dubas, C.~Downey, A.~Ferrera, J.~Frenje, E.~Fox-Widdows, R.~Gocht, G.~Gorini, R.~Granetz, M.~Greenwald, A.~Grieve, M.~Hanson, J.~Hawke, T.~Henderson, S.~Hicks, J.~Hillesheim, A.~Hoffmann, I.~Holmes, N.~Howard, A.~Hubbard, J.~W. Hughes, J.~Ilagan, J.~Irby, M.~Jean, G.~Kaur, R.~Kennedy, E.~Kowalski, A.~Q. Kuang, R.~Kulchy, M.~LaCapra, C.~Lafleur, M.~Lagieski, R.~Li, Y.~Lin, T.~Looby, R.~Zubieta~Lupo, S.~Mackie, E.~Marmar, S.~McKanas, A.~Moncada, R.~Mumgaard, C.~E. Myers, V.~Nikolaeva, M.~Nocente, S.~Normile, C.~Novoa, S.~Ouellet, E.~Panontin, C.~Paz-Soldan, J.~Pentecost, C.~Perks, M.~Petruzzo, M.~Quinn, J.~Raimond, P.~Raj, M.~Rebai, V.~Riccardo, D.~Rigamonti, J.~E. Rice, A.~Rosenthal, M.~Safabakhsh, A.~Saltos, J.~Shanahan, M.~Silva~Sa, I.~Song, J.~Souza, B.~Stein-Lubrano, I.~G. Stewart, R.~Sweeney, M.~Tardocchi, A.~Tinguely, D.~Vezinet,
  X.~Wang, and J.~Witham.
\newblock Overview of the early campaign diagnostics for the sparc tokamak (invited).
\newblock \emph{Review of Scientific Instruments}, 95\penalty0 (10):\penalty0 103518, 10 2024.
\newblock ISSN 0034-6748.
\newblock \doi{10.1063/5.0218254}.
\newblock URL \url{https://doi.org/10.1063/5.0218254}.

\bibitem[Ribeiro and Scott(2010)]{Riberio2010}
Tiago~Tamissa Ribeiro and Bruce~D. Scott.
\newblock Conformal tokamak geometry for turbulence computations.
\newblock \emph{IEEE Transactions on Plasma Science}, 38\penalty0 (9):\penalty0 2159--2168, 2010.
\newblock \doi{10.1109/TPS.2010.2056935}.

\bibitem[Ridders(1979)]{Ridders}
C.J.F. Ridders.
\newblock A new algorithm for computing a single root of a real continuous function.
\newblock \emph{IEEE Electronic Library (IEL) Journals}, 26\penalty0 (11):\penalty0 979--980, 1979.
\newblock \doi{10.1109/tcs.1979.1084580}.

\bibitem[Rodriguez-Fernandez et~al.(2022)Rodriguez-Fernandez, Creely, Greenwald, Brunner, Ballinger, Chrobak, Garnier, Granetz, Hartwig, Howard, Hughes, Irby, Izzo, Kuang, Lin, Marmar, Mumgaard, Rea, Reinke, Riccardo, Rice, Scott, Sorbom, Stillerman, Sweeney, Tinguely, Whyte, Wright, and Yuryev]{SPARCGain}
P.~Rodriguez-Fernandez, A.J. Creely, M.J. Greenwald, D.~Brunner, S.B. Ballinger, C.P. Chrobak, D.T. Garnier, R.~Granetz, Z.S. Hartwig, N.T. Howard, J.W. Hughes, J.H. Irby, V.A. Izzo, A.Q. Kuang, Y.~Lin, E.S. Marmar, R.T. Mumgaard, C.~Rea, M.L. Reinke, V.~Riccardo, J.E. Rice, S.D. Scott, B.N. Sorbom, J.A. Stillerman, R.~Sweeney, R.A. Tinguely, D.G. Whyte, J.C. Wright, and D.V. Yuryev.
\newblock Overview of the sparc physics basis towards the exploration of burning-plasma regimes in high-field, compact tokamaks.
\newblock \emph{Nuclear Fusion}, 62\penalty0 (4):\penalty0 042003, 03 2022.
\newblock \doi{10.1088/1741-4326/ac1654}.
\newblock URL \url{https://doi.org/10.1088/1741-4326/ac1654}.

\bibitem[Roeltgen et~al.(2024)Roeltgen, Shukla, Francisquez, Juno, Bernard, and Kotschenreuther2]{radiation}
J.~Roeltgen, A.~Shukla, M.~Francisquez, J.~Juno, T.~N. Bernard, and D.~R. Kotschenreuther2, M.~Hatch.
\newblock Analysis of a kinetic radiation operator for the gyrokinetic code gkeyll.
\newblock In \emph{26th PSI Conference}, 2024.

\bibitem[Roeltgen et~al.(2025)Roeltgen, Juno, Kotschenreuther, Bernard, Shukla, Francisquez, Hakim, Hammett, Power, and Hatch]{Roeltgen25}
Jonathan~Patrick Roeltgen, Jimmy Juno, Michael Kotschenreuther, Tess~N Bernard, Akash Shukla, Manaure Francisquez, Ammar Hakim, Gregory~W Hammett, Dominic Power, and David~R Hatch.
\newblock A kinetic line-driven radiation operator and its application to gyrokinetics.
\newblock \emph{Nuclear Fusion}, 2025.
\newblock URL \url{http://iopscience.iop.org/article/10.1088/1741-4326/adff28}.

\bibitem[Rognlien et~al.(1992)Rognlien, Milovich, Rensink, and Porter]{ROGNLIEN1992}
T.D. Rognlien, J.L. Milovich, M.E. Rensink, and G.D. Porter.
\newblock A fully implicit, time dependent 2-d fluid code for modeling tokamak edge plasmas.
\newblock \emph{Journal of Nuclear Materials}, 196-198:\penalty0 347--351, 1992.
\newblock ISSN 0022-3115.
\newblock \doi{https://doi.org/10.1016/S0022-3115(06)80058-9}.
\newblock URL \url{https://www.sciencedirect.com/science/article/pii/S0022311506800589}.
\newblock Plasma-Surface Interactions in Controlled Fusion Devices.

\bibitem[Rozhansky et~al.(2021)Rozhansky, Kaveeva, Senichenkov, Veselova, Voskoboynikov, Pitts, Coster, Giroud, and Wiesen]{Rozhansky2021}
V.~Rozhansky, E.~Kaveeva, I.~Senichenkov, I.~Veselova, S.~Voskoboynikov, R.A. Pitts, D.~Coster, C.~Giroud, and S.~Wiesen.
\newblock Multi-machine solps-iter comparison of impurity seeded h-mode radiative divertor regimes with metal walls.
\newblock \emph{Nuclear Fusion}, 61\penalty0 (12):\penalty0 126073, 12 2021.
\newblock \doi{10.1088/1741-4326/ac3699}.
\newblock URL \url{https://dx.doi.org/10.1088/1741-4326/ac3699}.

\bibitem[Sabo et~al.(2022)Sabo, Smolyakov, Yushmanov, and Putvinski]{Sabo2022}
A.~Sabo, A.~I. Smolyakov, P.~Yushmanov, and S.~Putvinski.
\newblock {Ion temperature effects on plasma flow in the magnetic mirror configuration}.
\newblock \emph{Physics of Plasmas}, 29\penalty0 (5):\penalty0 052507, 05 2022.
\newblock ISSN 1070-664X.
\newblock \doi{10.1063/5.0088534}.
\newblock URL \url{https://doi.org/10.1063/5.0088534}.

\bibitem[Schneider et~al.(2006)Schneider, Bonnin, Borrass, Coster, Kastelewicz, Reiter, Rozhansky, and Braams]{Schneider2006}
R.~Schneider, X.~Bonnin, K.~Borrass, D.~P. Coster, H.~Kastelewicz, D.~Reiter, V.~A. Rozhansky, and B.~J. Braams.
\newblock Plasma edge physics with b2-eirene.
\newblock \emph{Contributions to Plasma Physics}, 46\penalty0 (1-2):\penalty0 3--191, 2006.
\newblock \doi{https://doi.org/10.1002/ctpp.200610001}.
\newblock URL \url{https://onlinelibrary.wiley.com/doi/abs/10.1002/ctpp.200610001}.

\bibitem[Scholte et~al.(2025)Scholte, Al, Horsely, Iafrati, Manhard, Martelli, Morbey, Roccella, Vernimmen, and Morgan]{Scholte2025}
J.~G.~A. Scholte, R.~S. Al, D.~Horsely, M.~Iafrati, A.~Manhard, E.~Martelli, M.~Morbey, S.~Roccella, J.~W.~M. Vernimmen, and T.~W. Morgan.
\newblock Liquid metal droplet ejection through bubble formation under hydrogen plasma and radical exposure.
\newblock \emph{Journal of Fusion Energy}, 44:\penalty0 22, 2025.
\newblock \doi{10.1007/s10894-025-00492-5}.
\newblock URL \url{https://doi.org/10.1007/s10894-025-00492-5}.

\bibitem[Scott(1998)]{Scott1998}
B.~Scott.
\newblock Global consistency for thin flux tube treatments of toroidal geometry.
\newblock \emph{Physics of Plasmas}, 5\penalty0 (6):\penalty0 2334--2339, 06 1998.
\newblock ISSN 1070-664X.
\newblock \doi{10.1063/1.872907}.
\newblock URL \url{https://doi.org/10.1063/1.872907}.

\bibitem[Scott(2001)]{Scott2001}
B.~Scott.
\newblock Shifted metric procedure for flux tube treatments of toroidal geometry: Avoiding grid deformation.
\newblock \emph{Physics of Plasmas}, 8\penalty0 (2):\penalty0 447--458, 02 2001.
\newblock ISSN 1070-664X.
\newblock \doi{10.1063/1.1335832}.
\newblock URL \url{https://doi.org/10.1063/1.1335832}.

\bibitem[Scott(2010)]{Scott2010}
B.~Scott.
\newblock Derivation via free energy conservation constraints of gyrofluid equations with finite-gyroradius electromagnetic nonlinearities.
\newblock \emph{Physics of Plasmas}, 17\penalty0 (10):\penalty0 102306, 10 2010.
\newblock ISSN 1070-664X.
\newblock \doi{10.1063/1.3484219}.
\newblock URL \url{https://doi.org/10.1063/1.3484219}.

\bibitem[Seto et~al.(2019)Seto, Xu, Dudson, and Yagi]{Seto2019}
H.~Seto, X.~Q. Xu, B.~D. Dudson, and M.~Yagi.
\newblock Interplay between fluctuation driven toroidal axisymmetric flows and resistive ballooning mode turbulence.
\newblock \emph{Physics of Plasmas}, 26\penalty0 (5):\penalty0 052507, 05 2019.
\newblock ISSN 1070-664X.
\newblock \doi{10.1063/1.5086998}.
\newblock URL \url{https://doi.org/10.1063/1.5086998}.

\bibitem[Shi(2017)]{shithesis}
E.~L. Shi.
\newblock \emph{Gyrokinetic Continuum Simulation of Turbulence in Open-Field-Line Plasmas}.
\newblock PhD thesis, Princeton University, 2017.
\newblock URL \url{https://arxiv.org/abs/1708.07283}.

\bibitem[Shi et~al.(2017)Shi, Hammett, Stoltzfus-Dueck, and Hakim]{Shi17}
E.~L. Shi, G.~W. Hammett, T.~Stoltzfus-Dueck, and A.~Hakim.
\newblock Gyrokinetic continuum simulation of turbulence in a straight open-field-line plasma.
\newblock \emph{Journal of Plasma Physics}, 83\penalty0 (3):\penalty0 905830304, 2017.
\newblock \doi{10.1017/S002237781700037X}.

\bibitem[Shukla et~al.(2025{\natexlab{a}})Shukla, Roeltgen, Kotschenreuther, Juno, Bernard, Hakim, Hammett, Hatch, Mahajan, and Francisquez]{Shukla25}
A.~Shukla, J.~Roeltgen, M.~Kotschenreuther, J.~Juno, T.~N. Bernard, A.~Hakim, G.~W. Hammett, D.~R. Hatch, S.~M. Mahajan, and M.~Francisquez.
\newblock Direct comparison of gyrokinetic and fluid scrape-off layer simulations.
\newblock \emph{AIP Advances}, 15\penalty0 (7):\penalty0 075121, 07 2025{\natexlab{a}}.
\newblock ISSN 2158-3226.
\newblock \doi{10.1063/5.0268104}.
\newblock URL \url{https://doi.org/10.1063/5.0268104}.

\bibitem[Shukla et~al.(2025{\natexlab{b}})Shukla, Hakim, Juno, Hammett, and Francisquez]{Shukla2025Xpt}
Akash Shukla, Ammar Hakim, James Juno, Gregory Hammett, and Manaure Francisquez.
\newblock Constructing field aligned coordinate systems for gyrokinetic simulations of tokamaks in x-point geometries, 2025{\natexlab{b}}.
\newblock URL \url{https://arxiv.org/abs/2510.21676}.

\bibitem[Shukla et~al.(2025{\natexlab{c}})Shukla, Roeltgen, Kotschenreuther, Hatch, Francisquez, Juno, Bernard, Hakim, Hammett, and Mahajan]{shukla2025LR}
Akash Shukla, Jonathan Roeltgen, Michael Kotschenreuther, David~R. Hatch, Manaure Francisquez, James Juno, Tess~N. Bernard, Ammar Hakim, Gregory~W. Hammett, and Swadesh~M. Mahajan.
\newblock Gyrokinetic simulations of a low recycling scrape-off layer without a lithium target, 2025{\natexlab{c}}.
\newblock URL \url{https://arxiv.org/abs/2511.09437}.

\bibitem[Stangeby(2000)]{Stangeby2000}
P.~C. Stangeby.
\newblock \emph{The Plasma Boundary of Magnetic Fusion Devices}.
\newblock Plasma Physics Series. Taylor \& Francis / Institute of Physics Publishing, Bristol / New York, 2000.
\newblock ISBN 9780750305594.

\bibitem[Stangeby and McCracken(1990)]{Stangeby1990}
P.C. Stangeby and G.M. McCracken.
\newblock Plasma boundary phenomena in tokamaks.
\newblock \emph{Nuclear Fusion}, 30\penalty0 (7):\penalty0 1225, 07 1990.
\newblock \doi{10.1088/0029-5515/30/7/005}.
\newblock URL \url{https://doi.org/10.1088/0029-5515/30/7/005}.

\bibitem[Stegmeir et~al.(2016)Stegmeir, Coster, Maj, Hallatschek, and Lackner]{STEGMEIR2016139}
Andreas Stegmeir, David Coster, Omar Maj, Klaus Hallatschek, and Karl Lackner.
\newblock The field line map approach for simulations of magnetically confined plasmas.
\newblock \emph{Computer Physics Communications}, 198:\penalty0 139--153, 2016.
\newblock ISSN 0010-4655.
\newblock \doi{https://doi.org/10.1016/j.cpc.2015.09.016}.
\newblock URL \url{https://www.sciencedirect.com/science/article/pii/S0010465515003641}.

\bibitem[Stegmeir et~al.(2018)Stegmeir, Coster, Ross, Maj, Lackner, and Poli]{Stegmeir2018}
Andreas Stegmeir, David Coster, Alexander Ross, Omar Maj, Karl Lackner, and Emanuele Poli.
\newblock Grillix: a 3d turbulence code based on the flux-coordinate independent approach.
\newblock \emph{Plasma Physics and Controlled Fusion}, 60\penalty0 (3):\penalty0 035005, 1 2018.
\newblock \doi{10.1088/1361-6587/aaa373}.
\newblock URL \url{https://dx.doi.org/10.1088/1361-6587/aaa373}.

\bibitem[Stegmeir et~al.(2026)Stegmeir, Finkbeiner, Pitzal, Geiger, and Jenko]{STEGMEIR2026}
Andreas Stegmeir, Marion~E. Finkbeiner, Christoph Pitzal, Joachim Geiger, and Frank Jenko.
\newblock Grillix as unified fluid turbulence code for tokamaks and stellarators.
\newblock \emph{Computer Physics Communications}, 318:\penalty0 109874, 2026.
\newblock ISSN 0010-4655.
\newblock \doi{https://doi.org/10.1016/j.cpc.2025.109874}.
\newblock URL \url{https://www.sciencedirect.com/science/article/pii/S0010465525003765}.

\bibitem[Stroth(2022)]{Stroth2022}
U.~et~al. Stroth.
\newblock Progress from asdex upgrade experiments in preparing the physics basis of iter operation and demo scenario development.
\newblock \emph{Nuclear Fusion}, 62\penalty0 (4):\penalty0 042006, mar 2022.
\newblock \doi{10.1088/1741-4326/ac207f}.
\newblock URL \url{https://doi.org/10.1088/1741-4326/ac207f}.

\bibitem[Subba et~al.(2021)Subba, Coster, Moscheni, and Siccinio]{Subba2021}
F.~Subba, D.P. Coster, M.~Moscheni, and M.~Siccinio.
\newblock Solps-iter modeling of divertor scenarios for eu-demo.
\newblock \emph{Nuclear Fusion}, 61\penalty0 (10):\penalty0 106013, 09 2021.
\newblock \doi{10.1088/1741-4326/ac1c85}.
\newblock URL \url{https://dx.doi.org/10.1088/1741-4326/ac1c85}.

\bibitem[Tabarés et~al.(2016)Tabarés, Oyarzabal, Martin-Rojo, Tafalla, de~Castro, and Soleto]{Tabares2017}
F.L. Tabarés, E.~Oyarzabal, A.B. Martin-Rojo, D.~Tafalla, A.~de~Castro, and A.~Soleto.
\newblock Reactor plasma facing component designs based on liquid metal concepts supported in porous systems.
\newblock \emph{Nuclear Fusion}, 57\penalty0 (1):\penalty0 016029, nov 2016.
\newblock \doi{10.1088/0029-5515/57/1/016029}.
\newblock URL \url{https://doi.org/10.1088/0029-5515/57/1/016029}.

\bibitem[Tanabe et~al.(2003)Tanabe, Bekris, Coad, Skinner, Glugla, and Miya]{Tanabe2003}
T~Tanabe, N~Bekris, P~Coad, C.H Skinner, M~Glugla, and N~Miya.
\newblock Tritium retention of plasma facing components in tokamaks.
\newblock \emph{Journal of Nuclear Materials}, 313-316:\penalty0 478--490, 2003.
\newblock ISSN 0022-3115.
\newblock \doi{https://doi.org/10.1016/S0022-3115(02)01377-6}.
\newblock URL \url{https://www.sciencedirect.com/science/article/pii/S0022311502013776}.
\newblock Plasma-Surface Interactions in Controlled Fusion Devices 15.

\bibitem[Team(2026)]{Gkeyllwebsite}
Gkeyll Team.
\newblock Gkeyll 2.0 documentation.
\newblock \url{https://gkeyll.readthedocs.io}, 2026.
\newblock Accessed: 2024-12-08.

\bibitem[Wade and Leuer(2021)]{Wade2021}
M.~R. Wade and J.~A. Leuer.
\newblock Cost drivers for a tokamak-based compact pilot plant.
\newblock \emph{Fusion Science and Technology}, 77\penalty0 (2):\penalty0 119--143, 2021.
\newblock \doi{10.1080/15361055.2020.1858670}.
\newblock URL \url{https://doi.org/10.1080/15361055.2020.1858670}.

\bibitem[Waldon et~al.(2024)Waldon, Muldrew, Keep, Verhoeven, Thompson, and Kisbey-Ascott]{Waldon23}
Chris Waldon, Stuart~I. Muldrew, Jonathan Keep, Roel Verhoeven, Terry Thompson, and Mark Kisbey-Ascott.
\newblock Concept design overview: a question of choices and compromise.
\newblock \emph{Philosophical Transactions of the Royal Society A: Mathematical, Physical and Engineering Sciences}, 382\penalty0 (2280):\penalty0 20230414, 2024.
\newblock \doi{10.1098/rsta.2023.0414}.
\newblock URL \url{https://royalsocietypublishing.org/doi/abs/10.1098/rsta.2023.0414}.

\bibitem[WANG et~al.(2025)WANG, GU, HUA, WANG, BO, CHEN, SHI, XU, WANG, LIANG, and Team]{WANG2025}
Fuqiong WANG, Xiang GU, Jiankun HUA, Yumin WANG, Xiaokun BO, Bo~CHEN, Yuejiang SHI, Shuai XU, Erhui WANG, Yunfeng LIANG, and the EHL-1 Team.
\newblock Divertor heat flux challenge and mitigation in the ehl-2 spherical torus.
\newblock \emph{Plasma Science and Technology}, 27\penalty0 (2):\penalty0 024009, 02 2025.
\newblock \doi{10.1088/2058-6272/adadb8}.
\newblock URL \url{https://dx.doi.org/10.1088/2058-6272/adadb8}.

\bibitem[Wesson(2011)]{Wesson}
John Wesson.
\newblock \emph{Tokamaks}.
\newblock International Series of Monographs on Physics. Oxford University Press, Oxford, 4th edition, 2011.
\newblock ISBN 9780199592234.

\bibitem[Wiesen et~al.(2015{\natexlab{a}})Wiesen, Reiter, Kotov, Baelmans, Dekeyser, Kukushkin, Lisgo, Pitts, Rozhansky, Saibene, Veselova, and Voskoboynikov]{SOLPS}
S.~Wiesen, D.~Reiter, V.~Kotov, M.~Baelmans, W.~Dekeyser, A.S. Kukushkin, S.W. Lisgo, R.~A. Pitts, V.~Rozhansky, G.~Saibene, I.~Veselova, and S.~Voskoboynikov.
\newblock The new solps-iter code package.
\newblock \emph{Journal of Nuclear Materials}, 463:\penalty0 480--484, 2015{\natexlab{a}}.
\newblock ISSN 0022-3115.
\newblock \doi{https://doi.org/10.1016/j.jnucmat.2014.10.012}.
\newblock URL \url{https://www.sciencedirect.com/science/article/pii/S0022311514006965}.
\newblock PLASMA-SURFACE INTERACTIONS 21.

\bibitem[Wiesen et~al.(2015{\natexlab{b}})Wiesen, Reiter, Kotov, Baelmans, Dekeyser, Kukushkin, Lisgo, Pitts, Rozhansky, Saibene, Veselova, and Voskoboynikov]{WIESEN2015}
S.~Wiesen, D.~Reiter, V.~Kotov, M.~Baelmans, W.~Dekeyser, A.S. Kukushkin, S.W. Lisgo, R.A. Pitts, V.~Rozhansky, G.~Saibene, I.~Veselova, and S.~Voskoboynikov.
\newblock The new solps-iter code package.
\newblock \emph{Journal of Nuclear Materials}, 463:\penalty0 480--484, 2015{\natexlab{b}}.
\newblock ISSN 0022-3115.
\newblock \doi{https://doi.org/10.1016/j.jnucmat.2014.10.012}.
\newblock URL \url{https://www.sciencedirect.com/science/article/pii/S0022311514006965}.
\newblock PLASMA-SURFACE INTERACTIONS 21.

\bibitem[Wiesen et~al.(2015{\natexlab{c}})Wiesen, Reiter, Kotov, Baelmans, Dekeyser, Kukushkin, Lisgo, Pitts, Rozhansky, Saibene, Veselova, and Voskoboynikov]{Wiesen25}
S.~Wiesen, D.~Reiter, V.~Kotov, M.~Baelmans, W.~Dekeyser, A.S. Kukushkin, S.W. Lisgo, R.A. Pitts, V.~Rozhansky, G.~Saibene, I.~Veselova, and S.~Voskoboynikov.
\newblock The new solps-iter code package.
\newblock \emph{Journal of Nuclear Materials}, 463:\penalty0 480--484, 2015{\natexlab{c}}.
\newblock ISSN 0022-3115.
\newblock \doi{https://doi.org/10.1016/j.jnucmat.2014.10.012}.
\newblock URL \url{https://www.sciencedirect.com/science/article/pii/S0022311514006965}.
\newblock PLASMA-SURFACE INTERACTIONS 21.

\bibitem[Wiesenberger and Held(2024)]{Wiesenberger2024}
M~Wiesenberger and M~Held.
\newblock Effects of plasma resistivity in feltor simulations of three-dimensional full-f gyro-fluid turbulence.
\newblock \emph{Plasma Physics and Controlled Fusion}, 66\penalty0 (6):\penalty0 065003, apr 2024.
\newblock \doi{10.1088/1361-6587/ad3670}.
\newblock URL \url{https://doi.org/10.1088/1361-6587/ad3670}.

\bibitem[Wiesenberger et~al.(2017)Wiesenberger, Held, and Einkemmer]{WIESENBERGER2017}
M.~Wiesenberger, M.~Held, and L.~Einkemmer.
\newblock Streamline integration as a method for two-dimensional elliptic grid generation.
\newblock \emph{Journal of Computational Physics}, 340:\penalty0 435--450, 2017.
\newblock ISSN 0021-9991.
\newblock \doi{https://doi.org/10.1016/j.jcp.2017.03.056}.
\newblock URL \url{https://www.sciencedirect.com/science/article/pii/S0021999117302577}.

\bibitem[Wiesenberger et~al.(2018)Wiesenberger, Held, Einkemmer, and Kendl]{WIESENBERGER2018}
M.~Wiesenberger, M.~Held, L.~Einkemmer, and A.~Kendl.
\newblock Streamline integration as a method for structured grid generation in x-point geometry.
\newblock \emph{Journal of Computational Physics}, 373:\penalty0 370--384, 2018.
\newblock ISSN 0021-9991.
\newblock \doi{https://doi.org/10.1016/j.jcp.2018.07.007}.
\newblock URL \url{https://www.sciencedirect.com/science/article/pii/S0021999118304625}.

\bibitem[Wiesenberger and Held(2023)]{WIESENBERGER2023}
Matthias Wiesenberger and Markus Held.
\newblock A finite volume flux coordinate independent approach.
\newblock \emph{Computer Physics Communications}, 291:\penalty0 108838, 2023.
\newblock ISSN 0010-4655.
\newblock \doi{https://doi.org/10.1016/j.cpc.2023.108838}.

\bibitem[You et~al.(2022)You, Mazzone, Visca, Greuner, Fursdon, Addab, Bachmann, Barrett, Bonavolontà, Böswirth, Castrovinci, Carelli, Coccorese, Coppola, Crescenzi, {Di Gironimo}, {Di Maio}, {Di Mambro}, Domptail, Dongiovanni, Dose, Flammini, Forest, Frosi, Gallay, Ghidersa, Harrington, Hunger, Imbriani, Li, Lukenskas, Maffucci, Mantel, Marzullo, Minniti, Müller, Noce, Porfiri, Quartararo, Richou, Roccella, Terentyev, Tincani, Vallone, Ventre, Villari, Villone, Vorpahl, and Zhang]{YOU2022}
J.H. You, G.~Mazzone, E.~Visca, H.~Greuner, M.~Fursdon, Y.~Addab, C.~Bachmann, T.~Barrett, U.~Bonavolontà, B.~Böswirth, F.M. Castrovinci, C.~Carelli, D.~Coccorese, R.~Coppola, F.~Crescenzi, G.~{Di Gironimo}, P.A. {Di Maio}, G.~{Di Mambro}, F.~Domptail, D.~Dongiovanni, G.~Dose, D.~Flammini, L.~Forest, P.~Frosi, F.~Gallay, B.E. Ghidersa, C.~Harrington, K.~Hunger, V.~Imbriani, M.~Li, A.~Lukenskas, A.~Maffucci, N.~Mantel, D.~Marzullo, T.~Minniti, A.V. Müller, S.~Noce, M.T. Porfiri, A.~Quartararo, M.~Richou, S.~Roccella, D.~Terentyev, A.~Tincani, E.~Vallone, S.~Ventre, R.~Villari, F.~Villone, C.~Vorpahl, and K.~Zhang.
\newblock Divertor of the european demo: Engineering and technologies for power exhaust.
\newblock \emph{Fusion Engineering and Design}, 175:\penalty0 113010, 2022.
\newblock ISSN 0920-3796.
\newblock \doi{https://doi.org/10.1016/j.fusengdes.2022.113010}.
\newblock URL \url{https://www.sciencedirect.com/science/article/pii/S0920379622000102}.

\bibitem[Zakharov et~al.(2007)Zakharov, Blanchard, Kaita, Kugel, Majeski, and Timberlake]{Zakharov2007}
L.E. Zakharov, W.~Blanchard, R.~Kaita, H.~Kugel, R.~Majeski, and J.~Timberlake.
\newblock Low recycling regime in iter and the liwall concept for its divertor.
\newblock \emph{Journal of Nuclear Materials}, 363-365:\penalty0 453--457, 2007.
\newblock ISSN 0022-3115.
\newblock \doi{https://doi.org/10.1016/j.jnucmat.2007.01.230}.
\newblock URL \url{https://www.sciencedirect.com/science/article/pii/S002231150700089X}.
\newblock Plasma-Surface Interactions-17.

\bibitem[Zakharov(2019)]{Zakharov2019}
Leonid~E. Zakharov.
\newblock On a burning plasma low recycling regime with pdt = 23–26 mw, qdt = 5–7 in a jet-like tokamak.
\newblock \emph{Nuclear Fusion}, 59\penalty0 (9):\penalty0 096008, 07 2019.
\newblock \doi{10.1088/1741-4326/ab246b}.
\newblock URL \url{https://doi.org/10.1088/1741-4326/ab246b}.

\bibitem[Zhang et~al.(2024)Zhang, Sang, Zhao, Rognlien, Zhang, Wang, Bian, and Wang]{Zhang2024}
M.~Z. Zhang, C.~F. Sang, M.~L. Zhao, T.~D. Rognlien, C.~Zhang, Y.~L. Wang, Y.~Bian, and Y.~Wang.
\newblock Uedge modeling of plasma detachment of cfetr with iter-like divertor geometry by external impurity seeding.
\newblock \emph{Contributions to Plasma Physics}, 64\penalty0 (7-8):\penalty0 e202300135, 2024.
\newblock \doi{https://doi.org/10.1002/ctpp.202300135}.
\newblock URL \url{https://onlinelibrary.wiley.com/doi/abs/10.1002/ctpp.202300135}.

\bibitem[Zhao et~al.(2021{\natexlab{a}})Zhao, Jaervinen, Joseph, and Rognlien]{Zhao20212}
M.~Zhao, A.E. Jaervinen, I.~Joseph, and T.D. Rognlien.
\newblock Impact of ion temperature anisotropy on 2d edge-plasma transport.
\newblock \emph{Nuclear Materials and Energy}, 26:\penalty0 100881, 2021{\natexlab{a}}.
\newblock ISSN 2352-1791.
\newblock \doi{https://doi.org/10.1016/j.nme.2020.100881}.
\newblock URL \url{https://www.sciencedirect.com/science/article/pii/S2352179120301459}.

\bibitem[Zhao et~al.(2021{\natexlab{b}})Zhao, Rognlien, Jarvinen, and Joseph]{Zhao2021}
Menglong Zhao, Tom Rognlien, Aaro Jarvinen, and Ilon Joseph.
\newblock Ion temperature anisotropy in the tokamak scrape-off layer.
\newblock \emph{Plasma Physics and Controlled Fusion}, 63\penalty0 (12):\penalty0 125028, 11 2021{\natexlab{b}}.
\newblock \doi{10.1088/1361-6587/ac32e5}.
\newblock URL \url{https://dx.doi.org/10.1088/1361-6587/ac32e5}.

\bibitem[Zhao et~al.(2022)Zhao, Rognlien, Jarvinen, Joseph, and Umansky]{Zhao2022}
Menglong Zhao, Tom Rognlien, Aaro Jarvinen, Ilon Joseph, and Maxim Umansky.
\newblock Ion temperature anisotropy model with cross-field drifts in the scrape-off layer.
\newblock \emph{Contributions to Plasma Physics}, 62\penalty0 (5-6):\penalty0 e202100164, 2022.
\newblock \doi{https://doi.org/10.1002/ctpp.202100164}.
\newblock URL \url{https://onlinelibrary.wiley.com/doi/abs/10.1002/ctpp.202100164}.

\bibitem[Zohm(2010)]{Zohm10}
Hartmut Zohm.
\newblock On the minimum size of demo.
\newblock \emph{Fusion Science and Technology}, 58\penalty0 (2):\penalty0 613--624, 2010.
\newblock \doi{10.13182/FST10-06}.
\newblock URL \url{https://doi.org/10.13182/FST10-06}.

\end{thebibliography}


\begin{vita}
\theauthor\ was born in Aurora, Illinois and was raised in Houston, Texas. After attending Bellaire High School, he attended The University of Texas at Austin where he earned a BS in Physics and a BS in Electrical Engineering. During his undergraduate degree, he worked under Dr. David Hatch studying ITG turbulence. Following his undergraduate studies, he continued his education pursuing a doctoral degree under the supervision of Dr. David Hatch. During his PhD he worked in collaboration with Princeton Plasma Physics Laboratory to develop the Gkeyll code and study spherical tokamaks. During his PhD he also got married to his wife Maya Waterland.
\end{vita}

\end{document}